\documentclass[epj]{svjour}
\usepackage{graphics}
\usepackage{url}
\usepackage{color}
\usepackage{mathrsfs}
\usepackage{amssymb}
\usepackage{bm}
\usepackage[numbers]{natbib}
\usepackage{rotating}
\usepackage[normalem]{ulem}
\usepackage{epstopdf}
\usepackage{lipsum}
\usepackage{mathtools}
\usepackage{cuted}
\usepackage{enumitem}

\renewcommand{\v}[1]{\textbf{#1}}
\renewcommand{\d}{\mathrm{d}}

\def\esym{$E_{\rm sym}(\rho)$~}

\def\es0{$E_{\rm sym}(\rho_0)$}

\def\us0{$U_{\rm sym}(\rho_0,k_F)$~}

\def\lr{$L(\rho)$~}

\def\l0{$L(\rho_0)$~}

\begin{document}
\title{Towards Understanding Astrophysical Effects of Nuclear Symmetry Energy}
\author{Bao-An Li\inst{1}, Plamen G. Krastev\inst{2}, De-Hua Wen\inst{3} and Nai-Bo Zhang\inst{4} 
}                     
\mail{Bao-An.Li@Tamuc.edu}          
\institute{Department of Physics and Astronomy, Texas A$\&$M University-Commerce, Commerce, TX 75429, USA
\and Research Computing, Faculty of Arts and Sciences, Harvard University, Cambridge, MA 02138, USA
\and School of Physics and Optoelectronic Technology, South China University of Technology, Guangzhou 510641, P.R. China
\and Shandong Provincial Key Laboratory of Optical Astronomy and Solar-Terrestrial Environment, Institute of Space Sciences, Shandong University, Weihai, 264209, China}
\date{Received: date / Revised version: date}
%
\abstract{Determining the Equation of State (EOS) of dense neutron-rich nuclear matter is a shared goal of both nuclear physics and astrophysics. Except possible phase transitions, the density dependence of nuclear symmetry \esym is the most uncertain part of the EOS of neutron-rich nucleonic matter especially at supra-saturation densities. Much progresses have been made in recent years in predicting the symmetry energy and understanding why it is still very uncertain using various microscopic nuclear many-body theories and phenomenological models. Simultaneously, significant progresses have also been made in probing the symmetry energy in both terrestrial nuclear laboratories and astrophysical observatories. In light of the GW170817 event as well as ongoing or planned nuclear experiments and astrophysical observations probing the EOS of dense neutron-rich matter, we review recent progresses and identify new challenges to the best knowledge we have on several selected topics critical for understanding astrophysical effects of the nuclear symmetry energy. \\
}
\authorrunning{B.A Li, P.G. Krastev, D.H. Wen and N.B. Zhang}
\titlerunning{Astrophysical Effects of Nuclear Symmetry Energy}
\maketitle
\setcounter{tocdepth}{3} 

\tableofcontents
\section{Introduction}
To understand the nature and constrain the Equation of State (EOS) of dense neutron-rich nuclear matter is a major science goal \cite{NAP2011,NAP2012}
shared by many astrophysical observations, see, e.g. refs. \cite{Lat01,Watts16,Oz16a,Oertel17,ISAAC18,BUR18,Blas18,Bom18,Cos19} and terrestrial nuclear experiments, see, e.g. refs. \cite{Dan02,ditoro,Li2008,Lynch09,Trau12,Joe14,Bonasera,Garg18,Bor19,Ono19}. Realizing this goal is very important but rather challenging for many scientific reasons. The energy per nucleon $E(\rho ,\delta )$ in nuclear matter at density $\rho$ and isospin asymmetry $\delta\equiv (\rho_n-\rho_p)/\rho$ is the most basic input for calculating the EOS of neutron star matter regardless of the models used. It has a symmetry energy term $E_{\rm{sym}}(\rho )\cdot\delta^2$ which quantifies the energy needed to make nuclear matter more neutron rich. While much progress has been made over the last few decades in constraining mostly the EOS of symmetric nuclear matter (SNM) and the symmetry energy $E_{\rm{sym}}(\rho )$ around but mostly below the saturation density of nuclear matter $\rho_0\approx 2.8\times 10^{14}$ g/cm$^{3}$ (0.16 fm$^{-3}$), very little is known about the symmetry energy at supra-saturation densities. In fact, the high-density $E_{\rm{sym}}(\rho )$ has been broadly recognized as the most uncertain part of the EOS of dense neutron-rich nucleonic matter \cite{Steiner05,Tesym,Bal16,Li17,PPNP-Li}.

Because of the widely recognized importance of knowing precisely the density dependence of the nuclear symmetry energy in both astrophysics and nuclear physics, essentially all existing nuclear
many-body theories using all available nuclear forces have been used to predict the symmetry energy $E_{\rm{sym}}(\rho )$. Mostly by design, they all agree with existing constrains available around and below the saturation density. However, at supra-saturation densities their predictions diverge very broadly. The fundamental reason for the very uncertain high-density nuclear symmetry energy is our poor knowledge about the relatively weak isospin-dependence (i.e., the difference between neutron-proton interactions in the isosinglet and isotriple channels) of the two-body force as well as the spin-isospin dependence of the three-body and tensor forces at short distances in dense neutron-rich nuclear matter. Determining the high-density $E_{\rm{sym}}(\rho )$ using nuclear reactions induced by high-energy rare isotope beams has been identified as a major science thrust in both the 2015 American \cite{LRP2015} and 2017 European \cite{NuPECC} nuclear physics long range plans for the next decade.

Unlike the relatively small isospin effects in laboratory experiments, neutron stars (NSs) are the natural testing ground of the isospin-dependence of strong interactions and the corresponding EOS of cold neutron-rich matter at extremely high densities and isospin asymmetries. While recent analyses of astrophysical data including the radii and tidal deformability of canonical neutron stars already ruled out many model predictions up to about $2\rho_0$, huge uncertainties remain especially at higher densities. The proton fraction $x_p(\rho)$ in NSs is uniquely determined by the $E_{\rm{sym}}(\rho )$ through the $\beta$-equilibrium and charge neutrality conditions. Consequently, the composition, critical nucleon density $\rho_c$ (where $x_p(\rho_c)\approx 1/9$ in the $npe$ matter at $\beta$ equilibrium) above which the fast cooling of protoneutron stars by neutrino emissions through the direct URCA process can occur, and the crust-core transition density in NSs all depend sensitively on the $E_{\rm{sym}}(\rho )$. Moreover, the frequencies and damping times of various oscillations (especially the g-mode of the core and the torsional mode of the crust), quadrupole deformations of isolated NSs and the tidal deformability of NSs in inspiraling binaries also depend on the $E_{\rm{sym}}(\rho )$. Furthermore, there is a degeneracy between the EOS of super-dense neutron-rich matter and the strong-field gravity in understanding both properties of super-massive NSs and the minimum mass to form black holes. While understanding the nature of strong-field gravity has been identified as one of the eleven greatest physics questions for the new century by the U.S. National Research Council in 2003 \cite{11questions}.
Thus, a precise determination of the $E_{\rm{sym}}(\rho )$ especially at high densities has broad impacts in many areas of astrophysics, cosmology and nuclear physics.

In light of the GW170817 event as well as ongoing or planned nuclear experiments and astrophysical observations probing the EOS of dense neutron-rich nuclear matter using advanced facilities/detectors, we review here recent progresses and identify new challenges in understanding astrophysics effects of the nuclear symmetry energy. In particular, we examine \esym effects on several structural and dynamical properties of non-rotating, oscillating and rotating neutron stars as well as the associated gravitational wave (GW) signatures (e.g., strain amplitude, frequency and damping time). We organize the review into 8 sections with many subsection and subsubsections devoted to specific topics where the \esym plays a significant role. A summary is given at the end of each section. Finally, some concluding remarks and outlook are given in Section 9.

\section{The nuclear symmetry energy as we know it}
\subsection{The isospin dependence of nuclear EOS and single-nucleon potential in neutron-rich nuclear matter}
It is well known that the EOS of asymmetric nucleonic matter (ANM) of isospin asymmetry $\delta$ and density $\rho$ can be
written as
\begin{equation}\label{eos1}
E(\rho ,\delta )=E_0(\rho)+E_{\rm{sym},2}(\rho )\delta ^{2} +E_{\rm{sym},4}(\rho ) \delta ^{4} +\mathcal{O}(\delta^6)
\end{equation}
in terms of the energy per nucleon $E_0(\rho)\equiv E(\rho ,\delta=0)$ in symmetric nuclear matter (SNM),  the isospin-quadratic symmetry energy $E_{\rm{sym},2}(\rho )$ and the 
isospin-quartic (fourth-order) symmetry energy $E_{\mathrm{sym,4}}(\rho)$. In the literature, the $E_{\rm{sym},2}(\rho )$ is normally referred as the nuclear symmetry energy denoted often by $E_{\rm{sym}}(\rho )$ or $S$. In the following, we use the notations $E_{\rm{sym}}(\rho )$ or $E_{\rm{sym},2}(\rho )$ interchangeably for nuclear symmetry energy. The notation $E_{\rm{sym},2}(\rho )$ is mostly used when the symmetry energy appears in the same equation with the isospin-quartic symmetry energy $E_{\mathrm{sym,4}}(\rho)$. Specifically, they are defined as 
\begin{equation}
E_{\rm{sym}}(\rho )\equiv E_{\rm{sym},2}(\rho )\equiv\left.\frac{1}{2}\frac{\partial ^{2}E(\rho,\delta )}{\partial \delta ^{2}}\right|_{\delta=0}
\end{equation}
and 
\begin{equation}
E_{\mathrm{sym,4}}(\rho )\equiv \left.\frac{1}{24}\frac{\partial ^{4}E(\rho,\delta )}{\partial \delta ^{4}}\right|_{\delta=0} \label{Esyme4}.
\end{equation}
If the $E_{\mathrm{sym,4}}(\rho)$ is negligibly small, the Eq. (\ref{eos1}) is reduced to the so-called empirical parabolic approximation (PA) of nuclear EOS \cite{Bom91} and the
symmetry energy can be approximated by the difference between energy per nucleon in SNM and pure neutron matter (PNM)
\begin{equation}
E_{\rm sym}(\rho )\approx E(\rho,1)-E(\rho,0).
\end{equation}
It is also well known that the single-particle potential $U_{n/p}(k,\rho,\delta)$ for nucleons $\tau=n/p$ with $\tau_3=\pm$ at momentum $k$
in ANM can be written as
\begin{eqnarray}\label{sp}
&&U_{\tau}(k,\rho,\delta)=U_0(k,\rho)+\tau_3 U_{\rm sym,1}(k,\rho)\cdot\delta\nonumber\\
&&+U_{\rm sym,2}(k,\rho)\cdot\delta^2+\tau_3 U_{\rm sym,3}(k,\rho)\cdot\delta^3+\mathcal{O}(\delta^4)
\end{eqnarray}
in terms of the isoscalar $U_0(k,\rho)$ and $U_{\rm sym,2}(k,\rho)$ as well as the isovector $U_{\rm sym,1}(k,\rho)$ and $U_{\rm sym,3}(k,\rho)$ potentials, respectively.
Note that keeping only the zeroth and first order terms of this expansion, Eq. (\ref{sp}) reduces to the well known Lane potential \cite{Lan62}.

\subsection{Decomposition of the nuclear symmetry energy according to the Hugenholtz-Van Hove (HVH) theorem}\label{s-HVH}
The general Hugenholtz-Van Hove (HVH) theorem \cite{Hug58}
\begin{eqnarray}\label{HVH}
E_{\rm{F}}=\frac{d (\rho E)}{d \rho} =  E+ \rho\frac{d E}{d \rho} = E+ P/\rho
\end{eqnarray}
governs the relation between the Fermi energy $E_{\rm{F}}$ and the energy per nucleon $E$ in all Fermionic systems with pressure $P$ at zero temperature.
This fundamental theorem provides a link between the EOS of Eq. (\ref{eos1}) and the single-nucleon potentials of Eq. (\ref{sp}) at the Fermi momenta of neutrons and protons.
For more detailed discussions on this topic, we refer the reader to ref. \cite{PPNP-Li}.

\begin{figure*}[t]
\begin{center}
\includegraphics[width=0.8\linewidth]{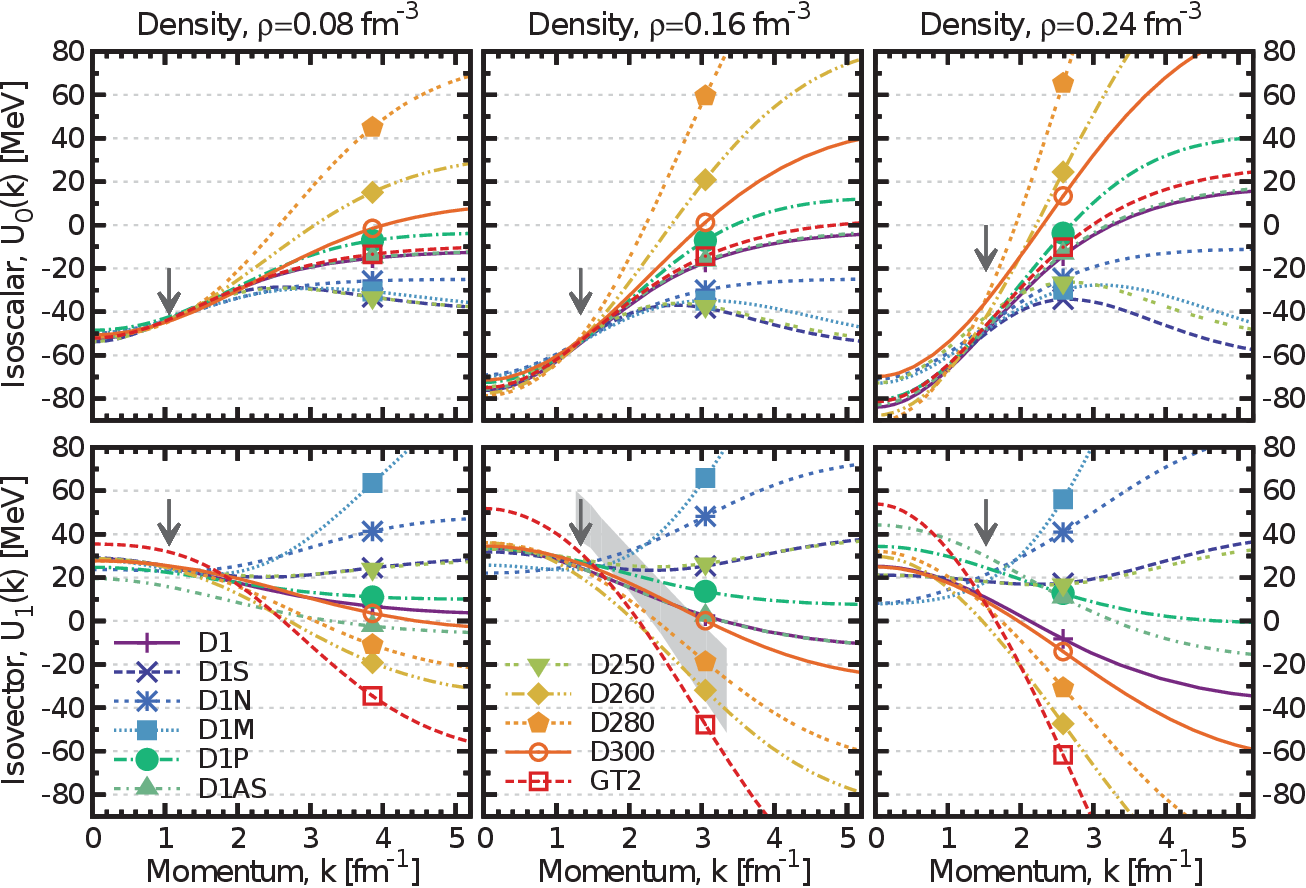}
\vspace{0.5cm}
\caption{(Color online) Isoscalar (top panels) and isovector (bottom panels) components of the single-particle potential as a function of momentum. Results for all 11 Gogny functionals are displayed at 3 densities: $\rho=0.08$ fm$^{-3}$ (left panels), $\rho=0.16$ fm$^{-3}$ (central panels) and $\rho=0.24$ fm$^{-3}$ (right panels) are displayed. The gray band in the bottom central panel is the allowed region of saturation isovector single-particle potentials obtained in Ref.~\cite{LiXH13}. The arrows mark the position of the Fermi momentum at each density. Taken from ref. \cite{Rios-Gogny}.}
\label{Rios-U}
\end{center}
\end{figure*}
\begin{figure}[t!]
\begin{center}
\includegraphics[width=0.8\linewidth]{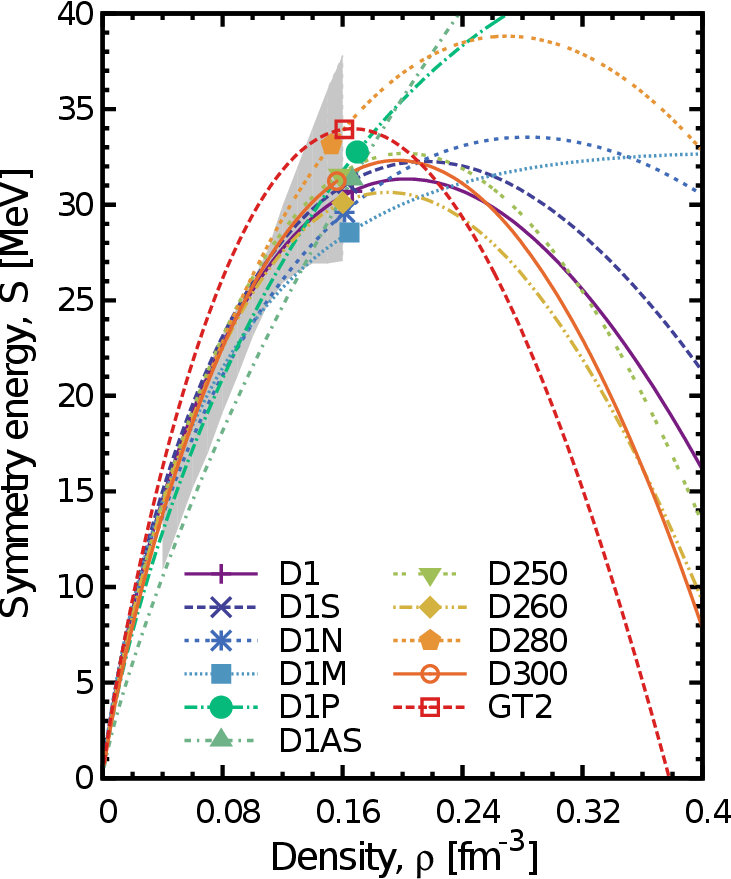}
\vspace{0.5cm}
\caption{(Color online) Symmetry energy as a function of density for all 11 Gogny functionals. The shaded region corresponds to the constraints arising from IAS of Ref.~\cite{Pawel14}.
Taken from ref. \cite{Rios-Gogny}.}
\label{Rios-esym}
\end{center}	
\end{figure}
\begin{figure*}
\begin{center}
\resizebox{0.8\textwidth}{!}{
 \includegraphics{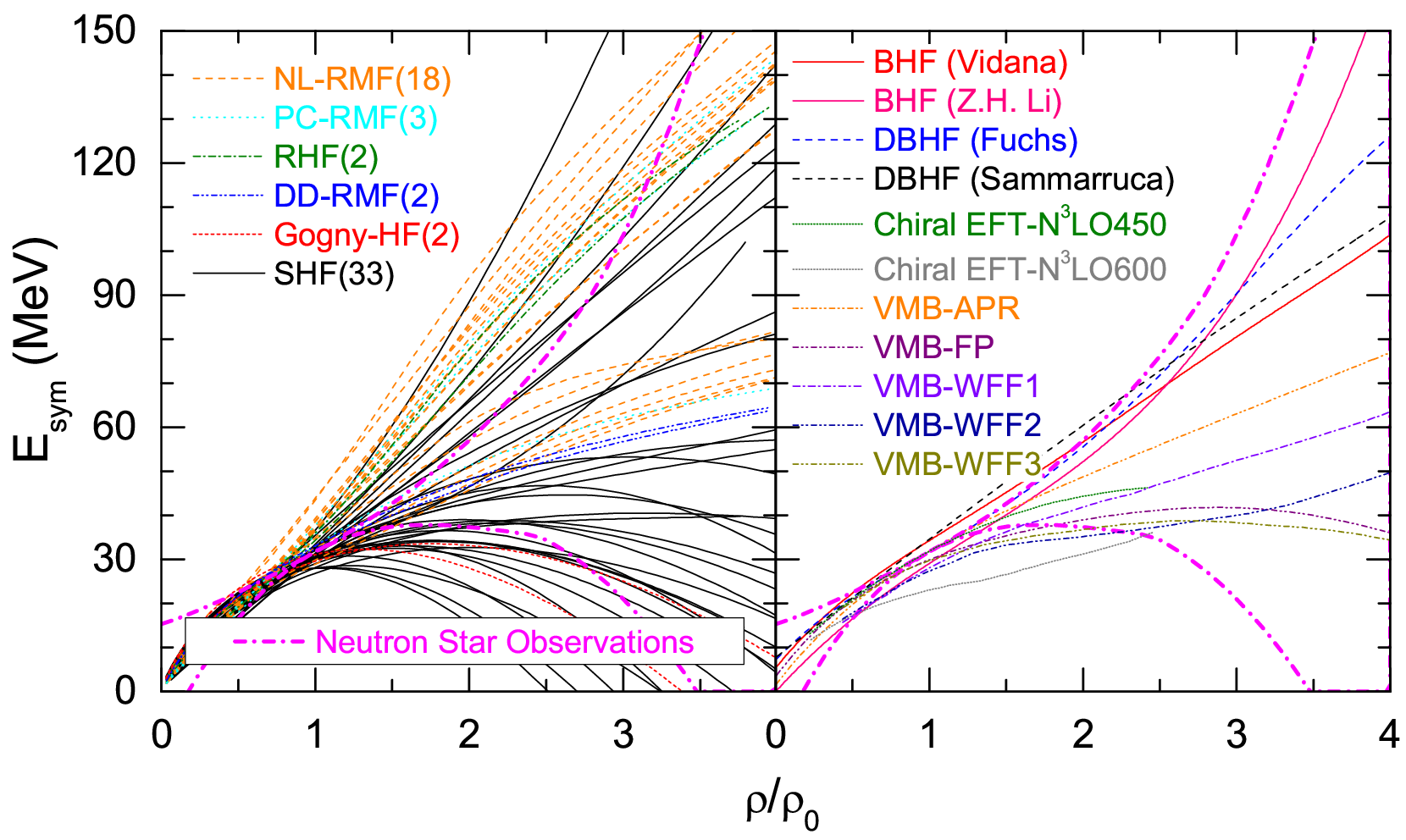}
}
\vspace{0.5cm}
\caption{(Color online) Examples of the nuclear symmetry energy \esym predicted by nuclear many-body theories using different interactions, energy density functionals and/or techniques
(made by amending a compilation in ref. \cite{Chen2017}) in comparison with the constraining boundaries (magenta dot-dashed lines) extracted from studying properties of neutron stars.
Taken from ref. \cite{Zhang2019EPJA}.}
\label{examples}
\end{center}
\end{figure*}
In terms of the components of the nucleon potentials, the quadratic symmetry energy can be written as
\cite{bru64,Dab73,FKW,XuC10,XuC11,Rchen,CXu14}
\begin{equation}
E_{\rm{sym},2}(\rho) =\frac{1}{3} \frac{k_{\rm{F}}^2}{2 m} +\frac{1}{2} U_{\rm{sym},1}(\rho,k_{\rm{F}})+\frac{k_{\rm{F}}}{6}\left(\frac{\partial U_0}{\partial k}\right)_{k_{\rm{F}}}-\frac{1}{6}\frac{k^4_{\rm{F}}}{2m^3}
\label{FKW}
\end{equation}
and the quartic symmetry energy $E_{\rm{sym},4}(\rho)$ can be written as\,\cite{XuC11}
\begin{eqnarray}\label{Esym4}
&&E_{\rm{sym},4}(\rho) = \frac{\hbar^2}{162m}
\left(\frac{3\pi^2}{2}\right)^{2/3} \rho^{2/3}\nonumber\\
&&+\Bigg[\frac{5}{324}\frac{\partial U_0(\rho,k)}{\partial k}
k - \frac{1}{108} \frac{\partial^2 U_0(\rho,k)}{\partial
k^2} k^2 +\frac{1}{648} \frac{\partial^3
U_0(\rho,k)}{\partial k^3}k^3\nonumber\\
&&-
\frac{1}{36} \frac{ \partial U_{\rm{sym},1}(\rho,k)}{\partial k}
k + \frac{1}{72} \frac{ \partial^2 U_{\rm{sym},1}(\rho,k)}{\partial
k^2} k^2\nonumber\\
&& + \frac{1}{12} \frac{\partial
U_{\rm{sym},2}(\rho,k)}{\partial k}k+ \frac{1}{4}
U_{\rm{sym},3}(\rho,k)\Bigg]_{k_{\rm{F}}}.
\end{eqnarray}
While the density slope L of the symmetry energy \esym
\begin{equation}
L(\rho) \equiv \left[3 \rho (\partial E_{\rm sym}/\partial \rho\right)]_{\rho}
\end{equation}
at an arbitrary density $\rho$ can be expressed generally as \cite{XuC11}
\begin{eqnarray}
&&L(\rho) = \frac{2}{3} \frac{\hbar^2 k_F^2}{2 m_0^*} + \frac{3}{2} U_{\rm sym,1}(\rho,k_F)
- \frac{1}{6}\Big(\frac{\hbar^2 k^3}{{m_0^*}^2}\frac{\partial m_0^*}{\partial k} \Big)|_{k_F} \nonumber\\
&&+\frac{\partial U_{\rm sym,1}}{\partial k}|_{k_F} k_F
+ 3U_{\rm sym,2}(\rho,k_F), \label{Lexp2}
\end{eqnarray}
where $k_F=(3\pi^2\rho/2)^{1/3}$ is the nucleon Fermi momentum and $m^*_0/m=(1+\frac{m}{\hbar^2k_{\rm F}}\partial U_0/\partial k)^{-1}|_{k_F}$ is the nucleon isoscalar effective mass of nucleons with free mass $m$. In terms of the $m^*_0$, the symmetry energy of Eq. (\ref{FKW}) can be rewritten as
\begin{equation}
E_{\rm{sym},2}(\rho) =\frac{1}{3} \frac{k_{\rm{F}}^2}{2 m^*_0(\rho,k_{\rm{F}})} +\frac{1}{2} U_{\rm{sym},1}(\rho,k_{\rm{F}}). \label{Esymexp2}
\end{equation}
The above decompositions of both the quadratic and quartic symmetry energies as well as the density slope $L(\rho)$
in terms of the density and momentum dependences of the isoscalar and isovector single-nucleon potentials reveal clearly the underlying microscopic physics at the non-relativistic mean-field level.
The above decompositions of \esym and \lr are transparent and model independent. They are useful for identifying the important underlying physics as we shall discuss next. The relativistic decompositions of \esym and $L$ in terms of the Lorentz covariant nucleon self-energies will be discussed in the Sect. \ref{s-fock}.

At this point, it is also worth mentioning that not in all many-body theories the average energy per nucleon can be explicitly separated into kinetic and potential contributions. Some techniques, such as the Hellmann-Feynman theorem used in the BHF approach  \cite{Vid11}, have to be used to separate the kinetic from the potential contribution to the symmetry energy. Moreover, in some other approaches, such as the self-consistent Green' s function (SCGF)
approach the energy expression involves the integration over the spectral function, see e.g. ref. \cite{SCG18}.  The Fermi energy, energy per nucleon and pressure from the SCGF approach have been shown to satisfy numerically the HVH theorem. However, it is unknown to us if it is possible to decompose the symmetry energy in terms of the kinetic contribution and single-particle potentials within this approach.

\begin{figure}[htb]
\hspace{-1.5cm}
\resizebox{0.7\textwidth}{!}{
  \includegraphics{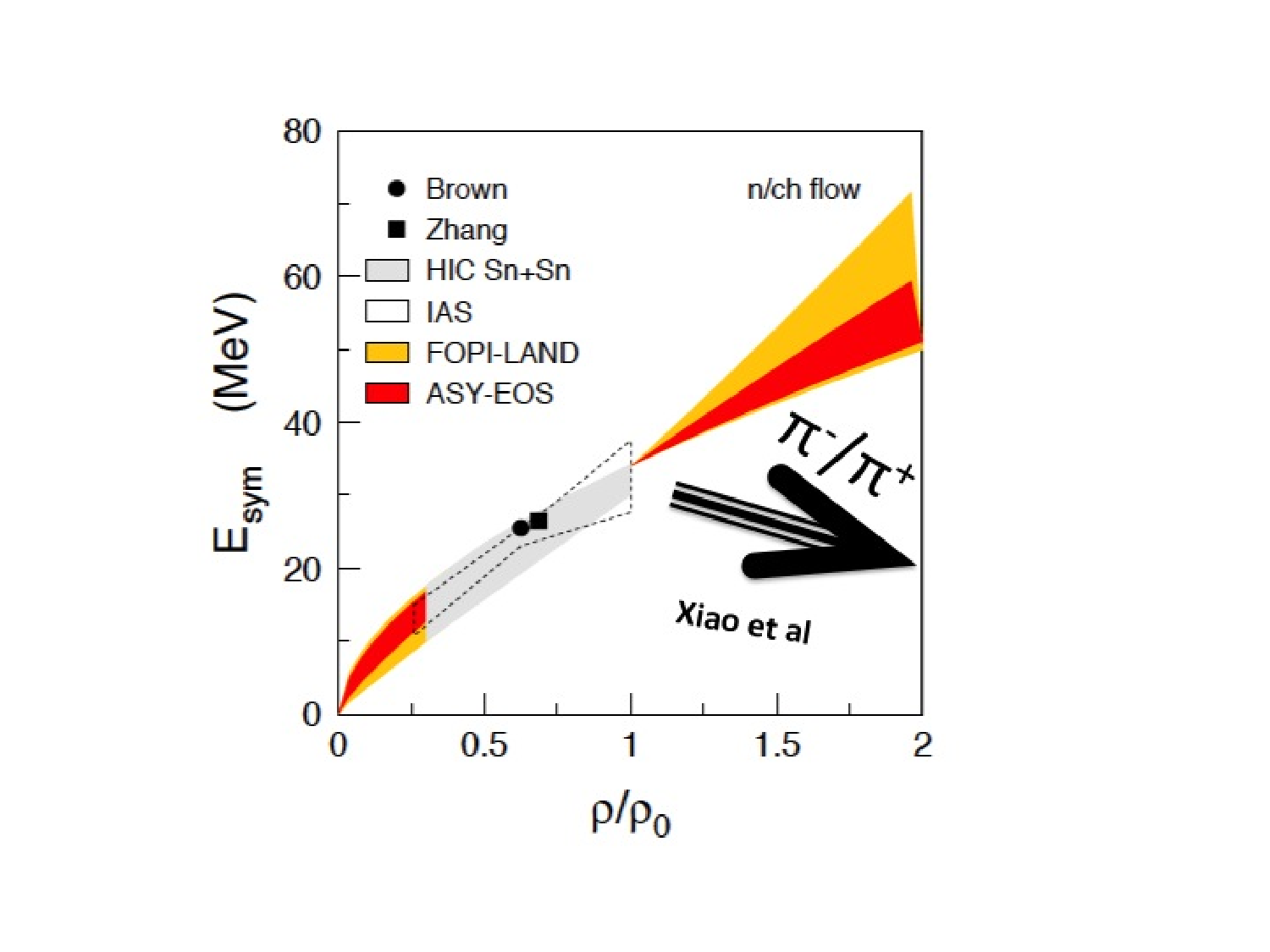}
  }
  \vspace{-1cm}
\caption{{\protect (Color online) Constraints on the density dependence of \esym using isospin diffusion data from MSU \cite{Tsang04}, world data of excitation energies of the isobaric analog states (IAS) \cite{Pawel14}, isospin-dependent flow measurements by the ASY-EOS Collaboration at GSI \cite{russ11,ASY-EOS} in comparison with the trend (arrow) of \esym from an earlier analysis of the FOPI/GSI pion data by Xiao et al. using the IBUU04 transport model \cite{XiaoPRL} as well as the constraints at $\rho_0$ from analyzing properties of double magic nuclei by Brown \cite{Brown13} and binding energies and neutron-skins by Chen and Zhang \cite{Zhang-skin}. Taken from ref. \cite{ASY-EOS}.}}\label{ASY}
\end{figure}

In non-relativistic frameworks,  the \esym has a kinetic term equivalent to $1/3$ the Fermi energy of quasi-nucleons with an isoscalar effective mass $m^*_0$ and a potential part equal to $1/2$ the isovector potential $U_{\rm sym,1}(\rho,k_{F})$ at the Fermi momentum $k_F$. The $L(\rho)$ has five terms depending, respectively, on (1) the isoscalar nucleon effective mass $m^*_0$, (2) the momentum dependence of $m^*_0$, (3) the isovector potential $U_{\rm sym,1}(\rho,k_F)$, (4) the momentum dependence of the isovector potential and (5) the second-order isoscalar potential $U_{\rm sym,2}(\rho,k_F)$. Analyses of the momentum dependence of the isoscalar and isovector potential at $\rho_0$ using experimental data from (p,n) charge-exchange and nucleon-nucleus elastic scatterings \cite{LiXH13,XuC10} indicate that the magnitude and the momentum-dependence of the symmetry potential $U_{\rm{sym},1}(\rho,k)$ play dominating roles but are also currently most uncertain.

Theoretical studies indicate that the density and momentum dependence of the isovector potential depends strongly on the interactions used \cite{Rios-Gogny,XuC11,Rchen,Das03,Zuo}. For example, the finite-range Gogny-type interactions and the resulting single-particle potentials are widely used in studying fruitfully nuclear structures and reactions.
Sellahewa and Rios conducted an extensive study of isovector properties of neutron-rich nuclear matter using all 11 parameter sets of Gogny interactions available in the literature \cite{Rios-Gogny}.
Shown in Fig.\ \ref{Rios-U} are the results of their Hatree-Fock calculations for the isoscalar (top) and isovector (bottom) components of the single-particle potential as a function of momentum at 3 densities of $\rho=0.08$ fm$^{-3}$ (left), $\rho=0.16$ fm$^{-3}$ (central) and $\rho=0.24$ fm$^{-3}$ (right) using the 11 parameter sets. The gray band in the bottom central panel is the allowed region of isovector single-particle potentials at saturation density obtained in Ref.~\cite{LiXH13} from analyzing the nucleon-nucleus optical potentials. The arrows mark the position of the Fermi momentum at each density. Largely be design, the different parameter sets give results consistent around the saturation density. However, especially at supra-saturation densities the predicted momentum dependence of especially the isovector single-particle potentials is rather different. The corresponding predictions for the \esym using all 11 Gogny functionals are shown in Fig.\ \ref{Rios-esym}. It is seen that the predicted \esym at supra-saturation densities diverge rather wildly, while in the sub-saturation region, most of the 11 calculations are generally consistent with the result from analyzing the
isobaric analog states in the shaded region~\cite{Pawel14}.
\begin{figure*}[htb]
\begin{center}
\resizebox{0.8\textwidth}{!}{
  \includegraphics{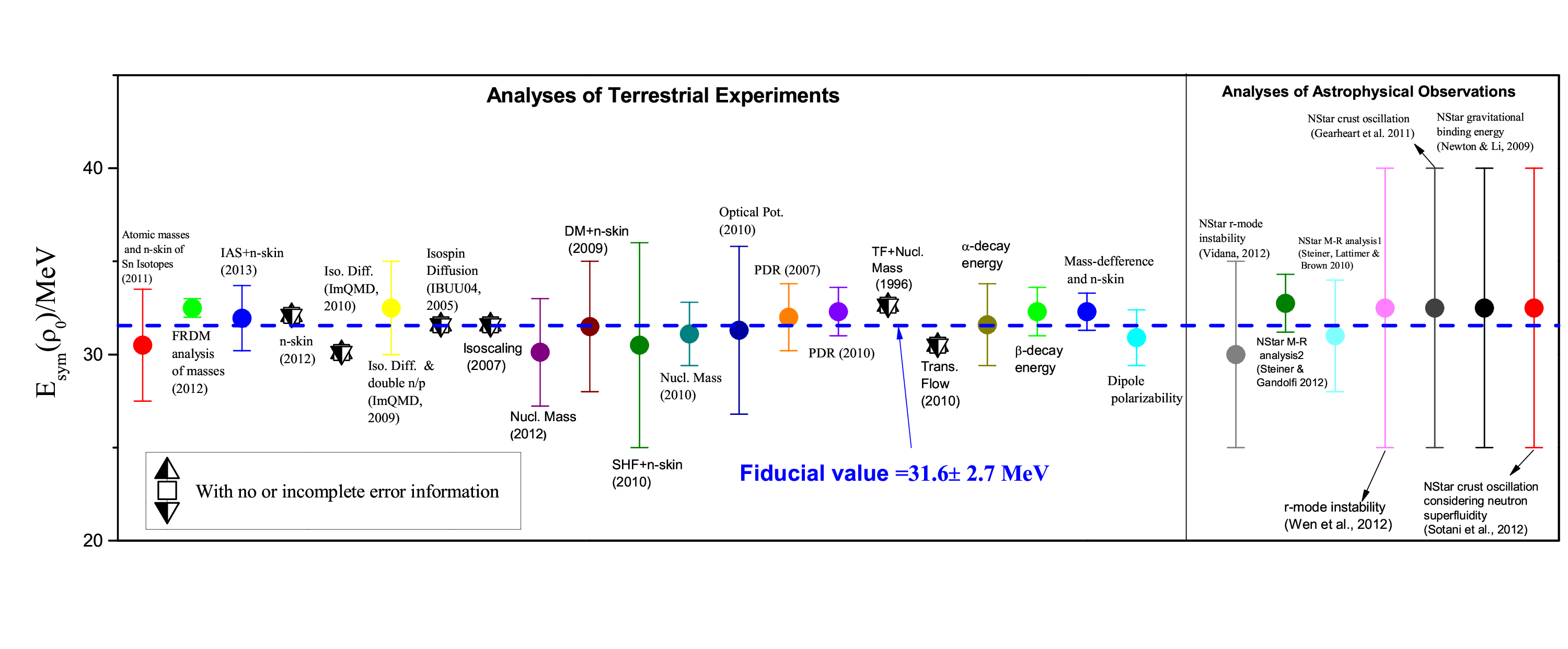}
  }
  \resizebox{0.8\textwidth}{!}{
  \includegraphics{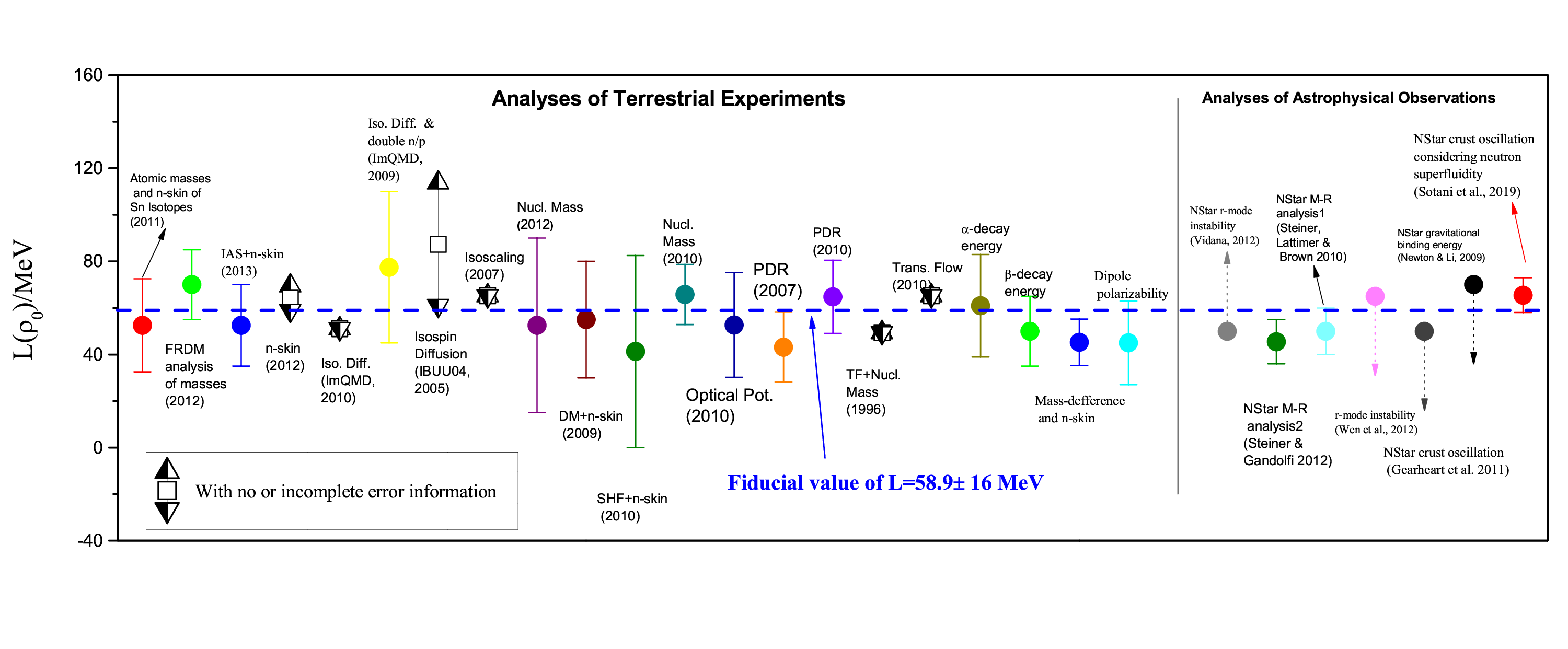}
}
\caption{{\protect (Color online) Central values of \es0 and $L(\rho_0)$ from 28 model analyses of terrestrial nuclear experiments and astrophysical observations. Modified from similar plots in ref. \cite{Li2013} by updating the result of Sotani et al in their analyses of the quasi-periodic oscillations of neutron stars \cite{Sotani2018,Sot19}.}}\label{Esym0-Li}
\end{center}
\end{figure*}

\subsection{What have we learned about the symmetry energy so far?}\label{lsys}
Over the last two decades, significant efforts have been devoted to understanding and constraining the \esym from both the theoretical and experimental/observational communities in
nuclear physics and astrophysics. Indeed, much progress has been made, see, e.g., refs.  \cite{ditoro,Li2008,Lynch09,Trau12,Steiner05,Tesym,Bal16,Li17,ibook01,Tsang12,Chuck14,Blaschke16,ireview98,Lattimer2012} for reviews.
To our best knowledge, basically all available nuclear many-body theories and forces have been used to predict the density dependence of nuclear symmetry energy.
 However, as illustrated using selected examples in Fig.\ \ref{examples}, model predictions spread over large regions at both sub-saturation and supra-saturation densities although they agree often by construction at $\rho_0$. What have we learned about the symmetry energy so far? There is probably no community consensus for the answer to this question. In the following we highlight a few points that might be biased and certainly incomplete. For a more comprehensive reviews, we refer the reader to some of the recent reviews mentioned above and the references cited therein.

Extensive efforts have also been devoted to extracting the \esym from sub-saturation to supra-saturation densities
from both terrestrial nuclear experiments and astrophysical observations. Aa an example, shown with the magenta dot-dashed lines in Fig. \ref{examples} are the boundaries of the \esym extracted from studying properties of neutron stars \cite{Zhang2019EPJA}. While the boundaries span over a large range above about twice the saturation density, they can already exclude some of the predictions. Similarly, laboratory nuclear experiments have also been used to constrain the symmetry energy. In particular,
various reaction observables and phenomena, energy and strength of various collective modes, various forms of isospin transport, hard photon production, ratios and differential flows of protons and neutrons as well as mirror nuclei in heavy-ion reactions at intermediate energies, pion, kaon and $\eta$ production in heavy-ion collisions up to 10 GeV/nucleon have been proposed as probes of the symmetry energy, see, e.g., reviews collected in ref. \cite{Tesym}. Most of these observables probe directly the density and momentum dependence of the isovector potential $U_{\rm sym,1}(\rho,k)$. Depending on the conditions of the reactions, these observables often probe the \esym over a broad density range.
For example, as shown in Fig. \ref{ASY} with the gray band and labeled as HIC Sn+Sn, transport model analyses of the isospin diffusion data in reactions involving several Sn isotopes taken by Tsang et al. \cite{Tsang04} have consistently extracted a constraining band on the \esym in the range of $\rho_0/3$ to about $\rho_0$. This band is also consistent with constraints obtained from analyzing other observables. At supra-saturation densities, however, the situation is quite different as indicated by the different trends obtained from analyzing
different observables or the same observables but using different models. In the examples shown in Fig. \ref{ASY}, the decreasing trend of \esym was found from analyzing the $\pi^-/\pi^+$ data from the FOPI/GSI collaboration \cite{XiaoPRL,FOPI} using a BUU-type (Boltzmann-Uehling-Uhlenbeck) transport model \cite{LiBA04a}. Such kind of decreasing \esym at supra-saturation densities is possible as indicated by some model predictions and is not ruled out by the constraints from analyzing neutron star observables as shown in Fig. \ref{examples}. However, the analysis of the pion ratio data is still quite model dependent. On the other hand, the ASY-EOS Collaboration found that the \esym increases continuously with density from analyzing their data on the relative flows of neutrons w.r.t. protons, tritons w.r.t. $^3$He and yield ratios of light isobars using two versions of the QMD-type (Quantum Molecular Dynamics) transport models \cite{russ11,ASY-EOS}. Thus, existing analyses of heavy-ion reaction experiments have not reached a consensus regarding the high-density behavior of the symmetry energy. Certainly, ongoing and planned new experiments coupled with better coordinated theoretical efforts using systematically tested reaction models will help improve the situation hopefully in the near future.  Comparing the theoretical predictions with constraints from analyzing observables of neutron stars and terrestrial nuclear experiments as shown in Fig. \ref{examples} and Fig.\ \ref{ASY}, it is seen clearly that the \esym at supra-saturation densities are still largely unconstrained. Certainly, the multi-messengers approach combing tools in nuclear theories, astrophysical observations and terrestrial experiments has the great potential of helping us resolve this longstanding challenge.

Within the parabolic approximation of the EOS, it is customary to either Taylor expand (e.g., in energy density functional theories) at densities near $\rho_0$ or simply parameterize up to certain supra-saturation densities (e.g., in conducting Bayesian inferences and directly solving numerically the inverse-structure problem of neutron stars especially at supra-saturation densities) the
 $E_0(\rho)$ and $E_{\rm{sym}}(\rho )$, respectively, according to
\begin{eqnarray}\label{E0para}
E_{0}(\rho)&=&E_0(\rho_0)+\frac{K_0}{2}(\frac{\rho-\rho_0}{3\rho_0})^2+\frac{J_0}{6}(\frac{\rho-\rho_0}{3\rho_0})^3,\\
E_{\rm{sym}}(\rho)&=&E_{\rm{sym}}(\rho_0)+L(\frac{\rho-\rho_0}{3\rho_0})+\frac{K_{\rm{sym}}}{2}(\frac{\rho-\rho_0}{3\rho_0})^2\nonumber\\
&+&\frac{J_{\rm{sym}}}{6}(\frac{\rho-\rho_0}{3\rho_0})^3.\label{Esympara}
\end{eqnarray}
For the EOS of SNM, extensive studies have determined the most probable incompressibility of symmetric nuclear matter as $K_0=230 \pm 20$ MeV \cite{Shlomo2006,Piekarewicz2010},
while the skewness parameter $J_0$ is only roughly known to be in the range of $-800 \leq J_{0}\leq 400$ MeV \cite{Tews17,Zhang2017}. For the symmetry energy,
past efforts in both nuclear physics and astrophysics have been most fruitful in constraining the magnitude and slope, i.e., $E_{\rm sym}(\rho_0)$ and $L(\rho_0)$, of the \esym around $\rho_0$. For example, shown in Figs.\ \ref{Esym0-Li} are the central values of the $E_{\rm sym}(\rho_0)$ and $L(\rho_0)$ from 28 analyses of some data from both nuclear experiments and astrophysical observations available before 2013 \cite{Li2013}. The fiducial values extracted from these data are $E_{\rm sym}(\rho_0)=31.6\pm 2.7$ MeV and $L(\rho_0)=58.9\pm 16$ MeV, respectively. More recent surveys of more data analyses found that the central values of $E_{\rm sym}(\rho_0)$ and $L(\rho_0)$ are $31.7\pm 3.2$ MeV and $58.7\pm 28.1$ MeV \cite{Oertel17}, respectively, consistent with the earlier findings \cite{Li2013}, while the two high-density parameters are poorly known. Mostly based on surveys of calculations using over 500 energy density functionals, the curvature and skewness of the symmetry energy have been predicted to be in the range of $-400 \leq K_{\rm{sym}} \leq 100$ MeV, $-200 \leq J_{\rm{sym}}\leq 800$ MeV, respectively. Given this situation, for many purposes, one can fix the $K_0$, $E_{\rm sym}(\rho_0)$ and $L(\rho_0)$ at their most probable values presently known while use the $J_0, K_{\rm{sym}}$ and $J_{\rm{sym}}$ as free high-density parameters.

\section{Why is the symmetry energy still so uncertain especially at supra-saturation densities?}
Of course, answers to the above question are likely model dependent and the different techniques used in solving nuclear many-body problems certainly contribute to the diverse behaviors of \esym at high densities. However,  a few common key physics ingredients can be identified based on the HVH decomposition of \esym and \lr presented earlier. In the following we discuss effects of a few such ingredients on the high-density behavior of nuclear symmetry energy.

\subsection{The most important but poorly known physics behind the nuclear symmetry energy}
In fact, lying under the \esym is the isospin-dependence of strong interactions. Our poor knowledge about the latter in dense neutron-rich matter is ultimately responsible for the uncertain \esym at high densities. As shown earlier, the \esym and its density slope \lr depend on the density and momentum dependence of both the isoscalar and isovector potentials.
Currently,  the density and momentum dependence of the isoscalar single-nucleon potential is much better known than that of the isovector potential. For example, even at the saturation density of nuclear matter $\rho_0$, the isovector (symmetry) potential $U_{\rm{sym},1}(\rho,k)$ around and above the Fermi momentum is very poorly known \cite{PPNP-Li}. Consequently, the density dependence of the \esym is still very uncertain especially at supra-saturation densities.

\begin{figure}[htb]
\begin{center}
\resizebox{0.5\textwidth}{!}{
  \includegraphics{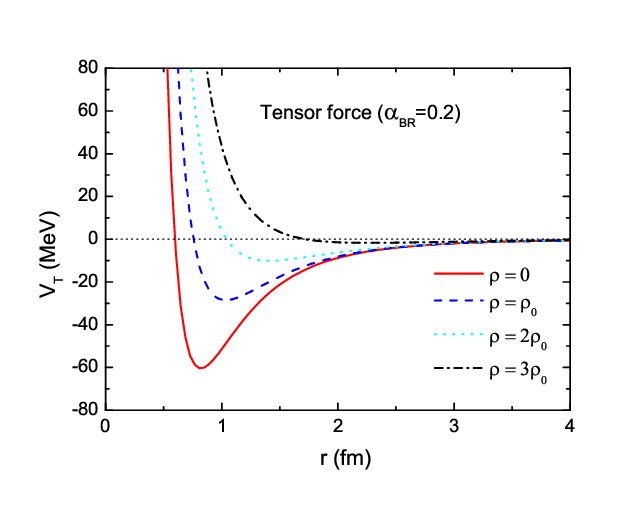}
  }
\caption{(Color online) The radial part of the tensor force due to pion and $\rho$ meson exchange
at densities of $\rho=0, \rho_0, 2\rho_0, \,$and$\, 3\rho_0$ with the in-medium $\rho$ mass of
$ m_{\rho}^{\star}/m_{\rho} =
1-0.2 \rho/\rho_0 $. Taken from ref. \cite{Xu-tensor}.}\label{Graph-Xu1}
\end{center}
\end{figure}
\begin{figure*}[htb]
\begin{center}
\resizebox{0.6\textwidth}{!}{
  \includegraphics{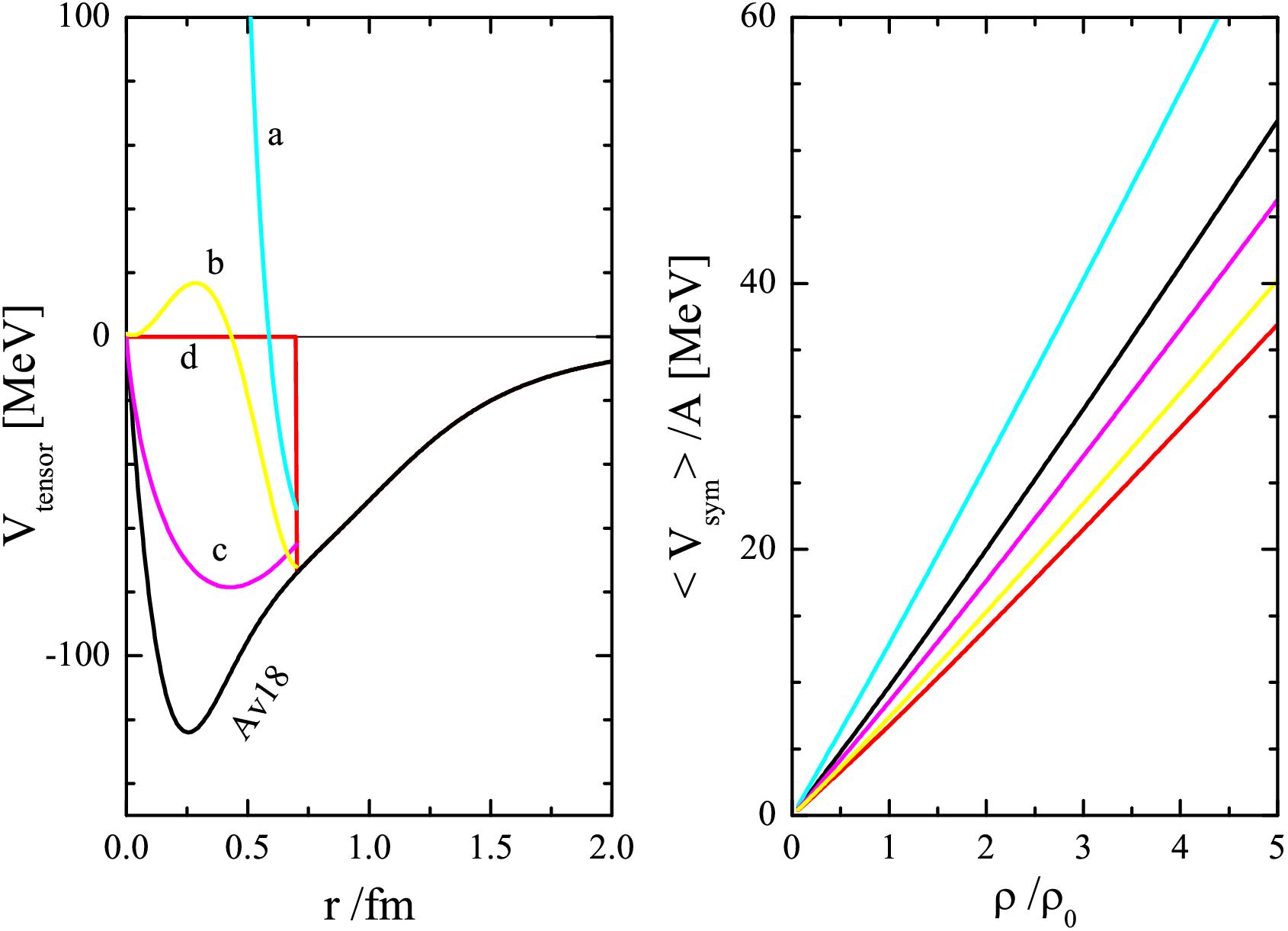}
  }
\caption{(Color online) Left panel: radial parts of the tensor interactions having different short-range behaviors but the same long-range ($r > 0.7$fm) part as the Av18,
Right panel: the corresponding potential symmetry energies. Taken from ref. \cite{LiLi}.} \label{f1}
\end{center}
\end{figure*}

To be more specific, it is instructive to inspect what is behind the $U_{\rm{sym},1}(\rho,k)$ within simple and transparent models.
For example, the Hartree term of the isovector potential at $k_F$ in the interacting Fermi gas model can be written as \cite{Xu-tensor,pre}
\begin{equation}
U_{\rm sym,1}(k_F,\rho)=
\frac{\rho}{4}\int [V_{T1}\cdot f^{T1}(r_{ij})-V_{T0}\cdot f^{T0}(r_{ij})]d^3r_{ij}
\end{equation}
in terms of the isosinglet (T=0) and isotriplet (T=1) nucleon-nucleon (NN) interactions $V_{T0}(r_{ij})$ and $V_{T1}(r_{ij})$, as well as the corresponding NN correlation functions $f^{T0}(r_{ij})$ and $f^{T1}(r_{ij})$, respectively. The charge independence of NN interaction requires that $V_{nn}=V_{pp}=V_{np}$ in the T=1 channel. However, because of the isospin dependence of the NN strong interaction, the $V_{np}$ interactions and the associated NN correlations in the T=1 and T=0 channels are not necessarily the same. As manifested in properties of the deuteron, the tensor force and the resulting short-range correlation in the T=0 channel is much stronger than that in the T=1 channel \cite{bethe,Ant93,Sub08,Hen14}.
While the $U_{\rm{sym},1}(\rho,k)$ from the Fock term using Gogny-type finite-range, isospin-dependent interactions \cite {Gogny} is often parameterized by using different strengths of interactions between like and unlike nucleon pairs \cite{Das03}. Therefore, the isospin dependence of strong NN interactions plays the key role in determining the density and momentum dependence of the isovector single-nucleon potential, and thus the density dependence of nuclear symmetry energy. Indeed, numerical calculations within microscopic many-body theories \cite{Bom91,Die03} have shown that the potential symmetry energy is dominated by the T=0 interaction channel while the T=1 contribution is almost zero.
Thus, the isospin dependence of NN interactions and correlations determines the potential symmetry energy. This general feature is qualitatively consistent with findings within microscopic nuclear many-body theories, see, e.g., ref. \cite{Eng98}.

While at the mean-field level, the tensor force has no contribution to the potential part of the symmetry energy, its secondary contribution is important. Moreover,
the isospin dependence of short-range correlation (SRC) induced by tensor forces leads to a high-momentum tail (HMT) in the single-nucleon momentum distribution \cite{bethe,Sub08,Hen14}. In neutron-rich matter, the HMT is highly isospin dependent. Namely, a larger fraction of protons than neutrons are in the HMT \cite{Hen14}. The isospin dependent HMT affects both the kinetic and potential parts of the symmetry energy when a momentum-dependent interaction is used. Furthermore, as illustrated in many studies in the literature, nuclear saturation properties can be well described by either using an in-medium tensor force or spin-isospin dependent three-body nuclear force \cite{Rap99,Don,Lee,Jeremy16}. Thus, in many studies only the latter is considered while effects of the tensor force are either completely ignored or mimicked by using the three-body force. In the following, we briefly discuss respective effects of the tensor force, SRC and the three-body force on the high-density behavior of nuclear symmetry energy.

\subsection{The role of the tensor force in the isosinglet nucleon-nucleon interaction channel}
The second-order tensor contribution to nuclear symmetry energy has been studied for a long time, see, e.g., refs. \cite{kuo65,bro81,sob,mac94}.
It is approximately
\begin{equation}\label{Mac}
<V_{\rm sym}> = \frac{12}{e_{\rm eff}}<V_t^2(r)>
\end{equation}
where $e_{\rm eff}\approx 200$ MeV and $V_t(r)$ is the radial part of the tensor force~\cite{mac94}. In the
one-boson-exchange picture, the tensor interaction results from
exchanges of the isovector $\pi$ and $\rho$ mesons. The tensor part of the one-pion exchange potential can be
written as~\cite{BR}
\begin{eqnarray}
 &&V_{t\pi}= -\frac{f_{\pi}^2}{4\pi}m_{\pi}(\tau_1\cdot\tau_2)S_{12}
\nonumber \\~~~~~&&
\cdot[\frac{1}{(m_{\pi}r)^3}+\frac{1}{(m_{\pi}r)^2}+\frac{1}{3m_{\pi}r}]\exp(-m_{\pi}r)
\label{pi}
\end{eqnarray}
where $r$ is the inter-particle distance and
\begin{equation}
S_{12}=3\frac{(\sigma_1 \cdot r)(\sigma_1 \cdot
r)}{r^2}-(\sigma_2\cdot\sigma_2)
\end{equation}
is the tensor operator. The
$\rho$-exchange tensor interaction $V_{t\rho}$ has the same
functional form but an opposite sign, namely, the $m_{\pi}$ is replaced
everywhere by $m_\rho$, and the $f_{\pi}^2$ by $-f_{\rho}^2$. The
magnitudes of both the $\pi$ and $\rho$ contributions grow quickly but in opposite directions
with decreasing $r$ as the density increases. The net result from the $\pi$ and $\rho$ exchanges depends strongly on the
poorly known $\rho$-nucleon coupling strength. Moreover, while there is still no solid experimental confirmation, it is possible that the in-medium $\rho$ meson mass $m_{\rho}$ is different from its free-space value. A density-dependent in-medium $\rho$ meson mass $m_{\rho}$
leads to very different tensor forces in dense medium~\cite{BR}, and thus different \esym at high densities~\cite{Xu-tensor,Don,Rho11,Xulili}.
As an illustration, shown in Fig.~\ref{Graph-Xu1} is the radial part of the total tensor force
$V_T=V_T^{\rho}(r)+V_T^{\pi}(r)$ at densities $\rho=0, \rho_0, 2\rho_0, \,$
and $3\rho_0$, respectively, with reduced in-medium $\rho$ meson mass according to $ m_{\rho}^{\star}/m_{\rho} =
1-0.2 \rho/\rho_0 $ \cite{Xu-tensor,BR}. As one expects, the total tensor force becomes
more repulsive in denser matter when the $\rho$ meson mass is
reduced. It was shown within the simple interacting Fermi gas model assuming all isosinglet neutron-proton pairs behave as bound
deuterons with $S_{12}=2$, the potential symmetry energy at supra-saturation densities is indeed found very sensitive to the strength of the tensor force~\cite{Xu-tensor}.

It is well known that various realistic nuclear potentials usually have widely different tensor components
at short range ($r\leq$0.8 fm). While the different behaviors of the tensor force at short-distance have no effect on nuclear structure calculations as often a common cut-off of about $r=0.8$ fm is normally used~\cite{ots05}, they do affect significantly the high-density behavior of the symmetry energy.
As an example, shown in the left panel of Fig.~\ref{f1} are the strengthes of several typical tensor forces widely used in the literature. They are the standard $\pi+\rho$ exchange (labelled as $a$), the G-Matrix (GM)~\cite{ots05} (labelled as $b$),
M3Y~\cite{m3y}(labelled as $c$) and the Av18~\cite{av18} (labelled
as Av18). They are rather differently at short distance but merge to the same Av18
tensor force at longer range. For comparisons, in the case $d$ the tensor force vanishes for $r\leq 0.7$ fm.
The corresponding potential symmetry energies evaluated using the Eq. (\ref{Mac}) are shown in the right panel of Fig.~\ref{f1}.
Clearly, the variation of the short-range tensor force leads to
significantly different potential symmetry energies at supra-saturation densities \cite{LiLi}.

It was predicted a long time ago based on an earlier version of the variational many body theory \cite{Pan72,Wiringa1988} that the tensor force in the isosiglet channel may lead to decreasing \esym at high densities when the repulsive $\rho$ contribution to the tensor dominates. In PNM effects of the tensor force is negligible while the tensor force in the isosinglet channel in SNM makes its potential energy increases very fast at high densities. Depending on the strength of the tensor force at short distance in dense matter, it is then possible that the energy in SNM increases much faster than that in PNM, leading to decreasing or even negative symmetry energies at supra-saturation densities. Possible astrophysical consequences of such kind of \esym have been explored in the literature, see, e.g.,
refs. \cite{XiaoPRL,Kut93,Kut94,Kut06,LiBA02,wen,Kho96,Bas07,Ban00}. However, whether or not the \esym can decrease or become negative at high densities remains a controversial issue.

\subsection{The role of the spin-isospin dependence of three-body forces}\label{3bf}
As we mentioned earlier, three-body nuclear forces play similar roles and thus are often used to mimic tensor force effects on saturation properties of nuclear matter. Effects of the spin-isospin dependence of the three-body force on the \esym are also better understood than those due to the tensor force. Within over 300 Skyrme Hatree-Fock \cite{Cha97,Sto03,Dut12,Pearson} and/or
Gogny Hartree-Fock \cite{dec} energy density functionals, a zero-range three-body force is often reduced to an effective two-body force \cite{dec,vau,Oni78,Gra89}
\begin{equation}\label{VD1}
V_{3}=t_0(1+x_0P_{\sigma})\rho^{\alpha}\delta(r),
\end{equation}
or its improved version involving separate densities $\rho_i$ and $\rho_j$ of the two interacting nucleons $i$ and $j$
\begin{equation}\label{VD}
V_{3ij}=t_0 (1+x_0 P_{\sigma}) [\rho_{i}(\textbf{r}_i) +
\rho_{j}(\textbf{r}_j)]^{\alpha} \delta(\textbf{r}_{ij})
\end{equation}
where $t_0$, $\alpha$ and $x_0$ are parameters while $P_{\sigma}$ is
the spin-exchange operator \cite{bru64,Dab73,koh,Dut2,Dut3,Dut4,Dut5,Dut6,Spr,Neg}.
To differentiate from calculations using the original MDI (Momentum-Dependent Interaction) with the three-body force of Eq. (\ref{VD1}), the MDI with the three-body force of Eq. (\ref{VD}) is referred as the Improved MDI (IMDI).
The parameter $x_0$ controls the relative contributions from the isosinglet and isotriple NN
interactions. More specifically, the potential energies due to the three-body force in the T=1
and T=0 NN interaction channels are, respectively \cite{dec},
\begin{equation}
V_d^{T1}=\frac{1-x_0}{2} \frac{3t_0}{8} \rho ^{\alpha +1}~\rm{and}~
V_d^{T0}=\frac{1+x_0}{2} \frac{3t_0}{8} \rho ^{\alpha +1}.
\end{equation}
One sees immediately that the terms containing $x_0$ cancel out in
calculating the EOS of SNM. While the three-body force contribution to the \esym is \cite{Xu10i}
\begin{equation}
E_{\rm sym}(V_3)=-(1+2x_0)\frac{t_0}{8}\rho ^{\alpha +1}.
\end{equation}
Thus, both the $x_0$ and $\alpha$ affect the symmetry energy but only the latter also affects the EOS of SNM.
Many energy density functionals have used either the $V_3$ or $V_{3ij}$ for the three-body force term with different values of $x_0$ and $\alpha$, leading to diverse predictions for
the \esym at supra-saturation densities.
\begin{figure}[htb]
\begin{center}
\resizebox{0.45\textwidth}{!}{
\includegraphics{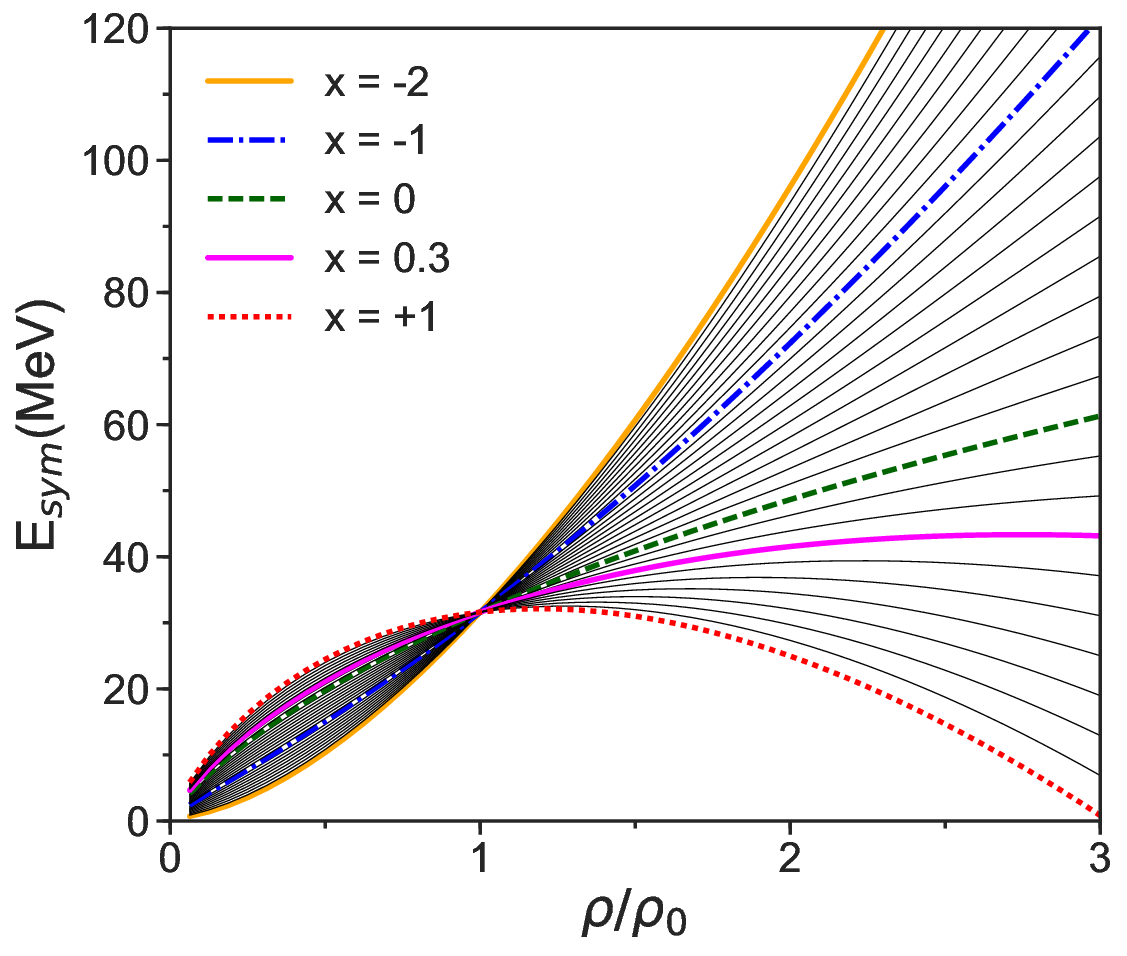}
}
\caption{(Color online) The density dependence of nuclear symmetry energy using the MDI interaction by varying the parameter $x$ from -2 to +1 in steps of $\Delta x = 0.1$ but using the same $\sigma=4/3$. Taken from ref.\ \cite{Plamen3}.}\label{Plamen18a}

\end{center}
\end{figure}

\begin{figure}[htb]
\begin{center}
\resizebox{0.45\textwidth}{!}{
\includegraphics{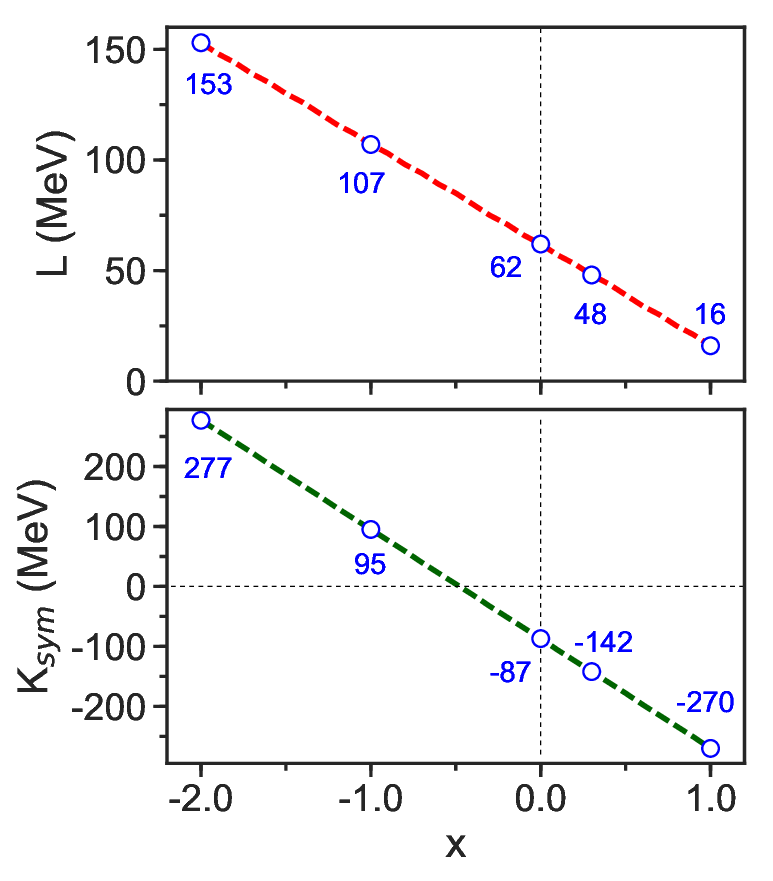}
  }
\caption{(Color online) The slope $L$ (upper panel) and curvature $K_{\rm sym}$ (lower panel) of the MDI symmetry energy shown in Fig. \ref{Plamen18a}. Taken from ref.\ \cite{Plamen3}.}\label{Plamen18b}
\end{center}
\end{figure}

Within a given energy density functional (EDF), when the $x_0$ parameter is varied other parameters have to be adjusted self-consistently to satisfy the same set of
constraints. As an illustration of three-body force effects on the symmetry energy, we quote in the following results of a study \cite{Xu10i} using the
MDI energy density functional \cite{Das03}. The latter has been used extensively in both simulating heavy-ion reactions \cite{LiBA04a,Chen05a,LiChen05} and
studying properties of neutron stars \cite{Li:2005sr,Li2006,Plamen2,JXu,Plamen1,Newton12}. It is developed from a modified Gogny-type interaction within the Hartree-Fock approach \cite{Das03}.
Using the MDI, the energy per nucleon $E(\rho,\delta)$ can be written as \cite{Cai18}
\begin{eqnarray}\label{EDF}
&&E(\rho,\delta)=\sum_{J=\rm{n,p}}\frac{1}{\rho_J}\int_0^{\infty}\frac{\v{k}^2}{2M}n_{\v{k}}^J(\rho,\delta)\d\v{k}\nonumber\\
&&+\frac{A_\ell(\rho_{\rm{p}}^2+\rho_{\rm{n}}^2)}{2\rho\rho_0}
+\frac{A_{\rm{u}}\rho_p\rho_n}{\rho\rho_0}
+\frac{B}{\sigma+1}\left(\frac{\rho}{\rho_0}\right)^{\sigma}(1-x\delta^2)\nonumber\\
&&+\sum_{J,J'}\frac{C_{J,J'}}{\rho\rho_0}\int\d\v{k}\d\v{k}'f_J(\v{r},\v{k})f_{J'}(\v{r},\v{k}')\Omega(\v{k},\v{k}').
\end{eqnarray}
The first term is the kinetic energy while the second to fourth terms are the usual zero-range 2-body and effective 3-body force contributions, respectively.
The parameters B and $\sigma$ are related to the $t_0$ and $\alpha$ in the Gogny effective
interaction via $t_0 = \frac{8}{3} \frac{B}{\sigma+1}
\frac{1}{\rho_0^{\sigma}}$, and $\sigma = \alpha + 1$,
respectively. The parameter $x$ is related to the $x_0$ via $x=(1+2x_0)/3$ \cite{Xu10i}.
As the $x$ parameter varies, the competition between the isosinglet and
isotriplet 2-body interactions is changed.

The last term in Eq. (\ref{EDF}) is the contribution from the
finite-range 2-body interactions characterized by the strength
parameter $C_{J,J}\equiv C_\ell$ for like and
$C_{J,\overline{J}}\equiv C_{\rm{u}}$ for unlike nucleon paris,
respectively, using the notations $\overline{\rm{n}}=\rm{p}$ and
$\overline{\rm{p}}=\rm{n}$. The $f_J(\v{r},\v{k})$ and
$n_{\v{k}}^J(\rho,\delta)$ are  the nucleon phase space distribution
function and momentum distribution function, respectively. In
equilibrated nuclear matter at zero temperature, they are related by
\begin{equation}
f_J(\v{r},\v{k})=\frac{2}{h^3}n_{\v{k}}^J(\rho,\delta)=\frac{1}{4\pi^3}n_{\v{k}}^J(\rho,\delta),~~\hbar=1.
\end{equation}
For example, in the free Fermi gas (FFG),  $n_{\v{k}}^J=\Theta(k_{\rm{F}}^J-|\v{k}|)$ with the
standard step function $\Theta$, then
\begin{equation}
f_J(\v{r},\v{k})=(1/4\pi^3)\Theta(k_{\rm{F}}^J-|\v{k}|).
\end{equation}
The regulating function $\Omega(\v{k},\v{k}')$ originating from the meson exchange theory of nuclear force normally
has the form of \cite{Das03,Gal87}
\begin{equation}\label{Ome}
\Omega(\v{k},\v{k}')=\left[1+\left(\frac{\v{k}-\v{k}'}{\Lambda}\right)^2\right]^{-1}
\end{equation}
where $\v{k}$ and $\v{k}'$ are the momenta ($\v{p}=\hbar \v{k}$) of two interacting
nucleons and $\Lambda$ is a parameter regulating the momentum
dependence of the single-particle potential.

\begin{figure}[htb]
\begin{center}
\resizebox{0.45\textwidth}{!}{
  \includegraphics{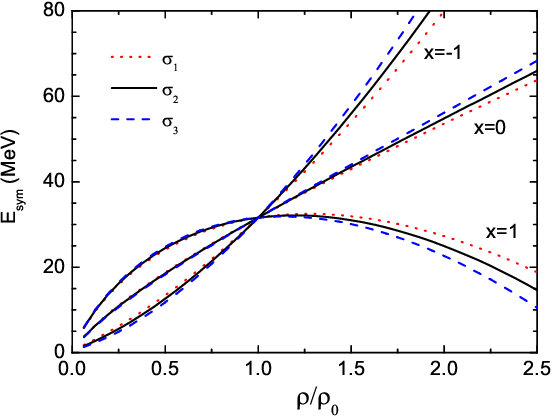}
  }
\caption{(Color online) The symmetry energy \esym obtained with the improved MDI interaction using the three-body force given in Eq. (\ref{VD}) with $x=1, 0$ and $-1$ and three $\sigma_{1,2,3}$ values of $\frac{4}{3}-\frac{1}{30}$, $\frac{4}{3}$, and
$\frac{4}{3}+\frac{1}{30}$, respectively. Taken from ref.\ \cite{Xu10i}.}\label{Esym-Xu2}

\end{center}
\end{figure}

Using the step function for the momentum distribution function, namely, neglecting the high momentum tail due to short-range correlation (its effects will be studied in the next subsection) at zero temperature, the symmetry energy can be expressed as
\begin{eqnarray}\label{Esym1}
&& E_{\mathrm{sym}}(\rho ) = \frac{\hbar ^2}{6m} (\frac{3\pi^2
\rho}{2})^{\frac{2}{3}}
\\
&&+ \frac{\rho }{4\rho _{0}}(A_{l}(x)-A_{u}(x))- \frac{Bx}{\sigma
+1}\left( \frac{\rho }{\rho _{0}}\right) ^{\sigma } \nonumber
\\
&&+\frac{C_{\ell}}{9\rho _{0}\rho }\left( \frac{4\pi
\Lambda}{h^{3}}\right)
^{2}\left[ 4p_{f}^{4}-\Lambda ^{2}p_{f}^{2}\ln \frac{%
4p_{f}^{2}+\Lambda ^{2}}{\Lambda ^{2}}\right] \nonumber
\\
&&+\frac{C_{\rm{u}}}{9\rho _{0}\rho }\left( \frac{4\pi
\Lambda}{h^{3}}\right) ^{2}\left[
4p_{f}^{4}-p_{f}^{2}(4p_{f}^{2}+\Lambda ^{2})\ln
\frac{4p_{f}^{2}+\Lambda ^{2}}{\Lambda ^{2}}\right].\nonumber
\end{eqnarray}
With the 3-body force $V_{3}$ of Eq. (\ref{VD1}), the parameters $A_\ell,
A_{\rm{u}}$ in the two-body forces have to be adjusted with the varying $x$ parameter via \cite{Das03,Chen05a}
\begin{eqnarray}\label{aual}
A_{u}(x)&=&-95.98-x\frac{2B}{\sigma+1},\nonumber\\
A_{l}(x)&=&-120.57+x\frac{2B}{\sigma +1}.
\end{eqnarray}%
Different values of $x$ can lead to widely different trends for the $E_{\rm sym}(\rho)$ without changing the
SNM EOS and the magnitude of the symmetry energy at saturation density. This is illustrated in Fig.~\ref{Plamen18a} which
displays representative examples of the $E_{\rm sym}(\rho)$ using Eq. (\ref{Esym1}) for values of $x$ in the interval between -2 and +1 but the same $\sigma=4/3$.
It needs to be emphasized that various values of $x$ correspond to various values of $L$ and $K_{\rm sym}$, i.e., varying $x$ changes both
parameters simultaneously as shown in Fig.~\ref{Plamen18b}.

While with the 3-body force $V_{3ij}$ of Eq. (\ref{VD}), the $A_\ell$ and
$A_{\rm{u}}$ have to be modified to \cite{Xu10i}
\begin{eqnarray}
A_u'(x)&=&-95.98 - \frac{2B}{\sigma+1}\left[1-2^{\sigma-1}(1-x)\right],\nonumber\\
A_l'(x)&=&-120.57 +\frac{2B}{\sigma+1}\left[1-2^{\sigma-1}(1-x)\right]
\end{eqnarray}
with B=106.35 MeV to reproduce the same saturation properties of nuclear matter and
$E_{\rm sym}(\rho_0)=30$ MeV at $\rho_0=0.16/fm^3$. We notice that, by design, with $x=1$, the above two parameters are identical to those in Eq. (\ref{aual}). The improved MDI then reduces to the original MDI interaction. The \esym for the improved MDI is given by \cite{Xu10i}
\begin{eqnarray}\label{Esym2}
&& E_{\mathrm{sym}}(\rho ) = \frac{\hbar ^2}{6m} (\frac{3\pi^2
\rho}{2})^{\frac{2}{3}} + \frac{\rho }{4\rho
_{0}}(A'_{l}(x)-A'_{u}(x))
\\
&&+ \frac{B}{\sigma +1}\left( \frac{\rho }{\rho _{0}}\right)
^{\sigma } \left[2^{\sigma-1}(1-x)-1\right] \nonumber
\\
&&+\frac{C_{l}}{9\rho _{0}\rho }\left( \frac{4\pi
\Lambda}{h^{3}}\right)
^{2}\left[ 4p_{f}^{4}-\Lambda ^{2}p_{f}^{2}\ln \frac{%
4p_{f}^{2}+\Lambda ^{2}}{\Lambda ^{2}}\right] \nonumber
\\
&&+\frac{C_{u}}{9\rho _{0}\rho }\left( \frac{4\pi
\Lambda}{h^{3}}\right) ^{2}\left[
4p_{f}^{4}-p_{f}^{2}(4p_{f}^{2}+\Lambda ^{2})\ln
\frac{4p_{f}^{2}+\Lambda ^{2}}{\Lambda ^{2}}\right] \nonumber.
\end{eqnarray}
To see the relative effects of the $X$ and $\sigma$ parameters, shown in Fig.\ \ref{Esym-Xu2} are the \esym obtained
using the improved MDI with $x=1, 0$ and $-1$ and three $\sigma_{1,2,3}$ values of
$\frac{4}{3}-\frac{1}{30}$, $\frac{4}{3}$, and
$\frac{4}{3}+\frac{1}{30}$, respectively. While the variation of $\sigma$ parameter alters appreciably the high density behavior of $E_{\rm sym}(\rho)$, it is the $x$ parameter that has the most dramatic influences on the \esym especially at supra-saturation densities.

In summary of this subsection, both the form and the spin-isospin dependence of the three-body force are important for determining the high-density behavior of the symmetry energy.

\begin{figure}[htb]
\begin{center}
\resizebox{0.5\textwidth}{!}{
   \includegraphics[width=5cm,angle=-90]{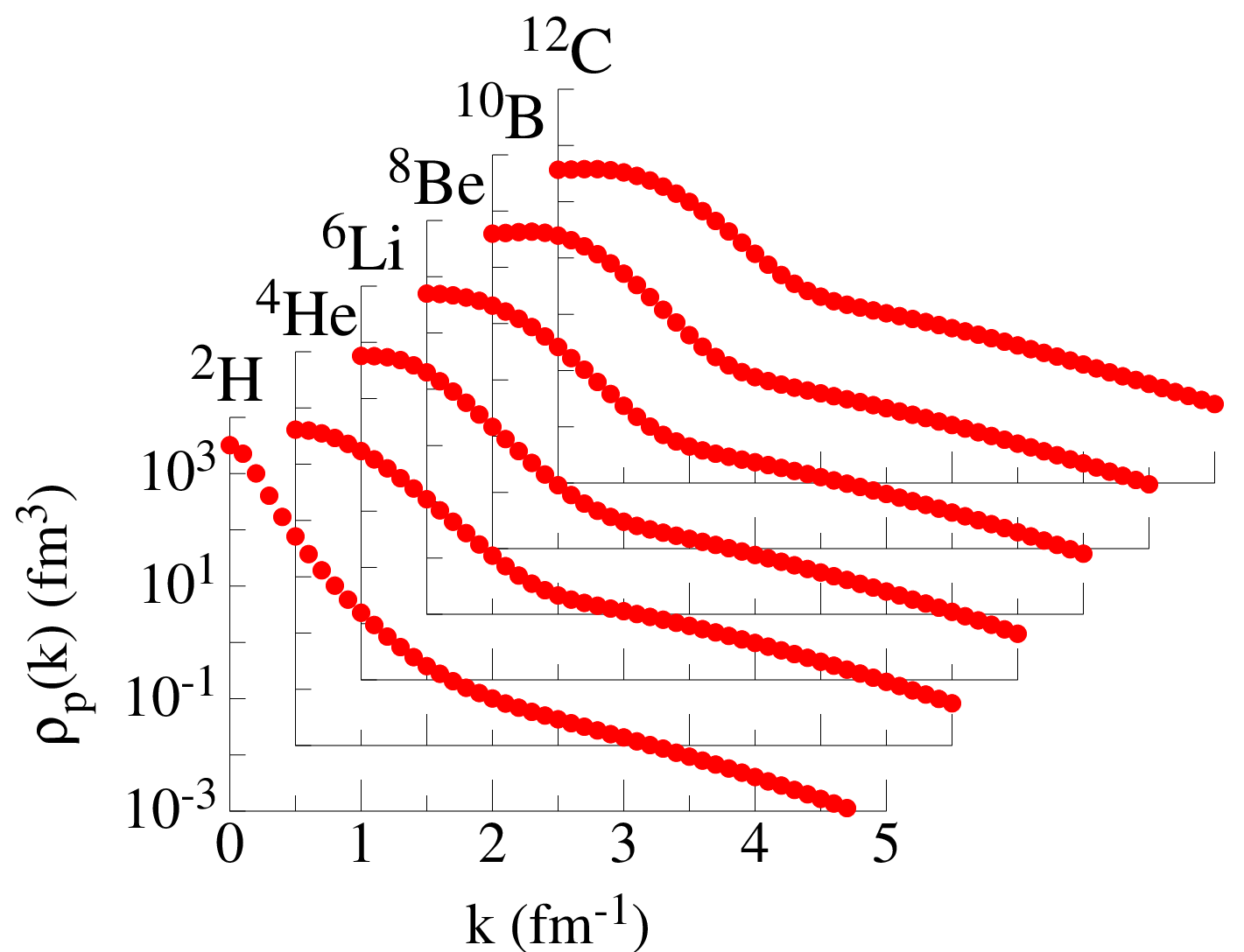}
  }
\caption{{\protect Color online) The proton momentum distributions in all $T$=0 nuclei from $A$=2--12 calculated by Wiringa et al using their variational Monte Carlo theory in ref. \cite{Bob1}.}}
 \label{Bob-fig}
\end{center}
\end{figure}

\begin{figure}[htb]
\begin{center}
\resizebox{0.4\textwidth}{!}{
  \includegraphics{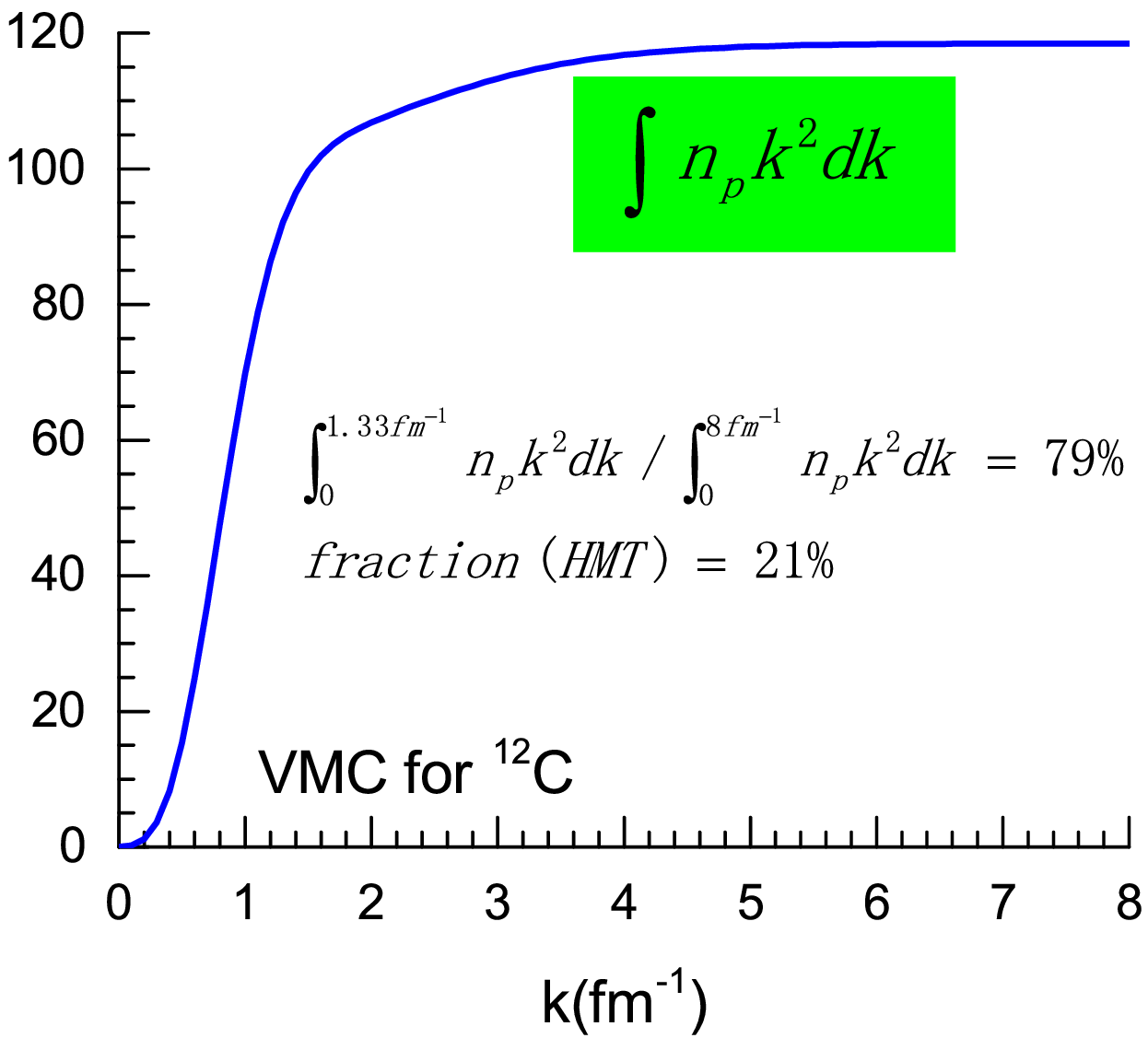}
  }
\caption{{\protect (Color online) The un-normalized population of nucleons in $^{12}$C up to momentum $k$ (blue) \cite{BCai15} from the variational Monte Carlo prediction shown in Fig. \ref{Bob-fig}.}}
  \label{Cai-fig}
\end{center}
\end{figure}
\subsection{The role of the isospin-dependent short-range correlation (SRC) induced by tensor forces}
Another major source of uncertainties of the high-density \esym is the poorly understood but very interesting isospin dependence of SRC in neutron-rich matter.
It is well known that nucleon-nucleon short-range repulsive core (correlations) and tensor force lead to a high (low) momentum tail (depletion) in the
single-nucleon momentum distribution above (below) the nucleon Fermi surface in both finite nuclei and nuclear matter \cite{Mig57,Lut60,Bethe,Ant88,Pan97,Mah92}.
As an example, shown in Fig.\,\ref{Bob-fig} are single nucleon momentum distributions in all T=0 nuclei with mass number between 2 to 12 from the variational Monte Carlo theory \cite{Bob1}.
One outstanding feature of these momentum distribution functions is the universal scaling of the high momentum tail (HMT) in all nuclei considered. Such scaling was actually seen in all nuclei from deuteron to infinite nuclear matter, see, e.g., ref. \cite{pan92}, indicating the shared short-range nature of the HMT in all systems. More quantitatively, the un-normalized population of momentum space up to the wave number $k$ is shown with the blue line in Fig. \ref{Cai-fig}. As indicated, the fractional occupation up to $k=1.33$ fm$^{-1}$ corresponding to the Fermi momentum at saturation density of nuclear matter is about 79\%, leaving about 21\% nucleons distributed above the Fermi momentum.

It is important to note here that systematic analyses of many experiments, see e.g., ref. \cite{Hen16x} for a recent review, indicate that about 25$\%$ nucleons are in the HMT in SNM,  while calculations are still model dependent. For example,  the Self-Consistent Green's Function
(SCGF) theory using the Av18 interaction predicts a 11-13\% HMT for SNM at saturation density $\rho_0$ \cite{Rio09,Rio14}, while Bruckner-Hartree-Fock calculations
predict a HMT between about 10\% using the N3LO450 to over 20\% using the Av18, Paris or Nij93 interactions \cite{Yin13,ZHLi}. Thus, there is a qualitative agreement but quantitative disagreement
regarding the size of the HMT even in SNM.

\begin{figure*}[t!]
  \begin{center}
    \includegraphics*[width=0.75\linewidth]{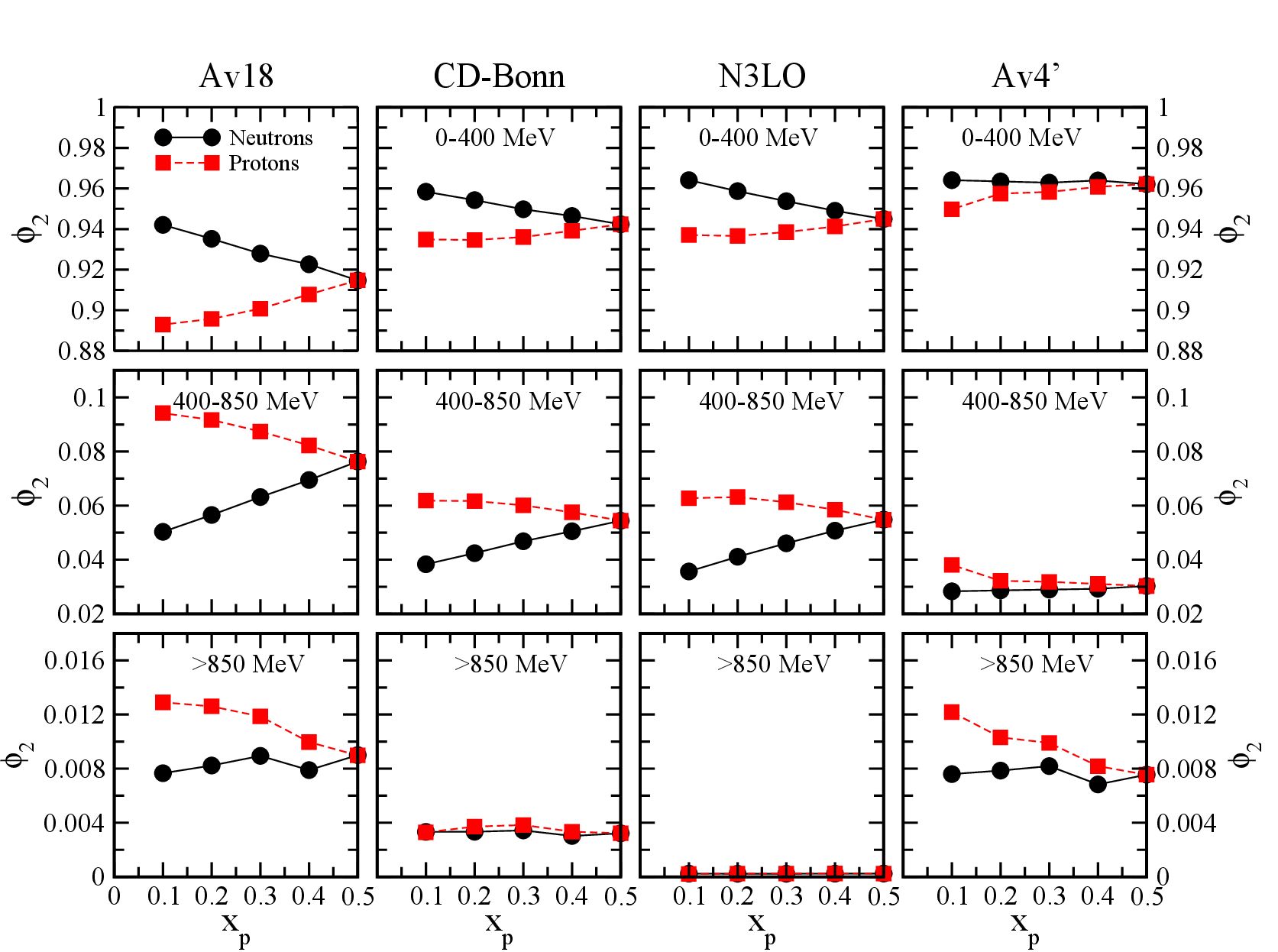}
    \caption{(Color online) Integrated single-particle strength for neutrons (circles) and protons (squares) in the three momentum regions using different NN interactions indicated in the figure. Taken from ref. \cite{Rio14}.}
    \label{fig:strength}
  \end{center}
\end{figure*}
Significant progresses have been made in understanding the source and features of the nucleon momentum distributions in finite nuclei especially from electron-nucleus scattering experiments during the last two decades \cite{Hen14,Arr12,Cio15,Egi06,Shn07,Wei11,Kor14,Fa17,Pia06,Wei15,Duer19}, albeit there are still controversies especially from those experiments using nuclear probes, see, e.g., refs. \cite{Gade,Bob06,Tsang-sf,Lee-Tsang,GSI}.
Theoretically, large uncertainties exist in quantifying the shape, size and isospin dependence of the HMT of single-nucleon momentum distributions in neutron-rich matter, for a recent review, see, e.g., ref. \cite{PPNP-Li}.
While some strong and consistent indications about the isospin dependence of the HMT have been found in electron scattering experiments \cite{Hen-nature}, quantitative predictions are still model dependent.  More specifically, based on the observation that the SRC strength of a neutron-proton pair is about 18-20 times that of two protons, the HMT in PNM was estimated to be about 1-2\% \cite{Hen15b}.  However, some theories predict a significantly higher HMT in PNM. For example, the SCGF predicted a 4-5\% HMT in PNM \cite{Rio09,Rio14}. More specifically, based on the ladder SCGF approach \cite{Rio14}, Rios, Polls and Dickhoff have shown clearly that the momentum distribution of neutrons with respect to that of protons depends strongly on the interactions used especially in neutron-rich matter.
The integrated strength defined as \cite{Rio14}
\begin{equation}
\phi_2(k_i,k_f) =  \frac{1}{\pi^2 \rho_\tau} \int_{k_i}^{k_f} \textrm{d} k \, k^{2} n_\tau(k)
\label{eq:strength}
\end{equation}
can be used to quantify the population in the momentum range between $k_i$ and $k_f$. Shown in Fig.~\ref{fig:strength} are the values of $\phi_2(k_i,k_f)$ in the three windows of momentum. As discussed in \cite{Rio14}, the low-momentum region, from $k_i=0$ to $k_f=400$ MeV,
includes depletion effects as well as the shifts in the Fermi momenta, the middle panels represent approximately the tensor-dominated region while
the bottom panels illustrate the remaining strength in the
very high-momentum region above $850$ MeV where three-body short-range correlations may play a significant role.

\begin{figure}[htb]
\begin{center}
\resizebox{0.9\textwidth}{!}{
  \includegraphics{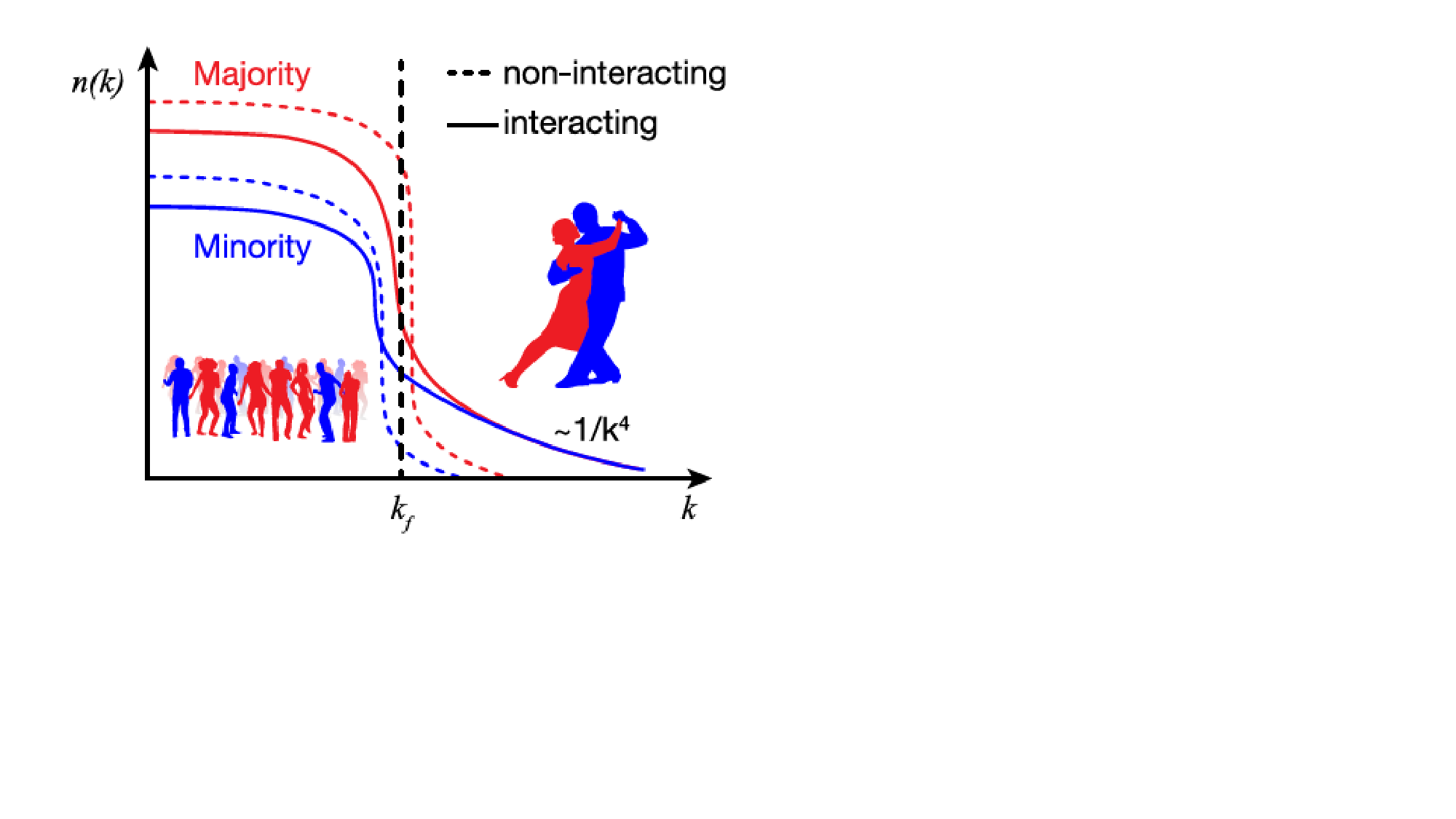}
  }
 \vspace{-3cm}
\caption{{\protect (Color online) The neutron-proton dominance picture of short-range correlations in isospin-asymmetric systems. Taken from ref.\,\cite{Hen14}.}}
  \label{dance}
\end{center}
\end{figure}

\begin{figure}[h!]
\centering
  \includegraphics[width=7.5cm]{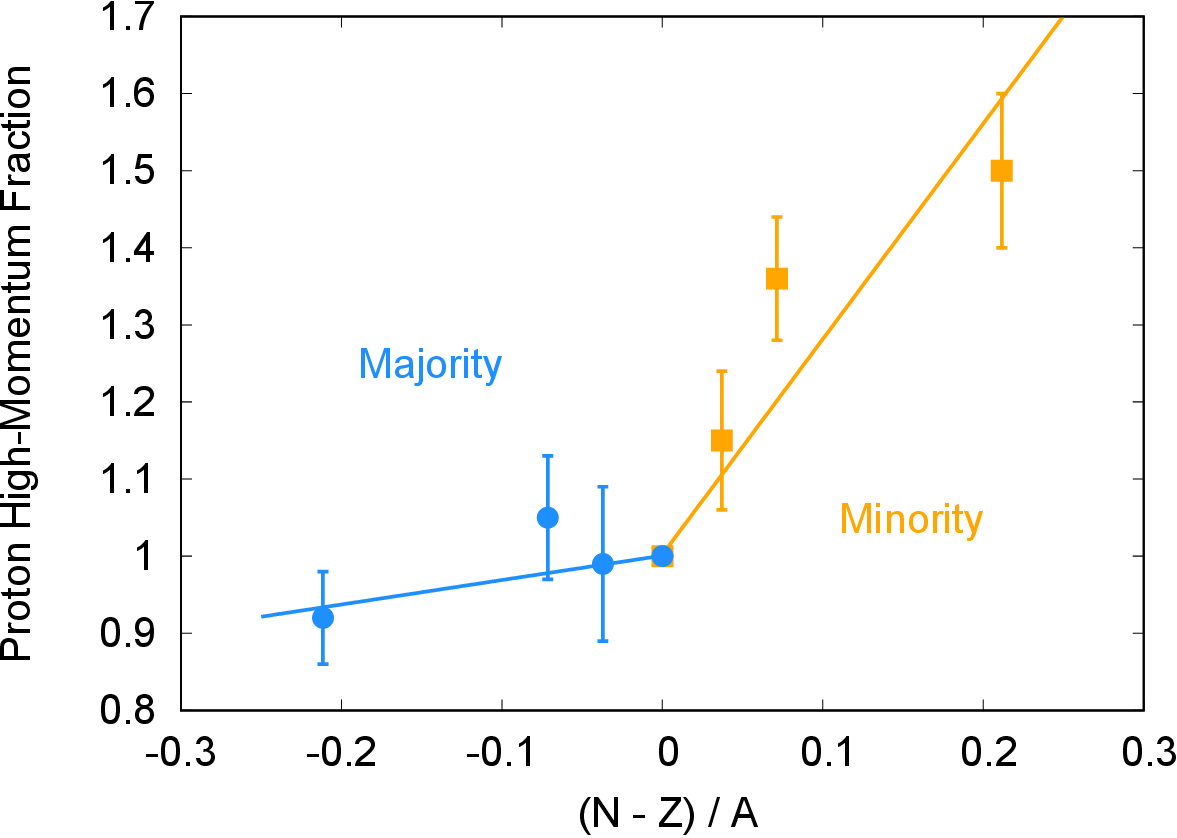}
   \caption{(Color online) The fraction of high-momentum protons relative to that of $^{12}$C as a function of the isospin-asymmetry $(N-Z)/A$ from data mining of electron-nucleus scattering experiments \cite{Hen-nature,Hen19}. Note that the values for negative isospin-asymmetry correspond to the fraction measured for neutrons for positive asymmetry. Taken from ref. \cite{Hen19} based on the measurements reported in ref.\ \cite{Hen-nature} .}
  \label{Hen-data}
\end{figure}

It is seen from Fig. \ref{fig:strength} clearly that as the isospin-asymmetry increases, higher fractions of protons populate the HMT while in the low-momentum region neutrons dominate. Most importantly, the integrated strength especially at high momentum is very interaction dependent. In particular, either turning off the tensor or high-momentum component of the interaction leads to significant reductions of the HMT. The relative populations of the HMT is also strongly interaction dependent. Moreover, within the SCGF approach it was also shown that
the shape of the HMT is also model dependent. The HMT does not always scale as
$1/k^4$ as predicted in some other models discussed in detail in ref. \cite{PPNP-Li}.
Nevertheless, the SCGF and all other models qualitatively confirm the deuteron-like neutron-proton dominance picture illustrated in Fig. \ref{dance} for the creation of the HMT.
In neutron-rich systems, neutrons are in the majority. The minority protons have a relatively larger chance of finding a neutron partner to form a T=0 pair for the tensor force to be active. Within the neutron-proton dominance approximation, there are equal number of neutrons and protons in the HMT. Thus, a larger fraction of protons is in the HMT.
An interesting example from data mining of electron-nucleus and proton-nucleus scattering experiments \cite{Hen-nature,Hen19} is shown in Fig. \ref{Hen-data}. The measured fraction of high-momentum protons relative to that of $^{12}$C is shown as a function of the isospin-asymmetry $(N-Z)/A$. It is seen clearly that from neutron-poor to neutron-rich systems the relative proton fraction in the HMT changes from less than 1 to significantly larger than 1 as one expects qualitatively based on both the schematic neutron-proton dominance picture and microscopic nuclear many-body calculations.

\begin{figure}[h!]
\centering
    \includegraphics[width=7.5cm,angle=0]{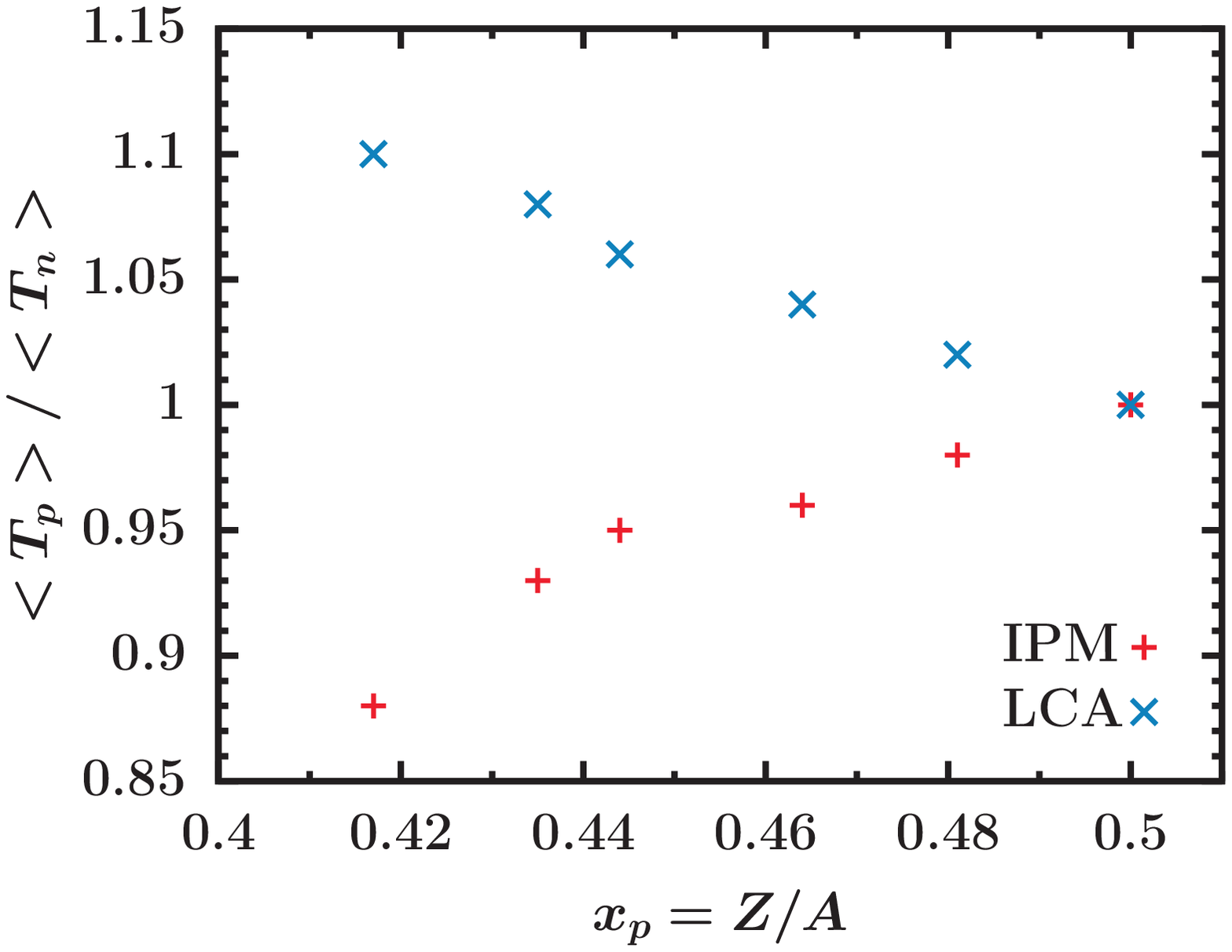}
  \caption{(Color online) Ratio of the average kinetic energy of protons to that of neutrons as a function of proton fraction $x_{\rm{p}}=Z/A$ within the independent particle model (IPM) and
the low order correlation operator approximation (LCA). Taken from ref.\,\cite{Ryc15}.}
  \label{Ryc}
\end{figure}

The isospin dependence of the HMT is expected to affect both the kinetic and potential parts of the energy density functionals, and thus the symmetry energy. In particular, because of the $k^4$ weighting in calculating the average kinetic energies of nucleons, a small change in the HMT may have a significant effect on the kinetic part of the symmetry energy \cite{Xulili}.
The kinetic EOS can be expanded in $\delta$ as
\begin{equation}
E^{\rm{kin}}(\rho,\delta)\approx E_0^{\rm{kin}}(\rho)+E_{\rm{sym}}^{\rm{kin}}(\rho)\delta^2+E_{\rm{sym,4}}^{\rm{kin}}(\rho)\delta^4+\mathcal{O}(\delta^6).
\end{equation}
For the free Fermi gas (FFG), it is well known that
\begin{equation}E^{\rm{kin}}_0(\rho)=\frac{3E_{\rm{F}}(\rho)}{5},~
E_{\rm{sym}}^{\rm{kin}}(\rho)=\frac{E_{\rm{F}}(\rho)}{3},~
E_{\rm{sym,4}}^{\rm{kin}}(\rho)=\frac{E_{\rm{F}}(\rho)}{81}
\end{equation}\label{FFG-e}
where $E_{\rm{F}}(\rho)=k_{\rm{F}}^2/2M$ is the nucleon Fermi energy.
Thus, in the FFG picture, the average kinetic energy of neutrons are higher than that of protons because of their higher Fermi energy. However, depending on the strength of the SRC the larger fraction of protons in the HMT may reverse this picture. As an example, shown in Fig. \ref{Ryc} are the ratio of average kinetic energy of protons over that of neutrons within the independent particle model (IPM) and the low order correlation operator approximation (LCA) \cite{Ryc15}, respectively. Clearly, the correlations make the protons more energetic. More recently, within the IPM and LCA approaches, a systematic study of the isospin composition and neutron/proton ratio (N/Z) dependence of the SRC
was carried out \cite{Ryc19}. It further confirmed that the minority species (protons) become increasingly more short-range correlated as the neutron/proton ratio increases.
The main feature shown in Fig. \ref{Ryc} is consistent with that found experimentally in systems from C and Pb \cite{Hen-nature}. Such change in kinematics of neutrons and protons in isospin asymmetric matter is expected to affect the nuclear symmetry energy. However, we note that the sign and magnitude of such phenomenon is still model and interaction dependent \cite{Rio14}.
\begin{figure*}[h!]
\centering
  \includegraphics[width=13.cm]{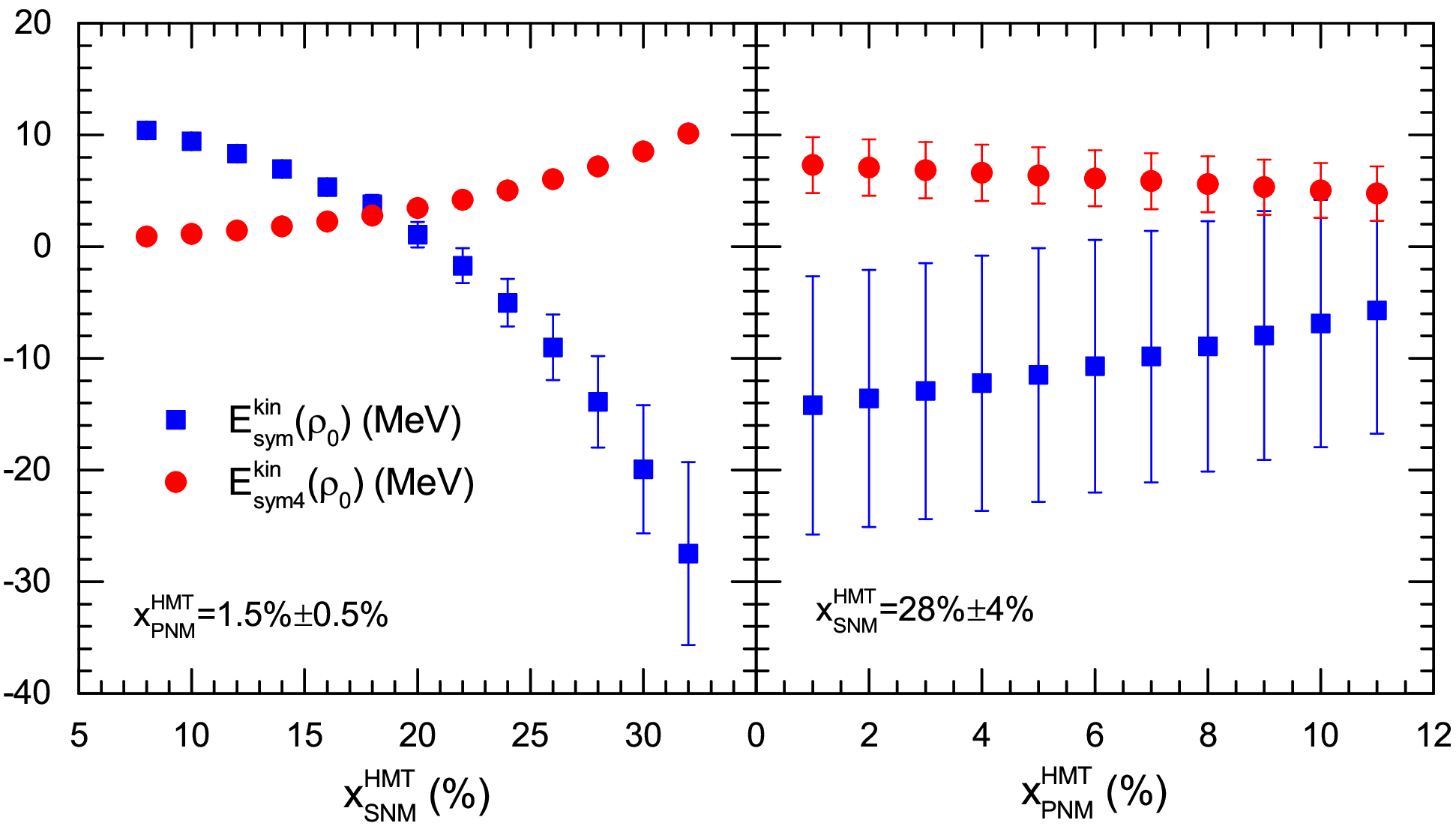}
  \caption{(Color online) Dependence of the kinetic symmetry energy and the isospin-quartic term on the fractions of high-momentum nucleons in symmetric nuclear matter ($x_{\rm{SNM}}^{\rm{HMT}}$) and pure neutron matter  ($x_{\rm{PNM}}^{\rm{HMT}}$). Taken from ref. \cite{PPNP-Li}.}
  \label{fig_E2E4withFraction}
\end{figure*}

\begin{figure}[h!]
\centering
  \includegraphics[width=7.5cm]{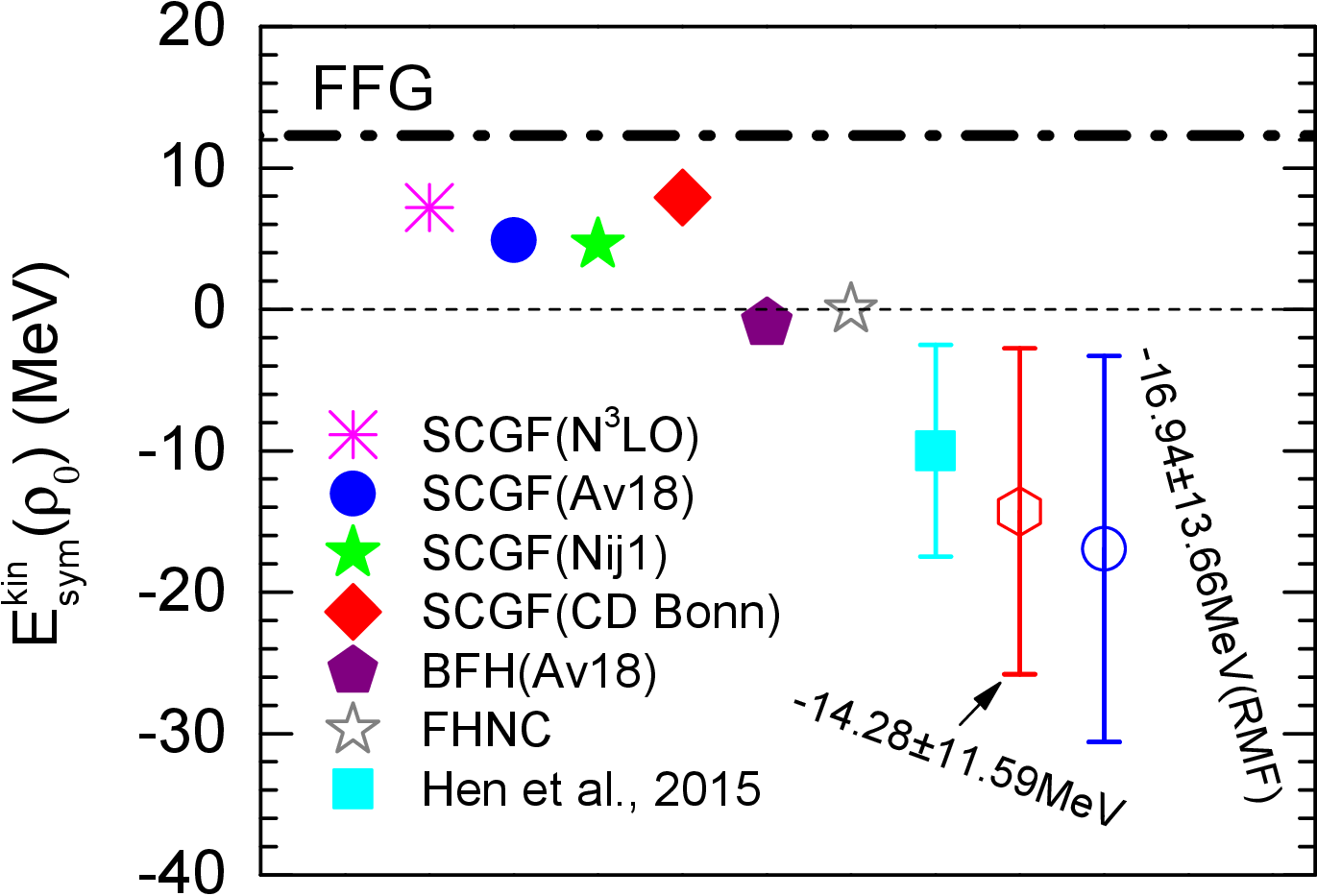}
  \caption{(Color online) SRC-induced reductions of nucleon kinetic symmetry energy from several models with respect to the value (12.3 MeV) for the free Fermi gas (FFG). Taken from ref. \cite{PPNP-Li}.} \label{fig_esym0}
\end{figure}

Effects of the HMT on the nuclear energy density functional can be examined by replacing the step function with the single-nucleon momentum distribution including the HMT at zero temperature. For example, within phenomenological models one can make the following substitution in both the kinetic and momentum dependent parts of the MDI EDF in Eq. (\ref{EDF})
\begin{eqnarray}\label{nHMT}
&&\int_0^{k_{\rm{F}}^J}n^J_{\v{k}}
(\rm{FFG~ step~function})\cdot d\v{k}\nonumber\\
&&\longrightarrow\int_0^{\phi_Jk_{\rm{F}}^J}n^J_{\v{k}}(\rho,\delta)
(\rm{with~ HMT})\cdot d\v{k}\nonumber
\end{eqnarray}
where $\phi_J$ is the HMT cut-off parameter \cite{Cai15a} in the single-nucleon
momentum distribution. The latter can be parameterized by \cite{Cai15a}
\begin{equation}\label{MDGen}
n^J_{\v{k}}(\rho,\delta)=\left\{\begin{array}{ll}
\Delta_J,~~&0<|\v{k}|<k_{\rm{F}}^J,\\
&\\
\displaystyle{C}_J\left(\frac{k_{\rm{F}}^{J}}{|\v{k}|}\right)^4,~~&k_{\rm{F}}^J<|\v{k}|<\phi_Jk_{\rm{F}}^J
\end{array}\right.
\end{equation}
where $\Delta_J$ is the depletion of the Fermi sphere at zero momentum while ${C}_J$ is the so-called ``contact" characterizing the size of the HMT.
While some of the parameters are constrained by the HMT data and normalization conditions, there are large uncertainties \cite{Cai15a}.
As an example, shown in Fig.\,\ref{fig_E2E4withFraction} are the kinetic symmetry energy $E_{\rm{sym}}^{\rm{kin}}(\rho)$ (left) and the isospin-quartic term $E_{\rm{sym,4}}^{\rm{kin}}(\rho)$ (right)
as a function of the HMT fractions in SNM $x_{\rm{SNM}}^{\rm{HMT}}$ and in PNM $x_{\rm{PNM}}^{\rm{HMT}}$, respectively. It is seen that the strength of HMT in SNM plays a leading role
in determining the kinetic symmetry energy, while a large quartic term is generated by a large difference between the $x_{\rm{SNM}}^{\rm{HMT}}$ and $x_{\rm{PNM}}^{\rm{HMT}}$.
Compared to their corresponding values for the FFG, it is seen that the isospin-dependent HMT decreases the
kinetic symmetry energy $E_{\rm{sym}}^{\rm{kin}}(\rho_0)$ from 12.3 MeV in FFG significantly to even negative values depending on the size of HMT.
Taking $x_{\rm{SNM}}^{\rm{HMT}}=(28\pm 4)\%$ and $x_{\rm{PNM}}^{\rm{HMT}}=(1.5\pm 0.5) \%$ consistent with the recent analyses of electron-nucleus scattering data at Jlab \cite{Hen14},
the resulting kinetic symmetry energy (red open circle with error bar) is compared in Fig. \ref{fig_esym0} with findings of several
studies using both phenomenological models \cite{Hen15b,Cai-RMF} and microscopic nuclear many-body theories\,\cite{Vid11,Rio14,Lov11,Car12,Car14}.
While the results are quantitatively different, they all consistently show significant reductions compared to the FFG value. Overall, the results cited above indicate clearly that the isospin dependent
SRC and the associated HMT affect significantly the kinetic symmetry energy of quasi-nucleons in isospin-asymmetric matter.

\begin{figure}[h!]
\centering
  \includegraphics[width=7.cm]{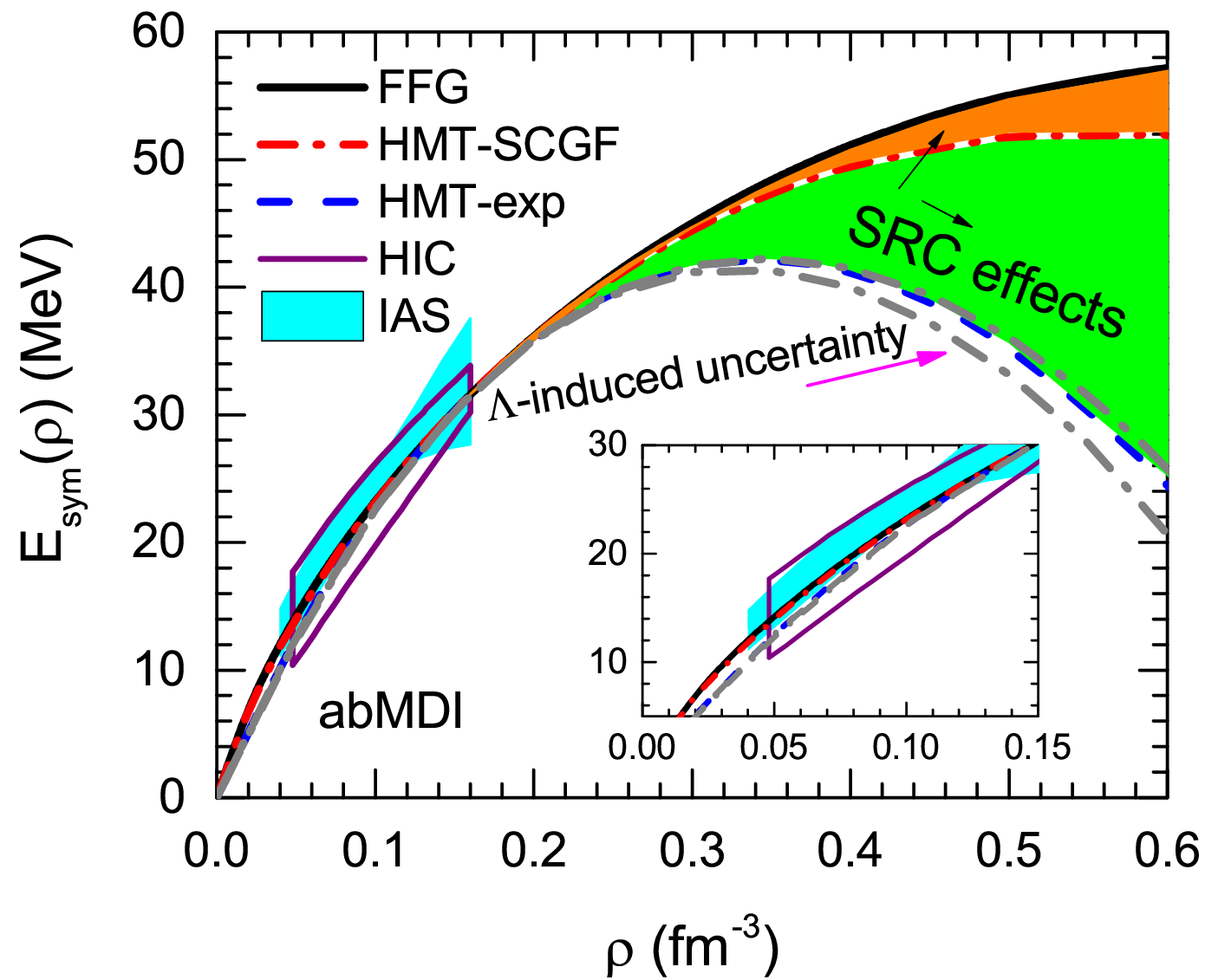}
  \caption{(Color online) Comparisons of the nuclear symmetry energy obtained within a modified Gogny energy density functional using the FFG, HMT-SCGF and HMT-exp parameter sets.
  Constraints on the symmetry energy from analyzing heavy-ion collisions (HIC)\,\cite{Tsang12} and the Isobaric Analog States (IAS)\,\cite{Pawel14} are also shown. Taken from ref.\,\cite{Cai18}.}
  \label{MDI-HMT}
\end{figure}

Effects of the HMT on the potential part or the total \esym can be studied within nuclear energy density functionals \cite{Cai18,Cai-RMF}. For example, making the replacement of
Eq. (\ref{nHMT}) in the modified MDI EDF in Eq. (\ref{EDF}), one can study how the HMT may affect both the kinetic and potential parts of the symmetry energy. This requires re-optimizing the
model parameters to reproduce all known constraints. Nevertheless, this approach can be considered as inconsistent because the quasi-nucleon momentum distribution with a HMT may not be produced self-consistently by the MDI interaction. Thus, this approach can only be viewed as a perturbative method for an orientation of the HMT effect. For comparisons, shown in Fig. \ref{MDI-HMT}
are the \esym from the modified Gogny EDF of Eq. (\ref{EDF}) using the following 3 parameter sets: (1) with the $n_{\v{k}}^J$ adopting a 28\% HMT in SNM and a 1.5\% HMT in
PNM (abbreviated as HMT-exp), (2) with the $n_{\v{k}}^J$ adopting a 12\% HMT in SNM and a 4\% HMT in PNM (abbreviated as HMT-SCGF) and (3) the original MDI interaction with the FFG
nucleon momentum distribution. As discussed in detail in ref. \cite{Cai18},  all three parameter sets have the same $E_{\rm{sym}}(\rho_0)$ and $L$ at $\rho_0$. While they all agree with the
constraints on the $E_{\rm{sym}}(\rho)$ around $\rho_0$ from intermediate energy heavy-ion collisions (HIC) \cite{Tsang12} and the isobaric analog states (IAS)\,\cite{Pawel14}, they have quite different
\esym especially at high densities. More quantitatively, the curvature coefficient
\begin{equation}
K_{\rm{sym}}\equiv 9\rho_0^2\d^2E_{\rm{sym}}(\rho)/\d\rho^2|_{\rho=\rho_0}
\end{equation}
of \esym was found to change from $-109$\,MeV in the FFG set to about
$-121\,\rm{MeV}$ and $-188$\,MeV in the HMT-SCGF and HMT-exp set,
respectively. This helps reproduce the experimentally measured
\begin{equation}
K_{\tau}=K_{\rm{sym}}-6L-J_0L/K_0
\end{equation}
where the skewness of SNM
\begin{equation}
J_0\equiv27\rho_0^3{\d^3E_0(\rho)}/{\d\rho^3}|_{\rho=\rho_0}
\end{equation}
is approximately $-381$, $-376$ and $-329\,\rm{MeV}$ in the FFG,
HMT-SCGF and HMT-exp set, respectively. The resulting $K_{\tau}$ was found to change from $-365\,\rm{MeV}$ in the FFG set to about
$-378\,\rm{MeV}$ and $-457\,\rm{MeV}$ in the HMT-SCGF and HMT-exp
set \cite{Cai18}, respectively, in better agreement with the best
estimate of $K_{\tau}\approx-550\pm 100$\,MeV from analyzing several
different kinds of experimental data currently
available\,\cite{Col14}.

Considering the SRC-induced reduction of kinetic symmetry energy of quasi-nucleons with respect to the FFG value given in Eq. (\ref{FFG-e}), a reduction factor $\eta$ was introduced to parameterize the $E_{\rm{sym}}(\rho)$ around the saturation density as \cite{LiG15}
\begin{eqnarray}\label{ffg-esym}
E_{\rm{sym}}(\rho)&=&\eta\cdot E_{\rm{sym}}^{\rm{kin}}(\textrm{FFG})(\rho)\nonumber\\
&+&\left[S_0-\eta\cdot E_{\rm{sym}}^{\rm{kin}}(\textrm{FFG})(\rho_0)\right]\left(\frac{\rho}{\rho_0}\right)^{\gamma}.
\end{eqnarray}
Normally, without considering the SRC effects one sets $\eta=1$ and varies the parameter $\gamma$ of the potential symmetry energy in transport model simulations of heavy-ion reactions. Taking both $\eta$ and $\gamma$ as free parameters, their correlation is determined by that between the $E_{\rm{sym}}(\rho_0)$ and $L$. Unfortunately, within the current uncertain ranges of $E_{\rm{sym}}(\rho_0)$, $L$ and $\gamma$, the $\eta$ can be anything between 0 and 1 \cite{PPNP-Li,LiG15}.  It is interesting to note that efforts are being made to better constrain the value of $\eta$ and investigate its effects on the finite temperature EOS for applications in astrophysics \cite{Arizona-EOS}.

In short, the isospin dependence of SRC induced by tensor force leads to different momentum distributions for neutrons and protons in neutron-rich matter. Subsequently, both the kinetic and potential parts of the symmetry energy may be affected. While there are still large uncertainties about the SRC physics, the \esym carries fundamental and interesting information about the isospin dependence of strong interaction at short distance in dense neutron-rich matter.

\subsection{The role of the Fock exchange terms in relativistic models}\label{s-fock}
As mentioned earlier, besides the isospin-dependence of nuclear forces and correlations another well-known reason limiting our current knowledge on the \esym is the long-standing challenge of treating accurately nuclear many-body problems. In this regard, perhaps one of the most important problems is the treatment of Fock (exchange) terms and the associated energy/momentum dependence of
nucleon self-energies. Very often, only the mean-field (Hartree) terms are included. Consequently, the resulting Schr\"{o}dinger-equivalent potential (SEP) from these models can not properly describe
the energy/momentum dependence of even the isoscalar nucleon optical potentials extracted from nucleon-nucleus scattering data. For the isovector optical potential and the corresponding neutron-proton effective mass splitting even at the saturation density $\rho_0$, the situation is much worse \cite{PPNP-Li}. Of course, these problems are well recognized by the community. Indeed, much efforts and progresses have been made in recent years. In particular, effects of the Fock terms on the symmetry energy have been studied in the Relativistic Hartree-Fock (RHF) model or covariant density functional theory~\cite{Sun:2008zzk,Zhao:2014bga,Sun:2016ley,Liu:2018far,Liu:2018kfv,Miy19}. In the following, a few main results of the very recent work by Tsuyoshi Miyatsu et al. \cite{Miy19} within their extended RHF (ERHF) approach are used to illustrate the main roles of the Fock exchange terms on the \esym and some remaining issues.
\begin{figure}[htb]
\centering
\resizebox{0.5\textwidth}{!}{
   \includegraphics{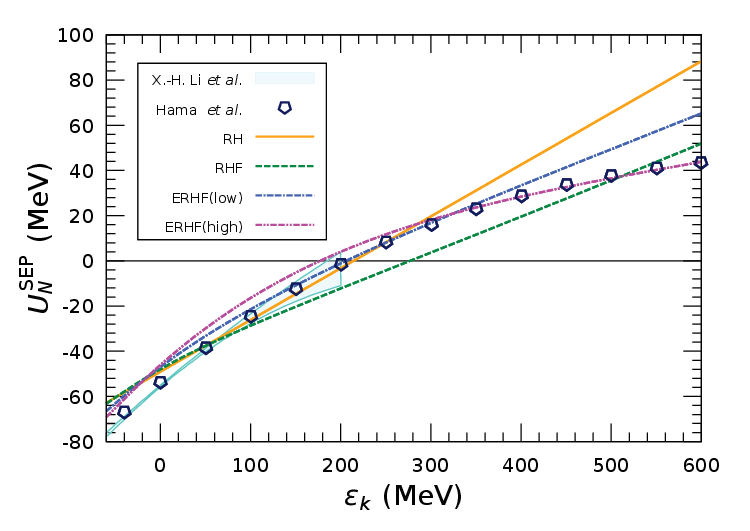}
     }
\caption{\label{USEP} (Color online) The isoscalar singe-nucleon potential $U_{N}^{\rm SEP}$ as a function of nucleon kinetic energy in symmetric nuclear matter at $\rho_{0}$. The shaded band shows the result of the nucleon-optical-model potential extracted from analyzing the nucleon-nucleus scattering data by X.-H.~Li {\it et al.}~\cite{XHLi1,XHLi2}.
The Schr\"{o}dinger-equivalent potential (SEP) obtained by by Hama {\it et al.}~\cite{Hama:1990vr} using the Dirac phenomenology for elastic proton-nucleus scattering data is shown
with the open diamonds. Taken from ref. \cite{Miy19}.}
\end{figure}
\begin{figure*}[htb]
\centering
\resizebox{1.\textwidth}{!}{
   \includegraphics{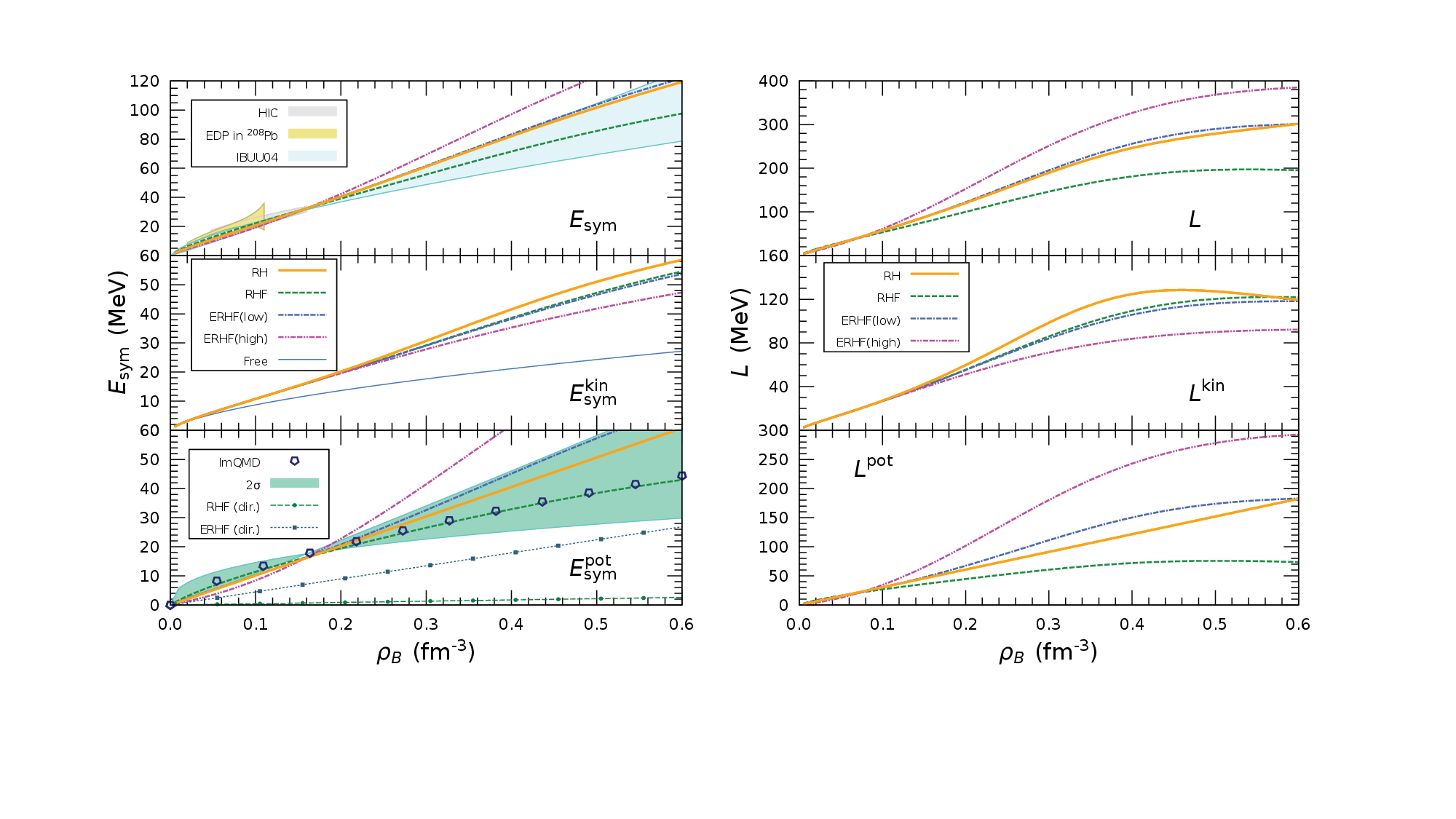}
     }
     \vspace{-2cm}
\caption{(Color online) The nuclear symmetry energy (left) and its slope $L$ (right) as functions of baryon density $\rho_{B}$ from the indicated four model calculations.
Constraints on the \esym from (1) analyzing the isospin diffusion data using the ImQMD (Improved Quantum Molecular Dynamics)~\cite{Tsang:2012se} for heavy-ion collisions (HIC), (2) analyzing the same isospin diffusion data using the isospin-dependent Boltzmann-Uehling-Uhlenbeck (IBUU04) transport model using the MDI interaction \cite{Chen2005,Chen05a} and (3) the electric dipole polarizability (EDP) in $^{208}$Pb \cite{Zhang:2015ava} are also shown. In the middle panel, the word ``free'' denotes $E_{\rm sym}^{\rm kin}$when the interactions are switched off.  In the
lower-left window, the potential part of the \esym used in the ImQMD within $2\sigma$ is also shown. Taken from ref. \cite{Miy19}.}
 \label{RHF-fig}
\end{figure*}

To our best knowledge, the HVH decomposition of \esym and $L$ based on the Lorentz-covariant nucleon self-energies in relativistic approaches was first given by Cai and Chen in ref. \cite{Cai-EL}.  Similar to their non-relativistic counterparts discussed in Sect. \ref{s-HVH},  the relativistic expressions for \esym and $L$ from the HVH decomposition were found useful \cite{PPNP-Li,Liu:2018kfv,Miy19}.  The self-energy  $\Sigma_{N}(k)$ of a nucleon with (three) momentum $\bm{k}$ can be written as
\begin{equation}
  \Sigma_{N}(k) = \Sigma_{N}^{s}(k) - \gamma_{0}\Sigma_{N}^{0}(k) + (\bm{\gamma}\cdot\hat{k})\Sigma_{N}^{v}(k)
  \label{eq:nucleon-self-engy}
\end{equation}
in terms of the scalar ($s$), time ($0$), and space ($v$) components of $\Sigma_{N}(k)$. The kinetic $E_{\rm sym}^{\rm kin}$ and potential $E_{\rm sym}^{\rm pot}$ parts of \esym in the RHF approximation are respectively given by \cite{Miy19}
\begin{align}
  E_{\mathrm{sym}}^{\rm kin} &= \frac{1}{6}\frac{k_{F}^{\ast}}{E_{F}^{\ast}}k_{F},
  \label{eq:Esym-kin} \\
  E_{\mathrm{sym}}^{\rm pot} &= \frac{1}{8}\rho_{B}\left(\frac{M_{N}^{\ast}}{E_{F}^{\ast}}\partial\Sigma_{\rm sym}^{s}
                              - \partial\Sigma_{\rm sym}^{0} + \frac{k_{F}^{\ast}}{E_{F}^{\ast}}\partial\Sigma_{\rm sym}^{v}\right),
  \label{eq:Esym-pot}
\end{align}
with $k_{F}=k_{F_{p}}=k_{F_{n}}$, $E_{F}^{\ast} = \sqrt{k_{F}^{\ast2}+M_{N}^{\ast2}}$, and
\begin{equation}
  \partial\Sigma_{\rm sym}^{s(0)[v]} \equiv \left(\frac{\partial}{\partial\rho_{p}}-\frac{\partial}{\partial\rho_{n}}\right)
                                            \left(\Sigma_{p}^{s(0)[v]}-\Sigma_{n}^{s(0)[v]}\right)\Bigg|_{\rho_p=\rho_n}.
  \label{eq:Esym-func}
\end{equation}
By construction, the direct part in $E_{\rm sym}^{\rm pot}$ is exactly the same as in the relativistic Hartree (RH) approximation~\cite{Chen:2007ih,Dutra2014}
\begin{equation}
  E_{\mathrm{sym}}^{\rm pot, dir} = \frac{1}{2}\frac{g_{\rho}^{2}}{m_{\rho}^{2}}\rho_{B}
  \label{eq:Esym-pot-dir}
\end{equation}
in terms of the $\rho-N$ coupling constant $g_{\rho}$.
The kinetic part of $L$ corresponding to Eq. (\ref{eq:Esym-kin}) is
\begin{eqnarray}
L^{\rm kin} &=& \frac{1}{6}k_{F}\left[\frac{k_{F}^{\ast}}{E_{F}^{\ast}}
               + \frac{k_{F}}{E_{F}^{\ast}}\left(\frac{M_{N}^{\ast}}{E_{F}^{\ast}}\right)^{2}\right]\nonumber\\       
               &+& \frac{1}{6}k_{F}\left[ \frac{k_{F}}{E_{F}^{\ast}}\frac{M_{N}^{\ast}}{E_{F}^{\ast}}\left(
                 \frac{M_{N}^{\ast}}{E_{F}^{\ast}}\frac{\partial\Sigma_{N}^{v}}{\partial\rho_{B}}
               - \frac{k_{F}^{\ast}}{E_{F}^{\ast}}\frac{\partial\Sigma_{N}^{s}}{\partial\rho_{B}}\right)\right]
\end{eqnarray}
while the potential part of $L$ corresponding to Eq. (\ref{eq:Esym-pot}) is
\begin{eqnarray}
&&L^{\rm pot} = 3E_{\rm sym}^{\rm pot}\\
              &+& \frac{3}{8}\rho_{B}\frac{\partial}{\partial\rho_{B}}\left(\frac{M_{N}^{\ast}}{E_{F}^{\ast}}\partial\Sigma_{\rm sym}^{s}
               - \partial\Sigma_{\rm sym}^{0} + \frac{k_{F}^{\ast}}{E_{F}^{\ast}}\partial\Sigma_{\rm sym}^{v}\right). \nonumber
\end{eqnarray}
Neglecting the exchange contributions, i.e., setting
\begin{eqnarray}
&&k_{F}^{\ast}=k_{F}, \partial\Sigma_{N}^{s}/\partial\rho_{B}=\partial M_{N}^{\ast}/\partial\rho_{B}, \partial\Sigma_{N}^{v}/\partial\rho_{B}=0, \nonumber\\
&&{\rm and}~\partial\left(\partial\Sigma_{\rm sym}^{s(0)[v]}\right)/\partial\rho_{B}=0,\nonumber
\end{eqnarray}
the $L^{\rm kin}$ and $L^{\rm pot}$ will be reduced to their counterparts in the RH approximation~\cite{Miy19,Dutra2014}
\begin{align}
  L^{\rm kin, dir} &= \frac{1}{3}\frac{k_{F}^{2}}{\sqrt{k_{F}^{2}+M_{N}^{\ast2}}}\nonumber\\
  &\cdot
                      \left[1-\frac{k_{F}^{2}}{2\left(k_{F}^{2}+M_{N}^{\ast2}\right)}
                      \left(1+\frac{2M_{N}^{\ast}k_{F}}{\pi^{2}}\frac{\partial M_{N}^{\ast}}{\partial\rho_{B}}\right)\right],
  \label{eq:slope-kin-dir} \\
  L^{\rm pot, dir} &= \frac{3}{2}\frac{g_{\rho}^{2}}{m_{\rho}^{2}}\rho_{B}.
  \label{eq:slope-pot-dir}
\end{align}

One goal of the RHF or covariant energy density functionals is to better describe the energy/momentum dependence of single-nucleon optical
potentials extracted from analyzing nucleon-nucleus scattering data by X.H. Li et al ~\cite{XHLi1,XHLi2} and Hama \cite{Hama:1990vr} without introducing extra density dependence in the meson-nucleon coupling as normally done in some RMF approaches \cite{Typ14}. However, there are still serious difficulties to describe properly the experimentally constrained nucleon SEP over large energy ranges. For example, Miyatsu et al. \cite{Miy19} used two parameter sets by adjusting the exchange terms
within their extended RHF (ERHF) model. As shown in Fig. \ref{USEP}, the ERHF low (high) is made to well reproduce the experimentally-constrained single-nucleon potential at kinetic energies below (above) about 300 MeV. It is also seen that neither the original RH nor RHF can describe the high-energy parts of the experimentally-constrained SEP. Nevertheless, comparisons of these calculations allows one to investigate how the momentum dependence in the nucleon self-energy due to the exchange contribution affects the nuclear symmetry energy and its slope parameter.
Shown in Fig. \ref{RHF-fig} are the kinetic and potential parts of \esym and $L$ in the four cases obtained by Miyatsu et al. \cite{Miy19}.  While quantitatively effects of the exchange terms depend on how the model parameters are fixed, several interesting qualitative observations were made \cite{Miy19}. In particular,  the Fock contribution was found to suppress the kinetic term of nuclear symmetry energy at densities around and beyond $\rho_{0}$. Moreover, not only the isovector-vector ($\rho$) meson but also the isoscalar mesons ($\sigma, \omega$) and pion have significant influence on the potential symmetry energy through the exchange diagrams. The exchange terms were also found to prevent the slope parameter from increasing monotonically at high densities.

Thus, the Fock exchange terms may influence significantly the density dependence of nuclear symmetry energy. However, there are still a lot of uncertainties. Besides better describing the isoscalar SEP extracted from laboratory experiments, it is also necessary to examine how the RHF models can describe the energy/momentum dependence of the isovector potential $U_{sym,1}(k,\rho)$, the corresponding neutron-proton effective mass splitting and the associated quartic symmetry energy $E_{sym,4}(\rho)$. These are more closely related to the finite-range parts of the isovector interaction. In fact, some systematics of $U_{sym,1}(k,\rho_0)$ from analyzing large sets of experimental data has been reported, see, e.g., refs. \cite{PPNP-Li,XHLi1,XHLi2}. However, these data have not been used to constrain the RHF models yet. Moreover, ongoing experiments with radioactive beams will provide more data useful for constraining the isovector potential up to high energies.  It is thus very hopeful that effects of the Fock exchange terms on the \esym can be better understood.

In summary of this section, the density dependence of nuclear symmetry energy especially at supra-saturation densities is still very uncertain mainly because of our poorly knowledge about the
isospin dependence of strong interactions and correlations at short distance in dense neutron-rich matter. Among the most uncertain but very interesting and new physics ingredients affecting strongly the high-density behaviour of nuclear symmetry energy are the isospin dependent tensor force and its resulting isospin dependence of SRC, the spin-isospin dependence of three-body forces as well as the finite-range interaction induced momentum-dependence of the isovector single-nucleon potential and the resulting neutron-proton effective mass splitting. Nuclear reactions induced by radioactive beams
and electron scatterings on heavy nuclei can help constrain some of these ingredients.

\section{Symmetry energy effects on the crust-core transition density and pressure in neutron stars}\label{cc-tran}
``{\it The physics of neutron star crusts is vast, involving many different research fields, from nuclear and condensed matter physics to general relativity" \cite{Chamel08}}. The extremely active research on the very rich nuclear physics involved in understanding properties of neutron star crusts and their astrophysical ramifications were summarized earlier in several comprehensive reviews, see, e.g., refs. \cite{Chamel08,Lattimer2007,Carlos12,Matt17}. Key to many of the interesting questions is the core-crust transition density and pressure. The latter affect directly the thickness, fractional mass and moment of inertia of the crusts and thus the interpretation of several still puzzling astrophysical phenomena \cite{Newton12,Newton14,Lida14,Pro14}. In this section, we focus on discussing the role of low-density \esym on determining the core-crust transition density and pressure.

\subsection{Some very useful lessions from earlier studies}
Since the pioneering work of Baym et al in 1971 \cite{Baym1971,Baym2}, essentially all available EOSs have been used to calculate the crust-core transition density and pressure. The most widely used
approaches are based on examining whether small density fluctuations will grow in uniform matter using the so-called dynamical method considering the surface and Coulomb effects of clusters or its
long-wavelength limit (the so-called thermodynamical method), see, e.g., refs. \cite{JXu,Pet84,Pet1,Kubis2007,Dou00,Mou1,Cai12,Mou2,Sei14,Atta17,Rou16,Duc1,Ava08,Gor10,Pea12,Sha15,Vid09,Cam10,Duc2,Bao1,Bao2,Gon17,Fang,Tsa,JP18,Ant19}, or the RPA \cite{Horowitz2001,Car03,Fat10}. The core-crust transition has also been studied by comparing free energies of clustered matter with the uniform matter either using various mass models with the Compressible Liquid Drop Mode of nuclei as the most popular one \cite{Newton12,Baym1971,Baym2,Pet84,Pet1,Dou00,Oyamatsu2007,Steiner08} or the 3D Hartree-Fock \cite{Newton09,Farrooh17} for nuclei on the Coulomb lattice within the Wigner-Seitz approximation.

As it has already been pointed out by J. Arpoen in 1972 \cite{Arp72} that the core-crust transition density and pressure are very sensitive to the fine details of the isospin and dense dependences of the nuclear EOS. This is because the determination of the core-crust transition requires both the first and second derivatives of energy with respect to the densities of neutrons and protons, respectively. It was first demonstrated by J. Arpoen that very small modifications in the details of the nuclear matter energy may lead to considerable differences in the resulting core-crust phase boundary \cite{Arp72}.
Similarly, it was concluded by Baym et al. that because the nucleon chemical potentials depend on derivatives of the EOS, therefore the rapidly varying terms which contribute little to the EOS could conceivably influence proton and neutron chemical potentials considerably; derivatives of nucleon chemical potentials used in determining the core-crust transition are even more sensitive measures of such effects \cite{Baym1971}. Indeed, many subsequent calculations have confirmed these earlier findings. Given the diversity of predicted EOSs up to different orders of the isospin asymmetry and the density expansions of their coefficients from different theories, it is not surprising that the predicted core-crust transition density and pressure are rather model dependent. Depending on the approaches used, some predictions suffer from systematic uncertainties. Nevertheless, it is very encouraging to see that efforts are being made to quantity the uncertainties of the core-crust transition density and pressure \cite{Ant19,Fra-crust2}. Their correlations with the model ingredients are also being quantified, albeit only within the isospin-parabolic approximation for the EOS of neutron-rich matter. Interestingly, some common features of the core-crust transition have been firmly identified thanks to the great efforts of many people in the community. In the following, we discuss some of these features that are closely related to the \esym and identify some remaining challenges.

\subsection{Important nuclear physics inputs for determining the crust-core transition density}
It is useful to recall here the main physics ingredients determining the core-crust transition by considering small density fluctuations~\cite{Baym1971,Baym2,Pet84,Pet1,Duc1}
\begin{equation}
\rho_q = \rho^0_q + \delta \rho_q
\end{equation}
for the particle $q \in \{n,p,e\}$. It can be decoupled into plane-waves
\begin{equation}
\delta \rho_q = A_q e^{i\vec k \cdot \vec r} + c.c.,
\end{equation}
of wave vector $\vec k$ and amplitude $A_q$. The resulting variation
of the free energy density can be written
as~\cite{Baym1971,Duc1}
\begin{equation}
\delta f={\tilde{A}^*} \mathcal C^f \tilde{A},
\end{equation}
where
\begin{eqnarray}
\label{dymethod} C^f &=& \left(
\begin{array}{ccc}
\partial\mu _{n}/\partial\rho _{n} & \partial\mu _{n}/\partial\rho _{p} & 0\\
\partial\mu _{p}/\partial\rho _{n} & \partial\mu _{p}/\partial\rho _{p} & 0\\
0 & 0 & \partial\mu _{e}/\partial\rho _{e}\\
\end{array}
\right) \nonumber \\
&+& k^2 \left(
\begin{array}{ccc}
D_{nn} & D_{np} & 0\\
D_{pn} & D_{pp} & 0\\
0 & 0 & 0\\
\end{array}
\right) + \frac{4\pi e^2}{k^2} \left(
\begin{array}{ccc}
0 & 0 & 0\\
0 & 1 & -1\\
0 & -1 & 1\\
\end{array}
\right)
\end{eqnarray}
is the free-energy curvature matrix.
The first term is the bulk term from the uniform $npe$ matter, the second term is from
the density-gradient part of the nuclear interactions with strength $D_{nn}, D_{pp}$ and $D_{np}$, respectively,
while the last term is from the Coulomb interaction induced
by the plane-wave charge distribution. The $D_{nn}, D_{pp}$ and $D_{np}$ are often estimated using the Skyrme-Hartree-Fock (SHF) model~\cite{Cha97}. For small density fluctuations to remain stable,
the necessary convexity of the curvature matrix $C^f$ is guaranteed by a positive effective interaction between protons \cite{Baym1971,Pet1,Duc1}
\begin{equation}\label{Vdyn}
V_{dyn}(k) \approx V_0 + \beta k^2 + \frac{4 \pi e^2}{k^2 +
k^2_{TF}}>0,
\end{equation}
where
\begin{eqnarray}
&&V_0 = \frac{\partial \mu_p}{\partial \rho_p} - \frac{(\partial
\mu_n / \partial \rho_p)^2}{\partial \mu_n / \partial \rho_n},
\label{v0}\\
&&\beta = D_{pp} + 2 D_{np} \zeta + D_{nn} \zeta^2,\\
&&\zeta =-\frac{\partial \mu_p / \partial \rho_n}{\partial \mu_n / \partial \rho_n}
\end{eqnarray}
and $k_{TF}= [\frac{4 \pi e^2}{\partial \mu_e / \rho_e}]^{1/2}$ is the inverse screening length of electrons.
It was also shown that the condition of Eq. (\ref{Vdyn}) ensures that the clustered matter has an energy higher than the uniform $npe$ matter \cite{Baym1971}.
Since the $V_{dyn}(k)$ has a minimal value at $k =[ (\frac{4 \pi e^2}{\beta})^{1/2} - k^2_{TF} ]^{1/2}$
~\cite{Baym1971}, the following condition is then used to determine the core-crust transition density
\begin{equation}\label{Vdynmin}
V_{dyn} = V_0 + 2 (4 \pi e^2 \beta)^{1/2} - \beta k^2_{TF}=0.
\end{equation}

The approach outlined above is the so-called dynamical approach. At the long wavelength limit ($k\rightarrow 0$) and neglecting the Coulomb energy,
the dynamical stability condition of Eq. (\ref{Vdynmin}) reduces to the thermodynamical stability condition of the $npe$ matter against the growth of small density fluctuations\cite{JXu,Duc1}.
This condition can be written as~\cite{Lattimer2007,Kubis2007}
\begin{equation}\label{ther1}
-\left(\frac{\partial P}{\partial v}\right)_\mu>0,~~~
-\left(\frac{\partial \mu}{\partial q_c}\right)_v>0.
\end{equation}
The second inequality is usually valid. It has been shown that the first condition is equivalent to requiring a positive value of \cite{Lattimer2007,Kubis2007}
\begin{eqnarray} \label{Vther}
V_{ther} &=& 2 \rho \frac{\partial E_b(\rho,x_p)}{\partial \rho} +
\rho^2 \frac{\partial^2 E_b(\rho,x_p)}{\partial \rho^2} \nonumber\\
&-&
\left(\frac{\partial^2 E_b(\rho,x_p)}{\partial \rho
\partial x_p}\rho\right)^2/\frac{\partial^2 E_b(\rho,x_p)}{\partial x_p^2}.
\end{eqnarray}
The condition $V_{ther}=0$ has to be solved together with the charge neutrality and $\beta$ equilibrium condition. As found in the earlier studies, since the above equation involves both the first and second derivatives of energy $E(\rho,\delta)$ with respect to density and proton fraction $x_p$, the core-crust transition density is very sensitive to the fine details of the EOS.
If one adopts the parabolic approximation (PA) for the EOS, i.e., assuming $E(\rho,\delta)=E_0(\rho)+E_{\rm sym}(\rho)\delta^2$,
the condition of Eq. (\ref{Vther}) can be rewritten explicitly in terms of the first and second derivatives of \esym with respect to density as
\begin{eqnarray}\label{Vtherpa}
&&V^{PA}_{ther} = \rho^2 \frac{d^2 E_0}{d \rho^2} + 2 \rho \frac{d
E_0}{d \rho}
+ (1-2x_p)^2 \\
&\cdot&
\left[ \rho^2 \frac{d^2 E_{\rm sym}}{d \rho^2}
+2 \rho \frac{d E_{\rm sym}}{d \rho} - 2 E^{-1}_{\rm sym}
\left(\rho \frac{d E_{\rm sym}}{d \rho}\right)^2\right]. \nonumber
\end{eqnarray}
This expression is widely used in the literature. However, as we shall discuss next it may lead to very different core-crust transition densities and pressures compared to calculation retaining high-order terms in the isospin asymmetry $\delta$ in expanding the $E(\rho,\delta)$. Nevertheless, it is interesting to see by examining the terms in the bracket on the second line that the minus sign between the last two terms reduces effects of the slope $L$, leaving the first term related to the curvature $K_{\rm sym}$ of the \esym strongly influences the core-crust transition. Of course, as shown on the first line, the curvature of the SNM EOS also affects the core-crust transition density. Fortunately, it is relatively well determined already.

\begin{figure}[t!]
\centering
\includegraphics[scale=0.5]{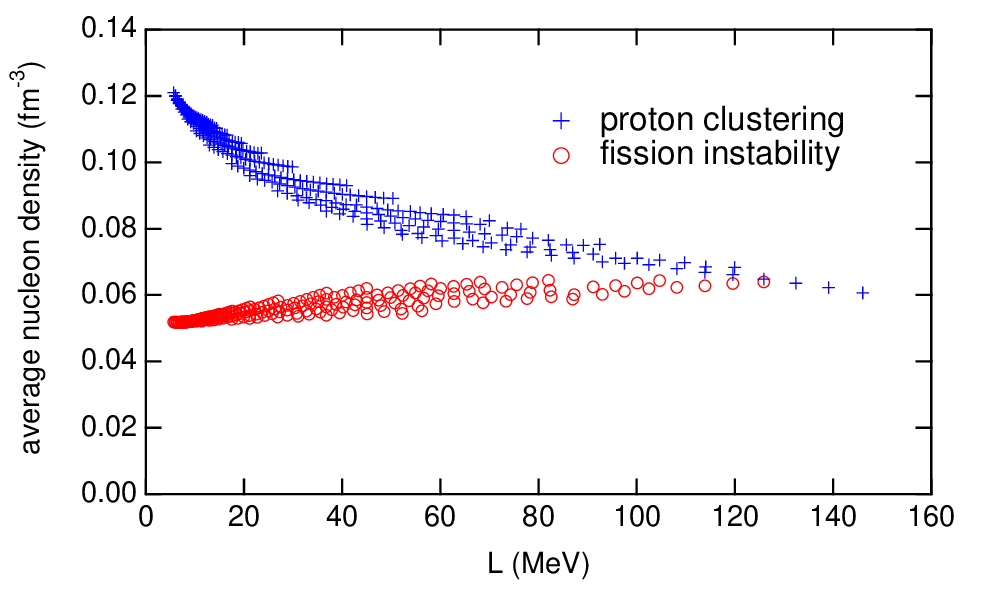}

\caption{{\protect\small (Color online) The onset (core-crust transition) density of proton clustering in uniform
nuclear matter and the lower boundary for pasta nuclei formation as functions of the $L$ parameter of nuclear symmetry energy.
Taken from ref.\ \cite{Oyamatsu2007}.}}
\label{nq}
\end{figure}

\begin{figure*}[t!]
\centering
\includegraphics[scale=0.5]{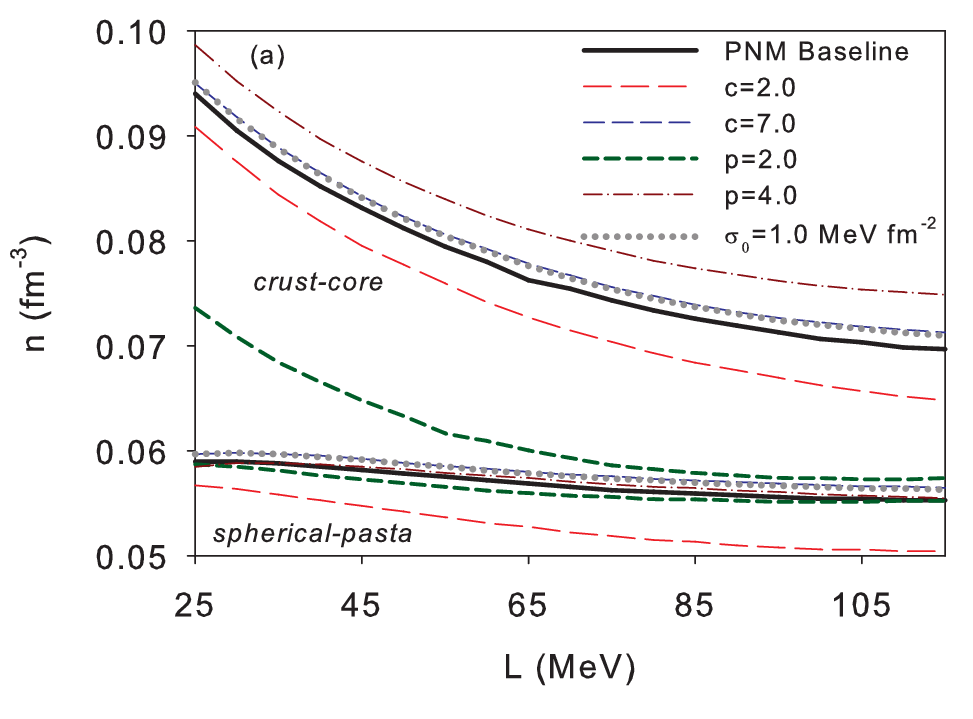}
\includegraphics[scale=0.5]{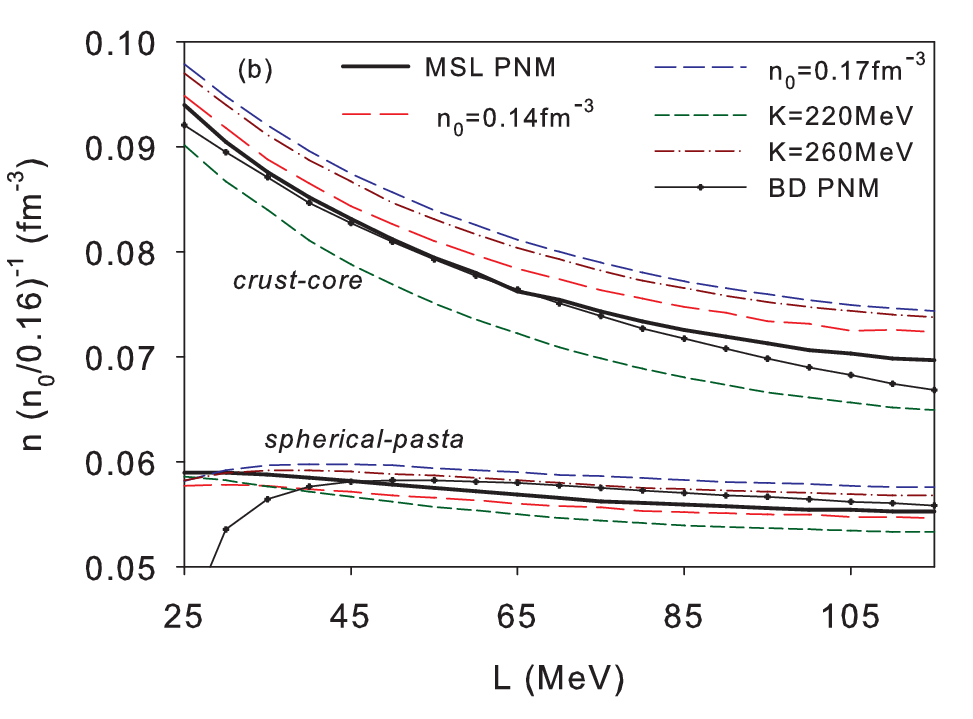}
\caption{{\protect\small (Color online) Crust-core and spherical nuclei-pasta transition densities versus $L$ for different parameters of the surface energy (left) and variations of parameters characterizing the EOS of symmetric nuclear matter (right) compared to the baseline (MSL and BD) EOSs constrained by low-density pure neutron matter (PNM) EOS predicted by microscopic many-body theories (thick solid lines). Taken from ref. \cite{Newton12}.}}
\label{NewtonFigs}
\end{figure*}

\subsection{Symmetry energy effects on the core-crust transition density and pressure based on the compressible liquid drop model}
Having outlined above the most widely used approaches for finding the core-crust transition point, we now turn to the main features of the results published by several groups in the literature in recent years. We focus on understanding effects of using different approaches and expanding the EOS to different powers of isospin asymmetry and density as well as identifying the parameters most influential on the core-crust transition density and pressure.

To our best knowledge, effects of the $L$ parameter on the core-crust and pasta formation were first studied by Kazuhiro Oyamatsu and Kei Iida using the Thomas-Fermi model for nuclei on the BCC lattice within the Wigner-Seitz approximation \cite{Lida14,Oyamatsu2007}. They used the parabolic approximation for the EOS and expanded the \esym only to the $L$ term. In their studies, the upper end of the pasta region is estimated by considering the proton clustering instability of uniform matter, i.e., the core-crust transition point. As shown in Fig.\ \ref{nq}, the transition density was found to decrease with increasing $L$ values. While the lower end of the pasta region can be understood from fission-like instability of spherical nuclei, they found that the baryon density for this boundary is of order 0.06 fm$^{-3}$ and is almost independent of the EOS models they used.

Similar to the work of ref. \cite{Oyamatsu2007} but keeping more terms in both $\delta$ and $\rho$ in expending the EOS and considering the
isospin dependence of the surface and curvature energy using the compressible liquid drop model, effects of all uncertain model parameters on the core-crust and pasta-spherical nuclei transitions
were studied extensively by Newton et al. \cite{Newton12} and also by Carreau et al. \cite{Fra-crust2} very recently.
The EOS of PNM $E_{\rm PNM}(\rho) \approx E_{\rm 0}(\rho) + E_{\rm sym}(\rho)$ can be calibrated at low densities by the available predictions from microscopic nuclear many-body theories \cite{Schwenk2005,Hebeler2010,Gandolfi2011}. In the work of Newton et al., the modified Skyrme-like (MSL) EOS \cite{MSL01} and the BD EOS originally developed by Bludman and Dover \cite{BD1981} and later modified by Oyamatsu and Iida \cite{Oya2003} were used. The MSL EOS constrained at low densities by the theoretical PNM EOS was used as the default baseline model while the constrained BD EOS was used for comparisons. For example, shown in Fig. \ref{NewtonFigs} are the core-crust and spherical nuclei-pasta transition densities versus $L$ for different parameterizations of the surface energy and variations in the SNM EOS compared to the two baseline models \cite{Newton12}. The surface energy depends on the isospin asymmetry of the surface region characterized by a parameter $p$. Another parameter $c$ was introduced to characterize how quickly the surface symmetry energy increases with the bulk symmetry energy $E_{\rm sym}(\rho_0)$.  A lower $c (p)$ corresponds to a higher surface energy at high (low) proton fractions. As seen in the left window, a stiff surface energy at low proton fractions ($p$=2) results in a notably lower core-crust transition density, highlighted by the thick, short dashed line. It is thus very clear that the isospin-dependence of surface energy of neutron-rich nuclei pays a very important role, which unfortunately is very poorly known. The recent study by Carreau et al. \cite{Fra-crust2} further explored the role of the parameter $p$ and quantified its importance in comparison with other model parameters. Their findings are in good agreement with that found by Newton et al.

In the right window of Fig. \ref{NewtonFigs}, by comparing calculations using different saturation densities of $n_0=0.14, 0.17$ fm$^{-3}$ and incompressibilities of $K_0=220,260$ MeV, it was found that decreasing (increasing) the incompressibility $K_0$ and the saturation density $n_0$ results in a decrease (increase) in the core-crust transition density $n_{\rm cc}$. While several details of the liquid drop model affect appreciably the core-crust transition density and the corresponding pressure, the negative correlation between the $L$ parameter and the core-crust transition density is a common feature shared by all models when the intrinsic correlation between $L$ and $K_{\rm sym}$ is used. We shall return to this point when we examine how the core-crust transition density depends individually on the $L$ and $K_{\rm sym}$ without considering any correlation between them.  Moreover, the week dependence of the pasta-spherical nuclei transition on the $L$ parameter was found consistently in refs. \cite{Newton12,Oyamatsu2007}.  A very recent study using a 3D Skyrme-Hartree-Fock approach
found that a variety of nuclear pasta geometries are present in the neutron star crust and the result strongly depends on the symmetry energy especially in neutron-rich systems \cite{Farrooh17}.

\subsection{Effects of high-order isospin and density dependences of the EOS on the core-crust transition density and pressure}
\begin{figure}[t!]
\centering
\includegraphics[scale=1.]{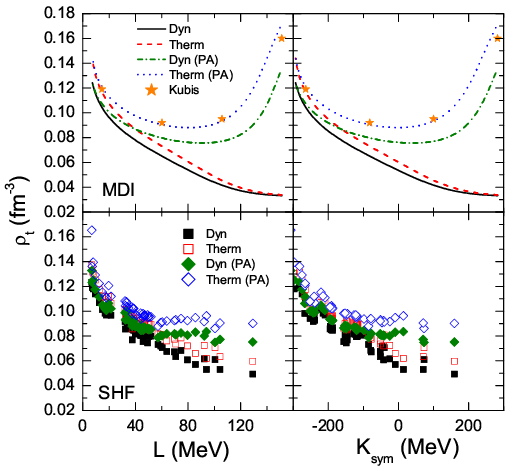}
\caption{{\protect (Color online) The core-crust transition density as
a function of $L$ (left) and $K_{\rm sym}$ (right) using
both the dynamical and thermodynamical approaches with the fully isospin-dependent EOS
and the parabolic approximation (PA), respectively. The MDI interactions with varying $x$ parameter (given in Sect. \ref{3bf}) and 51
Skyrme interactions are used in the upper and lower windows, respectively. Taken from ref. \cite{JXu}.}} \label{rhotLK}
\end{figure}

Different results of using both the dynamical and thermodynamical approaches with the fully isospin-dependent EOS and its parabolic approximation were demonstrated systematically by Xu et al. \cite{JXu}. For example, shown in the left and right panels of Fig.~\ref{rhotLK} are the core-crust transition density
$\rho_t$ as a function of $L$ and $K_{\rm sym}$, using both the MDI interaction with the varying $x$ parameter as discussed in Sect. \ref{3bf} and 51 Skyrme interactions, respectively.
Three important features can be identified: (1)  the dynamical approach predicts about 15\% smaller transition density compared to the thermodynamical calculations irrespective of the interactions used. This is well understood because due to the density gradients and the Coulomb terms included in the dynamical approach, system are more stable and thus lower the transition density. (2) The transition density decreases with increasing $L$ and $K_{\rm sym}$ consistent with other calculations. Again, as we shall discuss in
more detail that it is actually the $K_{\rm sym}$ that determines directly the transition density. The perceived dependence on the $L$ is mostly due to the intrinsic correlation between the $L$ and $K_{\rm sym}$ in the EOS models used. (3) Most strikingly, there are big differences in the transition densities obtained from calculations using the fully isospin-dependent EOS and its parabolic approximation. This basically verifies the findings from the earlier work of Baym et al \cite{Baym1971} and Arpoen \cite{Arp72}. It is useful to note that the curvature matrix elements involve the following first and second derivatives
\begin{eqnarray}
\partial E/\partial x_p&=& -4E_{\rm sym,2}(\rho)(1-2x_p) - 8E_{\rm sym,4}(\rho) (1-2x_p)^3 \nonumber\\
&+& \mathcal{O}(1-2x_p)^5,\nonumber \\
\partial^2 E/\partial x_p^2 &=& 8E_{\rm sym,2}(\rho) + 48E_{\rm sym,4}(\rho) (1-2x_p)^2 \nonumber\\
&+& \mathcal{O}(1-2x_p)^4.
\end{eqnarray}
The terms involving the $E_{\rm sym,4}(\rho)$ are not necessarily small
at $\beta$-equilibrium in both derivatives compared to the $E_{\rm sym}(\rho)$ terms since the $\delta=1-2x_p$ is normally not far from 1 and mathematically the higher-order derivative gains a larger multiplication factor.

\begin{figure}[t!]
\centering
\includegraphics[scale=1.]{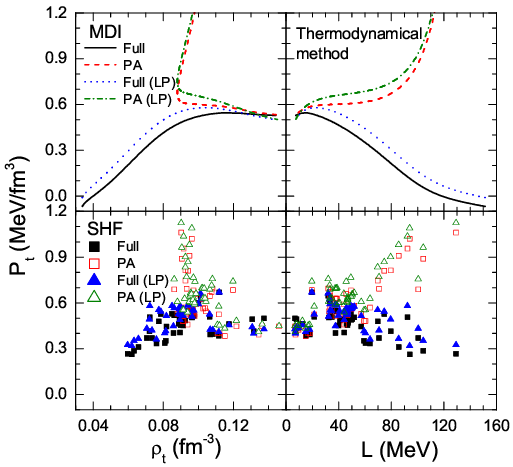}
\caption{{\protect (Color online) The transition pressure
$P_t$ as a function of $\rho_t$ (left) and $L$ (right) within the thermodynamical
method with the fully isospin-dependent EOS and its parabolic approximation (PA) using the
MDI (upper) and Skyrme (lower) interactions, respectively. Estimates using Eq. (\ref{lp}) given by Lattimer and Prakash (LP) with the transition density from Fig. \ref{rhotLK} are also shown for comparisons. Taken from ref. \cite{JXu}.}}
\label{PtrhotLther}
\end{figure}
\begin{figure}[t!]
 \centering
 \resizebox{0.45\textwidth}{!}{
  \includegraphics{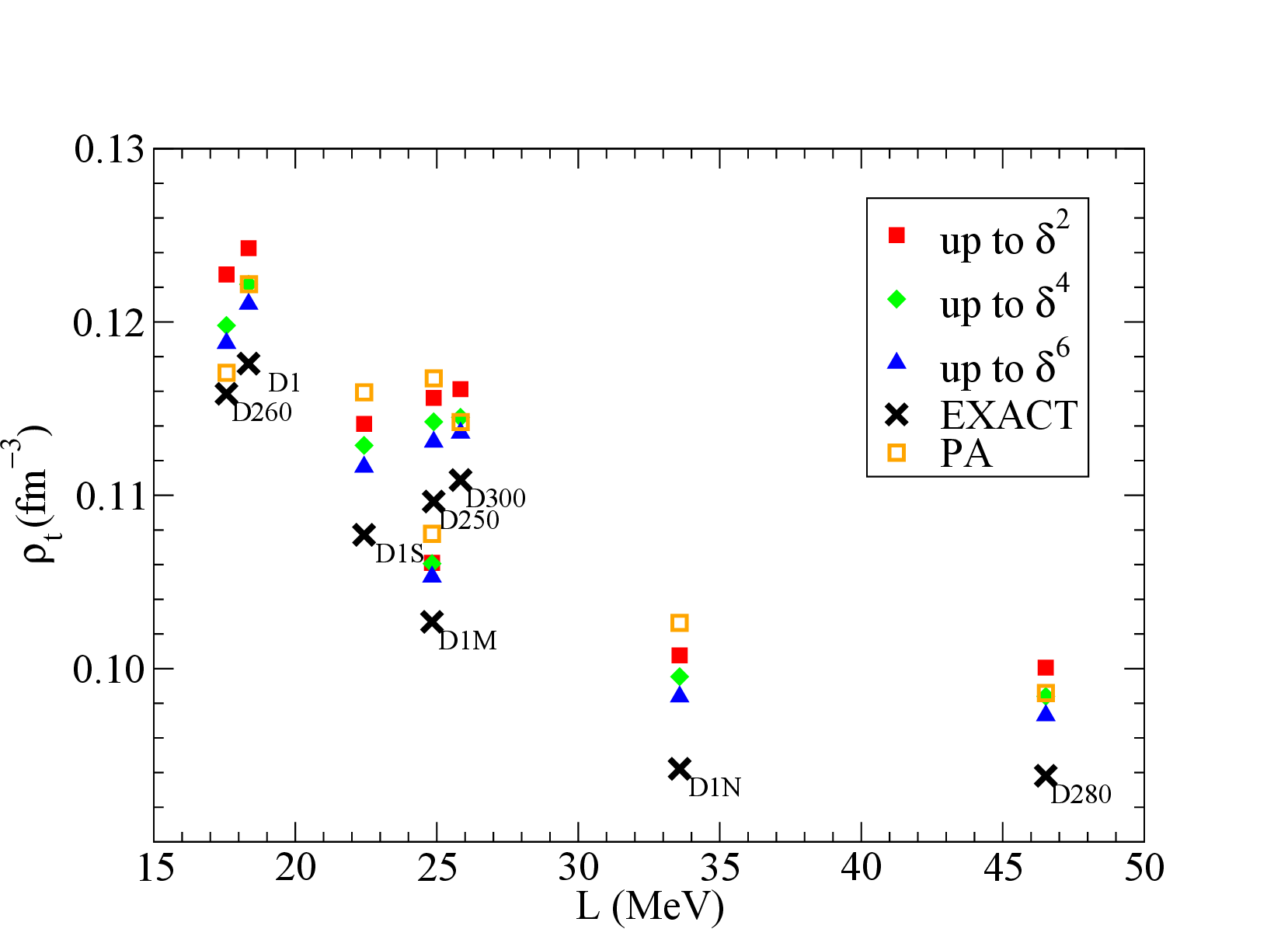}
  }
 \resizebox{0.45\textwidth}{!}{
  \includegraphics{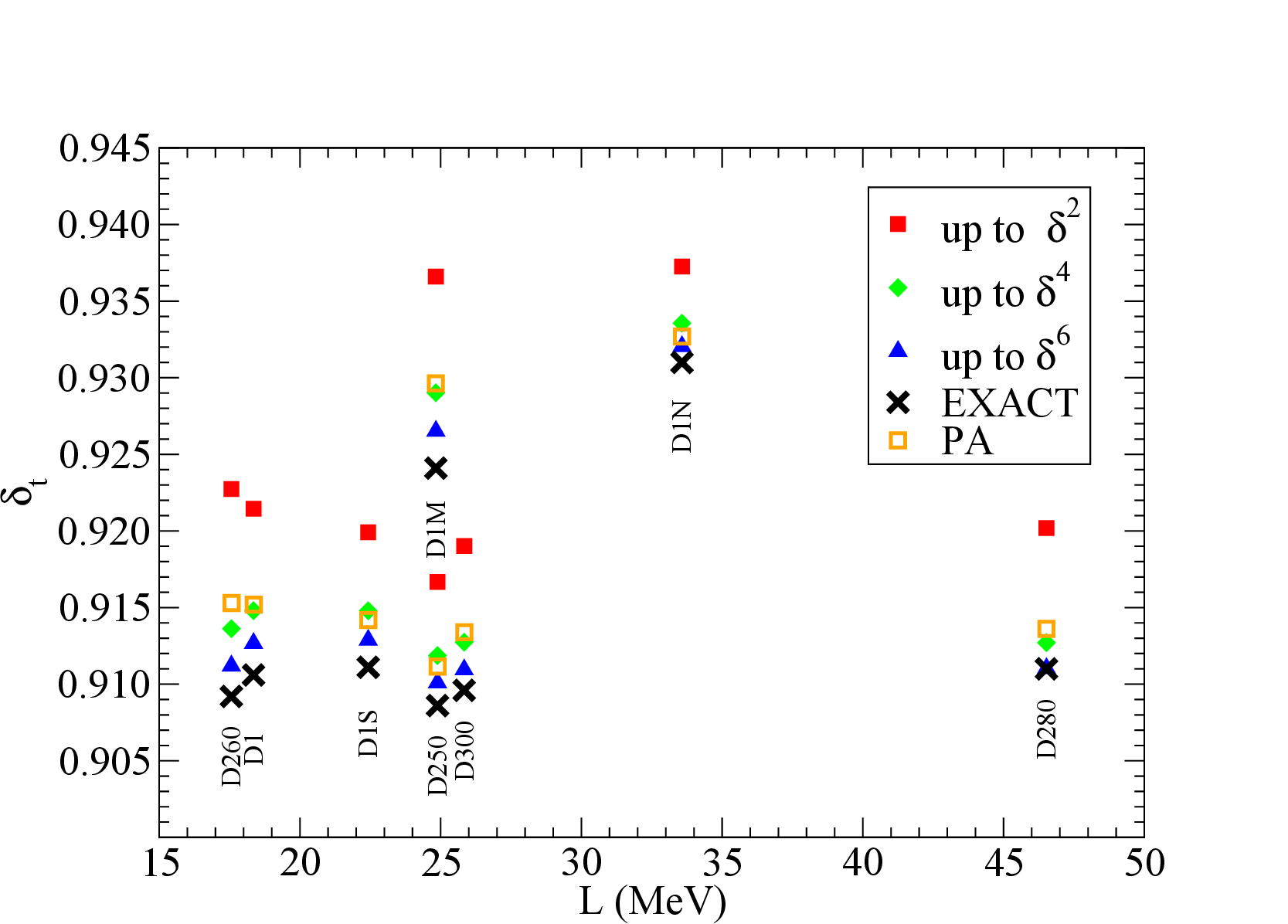}
  }
 \resizebox{0.45\textwidth}{!}{
  \includegraphics{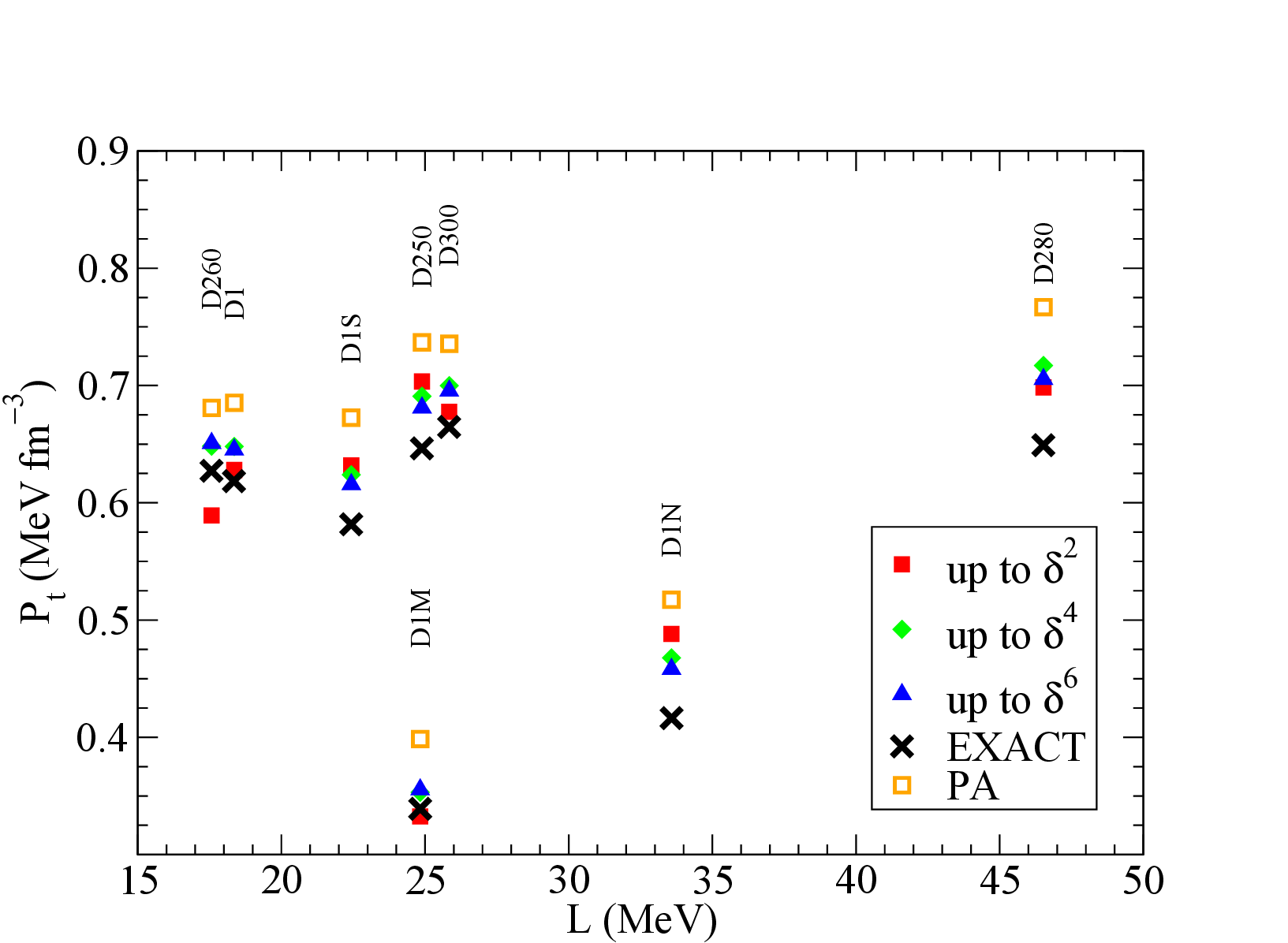}
  }
\caption{{\protect (Color online) Core-crust transition density $\rho_t$ (upper window), asymmetry $\delta_t$ (middle window) and pressure $P_t$ (lower window), as a function of the slope parameter $L$ calculated using the exact expression of the EoS (crosses), and the approximations up to second (solid squares), fourth (solid diamonds) and sixth order (solid triangles) with 11 Gogny interactions. The parabolic approximation is also included (empty squares). Taken from ref \cite{Gon17}.}}
 \label{RiosFigs}
\end{figure}

The core-crust transition pressure $P_t (\rho_t,\delta_t)$ can be calculated easily once the transition density $\rho_t$ and the isospin asymmetry there $\delta_t$ are known.
It relates directly with the crustal fraction of the moment of inertia that
can be measurable indirectly from observations of pulsar
glitches~\citep{Lattimer2007}. To see transparently effects of the \esym on the core-crust transition pressure, it is very instructive to recall the transition pressure estimated from
using the thermodynamical method with the PA by Lattimer and Prakash (LP)~\cite{Lattimer2007}
\begin{eqnarray}\label{lp}
P_t (\rho_t,\delta_t)
&&=
\frac{K_0}{9}\frac{\rho_t^2}{\rho_0}\left(\frac{\rho_t}{\rho_0}-1\right)\\
&&+\rho_t\delta_t
\left[\frac{1-\delta_t}{2}E_{\rm sym}(\rho_t)
+\left(\rho\frac{d
E_{\rm sym}(\rho)}{d \rho}\right)_{\rho_t}\delta_t\right].\nonumber
\end{eqnarray}
Obviously, the $P_t$ depends on the symmetry energy not only indirectly through the $\rho_t$ and
$\delta_t$,  but also explicitly through the value and slope
of the $E_{\rm sym}(\rho)$ at $\rho_t$.
The transition pressures $P_t$ using different interactions within the Lattimer-Prakash (LP) approximation are compared in Fig. \ref{PtrhotLther} with the full calculations of the core-crust transition pressure in the $npe$ matter at $\beta$ equilibrium. For ease of comparisons, the same transition densities from the thermodynamical calculations shown in Fig. \ref{rhotLK} are used.
It is seen that the Eq.~(\ref{lp}) predicts qualitatively
the same trend but quantitatively slightly higher values compared to the
original (full) expressions for the pressure with or without using the parabolic approximation (PA) for the EOS.
However, because the resulting core-crust transition densities change significantly when the PA is used, the corresponding transition pressures also change significantly.
More specifically, the $P_t$ essentially increases with
the increasing $\rho_t$ in calculations using the fully isospin-dependent EOS, while the use of PA may lead to a very
complex relation between the $P_t$ and $\rho_t$. Thus, the high-order isospin dependent terms of the EOS may influence significantly both the density and pressure at the
crust-core transition boundary.

The main features discussed above are in general agreement with those found in other recent studies, see, e.g., refs. \cite{Kubis2007,Dou00,Mou1,Cai12,Mou2,Sei14,Rou16,Duc1,Vid09,Cam10,Duc2,Bao1,Bao2,Gon17,Fang,Tsa,Ant19}, while quantitatively, the predicted core-crust transition densities and pressures are still model and interactions dependent for the reasons we mentioned above. Among all the available studies in the literature, it is very instructive to mention the rather systematic work in refs. \cite{Rios-Gogny,Gon17} using the same sets of Gogny interactions used in calculating the symmetry potential and energy shown in Fig. \ref{Rios-U} and Fig. \ref{Rios-esym}. Within the Gogny Hartree-Fock approach, all high-order coefficients of expanding the energy density functional in isospin asymmetry can be given analytically.  Shown in Fig.\ \ref{RiosFigs} are the core-crust transition density $\rho_t$, asymmetry $\delta_t$ and pressure $P_t$, as a function of the slope parameter $L$ calculated using the exact expression of the EOS (crosses), and the approximations up to second (solid squares), fourth (solid diamonds) and sixth orders (solid triangles) with the 11 Gogny interactions. For comparison, results of using the PA are shown with the empty squares. While it is well known that most of the Gogny interactions are too soft to support massive neutron stars of M$\ge 2.01$M$_{\odot}$ and may also predict $L\le 30$ MeV in obvious contradiction with the $L$ systematics discussed earlier in Sect. \ref{lsys}.

The results shown in Fig.\ \ref{RiosFigs} convey again the main message of this section: the physics of core-crust transition is unsettled. The numerical values of the transition density, isospin-asymmetry and pressure depend on the fine details of the interaction as well as the resulting isospin and density dependences of the EOS. Consequently, using these different core-crust transition properties in astrophysical models will lead to different predictions.
\begin{figure}
\begin{center}
 \resizebox{0.4\textwidth}{!}{
    \includegraphics{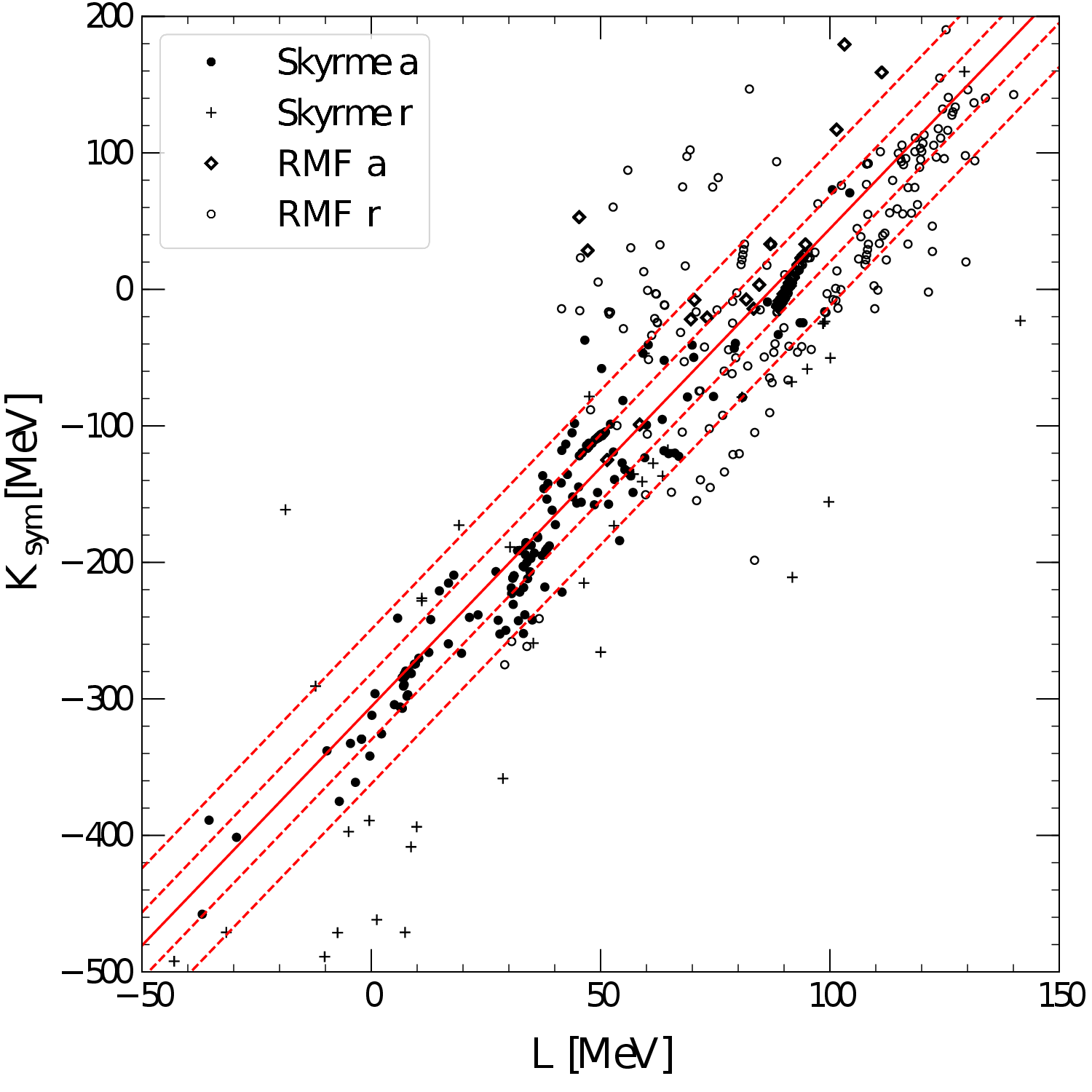}
  }
\\
 \resizebox{0.45\textwidth}{!}{
  \includegraphics{Isaac-KL.eps}
  }
\caption{(Color online) Correlation of $K_{\rm sym}$ and $K_{\tau}$ with $L$. Taken from refs. \cite{Tews17,Vid09}. }
\label{Isaac1}
\end{center}
\end{figure}
\begin{figure*}
\begin{center}
 \resizebox{0.9\textwidth}{!}{
  \includegraphics{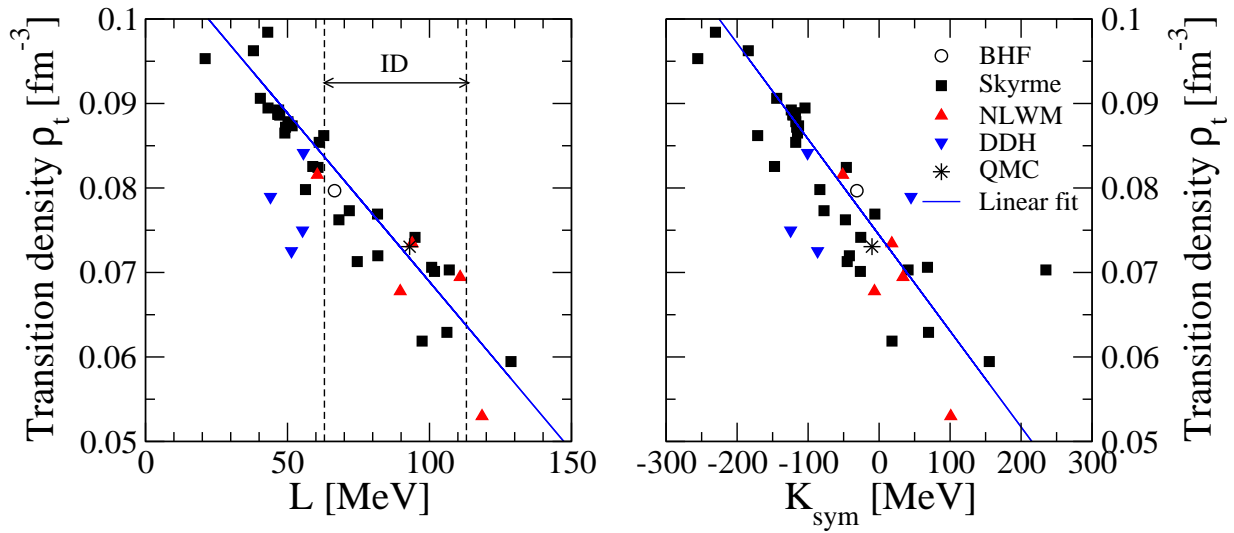}
  }
\caption{(Color online) Core-crust transition density as a function of $L$ (left) and $K_{\rm sym}$ (right). Taken from ref. \cite{Vid09}. }
\label{Isaac2}
\end{center}
\end{figure*}
\subsection{Individual roles of the slope $L$ and curvature $K_{\rm sym}$ of symmetry energy on the core-crust transition density}
As shown in essential all figures in the previous subsection, it is customary to plot the core-crust transition density as a function of $L$. In fact, the predicted transition densities are often parameterized as decreasing functions of increasing $L$, while the transition pressures are sometimes parameterized as functions of both $L$ and $K_{\rm sym}$ in the literature based on various calculations. A very comprehensive review on these parameterizations was given by Provid\^encia {\it et al.} \cite{Pro14}. Given the empirical correlation between $L$ and $K_{\rm sym}$ shown in several surveys of model predictions, plotting the transition density either as a function of $L$ or $K_{\rm sym}$ is indeed useful for presenting the model predictions. However, this general practice does not reveal the individual roles of the $L$ and $K_{\rm sym}$ on equal footing. In fact, as we shall show next, it actually overshadows the true role played by the $L$ and $K_{\rm sym}$ individually. The $K_{\rm sym}$ plays a dominating role and its increase makes the transition density $\rho_t$ decrease, while the $L$ plays a minor role and its increases at a fixed $K_{\rm sym}$ often makes the $\rho_t$ slightly increase.
Because of the empirical linear correlation between the $L$ and $K_{\rm sym}$ widely used in the literature, when the $\rho_t$ is plotted as a function of  $L$ only (by marginalizing the $K_{\rm sym}$), a false representation/impression that the increasing $L$ leads to the decreasing $\rho_t$ is then made.

Correlations among coefficients in expanding the $E_0(\rho)$ and $E_{\rm{sym}}(\rho)$ have been found based on surveys of predictions mostly from energy density functionals,
see, e.g., refs. \cite{Tews17,Dut12,Col14,Dutra2014,Maz13,India17}. For example, shown in Fig. \ref{Isaac1} are the $L$ and $K_{\rm sym}$ correlations from typical microscopic many-body theories and phenomenological models \cite{Tews17,Vid09}. More quantitatively, the results shown in the upper window are from calculations using 240 Skyrme forces \cite{Dut12} and 263 RMF forces \cite{Dutra2014} compiled by Dutra {\it et al.}. Excluding some of the unrealistic forces, Tews et al. \cite{Tews17} found that
$K_{\rm{sym}}=3.501L-(305.67\pm24.26)$ with a correlation coefficient of r=0.96 at 68\% confidence level. In the lower window, the results from the Brueckner-Hartree-Fock (BHF) theory, Skyrme Hartree-Fock (SHF) theory, Relativistic Mean-Field (RMF) theory, non-linear Walecka model (NLWM) with constant couplings, density dependent hadronic (DDH) model and the quark-meson coupling (QMC) model can be fitted with a straight line in general agreement with the results in the lower window. The core-crust transition densities obtained using all these modes are shown in Fig. \ref{Isaac2} as functions of $L$ and $K_{\rm sym}$, respectively. Consistent with many other studies, the transition density decreases with both increasing $L$ and $K_{\rm sym}$. Interestingly, a recent Bayesian analysis found that the $\rho_t$ is almost equally correlated with the $L$ and $K_{\rm sym}$ \cite{Ant19}. However, these results alone do not tell whether the apparent $L$ dependence is real or if it is simply due to the correlation between the $L$ and $K_{\rm sym}$. Of course, one may also ask if the apparent $K_{\rm sym}$ dependence is real or not, or both the $L$
and $K_{\rm sym}$ make the transition density decrease. These questions arise because the correlations among the $\rho_t$, $L$ and $K_{\rm sym}$ revealed in the surveys of multiple predictions are studied in an imaginary giant theory containing multiple models. The latter contain multiple parameters that do not necessarily communicate with each other across models. Namely, the macroscopic observables $\rho_t$, $L$ and $K_{\rm sym}$ depend differently on microscopic parameters that are not necessarily present in all models considered in the surveys. Even in the same kind of models, such as the SHF, from one set of model parameters to another both the $L$ and $K_{\rm sym}$ changes. While there must be some true physics behind the observed $L$-$K_{\rm sym}$ correlation, it is not clear which one of them causes the apparent decreasing
$\rho_t$ with increasing $L$ and $K_{\rm sym}$ from the surveys of multiple model predictions.

\begin{figure*}
\begin{center}
 \resizebox{0.45\textwidth}{!}{
  \includegraphics{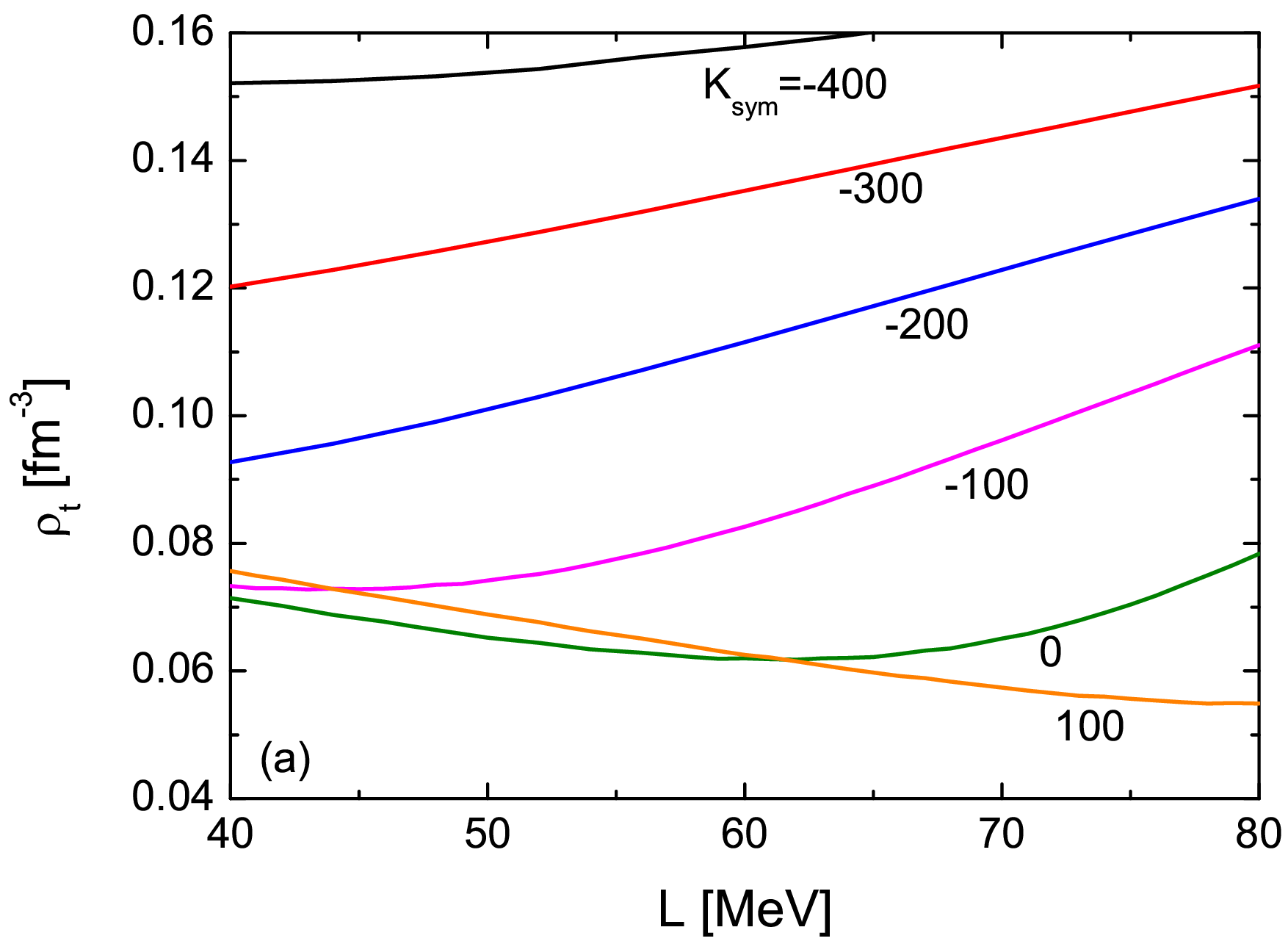}
  }
 \resizebox{0.45\textwidth}{!}{
  \includegraphics{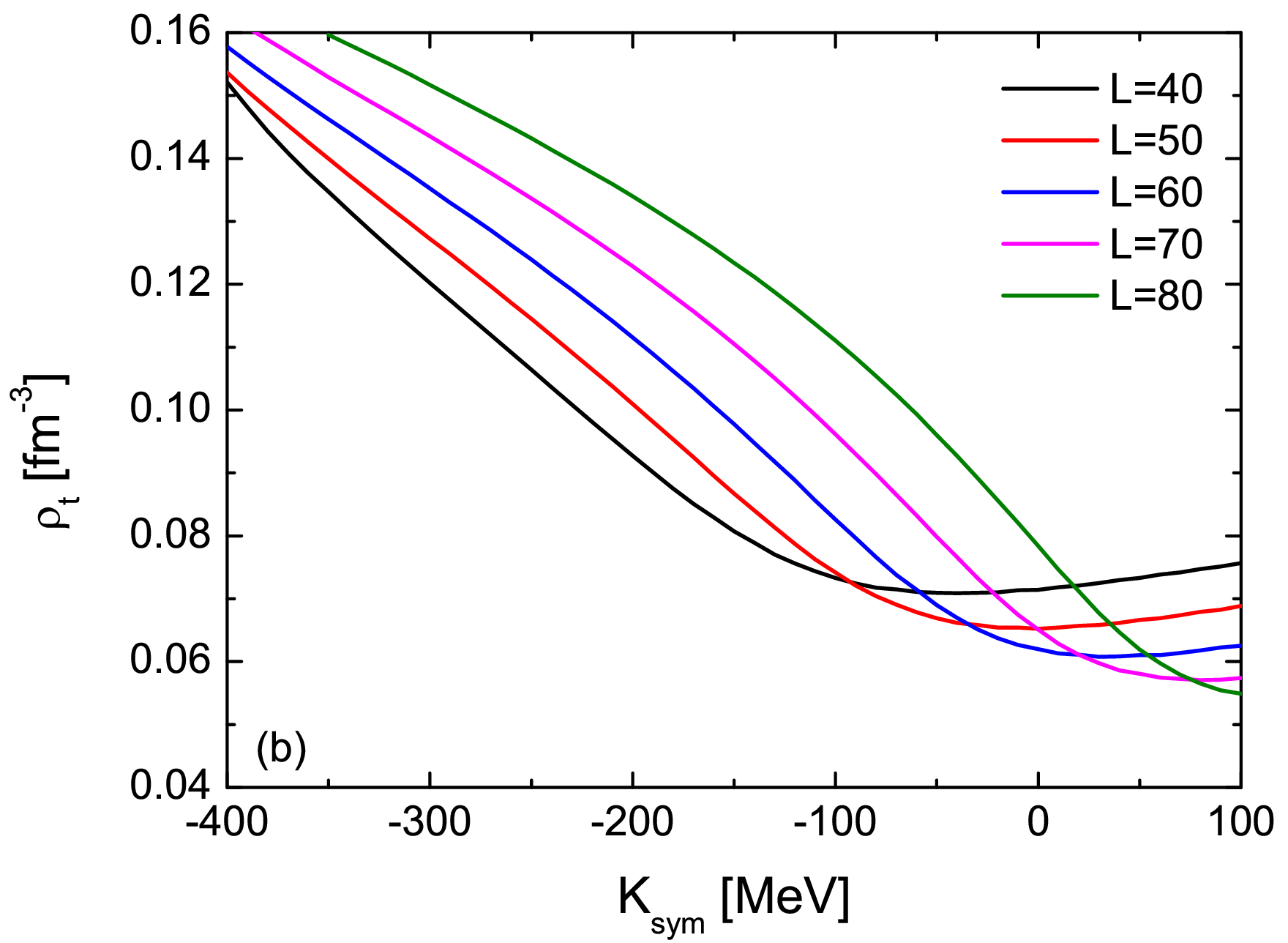}
  }
\caption{(color online) The core-crust transition density $\rho_t$ as a function of $L$ (left panel) with $K_{\rm{sym}}$ fixed at -400, -300, -200, -100, 0, and 100 MeV, and $K_{\rm{sym}}$ (right panel) with $L$ fixed at 40, 50, 60, 70, and 80 MeV, respectively. Taken from refs. \cite{Zhang2018}. }
\label{NBZ-lk}
\end{center}
\end{figure*}
To answer the questions raised above, one has to investigate independently the individual roles of $L$ and $K_{\rm sym}$ without using any correlation between them. Such a study was carried out in ref. \cite{Zhang2018} by using simply a parameterized EOS independent of any model within the thermodynamical approach. The parametric EOS used in
ref. \cite{Zhang2018} is the same in spirit as the so-called meta-modelling formulation of the EoS by Margueron, Casali and Gulminelli \cite{MM1,MM2,MM3}. Relevant to the
discussions in this section on the crust-core transition properties, it is very interesting to mention that they found that
some of the existing correlations among different empirical parameters of the nuclear EOS can be understood from basic physical constraints imposed on the Taylor expansions of the SNM EOS and symmetry energy around the saturation density \cite{MM3}. In particular, a huge dispersion of the correlations among low-order empirical parameters is induced by the unknown higher-order empirical parameters. For example, the correlation between $E_{\rm sym}(\rho_0)$ and $L$ depends strongly on the poorly known $K_{\rm sym}$, while the correlation between $L$ and $K_{\rm sym}$ is strongly blurred by the even more poorly known third-order parameters $J_0$ and $J_{\rm sym}$. Moreover, some of the
perceived correlations especially those involving the high-order parameters from combining different models may be spurious. These findings further call for cautions in using the
correlations of low-order parameters in determining the crust-core transition properties. 

Using $L$ and $K_{\rm{sym}}$ as two independent free parameters, shown in Fig. \ref{NBZ-lk} are the $\rho_t$ versus $L$ with fixed values of $K_{\rm{sym}}$ in the left window and versus $K_{\rm{sym}}$ with fixed values of $L$ in the right window, respectively. The $\rho_t$ changes much more dramatically with $K_{\rm{sym}}$ than $L$ in their respective uncertainty ranges. Most importantly, in most regions on the $L-K_{\rm sym}$ parameter plane, the transition density $\rho_t$ actually increases with increasing $L$ when the $K_{\rm{sym}}$ is fixed while it decreases with increasing $K_{\rm{sym}}$ when the $L$ is fixed.
Thus, one may easily get the impression that the $\rho_t$ decreases with increasing $L$ when it is only plotted as a function of $L$ while both the $L$ and $K_{\rm{sym}}$ are varied from model to model.

\begin{figure*}
\begin{center}
 \resizebox{0.45\textwidth}{!}{
  \includegraphics{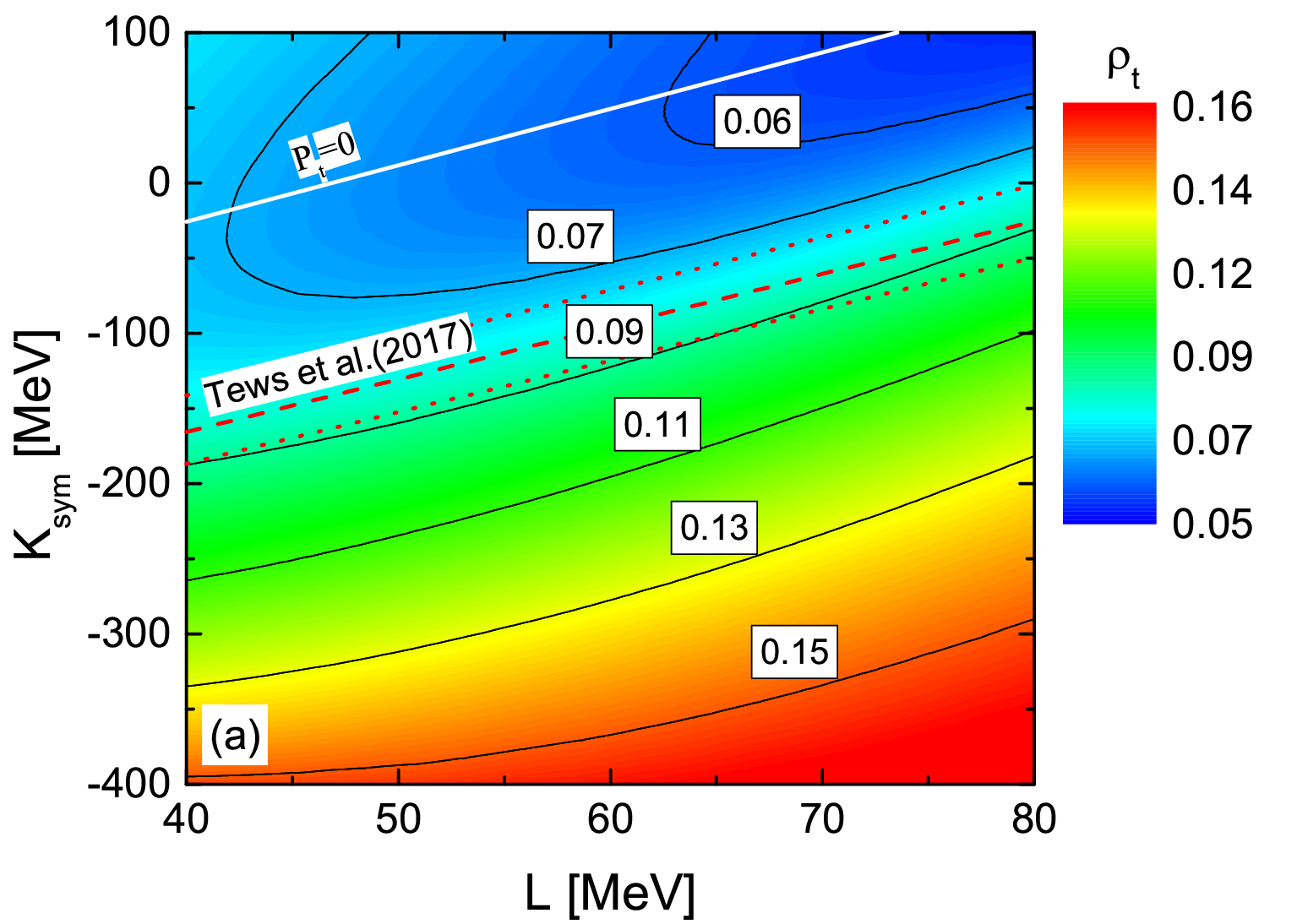}
  }
 \resizebox{0.45\textwidth}{!}{
  \includegraphics{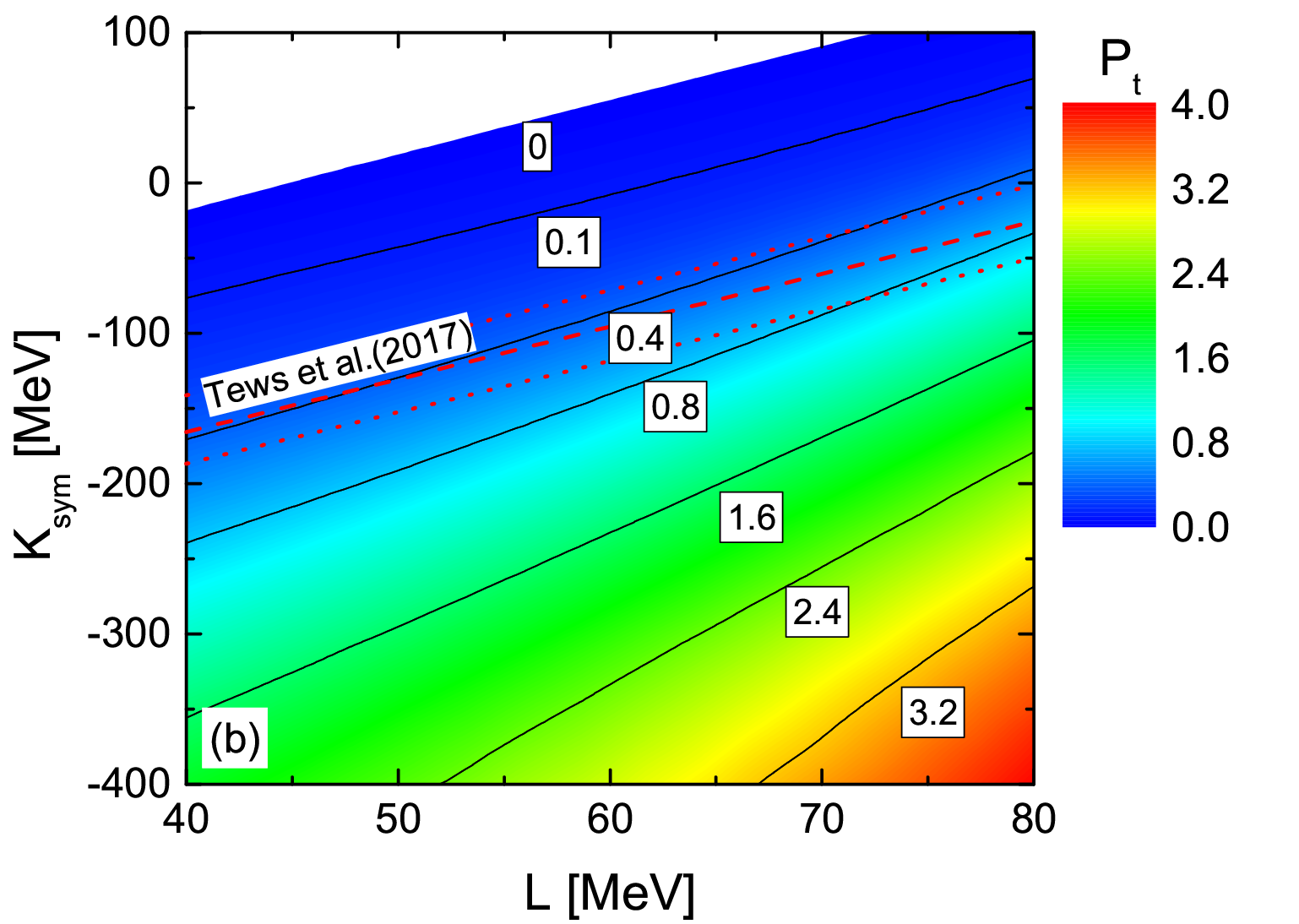}
  }
   \caption{(color online) Contours of the core-crust transition density $\rho_t$ in fm$^{-3}$ (left window) and the corresponding pressure $P_t$ in MeV fm$^{-3}$ (right window) in the $L-K_{\rm{sym}}$ plane. Lines with fixed values of transition densities and pressures are labeled. The red dashed lines are the results applying the $K_{\rm{sym}}$-$L$ correlation within 68\% confidence from the systematics given in ref. \cite{Tews17}. The white region in the right window is where the transition pressure vanishes. The $P_t=0$ boundary along the line $K_{\rm{sym}}=3.64 L-163.96~(\rm{MeV})$ marks the mechanical stability boundary. The corresponding boundary of $P_t=0$ is shown with the while line in the left window. Taken from refs. \cite{Zhang2018}.}
 \label{NBZ-dp}
\end{center}
\end{figure*}
All features shown in Fig. \ref{NBZ-lk} can be easily understood qualitatively from inspecting the competitions of the three terms in the bracket of line two in Eq. (\ref{Vtherpa}), i.e.,
\begin{equation}\label{form1}
\left[ \rho^2 \frac{d^2 E_{\rm sym}}{d \rho^2}
+2 \rho \frac{d E_{\rm sym}}{d \rho} - 2 E^{-1}_{\rm sym}
\left(\rho \frac{d E_{\rm sym}}{d \rho}\right)^2\right].
\end{equation}
The three terms are proportional to $K_{\rm{sym}}$, $L$ and $L^2/E_{\rm sym}$ at density $\rho$, respectively. The minus sign between the last two terms significantly reduces effects of $L$, making the $K_{\rm{sym}}$ dominate the variation of
$\rho_t$ which is a solution of setting Eq. (\ref{Vtherpa}) to zero. When the $K_{\rm{sym}}$ increases from big negative to small positive values, the necessary $\rho_t$ to make the total value of the first two terms on the first line of  Eq. (\ref{Vtherpa}), i.e.,  $\rho^2 \frac{d^2 E_0}{d \rho^2} + 2 \rho \frac{dE_0}{d \rho}$,
large enough decreases, regardless of the $L$ value. When the value of $K_{\rm{sym}}$ is fixed at large negative values, the competition of the last two terms in the bracket (\ref{form1})
with increasing $L$ makes the total value of the bracket even more negative, leading to an increasing $\delta_t$ with increasing $L$. However, when the $K_{\rm{sym}}$ is fixed at positive values,
the $\rho_t$ decreases with $L$. We note that while the value of $L$ is already relatively well constrained as we discussed earlier, the $K_{\rm{sym}}$ is rather poorly known. The discussions above clearly indicate the critical needs to better constrain the $K_{\rm{sym}}$.

Effects of $K_{\rm{sym}}$ and $L$ on the transition density and pressure are examined on equal footing in Fig. \ref{NBZ-dp} where contours of constant transition densities $\rho_t$ and pressures $P_t$ are shown. Obviously, the variation of $K_{\rm{sym}}$ has a dominating effect. Depending on the $K_{\rm{sym}}$ and $L$ values, both the transition density and pressure span large ranges. The marginalized dependence on either $K_{\rm{sym}}$ or $L$ would strongly depend on the intrinsic $K_{\rm{sym}}$-$L$ correlation in a given model. For instance, keeping a constant $L=60$ MeV would lead the $P_t$ to decrease quickly with increasing $K_{\rm{sym}}$, while keeping a constant $K_{\rm{sym}}=-350$ MeV would lead the $P_t$ to increase quickly with increasing L, qualitatively consistent with the observations in calculations using various SHF EOSs \cite{Newton12,Duc2}. Most strikingly, if one applies the ``universal" empirical $K_{\rm{sym}}$-$L$ correlation discussed earlier, such as the
red dashed lines from applying the $K_{\rm{sym}}$-$L$ correlation within 68\% confidence interval from the systematics given by Tews et al in ref. \cite{Tews17}, the $\rho_t$ is found to be a constant of about $0.08\pm 0.01$ fm$^{-3}$ essentially in the whole parameter plane, while the transition pressure has a larger variation around $P_t=0.40$ MeV fm$^{-3}$. Ironically, the constant $\rho_t$ is very close to the fiducial value of $0.07$ fm$^{-3}$, see, e.g., ref. \cite{Lat01},  $0.08$ fm$^{-3}$, see, e.g., ref. \cite{Dou00}, frequently used in the literature.

In the recent Bayesian analyses of the core-crust transition density and pressure \cite{Ant19,Fra-crust2} also within the PA, independent and flat priors (prior probability density distribution) for the $K_{\rm{sym}}$ and $L$ were used. The importance of both $K_{\rm{sym}}$ and $L$ is clear revealed. However, their relative roles are inconclusive, depending on the model used. Moreover, it was found that only if the surface tension is fixed to a reasonable but somewhat arbitrary value, strong correlations with the symmetry energy parameters ($L$ and $K_{\rm sym}$) are recovered. More quantitatively, a value of the surface isospin-dependence parameter $p=3$ is needed to reproduce the estimated average value of $\rho_t= 0.072\pm 0.011$ fm$^{-3}$ and the transition pressure as $P_t=0.339\pm0.115$ MeV fm$^{-3}$ obtained from many studies using the dynamical approach, while an even larger $p$ is required to reproduce the average results from many studies using the thermodynamical approach \cite{Fra-crust2}.

In summary of this section, as indicated already in the pioneering work in this field, the determination of core-crust transition point involves many interesting isospin physics. The core-crust transition point can be reached from both sides and locating it may thus require different techniques and nuclear physics inputs. The most widely used approach is by investigating the dynamical instabilities of uniform mater against the formation of clusters taking into accounts their surface and Coulomb energy. Because the curvature matrix of density fluctuations involves both the first and second derivatives of energy with respect to isospin asymmetry and density, the transition density and pressure are very sensitive to the fine details of nuclear EOS. The curvature $K_{\rm{sym}}$ of symmetry energy is the most critical but poorly known physics quantity determining the core-crust transition. The popular practice of plotting or parameterizing the transition density as a function of $L$ only hides the true dominating physics agent $K_{\rm{sym}}$ at work. Surprisingly, if one applies the empirical correlation between $K_{\rm{sym}}$ and $L$ found in extensive surveys of predictions by microscopic and phenomenological nuclear many-body theories to the transition density contours in the $K_{\rm{sym}}$ versus $L$ plane obtained using a parameterized EOS, the $\rho_t$ is found to be a constant of about $0.08\pm 0.01$ fm$^{-3}$.
Another outstanding quantity affecting significantly the core-crust transition point especially in models approaching the transition boundary from the cluster side is the parameter $p$ characterizing the isospin-dependence of surface energy. Obviously, much more work is still needed for the community to pin down the core-crust transition density and pressure.

Once the core-crust transition density and pressure are available, their astrophysical impacts can be examined. Most direct impacts are related to the crustal fraction of the moment of inertia affecting the glitch phenomena, crust thickness affecting the radii of neutron stars, sheer modulus, sheer viscosity and proton fraction in the crust affecting the torsional oscillation frequencies and r-mode damping rate. For earlier reviews of \esym effects on these phenomena we refer the reader to refs. \cite{Lat01,Lattimer2007}, while for the latest reviews it is very interesting to read the discussions in refs. \cite{Newton14,Tsa}.

\section{Symmetry energy effects on global properties of non-rotating neutron stars}
In this section, we discuss observables of neutron stars that can be used to probe the density dependence of nuclear symmetry energy.  Radii of neutron stars have long been considered as the most sensitive probe of nuclear symmetry energy at least since the earlier work of Prakash et al. in ref. \cite{Prakash1988}. More specifically, it was shown that the radius of a neutron star is most sensitive to the symmetry energy around $2\rho_0$~\cite{Lattimer2000,Lattimer2001}. While observational efforts of determining precisely the radii of neutron stars using x-ray data have been quite fruitful, no clear consensus was reached. The neutron star merger event GW170817 has triggered a welcomed flood of new studies extracting the radii of neutron stars from the tidal deformability of neutron stars involved in GW170817. Together with earlier results from analyzing the x-ray data, a narrow range for the radii of canonical neutron stars has emerged recently. It is then probably natural to ask now what we have learned so far from all the available astrophysical data about the \esym especially its high density behavior. In this section, we discuss this and several related issues.

\subsection{Solving the inverse-structure problem of neutron stars in a multi-dimensional high-density EOS parameter space}\label{inversion}
For completeness and ease of the following discussions, we first recall here the differential equation that has to be solved simultaneously with the Tolman-Oppenheimer-Volkov (TOV) equations
to investigate the tidal deformability $\lambda$ ~\cite{Flanagan2008,Hinderer2008,Binnington2009,Damour2009,Damour2010,Hinderer2010,Postnikov2010,Baiotti2010,Baiotti2011,Lackey2012,Pannarale2011,Damour2012,Fattoyev2013,Fattoyev2014}.  The $\lambda$ is related to the tidal Love number $k_2$ and radius $R$ via ~\cite{Flanagan2008,Damour2009,Damour2010,Hinderer2010}
\begin{equation}\label{lambda}
  \lambda=\frac{2}{3}k_2R^5.
\end{equation}
The tidal Love number $k_2$ depends on the stellar structure and can be calculated using the following expression~\cite{Hinderer2008,Postnikov2010}
\begin{eqnarray}\label{k2}
  k_2&=&\frac{1}{20}(\frac{R_s}{R})^5(1-\frac{R_s}{R})^2[2-y_R+(y_R-1]\frac{R_s}{R} \\\nonumber
   &\times&\{(6-3y_R+\frac{3R_s}{2R}(5y_R-8)+\frac{1}{4}(\frac{R_s}{R})^2[26  \\\nonumber
   &-&22y_R+(\frac{R_s}{R})(3y_R-2)+(\frac{R_s}{R})^2(1+y_R)])  \\\nonumber
   &+&3(1-\frac{R_s}{R})^2[2-y_R+(y_R-1)\frac{R_s}{R}]  \\\nonumber
   &\times&{\rm log}(1-\frac{R_s}{R})\}^{-1},
\end{eqnarray}
where $R_s\equiv2M$ is the Schwarzschild radius of the star, and $y_R\equiv y(R)$ can be calculated by solving the following first-order differential equation:
\begin{equation}\label{yr}
  r\frac{dy(r)}{dr}+y(r)^2+y(r)F(r)+r^2Q(r)=0,
\end{equation}
with
\begin{equation}\label{fr}
  F(r)=\frac{r-4\pi r^3(\varepsilon(r)-P(r))}{r-2M(r)},
\end{equation}
\begin{eqnarray}
  Q(r)&=&\frac{4\pi r(5\varepsilon(r)+9P(r)+\frac{\varepsilon(r)+P(r)}{\partial P(r)/\partial\varepsilon(r)}-\frac{6}{4\pi r^2})}{r-2M(r)}\\\nonumber
   &-&4[\frac{M(r)+4\pi r^3P(r)}{r^2(102M(r)/r)}]^2.
\end{eqnarray}
As the above equations are related to $M(r)$ and $r$, they thus should be solved together with the TOV equations~\cite{Oppenheimer39,Tolman1934}
\begin{equation}\label{TOVp}
\frac{dP}{dr}=-\frac{G(m(r)+4\pi r^3P/c^2)(\epsilon+P/c^2)}{r(r-2Gm(r)/c^2)},
\end{equation}
\begin{equation}\label{TOVm}
\frac{dm(r)}{dr}=4\pi\epsilon r^2
\end{equation}
by adopting the following boundary conditions: $y(0) = 2$, $P(0)=P_c$, and $M(0) = 0$.

\begin{figure}
\begin{center}
\resizebox{0.48\textwidth}{!}{
 \includegraphics{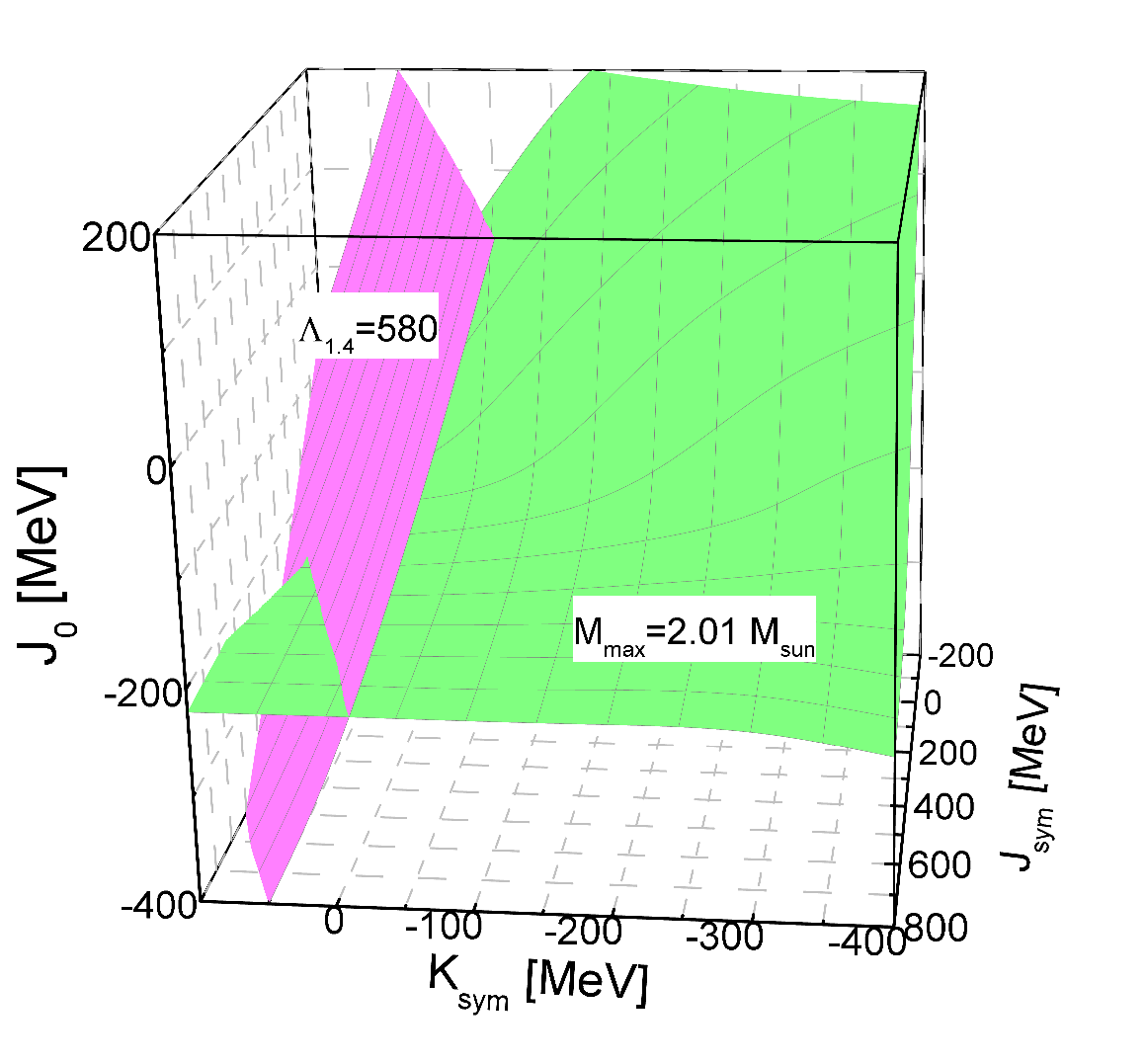}}
\caption{(Color online) Examples to show how the inverse-structure problem is solved. The green and magenta surface correspond to $M_{\rm max}=2.01$ M$_\odot$ and $\Lambda_{1.4}=580$, respectively.}
\label{example1}
\end{center}
\end{figure}
\begin{figure*}
\begin{center}
\resizebox{0.9\textwidth}{!}{
  \includegraphics{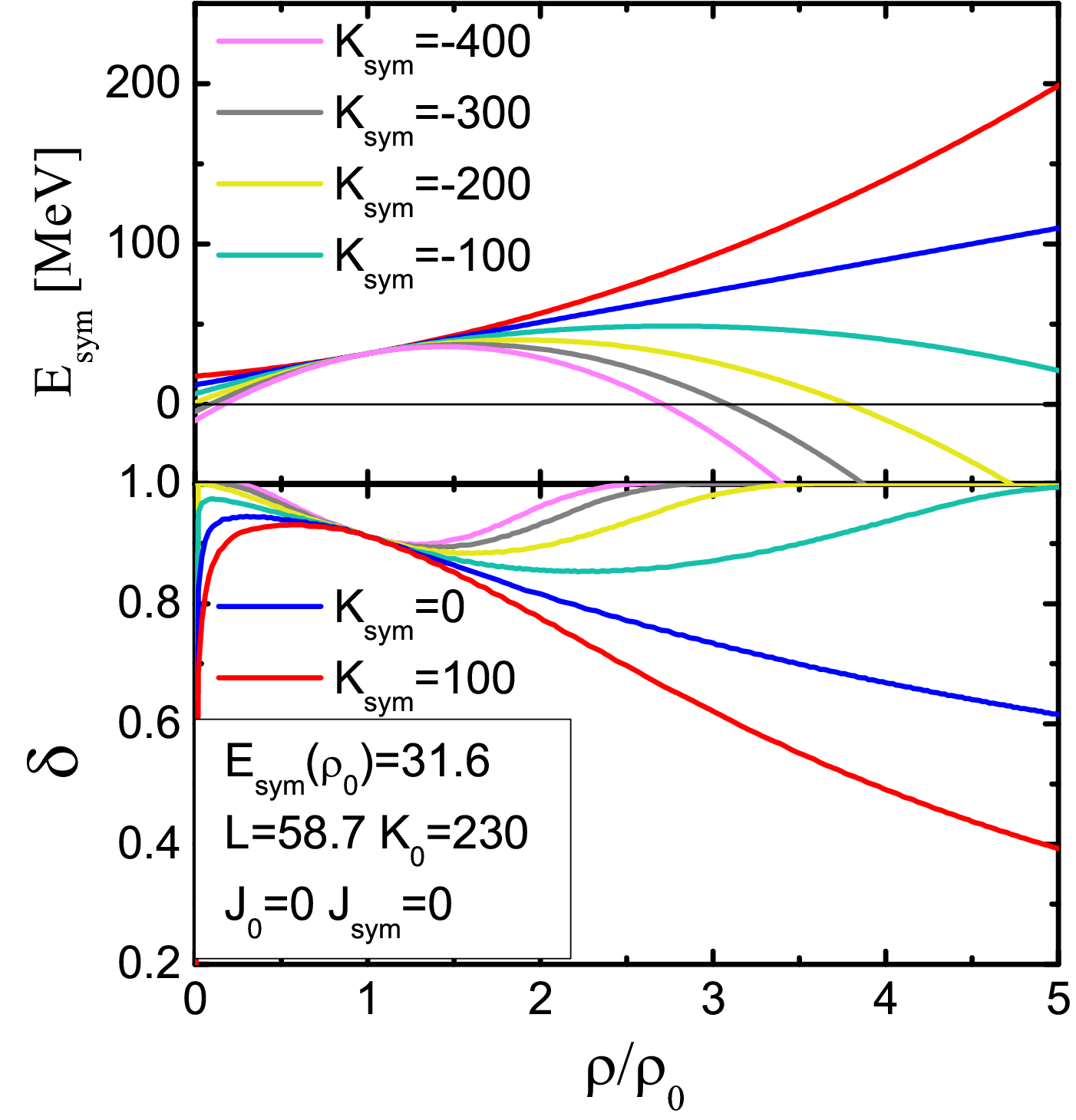}
  \includegraphics{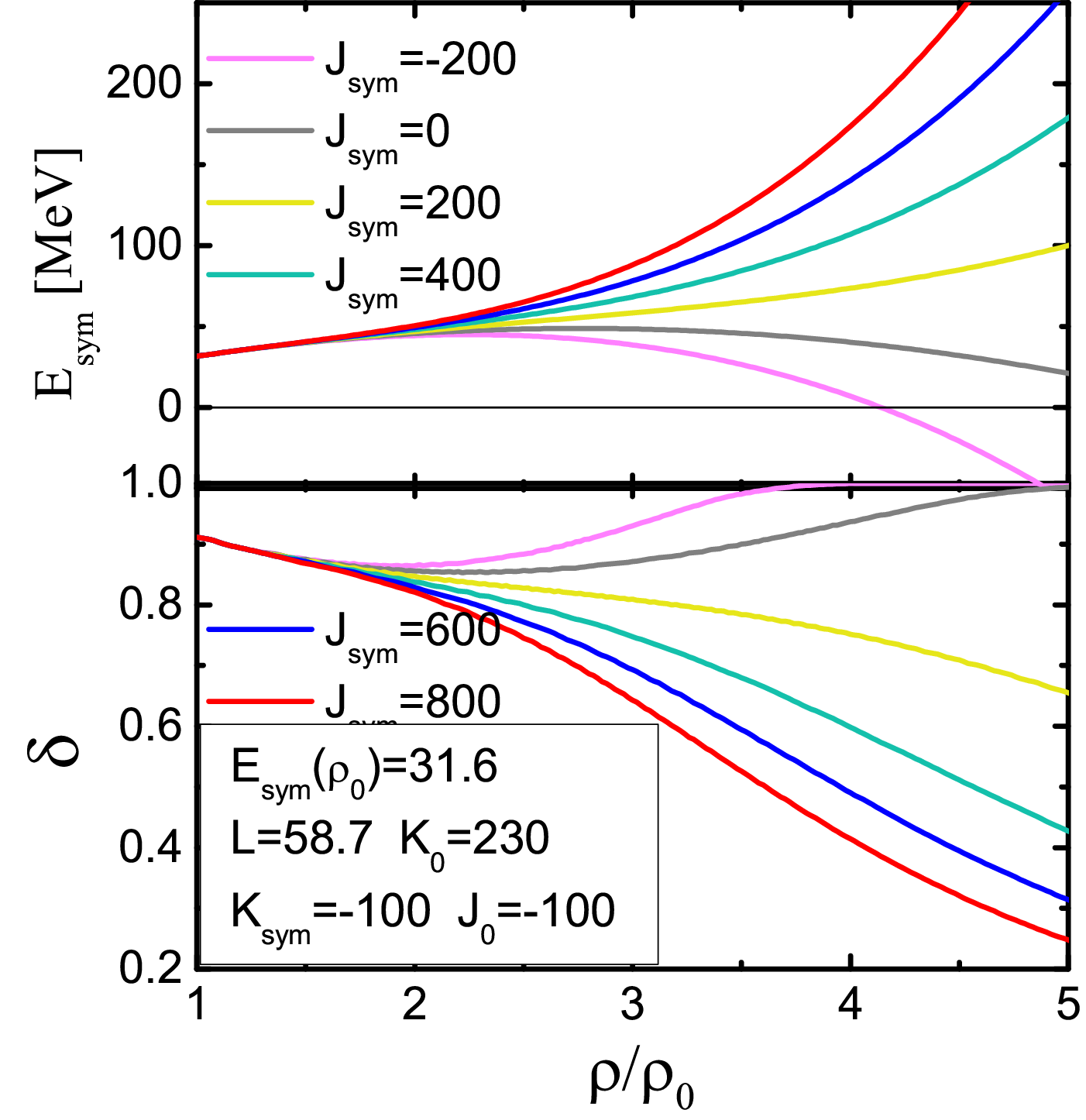}
  }
  \caption{(color online) The symmetry energy $E_{\rm{sym}}(\rho)$ and isospin asymmetry profile $\delta(\rho)$ in neutron star matter at $\beta$-equilibrium as a function of the reduced density $\rho/\rho_0$ for $K_{\rm{sym}}=-400$, -300, -200, -100, 0, and 100 MeV (left), and $J_{\rm{sym}}=-200$, 0, 200, 400, 600, and 800 MeV (right), respectively, while all other parameters are fixed as the specified values. Taken from ref. \cite{Zhang2019EPJA}}\label{Ksymeffect}
  \end{center}
\end{figure*}

The traditional way of investigating the relationships between properties of neutron stars and the EOS is straightforward: given a EOS, solve the TOV possibly coupled with other equations to find the mass-radius and possibly other observables, such as the tidal deformability and binding energy. On the other hand, inferences from the observational data information abut the underlying EOS parameters have also been very successful. In particular, the Bayesian statistical tools have been widely used to infer the posterior probability density distribution functions of EOS parameters. For an earlier review on the Bayesian inference, please see ref. \cite{Steiner16} and for a very recent one please see ref. \cite{Holt19}.
Another way to infer the EOS directly from observations is to solve the long-standing inverse-structure problem of neutron stars:
given an observable or a set of observables, such as the radii of several neutron stars with known masses, infer the necessary EOSs from the observational data. This approach was pioneered by Lindblom \cite{Lee00}, for a recent review, please see ref. \cite{Lee2018}. Some of us have recently made significant efforts in the latter approach \cite{Zhang2019EPJA,Zhang2018,Zhang2018JPG}. By using an explicitly isospin-dependent parameterization of the EOS determined by the Eq. (\ref{eos1}), Eq. (\ref{E0para}) and Eq. (\ref{Esympara}) together,
the TOV and related equations can be solved within multiple do-loops running through the multi-dimensional EOS parameter space.

It is worth noting that the most popular way of parameterizing the high-density EOS is using pieceweise analytical functions in each of n density/pressure domains \cite{Lee00}, e.g., the piecewise polytropic EOSs. While they are sufficient for solving the TOV equations, they have no explicit isospin-dependence and are thus unable to reveal directly information about the nuclear symmetry energy. Also, in many studies one assumes that neutron stars are made of only PNM. Using parameterized PNM EOSs one can also obtain some useful information about the high-density symmetry energy from the polytropes extracted from analyzing astrophysical observations. However, as we shall discuss later, the latest analyses considering all available constraints indicate that the proton fraction in neutron stars can be as high as about 30\% \cite{Zhang2019EPJA}.
The isospin-dependent EOS parameterized by the Eq. (\ref{eos1}), Eq. (\ref{E0para}) and Eq. (\ref{Esympara}) facilitates the inversion of the TOV equations. Around the saturation density, the EOS parameters naturally approach their empirical values, e.g., $E_0(\rho_0)=-16$ MeV, $E_{\rm sym}(\rho_0)=31.7$ MeV~\cite{Oertel17,Li2013}, and $K_0=230$ MeV~\cite{Shlomo2006,Piekarewicz2010,Khan2012}.
Below the core-crust transition density $\rho_t$,  a crust EOS, such as the NV~\cite{Negele1973} and BPS~\cite{Baym1971} EOSs may be used.

Depending on the specific purposes of solving the NS inverse-structure problems, one can select different high-density parameters as variables. Here we present two examples from refs. \cite{Zhang2019EPJA}. In the first example, shown in Fig. \ref{example1}, by setting the $L$ at its currently known most probably value of $L(\rho_0)=58.7$ MeV, the maximum mass $M_{\rm max}=2.01$ M$_\odot$ and the dimensionless tidal deformability $\Lambda_{1.4}=580$ were inverted as two independent observables in the 3D high-density EOS parameter space of $K_{\rm{sym}}-J_{\rm{sym}}-J_0$ spanning the regions of $-800\leq J_0\leq 400$ MeV, $-400 \leq K_{\rm{sym}} \leq 100$ MeV and $-200 \leq J_{\rm{sym}} \leq 800$ MeV~\cite{Zhang2017,Dut12,Dutra2014, Chen2009,Colo2014}.
Within these parameter ranges, the diverse high-density behaviors of \esym can be sampled by the parameterization of Eq. (\ref{Esympara})
as shown in the upper panels of Fig. \ref{Ksymeffect}. The lower panels are the resulting isospin asymmetry profile $\delta(\rho)$ calculated consistently for the $npe\mu$ matter at $\beta$ equilibrium \cite{Zhang2019EPJA}. With the slope $L$ fixed at 58.7 MeV, the $K_{\rm{sym}}$ and $J_{\rm{sym}}$ control the high-density symmetry energy. In the left panels, $K_{\rm{sym}}=-400$, -300, -200, -100, 0, and 100 MeV, while in the right panels $J_{\rm{sym}}=-200$, 0, 200, 400, 600, and 800 MeV, with all other parameters fixed at the values specified in the respective panels, respectively. As the \esym varies broadly at high-densities, the resulting $\delta(\rho)$ at $\beta$ equilibrium changes from values for very neutron-poor matter obtained with very stiff \esym to pure neutron matter when the $E_{\rm{sym}}(\rho)$ becomes super-soft obtained when the $K_{\rm{sym}}$ and/or $J_{\rm{sym}}$ are very small or negative.

We also notice that the parameterization Eq. (\ref{Esympara}) for \esym may not approach zero as $\rho\rightarrow 0$ when its parameters are completely freely varied. This was controlled by introducing additional parameters at nearly zero densities in the meta-modeling of EOS in refs. \cite{MM1,MM2}. While in the 
approach of refs. \cite{Zhang2019EPJA,Zhang2018}, the additional low-density terms are unnecessary as the core-crust transition point was calculated from the uniform liquid-core side and the NV~\cite{Negele1973} and BPS~\cite{Baym1971} EOSs are used for the crust of clustered matter. Nevertheless, this obviously different approaches of handling the low-density limit of nuclear EOS require us to make a few comments about the limitation of the EOS in Eq. (\ref{eos1}) and some issues regarding the symmetry energy (if it can be defined properly) of clustered matter. Firstly, the Eq. (\ref{eos1}) and the associated \esym are for uniform isospin-asymmetric nucleonic matter. It is well known that at low densities below the so-called Mott points, various clusters start forming \cite{Hag12}. One thus has to consider correlations/fluctuations and in-medium properties of clusters in constructing the EOS of stellar matter for astrophysical applications, see, e.g., refs. \cite{Lat-Eos,Shen,Hor06,Joe10,Rop13,Typ14}. Then the EOS in Eq. (\ref{eos1}) is obviously no longer valid. Moreover, there seems to be no need to introduce a symmetry energy of clustered matter for describing its EOS. As discussed in ref. \cite{Li17}, for the clustered matter, because of the different binding energies of mirror nuclei, Coulomb interactions, different locations of proton and neutron drip lines in the atomic chart, the system no longer possesses a proton-neutron exchange symmetry. Also, different clusters have their own local internal isospin asymmetries and average internal densities close to the saturation density of uniform nuclear matter. In fact, in terms of the average density $\rho_{av}$ and the average isospin asymmetry $\delta_{av}$ of the whole non-uniform system, the EOS of clustered matter has been found to have odd terms in $\delta_{av}$ that are appreciable compared to the $\delta^2_{av}$ term \cite{Agr14,Fan14,LWChen}. Thus, in our opinion, it is conceptually ambiguous to define a symmetry energy for clustered matter. Nonetheless, in practice, either the second-order derivative of energy per nucleon $e_{\rm{cluster}}(\rho_{av},\delta_{av})$ in clustered matter with respect to $\delta_{av}$, i.e., $E^{\rm cluster}_{\rm sym}(\rho_{av})\equiv\frac{1}{2}[\partial ^{2}e_{\rm cluster}/\partial \delta_{av} ^{2}]_{\delta_{av} =0}$, or the quantity $E^{\rm cluster}_{\rm sym}(\rho_{av})\equiv 1/2[e_{\rm{cluster}}(\delta_{av}=1)+ e_{\rm{cluster}}(\delta_{av}=-1)-2e_{\rm{cluster}}(\delta_{av}=0)]$ as if the EOS is parabolic in $\delta_{av}$, has been used in extracting the symmetry energy of clustered matter in the literature. This quantity stays finite at the limit of zero average density \cite{Joe10,Rop13,Typ14}. 

\begin{figure*}
\begin{center}
\resizebox{0.49\textwidth}{!}{
 \includegraphics{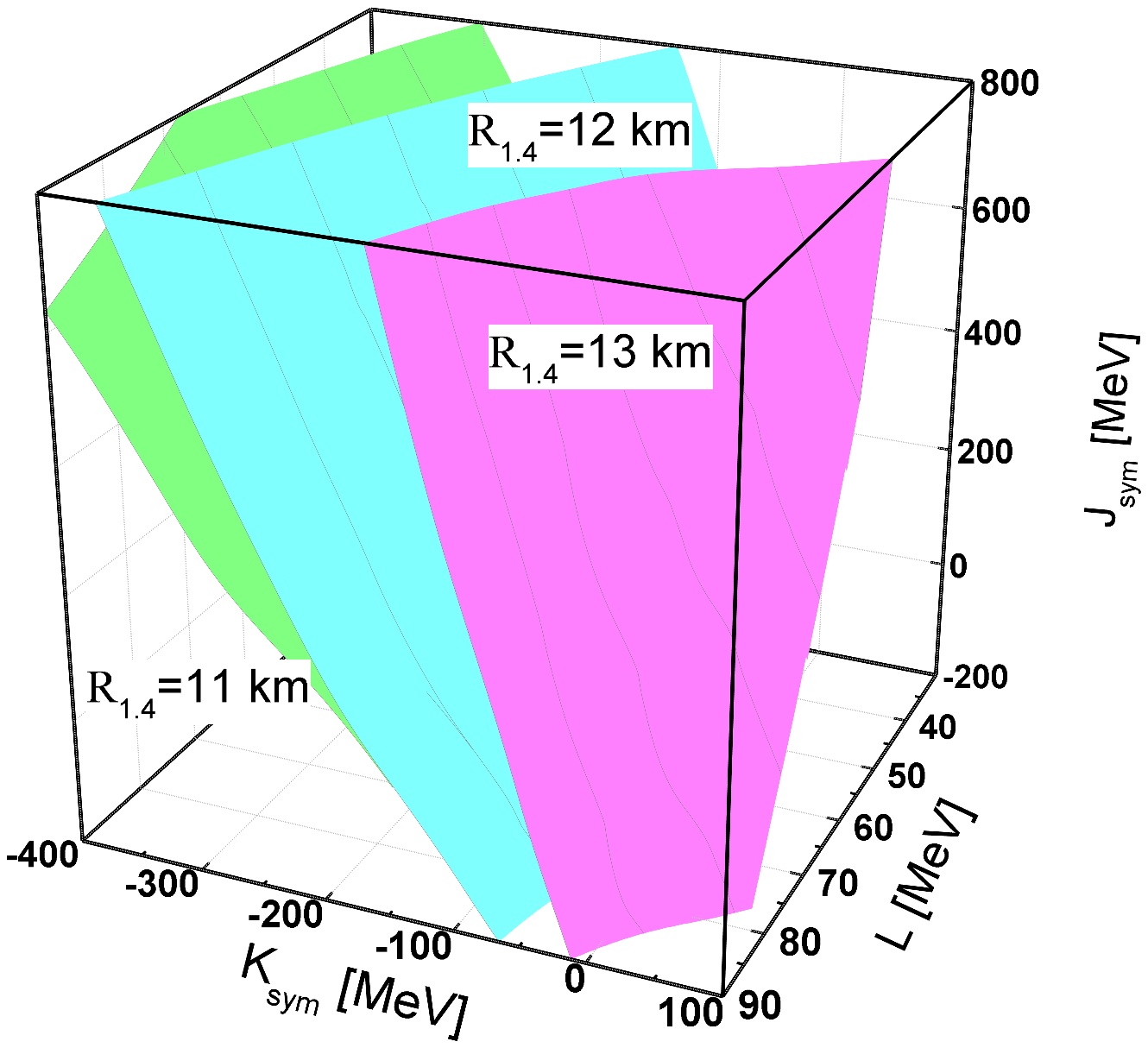}}
\resizebox{0.49\textwidth}{!}{
 \includegraphics{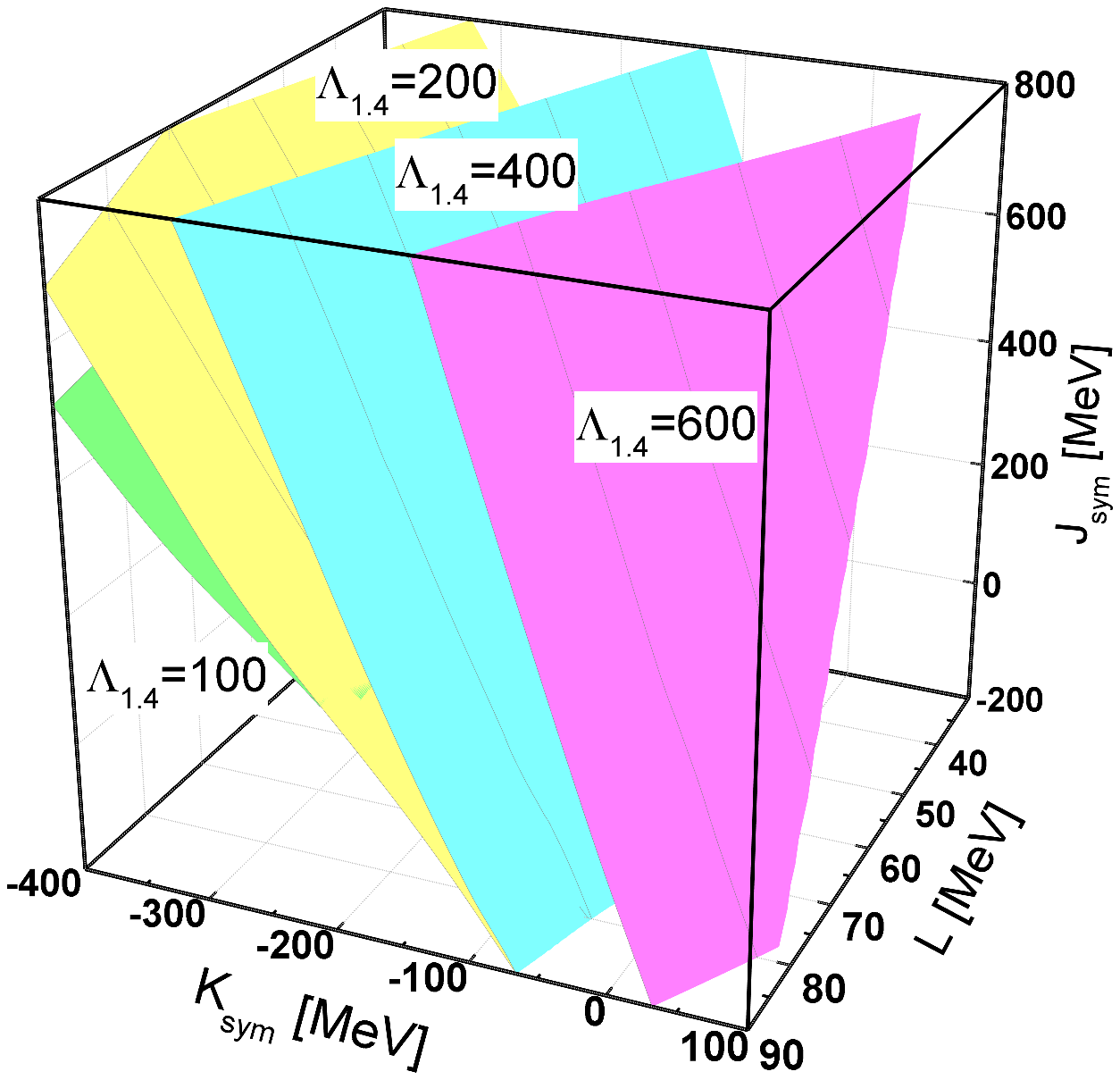}}
\caption{(Color online) Constant surfaces of radius $R_{1.4}$ and tidal deformability $\Lambda_{1.4}$ in the symmetry energy parameter $L-K_{\rm sym}-J_{\rm sym}$ space. Taken from ref.~\cite{Zhang2018JPG}.}
\label{JPGsurface}
\end{center}
\end{figure*}

In the inversion process, the TOV equations and the equations for the tidal deformability were solved inside the
three do-loops. Take $M=2.01$ M$_\odot$ as an example, starting at the point $K_{\rm{sym}}=-400$ MeV and $J_{\rm{sym}}=-200$ MeV, the do-loop in $J_0=J_0+\Delta J_0$ checks if the EOS with a specific value of $J_0$ can be found to produce the maximum mass of $M_{\rm max}=2.01$ M$_\odot$. If the answer is yes, then a point in the $K_{\rm sym}-J_{\rm sym}-J_0$ 3D space is found. Then the do-loops in $K_{\rm{sym}}=K_{\rm{sym}}+\Delta K_{\rm{sym}}$ and $J_{\rm{sym}}=J_{\rm{sym}}+\Delta J_{\rm{sym}}$ are performed subsequently. After running through all possible loops and collecting all points giving $M_{\rm max}=2.01$ M$_\odot$, a constant surface of $M_{\rm max}=2.01$ M$_\odot$ is obtained in the $K_{\rm sym}-J_{\rm sym}-J_0$ 3D parameter space, shown as the green surface in Fig. \ref{example1}. All points on the constant surface corresponding to different EOSs give the same maximum mass of $M_{\rm max}=2.01$ M$_\odot$. In the super-soft \esym regions where the values of $K_{\rm{sym}}$ and/or $J_{\rm{sym}}$ are small/negative, the required value of $J_0$ is high to make the EOS stiff enough to support neutron stars as massive as  $2.01$ M$_\odot$. Similarly, all points on the magenta surface lead to different EOSs giving the same $\Lambda_{1.4}=580$. The surface is rather vertical as
the value of $J_0$ has little effect on the radii and/or tidal deformability. Since $\Lambda_{1.4}=580$ is the maximum value of tidal deformability from the improved analysis by the LIGO and Virgo Collaborations, the vertical surface of $\Lambda_{1.4}=580$ sets an observational boundary from the left for the 3D high-density EOS parameter space, while the constant surface of $M_{\rm max}=2.01$ M$_\odot$ limit the parameter space from below. Their cross line determines a correlation among the three parameters along the south-west boundary of the EOS parameter space.

Since the high-density SNM EOS parameter $J_0$ has little effect on the radii and tidal deformability, by fixing it at a value large enough to support $M_{\rm max}=2.01$ M$_\odot$, one may
explore constant surfaces of radius $R_{1.4}$ and tidal deformability $\Lambda_{1.4}$ in the 3D symmetry energy parameter space $L-K_{\rm sym}-J_{\rm sym}$ as shown in Fig.~\ref{JPGsurface}.
These surfaces were obtained in the same way as in Fig.~\ref{example1} but with a fixed $J_0=-180$ MeV.
Each point on a given surface can generate a EOS satisfying the specified observational constraint. As one expects, there are large degeneracies. A given value of the observable $R_{1.4}$ or $\Lambda_{1.4}$ is not sufficient to completely determine the three parameters but can constrain their combinations to a surface. Indicated by the largely vertical orientations of the constant surfaces of $R_{1.4}$ or $\Lambda_{1.4}$, the high-order \esym parameter $J_{\rm sym}$ plays little role in determining the radius and tidal deformability of canonical neutron stars. Similar orientations of the constant surfaces of $R_{1.4}$ and $\Lambda_{1.4}$ in the same region of 3D \esym parameter space indicate that they are strongly correlated as we shall discuss later.

Not only observables, physical requirements, such as the causality condition, can also be inverted in the 3D EOS parameter space. While some perturbative QCD theories have predicted that the speed of sound $(v_s/v_c)^2\leq1/3$ at extremely high densities~\cite{Cherman2009,Kurkela2010,Borsanyi2012,Bedaque2015,Ecker2017,Tews2018b}, the causality condition $(v_s/v_c)^2\leq1$ is widely used in constraining the EOS of neutron star matter, see, e.g., ref. \cite{Larry} as one of the latest examples. As we shall discuss later, the inversion technique also helps determine the absolutely maximum mass of neutron stars by examining the maximum mass obtained on the causality surface.

\subsection{Predicted effects of nuclear symmetry energy on properties of neutron stars in the dawn of gravitational wave astronomy}
It is useful to first review a few relevant earlier work on this topic. In particular, using the traditional approach, i.e, given an EOS, one solves the TOV equations to predict a sequence of mass-radius relation and tidal deformability that can be compared to observations. While almost every available interaction/EOS has been used in such a way, some predictions using EOSs constrained by terrestrial data are particularly useful. Moreover, by studying correlations of the mass-radius relations with the EOS parameters used, some interesting predictions were made.
\begin{figure}[tbh]
\begin{center}
\resizebox{0.47\textwidth}{!}{
\includegraphics[angle=0]{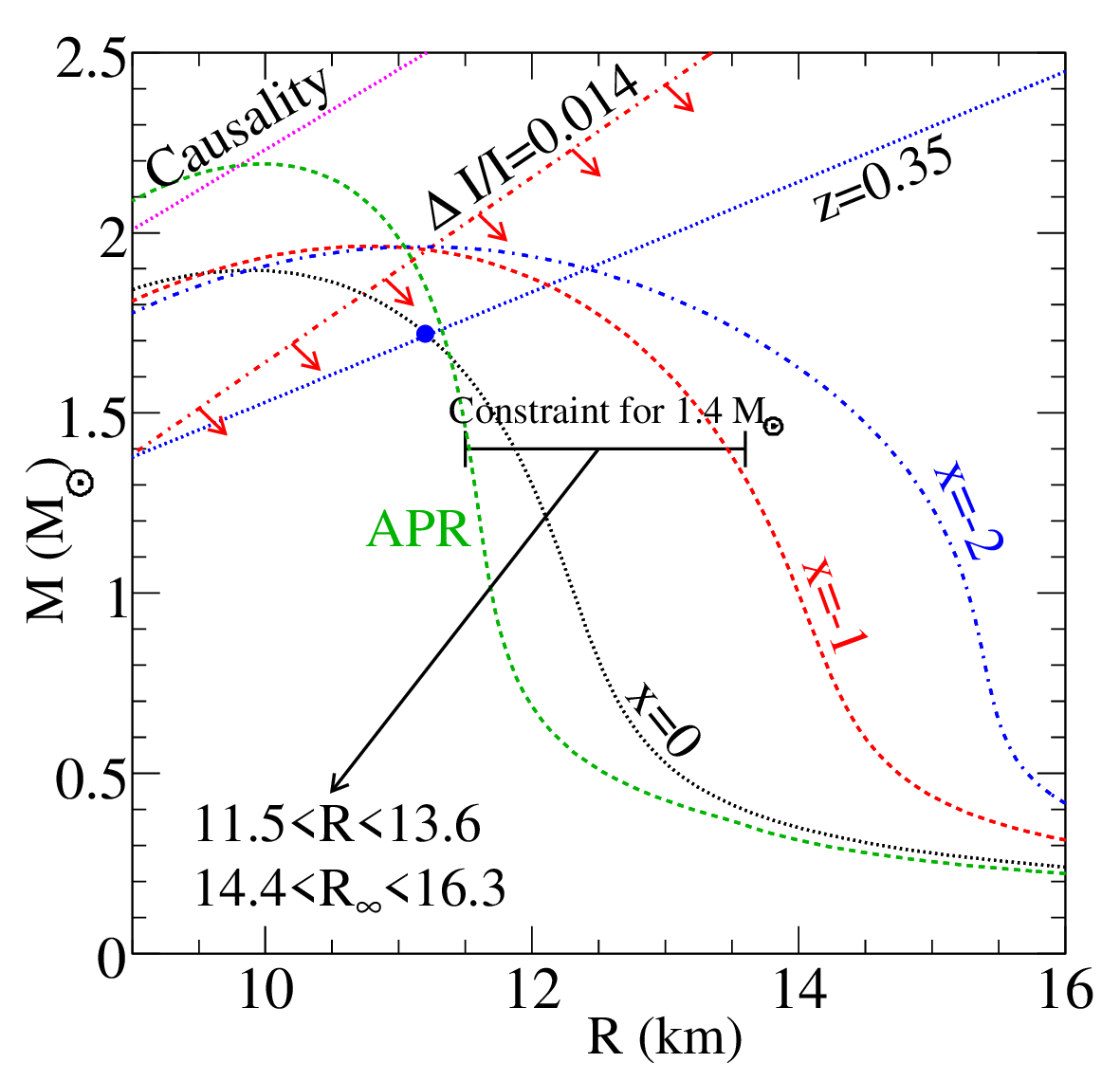}
}
\caption{{\protect (Color online) The mass-radius curves using the MDI interactions with $x=0,-1,$ and $-2$
and the APR EOS. Analyses of isospin diffusion in heavy-ion reactions and the size of neutron skin in $^{208}$Pb available up to 2006 limits the x-parameter to between $x=0$ and $x=-1$. The inferred radius of a 1.4 solar mass neutron star is between 11.5 and 13.6 km (the corresponding radiation radius $R_{\infty}$ between 14.4 and 16.3 km). Taken from ref. \cite{Li2006}. }}\label{LiSteiner}
\end{center}
\end{figure}
\begin{figure}[tbh]
\begin{center}
\resizebox{0.47\textwidth}{!}{
\includegraphics[angle=-90]{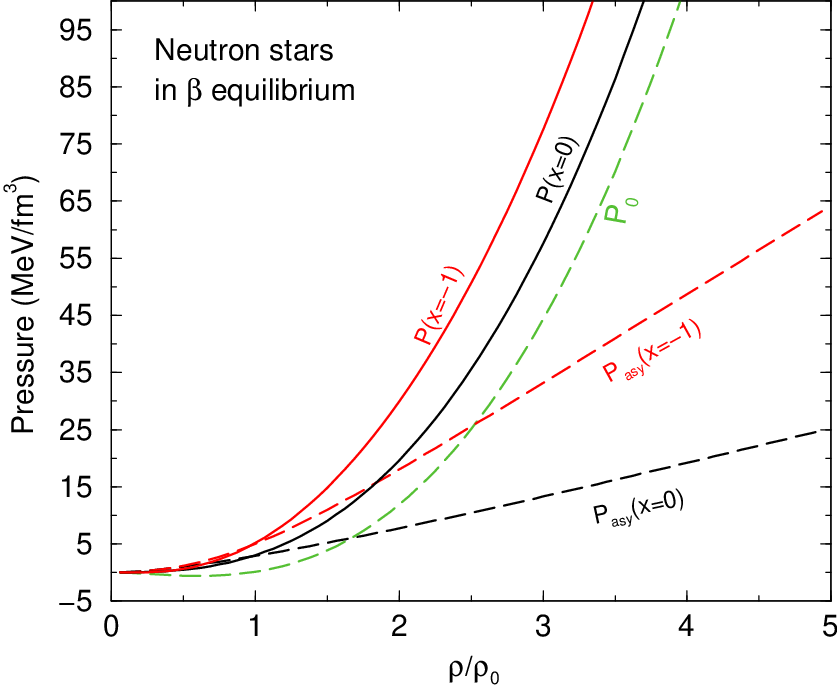}
}
\caption{{\protect (Color online) The pressure $p_0$ of symmetric nuclear matter and symmetry energy contribution $p_{asy}$ to the total pressure in neutron stars at $\beta$ equilibrium using the MDI interaction with the x-parameter $x=0$ (black) and $x=-1$ (red), respectively. Taken from ref. \cite{Li2006p}.
}}\label{Li-Mexico}
\end{center}
\end{figure}
\subsubsection{Predicted radii of canonical neutron stars using \esym constrained by isospin diffusion experiments in terrestrial nuclear laboratories}\label{MDI-p}
Essentially all available EOSs have been used in one way or another to predict the mass-radius correlation of neutron stars. Most of the earlier studies have focused on exploring effects of varying the saturation properties of SNM EOS, new degrees/particles or phase transitions at high densities. Effects of varying the \esym have also been studied extensively. As an example from 2006 \cite{Li2006},
shown in Fig. \ref{LiSteiner} are the mass vs. radius curves using the MDI EOSs with $x=0, -1$ and $x=-2$ having the same
compressibility ($K_0=211$ MeV) but different \esym shown in Fig. \ref{Esym-Xu2}. The APR EOS has a
compressibility of $K_0=269$ MeV but almost the same symmetry energy
as with $x=0$ is also shown for comparisons. As shown in Fig. \ref{RHF-fig} and the associated discussions earlier, the isospin diffusion data from NSCL/MSU together with the fiducial value of
$R_{skin}\equiv\Delta r_{np}=0.2\pm 0.04$ fm for the neutron-skin in $^{208}$Pb limit the x value to between $x=0$ and $x=-1$. This then restricts the radius for a 1.4 M$_\odot$ neutron star (and the corresponding radiation radius $R_{\infty}$) to the range of 11.5 km and 13.6 km (or $R_{\infty}$ between 14.4 km and 16.3 km).

The results shown in Fig. \ref{LiSteiner} clearly indicate that the observed maximum mass of neutron star constrain mostly the SNM EOS while the radii of neutron star constrain mostly the density dependence of nuclear symmetry energy. This finding was explained in ref. \cite{Li2006p} by studying the relative contributions from the SNM EOS and symmetry energy to the total pressure in neutron stars at $\beta$ equilibrium. The latter in $npe$ matter before muons appear in NSs can be written as
\begin{equation}\label{pre-npe}
  P(\rho, \delta)=\rho^2[\frac{dE_0(\rho)}{d\rho}+\frac{dE_{\rm{sym}}(\rho)}{d\rho}\delta^2]+\frac{1}{2}\delta (1-\delta)\rho E_{\rm{sym}}(\rho)
\end{equation}
where the density profile of isospin asymmetry $\delta(\rho)$ at $\beta$ equilibrium is uniquely determined by the
density dependence of symmetry energy \esym as we have discussed earlier.
The pressure $P_0$ of SNM from the first term in Eq. (\ref{pre-npe}) and the isospin-asymmetric pressure $P_{\rm asy}$ from the last two terms as well as their sum $P$ are shown in Fig. \ref{Li-Mexico} for the MDI interaction with the x-parameter $x=0$ (black) and $x=-1$ (red), respectively.
It is seen that in the density region around $\rho_0\sim 2.5\rho_0$, the isospin-dependent pressure $P_{\rm asy}$ dominates over the $P_0$ from SNM EOS, while the exact transition of dominance from $P_{\rm asy}$ to $P_0$ depends on the stiffnesses of both the SNM EOS and the symmetry energy.
The radii of neutron stars are known to be determined by the pressure at densities around $\rho_0\sim 2\rho_0$ \cite{Lat01}. They are thus sensitive to the density dependence of nuclear symmetry energy in this density region. At higher densities, while the $P_{\rm asy}$ may still contribute significantly, the total pressure is dominated by the $P_0$ from SNM EOS. The latter at densities reached in the core determine the maximum mass of neutron stars that can be supported. Thus, the observed maximum mass of neutron stars constrain mostly the stiffness of SNM EOS. Nevertheless, because the symmetry energy not only affects directly the pressure but also the composition profile $\delta(\rho)$ of neutron star matter, as we shall discuss in more detail, the observed maximum mass of neutron stars can also help constrain the high-density behavior of nuclear symmetry energy.

\begin{figure}
\begin{center}
\resizebox{0.5\textwidth}{!}{
  \includegraphics{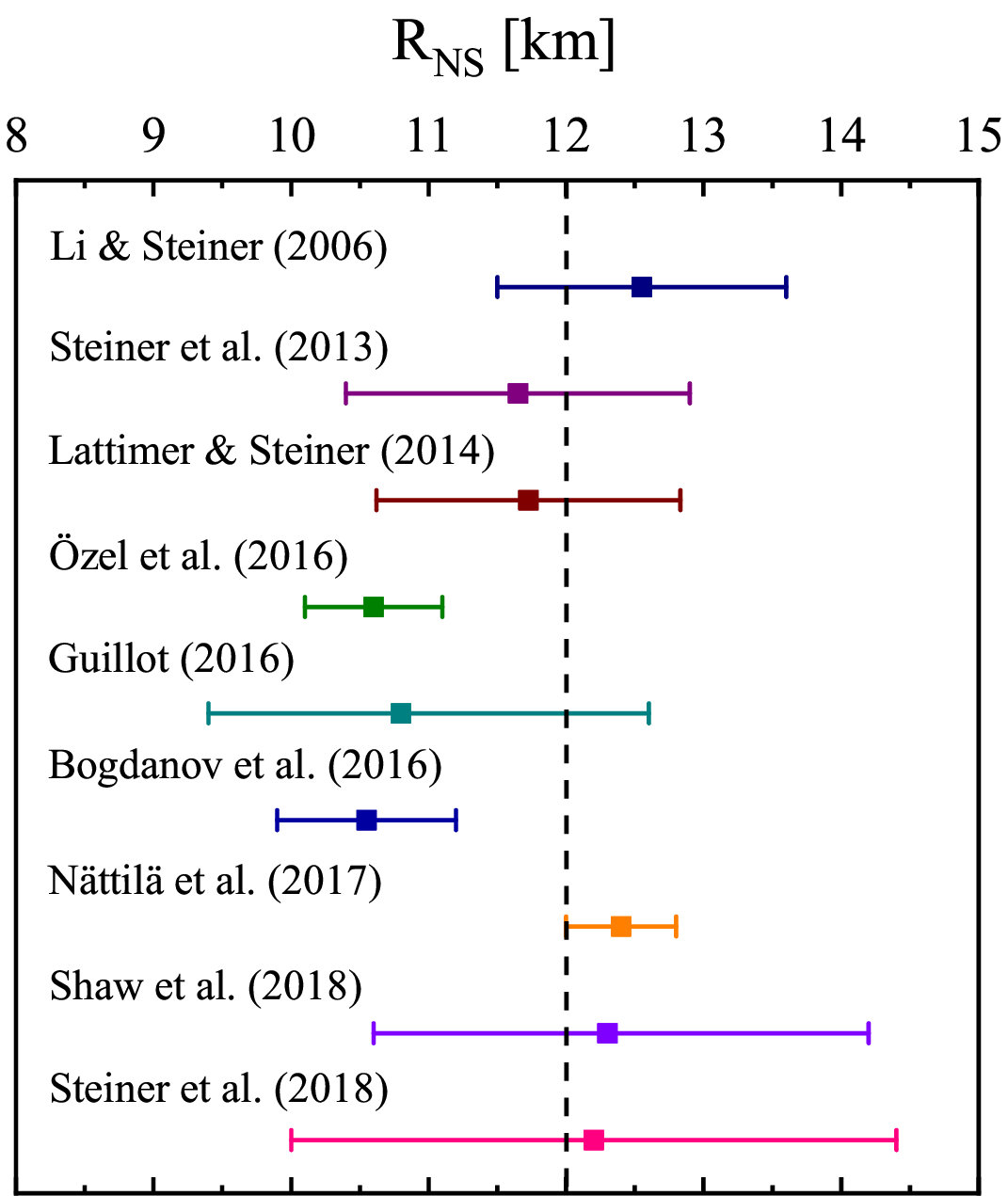} }
  \vspace{0.5 cm}
  \caption{(color online) The radii of canonical neutron stars $R_{\rm NS}$ extracted from analyses of LMXB~\cite{Lattimer2014,Ozel2016,Guillot2013,Steiner2013,Guillot2014,Guillot2016,Bogdanov2016,Shaw2018,Steiner2018} in comparison with a prediction based on the MDI EOS constrained by terrestrial nuclear laboratory data~\cite{Li2006}. }\label{radiusLMXB}
  \end{center}
\end{figure}
The predicted ``nuclear" limit on the radii in in Fig. \ref{LiSteiner} is consistent with most of the results extracted from analyzing both X-ray and gravitational wave data since 2006. The X-ray bursts from accreting neutron stars in low-mass X-ray binary (LMXB) systems provide potential possibilities to constrain the mass and radius simultaneously~\cite{Lattimer2014,Lewin1993,Strohmayer2006,Miller2016,Suleimanov2016,Nattila2017}.
Two methods have been used to infer the mass and radius from X-ray bursts of LMXB: the touchdown method \cite{Ozel2016,Ozel2006,Ozel2009,Guver2010} and cooling tail method~\cite{Suleimanov2011,Poutanen2014,Nattila2016,Suleimano2017}. In analyzing the observed spectral fluxes, there are several well-known challenges. Consequently, there are still large uncertainties associated with modeling the neutron star atmosphere especially its composition, determining the distance, the column density of X-ray absorbing materials, and the surface gravitational redshift. They all contribute to the uncertainties in determining the radii of neutron stars. Nevertheless, as summarized in ref.~\cite{Lukasik2018} and Fig.~\ref{radiusLMXB}, several constraints on the radii of neutron stars have been put forward in recent years: $10.4\leq R_{1.4}\leq12.9$ km~\cite{Steiner2013}, $10.62\leq R_{1.4}\leq12.83$ km~\cite{Lattimer2014}, $10.1\leq R_{1.5}\leq11.1$ km~\cite{Ozel2016}, $9.0\leq R_{\rm NS}\leq12.2$ km~\cite{Guillot2013,Guillot2014,Guillot2016}, $9.9\leq R_{1.5}\leq11.2$ km~\cite{Bogdanov2016}, $10.6\leq R_{1.4}\leq14.2$ km~\cite{Shaw2018}, $10\leq R_{1.4}\leq14.4$ km~\cite{Steiner2018}, and the radius of 4U 1702-429 $R = 12.4 \pm 0.4$ km~\cite{Nattila2017}.
For comparisons, the prediction of $11.5\leq R_{1.4}\leq13.6$ km \cite{Li2006} using the EOS constrained by the nuclear laboratory data is also shown in Fig. ~\ref{radiusLMXB}. It is seen that the predicted radius is larger than the results of some early X-ray bursts analyses but consistent with the two latest analyses. As we will see in the next subsection, the  prediction is in very good agreement with the majority of post-GW170817 analyses.

\begin{figure}[ht]
\begin{center}
\resizebox{0.55\textwidth}{!}{
  \includegraphics{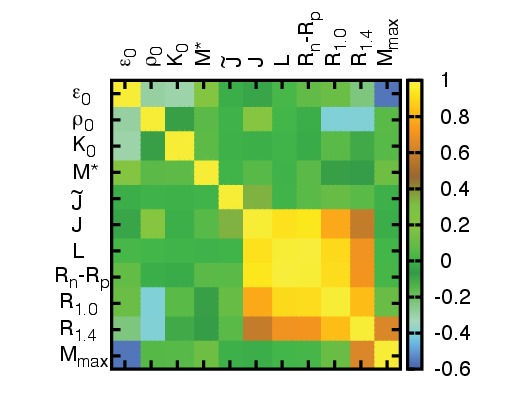}
  }
\caption{(Color online) Color-coded plot of 55 independent correlation coefficients among 11 physical variables relevant to the structure of neutron stars within the RMF model with the FSUGold interaction. Taken from ref. \cite{Farrooh11}.}
\label{F-Fig1}
\end{center}
\end{figure}

\begin{figure*}[ht]
\resizebox{0.49\textwidth}{!}{
  \includegraphics{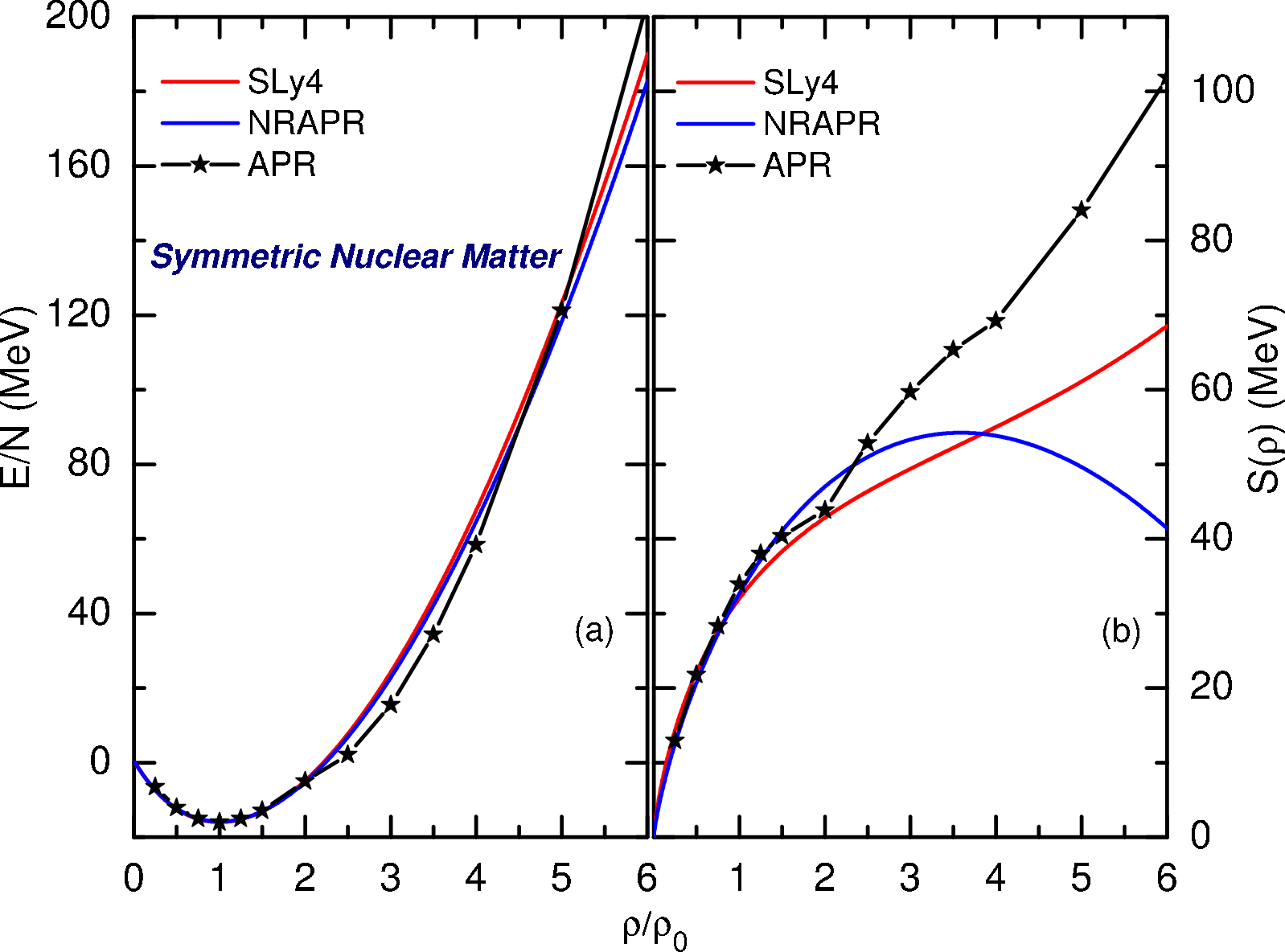}
  }
  \resizebox{0.49\textwidth}{!}{
  \includegraphics{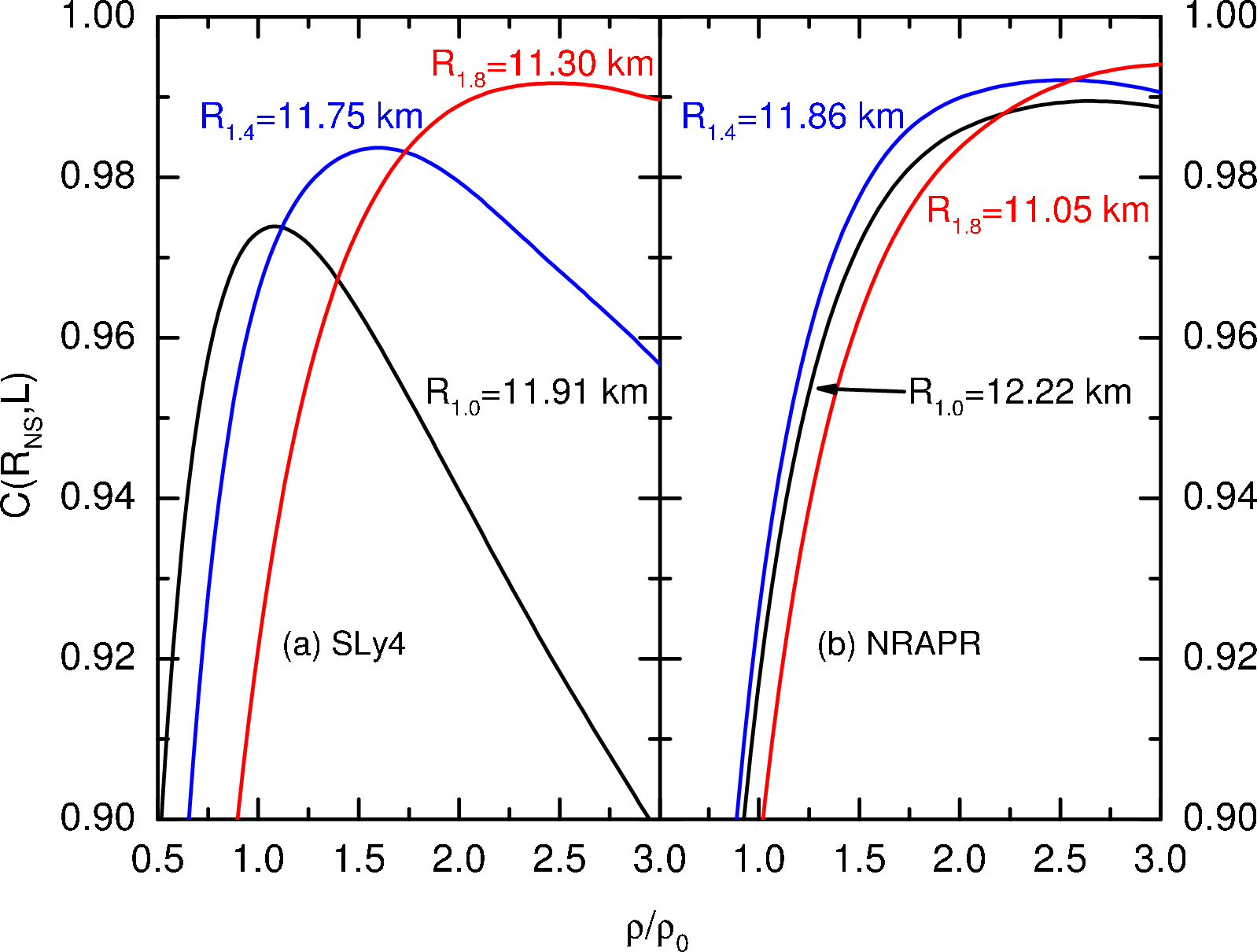}
}
\caption{(Color online) (Left) The EOS in SNM (a) and the symmetry energy (b) as functions of the reduced density $\rho/\rho_0$ for SLy4 and NRAPR Skyrme energy density functionals. The APR EOS and \esym are also shown as references. (Right) Pearson's correlation coefficients between $L(\rho)$ and the radii of neutron stars with masses of $1.0$-, $1.4$-, and $1.8$-solar mass as a
function of density calculated using the two EOSs shown in the left 2 windows. Taken from ref. \cite{Farrooh-skin}.}
\label{F-Fig2}
\end{figure*}

\subsubsection{Predicted correlation strength between the radii of neutron stars and the symmetry energy from low to high densities}\label{MDI-p}
It is well known that the mass-radius relation may be affected by several components of the EOS. A typical plot, such as the one in Fig. \ref{LiSteiner}, of mass versus radius with different EOSs does not tell accurately which part/property of the EOS is at work. Interestingly, covariant analyses with multi-parameters and multi-observables have been advanced to identify the most important model parameters and quantify the uncertainties of observables as well as their correlations \cite{PGR}. Whiile these analyses are not completely model independent, they are very informative. For example,
shown in Fig. \ref{F-Fig1} are color-coded 55 correlation coefficients for
11 variables from a study by Fattoyev and Piekarewicz \cite{Farrooh11} using their RMF model with the FSUGold interaction \cite{Gold}.
Among the 11 variables, the following 7 characterizing the EOS are directly related to their model parameters:
the binding energy $\varepsilon_{0}$, incompressibility $K_{0}$, nucleon Dirac effective mass $M_{0}^{\star}$ in SNM, the magnitude $J$ and slope $L$ of symmetry energy at saturation
density $\rho_{0}$ and the symmetry energy $\tilde{J}$ evaluated at $\rho\!\approx\!0.1\,{\rm fm}^{-3}$, while the following 4 are predicted observables:  the neutron-skin $R_{\rm skin}$ in $^{208}$Pb, the maximum neutron-star mass $M_{\rm max}$, radii $R_{1.0}$ and $R_{1.4}$ of neutron stars with masses of 1.0M$_{\odot}$ and 1.4M$_{\odot}$, respectively. Interestingly, the radii of neutron stars show the well-known and strong correlation with the $R_{\rm skin}$ \cite{Horowitz2001}.

Among the 3 variables characterizing the \esym below and around the saturation density, the slope $L$ has the strongest correlation with both
$R_{1.0}$ and $R_{1.4}$. While the magnitude $J$ of \esym at $\rho_0$ also has strong correlation with $R_{1.0}$ and $R_{1.4}$, the low density symmetry energy $\tilde{J}$ at $\rho\!\approx\!0.1\,{\rm fm}^{-3}$ has little influence on the radii. It is also seen that the maximum mass $M_{\rm max}$ has a very weak correlation with the parameters characterizing the symmetry energy near and below the saturation density. While these findings are very useful, up to which points these correlations are model independent remains an open question. Generally speaking, the quantitative values of the correlation matrix elements shown are likely model dependent, while the general trends are likely model independent. As shown in ref. \cite{MM3}, both the values and correlations of the low-order parameters in expanding the EOS in Taylor series depend strongly on the very poorly known high-order parameters. Since the later are largely unconstrained experimentally and are known to be model dependent, the correlations among neutron star observables and the low-order EOS parameters are therefore expected to be model dependent quantitatively. 

It is known qualitatively that the radii of neutron stars are sensitive to the behavior of \esym from around $\rho_0$ to $2\rho_0$ \cite{Lat01}. While very informative, the correlation coefficients shown in Fig. \ref{F-Fig1} do not tell at what densities the \esym for neutron stars of different masses is most important for determining their radii. Thus, extending the study of ref. \cite{Farrooh11}, Fattoyev {\it et al.} have later studied the Pearson's correlation coefficient between the neutron star radii and the slope $L(\rho)$ of \esym as a function of density $\rho$
using the SLy4 and NRAPR Skyrme energy density functionals \cite{Farrooh-skin}. As shown in the left 2 windows of Fig. \ref{F-Fig2}, the two interactions with the default values of their parameters lead to almost identical EOS for SNM and symmetry energy \esym for densities up to about $1.5\rho_0$. However, they have quite different behaviors at higher densities reachable in the cores of neutron stars. The Pearson's correlation coefficients between the neutron star radii and the $L(\rho)$ as a function of density were calculated by allowing the isovector effective mass $m_{\rm v}^{\ast}(\rho_0)$ and the symmetry-gradient coefficient $G_{\rm v}$ to have a 20\% theoretical error-bars while fixing all isoscalar parameters at their default values~\cite{Farrooh-skin}.
In the case of SLy4, it is seen that the radius of a 1.0M$_{\odot}$ neutron star has a strong correlation with the $L(\rho_0)$. For heavier neutron stars, the strongest correlation shifts to the $L(\rho)$ at higher densities, {\sl e.g.} at $1.5\rho_0$ for a 1.4M$_{\odot}$ neutron star, and at $2.5\rho_0$ for a 1.8M$_{\odot}$ neutron star. Moreover, the correlation coefficient remains almost
flat for higher densities in a 1.8M$_{\odot}$ neutron star \cite{Farrooh-skin}. Comparing the results from using the  SLy4 and NRAPR interactions, it is seen that the different high density behaviors of \esym affect the correlation coefficient for more massive neutron stars. In the case of NRAPR, the radius-$L(\rho)$ correlation depends weakly on the mass. There is no strongly pronounced peak in the correlation coefficient. Instead, the \esym in a broad range above about $1.5\rho_0$ has approximately equally important influence on the radii of all neutron stars \cite{Farrooh-skin}. It is also interesting to mention that the $L(\rho)$ can be decomposed into several components related to the fundamental properties of nuclear interaction according to the HVH theorem as we discussed in Sect. \ref{s-HVH}.
Indeed, it was found that the strongest contributions to the radius-$L(\rho)$ correlation come from the magnitude and momentum dependence of the symmetry potential due to the finite-range of isovector
interactions \cite{Farrooh-skin}.

In short, it is well known that the radii of neutron stars are strongly correlated with the density dependence of nuclear symmetry energy. Covariance analyses have quantified the strength and identified relevant density range of this correlation. Almost all EOSs of neutron-rich matter available in the literature have been used to predict the mass-radius correlations. Interestingly, some earlier predictions for the radius of canonical neutron stars using constraints on the \esym provided by terrestrial nuclear laboratory data have found strong supports in analyses of both X-ray and gravitational wave data as we shall discuss in detail next.

\begin{figure*}
\begin{center}
\resizebox{0.51\textwidth}{!}{
   \includegraphics{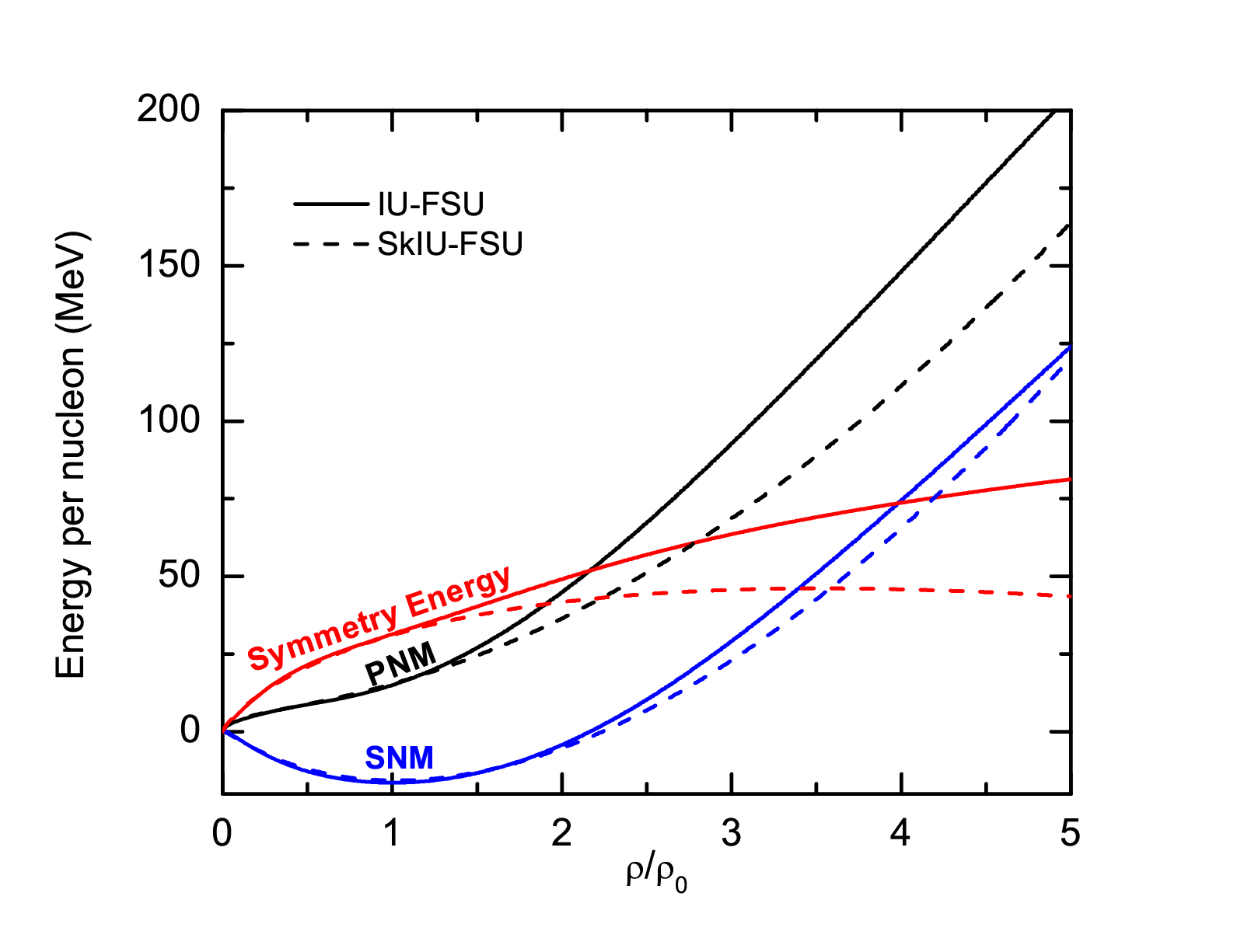} }
  \resizebox{0.445\textwidth}{!}{
   \includegraphics{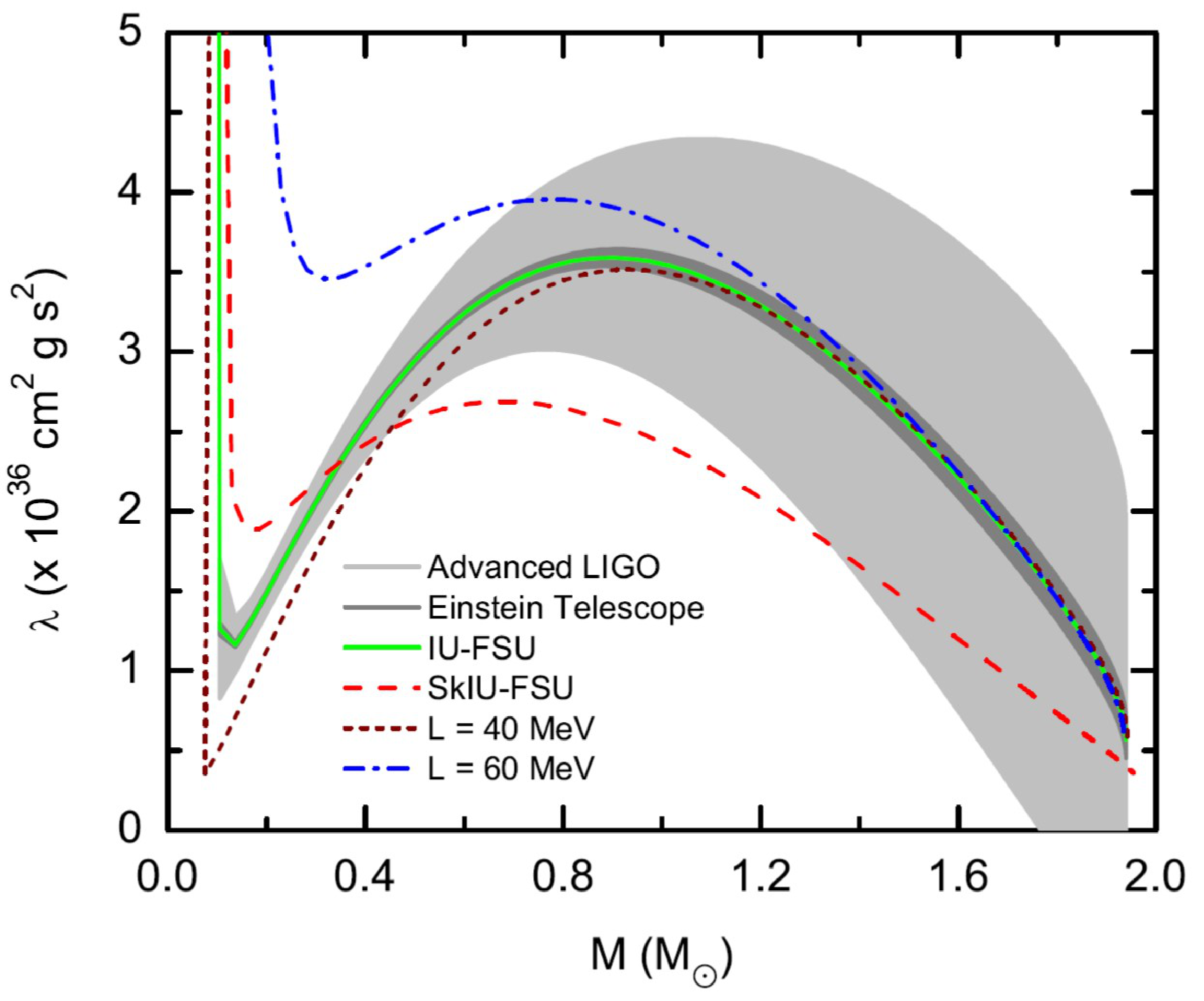} }
  \caption{(color online) Left: The EOS of SNM and PNM as well as the
symmetry energy as a function of density obtained within the IU-FSU
RMF model and the SHF approach using the SkIU-FSU parameter set.
  Right: Tidal deformability $\lambda$ of a single neutron star as a function of mass with the EOSs and symmetry energy shown in the left window. The shaded light-grey area and dark-grey area show a crude estimate of uncertainties in measuring $\lambda$ for equal mass binaries at a distance of $D = 100$ Mpc for the Advanced LIGO and the Einstein Telescope, respectively. Taken from ref.~\cite{Fattoyev2013}. }\label{fatt}
  \end{center}
\end{figure*}
\subsubsection{Predicted effects of the symmetry energy on the tidal deformability of neutron stars}
Once the EOS is given, the tidal deformability can be calculated. In turn, accurately measured tidal deformability can constrain the EOS and related symmetry energy. For example,
Fattoyev et al. \cite{Fattoyev2013} studied in 2013 effects of different parts of \esym on the tidal deformability. They constructed two
EOSs using the IU-FSU RMF model and SkIU-FSU SHF model. As shown in the left window of Fig.~\ref{fatt},  the two models have the same EOS for SNM
and PNM around and below $\rho_0$, thus the same \esym at and below $\rho_0$. However, they have very different behaviors
above about $1.5\rho_0$ with the IU-FSU leading to a much stiffer \esym at high densities. More quantitatively, the \esym of IU-FSU is $40-60\%$ higher in the density range of $\rho/\rho_0=3-4$.
The right window shows the resulting tidal deformability $\lambda$ of a single neutron star as a function of mass. The shaded light-grey area and dark-grey area show a crude estimate of uncertainties in measuring $\lambda$ for equal mass binaries at a distance of $D = 100$ Mpc for the Advanced LIGO and the Einstein Telescope, respectively. It is interesting to see that the $\lambda$ of canonical and more massive neutron stars is quite sensitive to the high-density behavior of the symmetry energy. However, only the $\lambda$ of light neutron stars is sensitive to the variation of the $L$ parameter.

As we shall discuss next, after the GW170817 event, much efforts have been devoted to constraining the EOS or related model parameters by comparing various calculations first with the upper limit from the  original analysis than later the range of tidal deformability from the improved analyses reported by LIGO and Virgo Collaborations. A number of these studies have examined effects of symmetry energy. Some of them have extracted constraints on the parameter $L$. We emphasize here that the $L$ only describes the \esym around saturation density. While the $L$ is often used to label the density dependence of nuclear symmetry energy, the results shown in  Fig.~\ref{fatt} indicates that there is likely a large degeneracy between the $L$ and the tidal deformability of canonical neutron stars. Indeed, as shown in Fig. \ref{JPGsurface}, many combinations of $L$, $K_{\rm sym}$ and $J_{\rm sym}$ can give the same tidal deformability. Thus, conclusions about $L$ drawn from comparing calculated and observed tidal deformability are all conditional depending on especially the $K_{\rm sym}$ used and thus should be taken cautiously.

\begin{figure}[h!]
\begin{center}
\resizebox{0.55\textwidth}{!}{
  \includegraphics{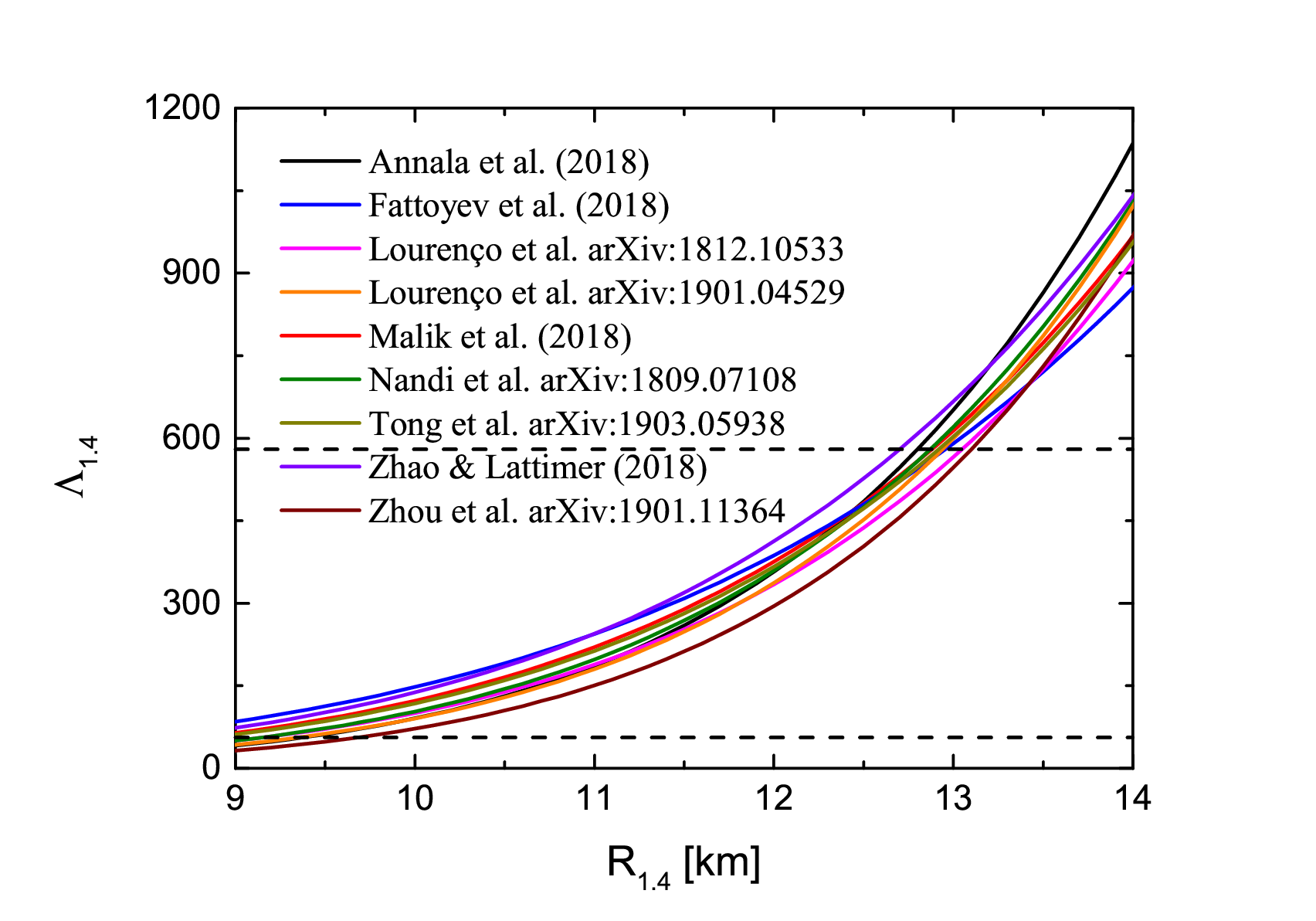}}
  \caption{(color online) Constraints on the relationship between $R_{\rm 1.4}$ and $\Lambda_{\rm 1.4}$. The line labeled by Zhao \& Lattimer (2018) is the suggested relationship between $R_{1.4}$ and $\tilde{\Lambda}$~\cite{Zhao2018}. The refined limits $70\leq\Lambda\leq580$ from LIGO and Virgo Collaborations in ref.~\cite{LIGO2018} are shown as dashed horizontal lines.}\label{Lambdaradius}
  \end{center}
\end{figure}
\subsection{Post-GW170817 analyses of tidal deformability and radii of neutron stars as well as constraints on the nuclear EOS and symmetry energy}
The GW170817 event~\cite{LIGO2018,LIGO2017} opened the era of gravitational wave astronomy. Stimulated by the observational data of GW170817, many interesting studies have been carried out to extract the radii of neutron stars and constraints on the EOS. Here we focus on a few aspects most relevant to constraining the density dependence of nuclear symmetry energy.
For ease of the following discussions, we recall here the notations used in the literature for the tidal deformability.
The dimensionless tidal deformability $\Lambda$ is related to
the compactness parameter $\beta\equiv R/M$ and the Love number $k_2$ through
\begin{equation}\label{LLL}
\Lambda = \frac{2}{3}\frac{k_2}{\beta^5}.
\end{equation}
And the mass-weighted (dimensionless) tidal deformability in an inspiraling binary system is given by
\begin{equation}
\tilde{\Lambda} = \frac{16}{13}\frac{(M_1+12M_2)M_1^4\Lambda_1+(M_2+12M_1)M_2^4\Lambda_2}{(M_1+M_2)^5}
\end{equation}
where $\Lambda_1=\Lambda_1(M_1)$ and $\Lambda_2=\Lambda_2(M_2)$ are the tidal
deformabilities of the individual binary components.

\subsubsection{GW170817 implications on the radii of neutron stars}\label{GW-LR}
Special efforts and attention have been devoted to extracting the radius $R_{\rm 1.4}$ from the reported range of tidal deformability of canonical neutron stars based on the analyses of GW170817 event~\cite{LIGO2018,Nandi2018,Lourenco2019,Raithel2018,Fattoyev2018,Malik2018,Zhou2019,Most,Annala2018,Lim2018,Radice2018,Tews2018,De2018,Bauswein2017,PKU-Meng}.
According to the definition of tidal deformability, i.e., Eq.~(\ref{lambda}) or Eq. (\ref{LLL}), an underlying relationship exists between $\Lambda$ and $R$ when the mass is fixed. However, the exact relationship is EOS dependent. This is because the $k_2$ also depends on the radius. The constraints on the relationship between $R_{\rm 1.4}$ and $\Lambda_{\rm 1.4}$ from many studies in the literature are summarized in Fig.~\ref{Lambdaradius}. The refined constraints $70\leq\Lambda\leq580$ from ref.~\cite{LIGO2018} are shown as dashed horizontal lines.
We notice here that a very recent study in ref.~\cite{Kiuchi2019} reevaluated the lower boundary of the tidal deformability $\tilde{\Lambda}$ by using information from the kilonova, AT 2017gfo, and found that $\tilde{\Lambda}\leq242$ if the mass ejection from the remnant is $0.05$ M$_\odot$. Indeed, several other studies have also emphasized the importance of better determining the lower boundary of the tidal deformability.

Overall, the reported $R_{\rm 1.4}$ versus $\Lambda_{\rm 1.4}$ relations spread within a narrow band. The upper ($\Lambda^{\mathrm{UL}}_{1.4}$) and lower ($\Lambda^{\mathrm{LL}}_{1.4}$) boundaries of this band can be well approximated by
\begin{eqnarray}
\Lambda^{\mathrm{UL}}_{1.4}&=&1.06\times10^{-6}\cdot R_{1.4}^{7.85}+71.62\\
\Lambda^{\mathrm{LL}}_{1.4}&=&1.24\times10^{-5}\cdot R_{1.4}^{6.87}-20.49.
\end{eqnarray}
More quantitatively about the typical analyses, using a large number of EOSs from RMF and SHF theories, studies in refs.~\cite{Nandi2018,Lourenco2019,Fattoyev2018,Malik2018,Zhou2019,Lourenco2018,PKU-Meng} obtained the following relationship: $\Lambda_{\rm 1.4}=1.53\times10^{-5}R_{\rm 6.83}^{7.5}$, $\Lambda_{\rm 1.4}=5.87\times10^{-6}R_{\rm 1.4}^{7.19}$, $\Lambda_{\rm 1.4}=7.76\times10^{-4}R_{\rm 1.4}^{5.28}$, $\Lambda_{\rm 1.4}=9.11\times10^{-5}R_{\rm 1.4}^{6.13}$, $\Lambda_{\rm 1.4}=1.41\times10^{-6}R_{\rm 1.4}^{7.71}$, $\Lambda_{\rm 1.4}=2.65\times10^{-5}R_{\rm 1.4}^{6.58}$, and $\Lambda_{\rm 1.4}=7.29\times10^{-5}R_{\rm 1.4}^{6.21}$, respectively. Similarly, studies in ref.~\cite{Annala2018} suggested the relationship $\Lambda_{\rm 1.4}=2.88\times10^{-6}R_{\rm 1.4}^{7.5}$ based on the parameterized EOSs that interpolate between theoretical results at low and high baryon densities. In addition, the relationship $R_{1.4}=(13.4\pm0.1)(\tilde{\Lambda}/800)^{1/6}$ was suggested in ref.~\cite{Zhao2018}. However, a linear relationship between $R_{\rm 1.4}$ and $\Lambda_{\rm 1.4}$ appears when the parameters of \esym are restricted to their current uncertainty ranges in ref. \cite{Zhang2018JPG}. Moreover, as shown in Fig.~6 of ref.~\cite{Nandi2018}, the $R_{\rm 1.4}\sim\Lambda_{\rm 1.4}$ relationship also differs for EOSs with or without the hadron-quark phase transition.

\begin{figure}
\begin{center}
\resizebox{0.5\textwidth}{!}{
  \includegraphics{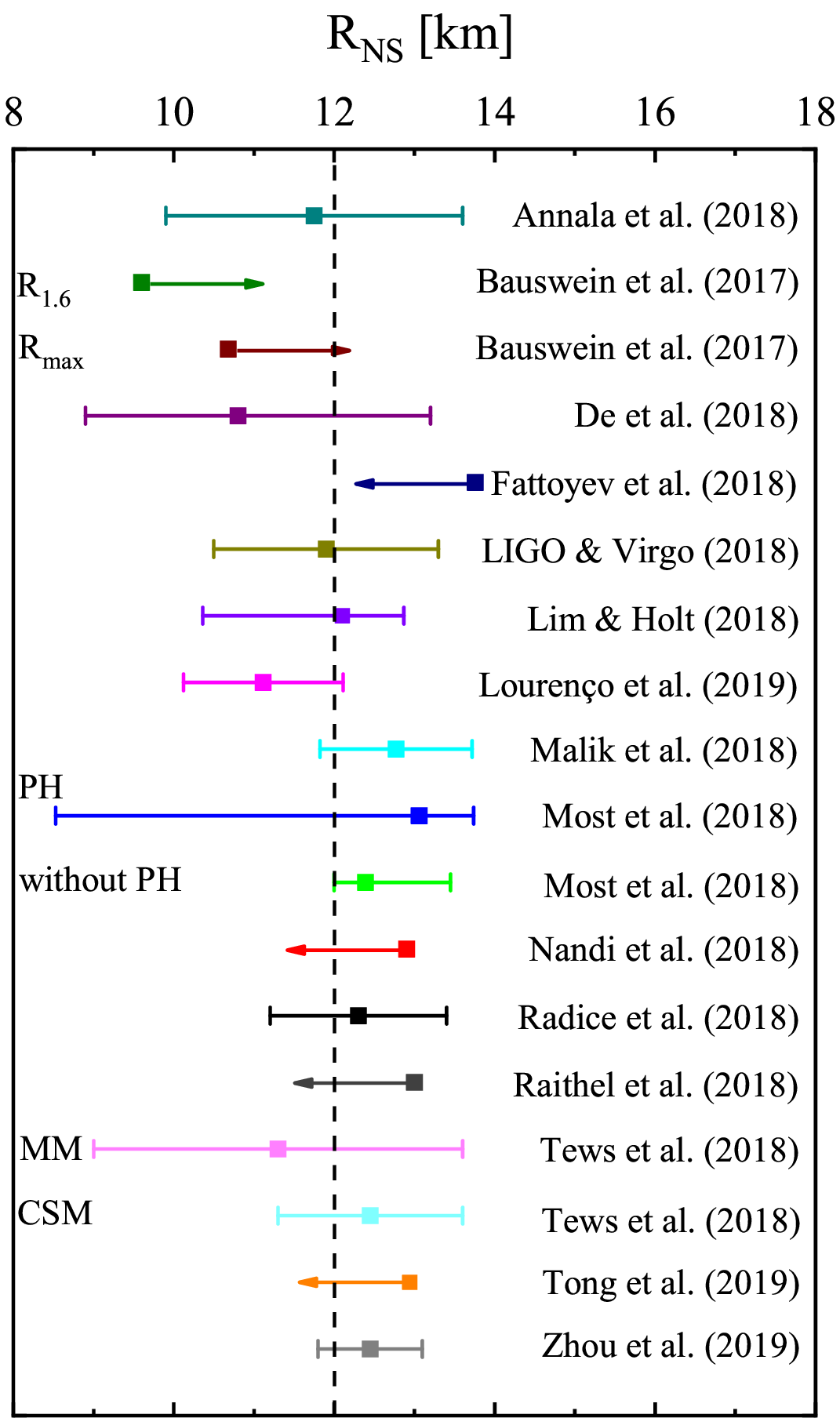} }
  \hspace{2cm}
  \caption{(color online) Constraints on the radii of neutron stars $R_{\rm NS}$ based on studies of the GW170817 event~\cite{LIGO2018,Nandi2018,Lourenco2019,Raithel2018,Fattoyev2018,Malik2018,Zhou2019,Most,Annala2018,Lim2018,Radice2018,Tews2018,De2018,Bauswein2017,PKU-Meng}. The $R_{\rm NS}$ represent the radii of neutron stars with $1.4$ M$_\odot$ ($R_{1.4}$), $1.6$ M$_\odot$ ($R_{1.6}$), and maximum mass ($R_{\rm max}$) from different studies. The ``PH'', ``MM'', and ``CSM'' on the left sides are used to distinguish the results in the same paper and denote phase transition (PH), minimal model (MM), and speed of sound model (CSM), respectively. See text for details.}\label{radius}
\end{center}
\end{figure}
Comparisons of the upper limit or range of $\Lambda_{\rm 1.4}$ from GW170817 with the calculated relationships between $R_{\rm 1.4}$ and $\Lambda_{\rm 1.4}$ have allowed the extraction of neutron
star radii. Summarized in Fig.~\ref{radius} are the radii of neutron stars with $1.4$ M$_\odot$ ($R_{1.4}$), $1.6$ M$_\odot$ ($R_{1.6}$), and the maximum mass ($R_{\rm max}$) based on different studies of the GW170817 event. The ``PH'', ``MM'', and ``CSM'' on the left sides are used to distinguish results in the same paper. They denote the model with phase transition (PH), minimal model (MM), and the speed of sound model (CSM), respectively. To infer the radius from observations of GW170817, both the crust and core EOSs may be crucial. However, as concluded in refs.~\cite{Piekarewicz2018,Gamba2019}, the mass or tidal deformability are almost independent of the crust EOS, although the $y(r)$ and $k_2$ strongly depend on it. Thus, the core EOS dominates the determination of tidal deformability.

Based on the constraints on the dimensionless tidal deformability $\Lambda$ or the mass-weighted dimensionless tidal deformability $\tilde{\Lambda}$ from the LIGO and Virgo Collaborations~\cite{LIGO2017,LIGO2018}, studies in refs. \cite{Nandi2018,Lourenco2019,Raithel2018,Fattoyev2018,Malik2018,Zhou2019,PKU-Meng} deduced that $R_{\rm 1.4}\leq12.9$ km, $10.12\leq R_{\rm 1.4}\leq12.11$ km, $R_{\rm 1.4}\leq13$ km, $R_{\rm 1.4}\leq13.76$, $11.82\leq R_{\rm 1.4}\leq13.72$ km, $11.8\leq R_{\rm 1.4}\leq13.1$ km, and $R_{\rm 1.4}\leq12.94$ km, using EOSs from relativistic mean field (RMF), Skyrme Hartree-Fock (SHF), or microscopic theories, respectively. By producing millions of parameterized EOSs, studies in ref.~\cite{Most} constrained the $R_{\rm 1.4}$ to $12.00\leq R_{1.4}\leq13.45$ ($8.53\leq R_{1.4}\leq13.74$) km for neutron stars without (with) phase transition (the most likely value is $R_{1.4}=12.39$ ($13.06$) km) using $400\leq\tilde{\Lambda}_{1.4}\leq800$ and $2.01\leq M_{\rm max}\leq2.16$ M$_\odot$. While studies in ref.~\cite{Annala2018} obtained $9.9\leq R_{1.4}\leq13.3$ km using $\Lambda_{1.4}\leq800$ and $M_{\rm max}\geq2.01$ M$_\odot$. Using Bayesian statistical analysis, studies in refs.~\cite{Lim2018,Radice2018} inferred that $10.36\leq R_{1.4}\leq12.87$ km (the most likely value is $R_{1.4}=11.89$ km) and $11.2\leq R_{\rm 1.4}\leq13.4$ km. In turn, studies in ref.~\cite{Tews2018} concluded that the $\tilde{\Lambda}$ of the two neutron stars in GW170817 are $80\leq\tilde{\Lambda}\leq580$ and $280\leq\tilde{\Lambda}\leq480$, while the $R_{1.4}$ are $9.0\leq R_{1.4}\leq13.6$ km and $11.3\leq R_{1.4}\leq13.6$ km using the speed of sound (CSM) and minimal models (MM), respectively. Moreover, besides the constraints on $R_{1.4}$, studies in ref.~\cite{De2018} suggested that the common areal radius of neutron stars satisfy $8.9\leq\hat{R}\leq13.2$ km with a mean value of $\langle\hat{R}\rangle=10.8$ km. Assuming the remnant was stable for at least $10$ ms to yield the observed ejecta properties, studies in ref.~\cite{Bauswein2017} found that the $R_{\rm 1.6}$ are larger than $10.68$ km and the $R_{\rm max}$ are larger than $9.60$ km. While LIGO \& Virgo Collaborations suggested that $10.5\leq R_{1,2}\leq13.3$ km for the two companions before their merger \cite{LIGO2018}. Besides the above constraints summarized in Fig.~\ref{radius}, some other studies also suggested constraints with limited EOSs, see, e.g., ref.~\cite{Kim2018}.

Given the diverse approaches used in analyzing albeit the same data from GW170817, the extracted radii shown in Fig.~\ref{radius} are remarkably consistent. Applying the democratic principle that is probably not so sound scientifically, a fiducial value of $R_{1.4}=12.42$ km with the lower and upper limits of $10.95$ km and $13.21$ km, respectively, can be extracted from the radii shown in
Fig.~\ref{radius}. Interestingly, these values are consistent with those from analyzing the X-ray data shown in Fig. ~\ref{radiusLMXB}. Ironically, these results from analyzing astrophysical observations are
in good agreement with the prediction of 11.5 km $\leq R_{1.4}\leq13.6$ km using the EOS constrained by terrestrial nuclear laboratory data \cite{Li2006}.

\begin{figure}
\begin{center}
\resizebox{0.48\textwidth}{!}{
 \includegraphics{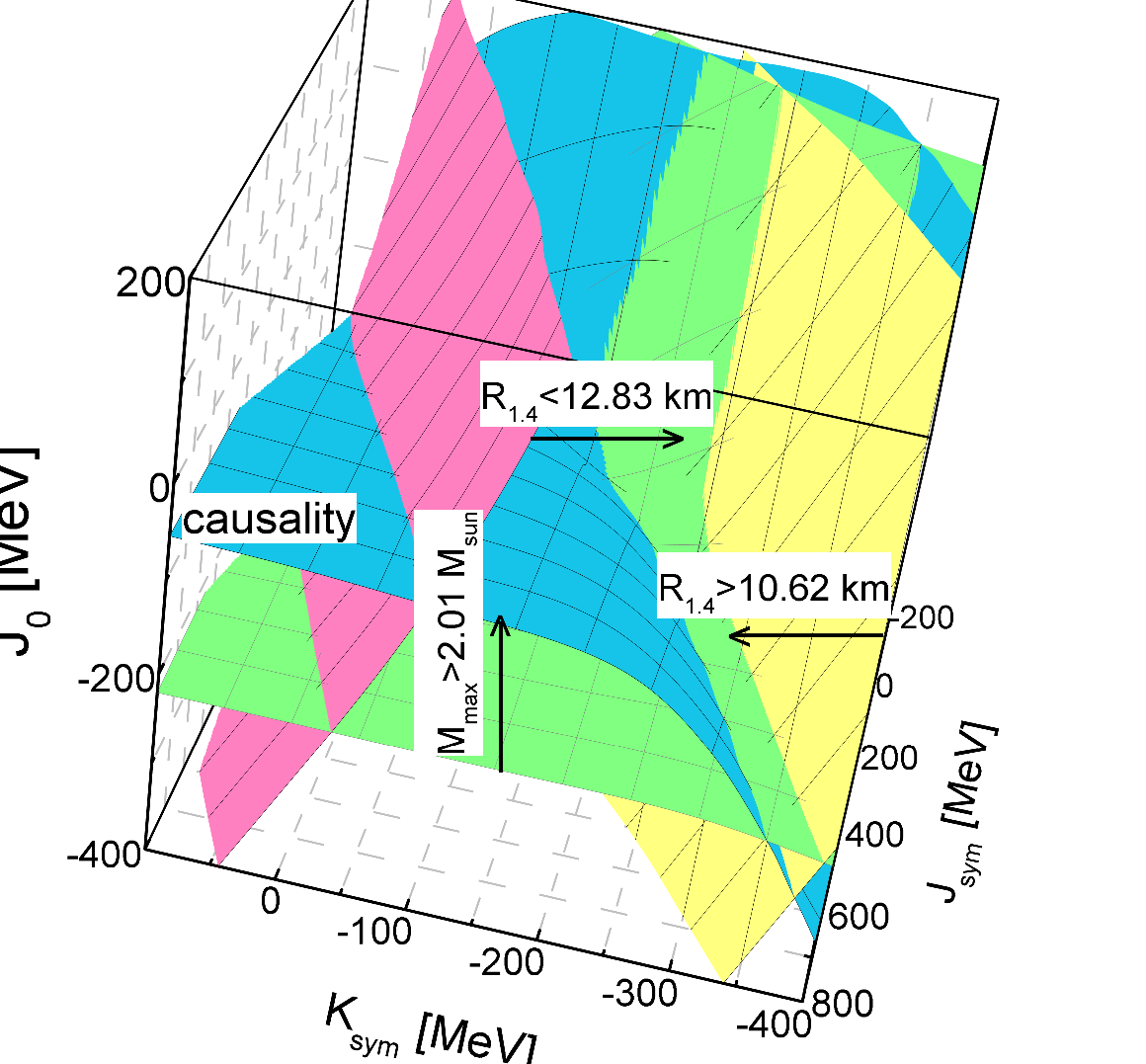}}
\caption{(Color online) Observational constraints of mass, radius, and causality in the $K_{\rm{sym}}-J_{\rm{sym}}-J_0$ 3D parameter space. The green, magenta, yellow, and blue surfaces represent $M_{\rm max}=2.01$ M$_\odot$, $R_{1.4}=12.83$ km, $R_{1.4}=10.62$ km, and the causality surface, respectively. The arrows show the directions satisfying corresponding observations. The physical constraint that transition pressure $P_t$ is larger than 0 MeV fm$^{-3}$ is demanded. Taken from ref.~\cite{Zhang2019EPJA}.}
\label{EPJAJ0}
\end{center}
\end{figure}
\begin{figure*}
\begin{center}
\begin{center}
\resizebox{0.42\textwidth}{!}{
  \includegraphics{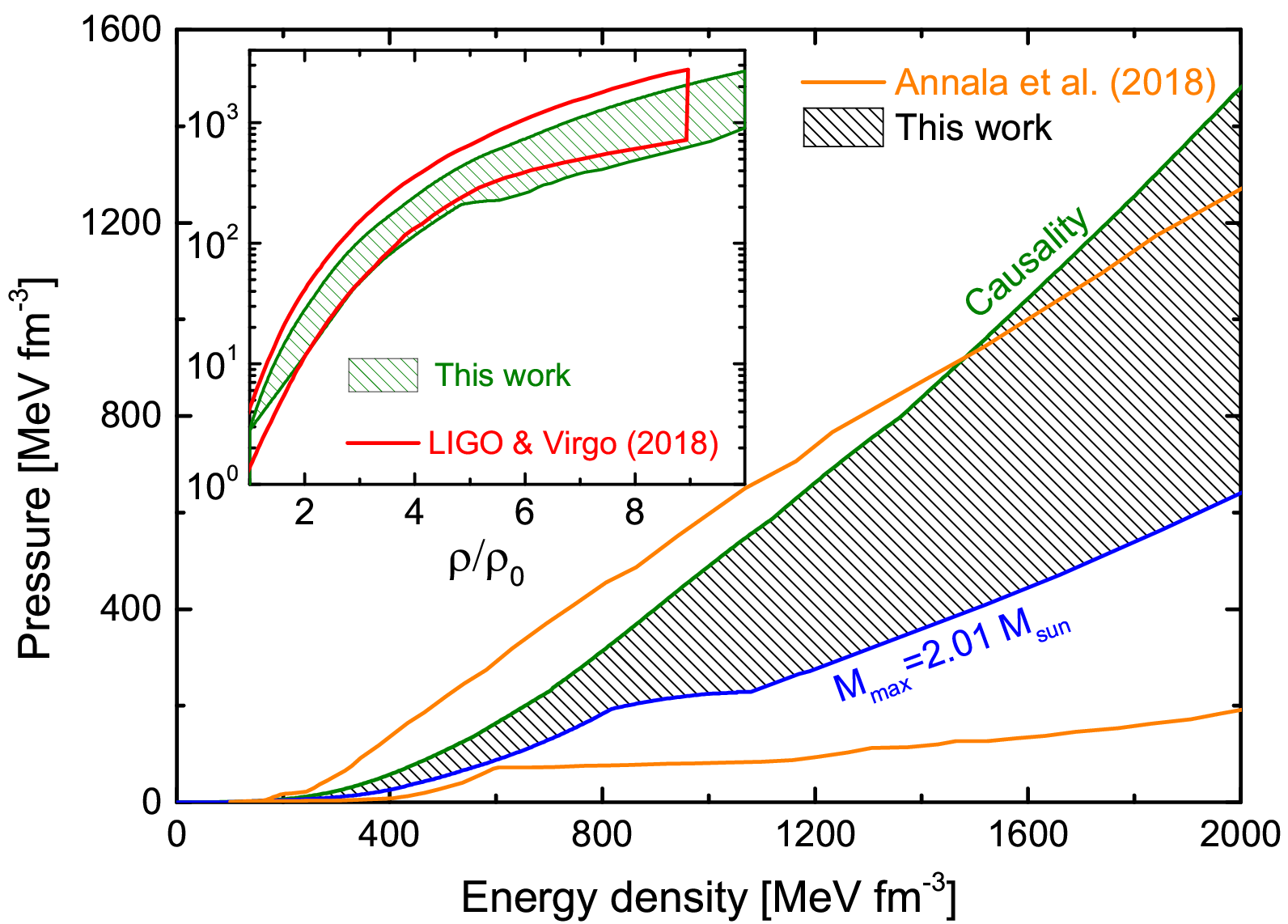}
  }
  \resizebox{0.42\textwidth}{!}{
  \includegraphics{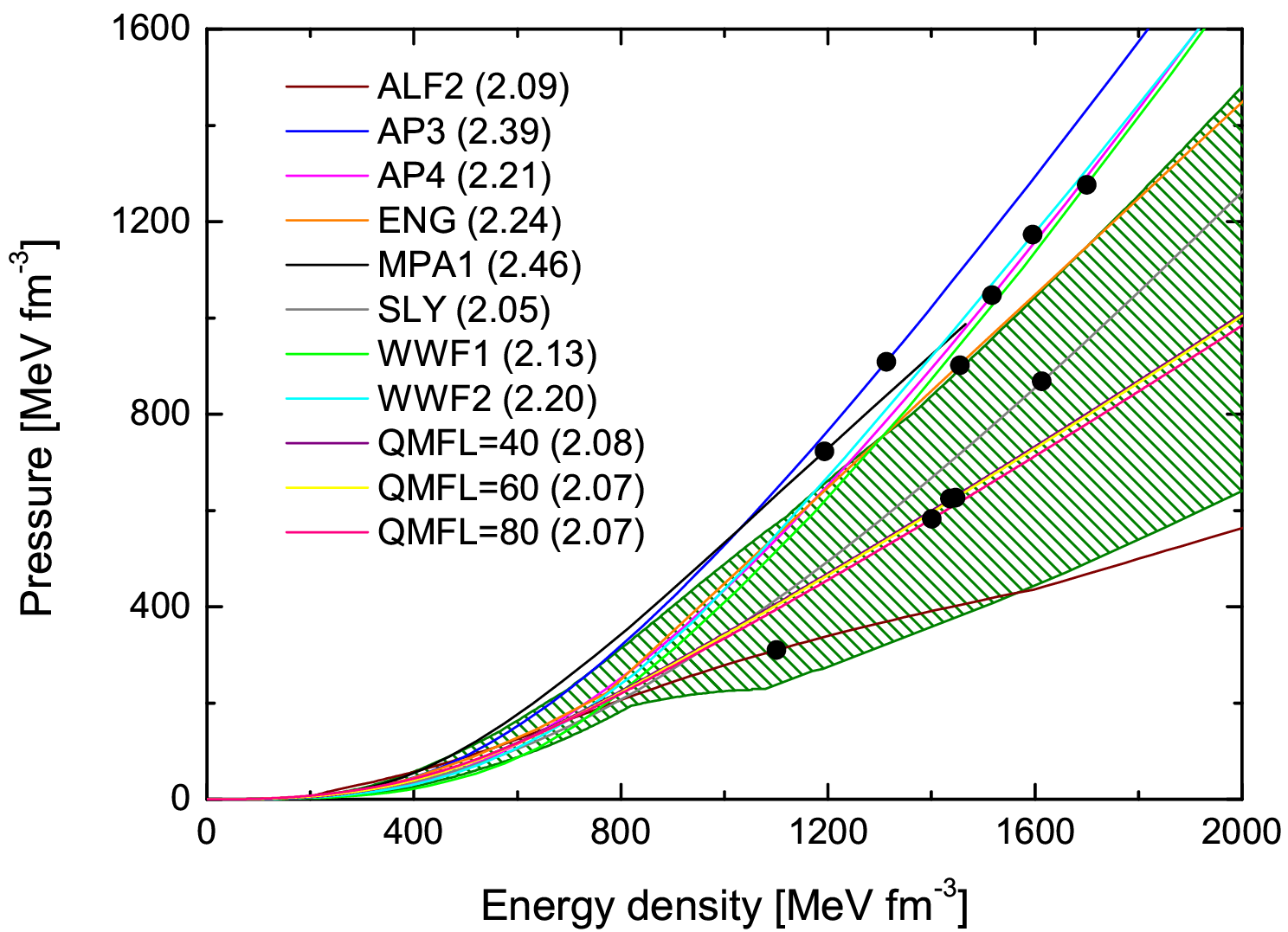}
}
\end{center}
  \caption{(color online) ) Left: Comparing the constrained EOSs with result of ref.~\cite{Annala2018} (orange line) and result of LIGO and Virgo Collaborations~\cite{LIGO2018} (red line) using parameterized EOSs interpolating between state-of-the-art theoretical results at different baryon densities and the analysis of GW170817 event, respectively. Right: Application of the extracted EOSs on several EOS models predicting maximum masses higher than $2.01$ M$_\odot$. The corresponding maximum masses are labeled after the model names. The black symbols indicate the maximum pressure and energy density reached at the maximum mass for each model. Modified from two figures in ref. \cite{Zhang2019EPJA}.}\label{EPJAEOS}
  \end{center}
\end{figure*}
\subsubsection{GW170817 implications on the EOS of neutron star matter}
The inversion technique demonstrated earlier in subsection \ref{inversion} can be used simultaneously to multiple observables to infer the underlying EOS. So far, the following observables and physical requirements have been used: 1) the observed maximum mass around $2.0$ M$_\odot$ for the two pulsars J1614-2230 \cite{Demorest2010} and J0348+0432 \cite{Antoniadis2013}; 2) the radius inferred from the X-ray bursts of LMXB: $10.62\leq R_{1.4}\leq12.83$ km~\cite{Lattimer2014}; 3) the tidal deformability $70\leq\Lambda_{1.4}\leq580$~\cite{LIGO2018} extracted by the LIGO and Virgo Collaborations. Except the constant surfaces of $\Lambda_{1.4}=70$ and $\Lambda_{1.4}=580$, the upper and lower limits of the above observables together with the causality surface are shown in
Fig. \ref{EPJAJ0} in the $K_{\rm{sym}}-J_{\rm{sym}}-J_0$ 3D parameter space. All other parameters (around the saturation densities) are taken as their currently known most probable values, e.g., $L=58.7$ MeV. The constant surfaces of tidal deformability of GW170817 are left out to make an unbiased comparison of the EOS extracted from inverting the observables and requirements specified in the figure with the EOS independently extracted earlier by the LIGO and Virgo Collaborations.
The surfaces of $M_{\rm max}=2.01$ M$_\odot$ and causality constrain the $J_0$ from the bottom and top sides, respectively. The lower and upper limits of $K_{\rm sym}$ are obtained by the intersecting line between the surfaces of $M_{\rm max}=2.01$ M$_\odot$ and $R_{1.4}=12.83$ km, as well as the surfaces of $M_{\rm max}=2.01$ M$_\odot$ and causality surface, respectively.  The $J_{\rm sym}$ can not be narrowed down further than its current uncertainty range by the observables considered. The upper and lower boundaries of the pressure allowed by these observables and physical requirement were extracted in ref.~\cite{Zhang2019EPJA}. As shown in Fig.~\ref{EPJAEOS}, this pressure as a function of baryon density (green band) is in very good agreement with the one (red lines in the inset) from analyzing the tidal deformability of GW170817~\cite{LIGO2018}. Moreover, the study in ref.~\cite{Annala2018} constrained the energy density as a function of pressure based on the finding $\Lambda\leq800$ first reported by LIGO and VIRGO collaborations and $M_{\rm max}\geq2.01$ M$_\odot$. Their result (orange lines) is compared with the one from the inversion approach in the main frame of Fig.~\ref{EPJAEOS}. It is seen that their constraints on the pressure are consistent but spread to a larger uncertainty region at high densities.

To check the impact of the constrained EOS from the inversion of astrophysical observables and physical requirements discussed above, several EOS models for neutron star matter that all predict maximum masses higher than $2.01$ M$_\odot$ are shown in the right panel of figure~\ref{EPJAEOS}. They are the  ALF2 for hybrid stars~\cite{ALF2}, APR3 and APR4~\cite{AP34}, ENG~\cite{ENG}, MPA1~\cite{Muther1987}, SLy~\cite{SLy}, WWF1 and WWF2~\cite{Wiringa1988}, QMFL40, QMFL60 and QMFL80~\cite{Zhu2018}. The corresponding maximum masses are labeled after the model names and shown as black symbols. It is seen that several EOSs go out of the constrained pressure band before reaching the maximum mass and thus can be excluded.

\begin{figure*}
\begin{center}
  \resizebox{0.42\textwidth}{!}{
 \includegraphics{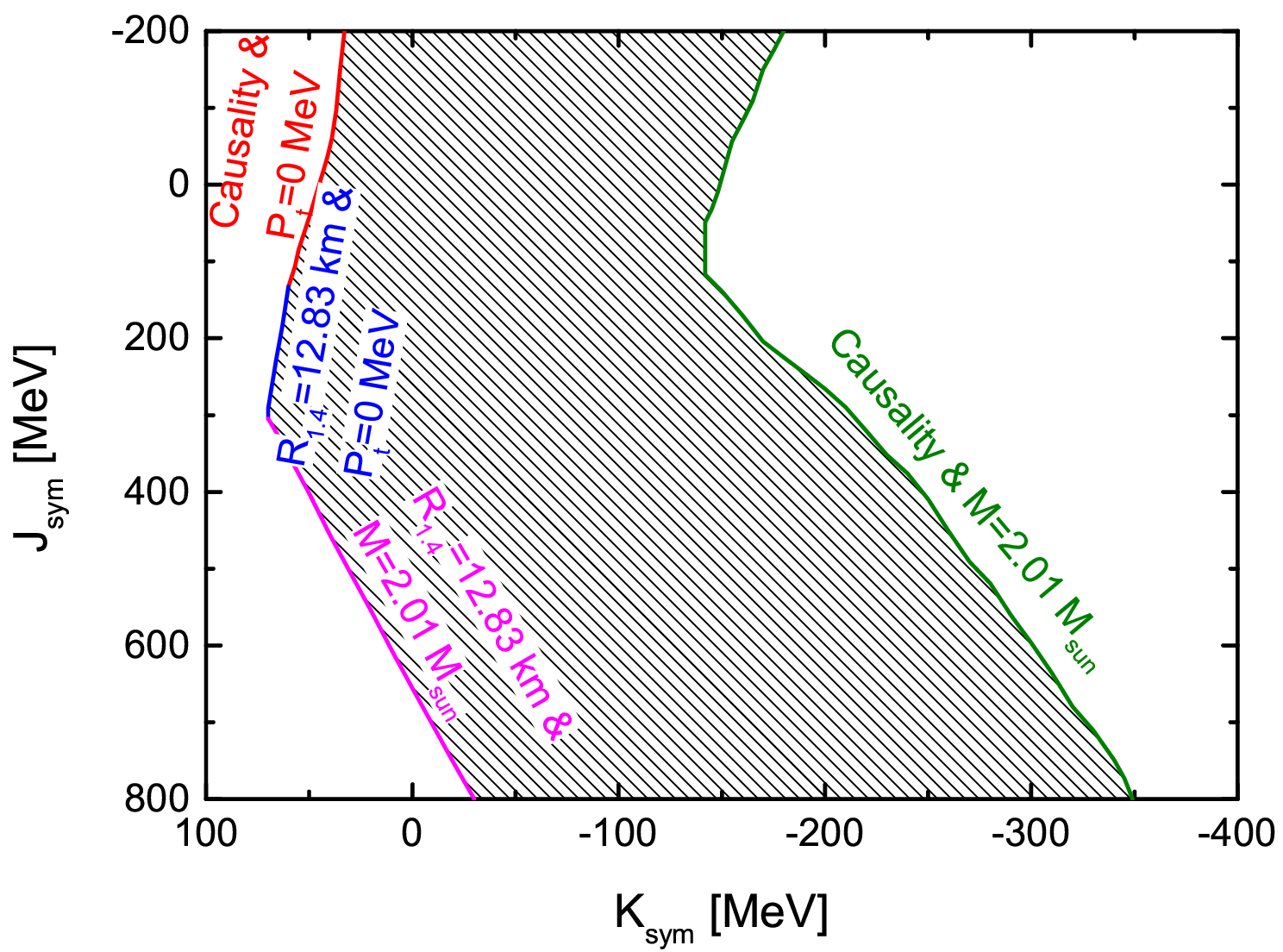}
}
  \resizebox{0.42\textwidth}{!}{
   \includegraphics{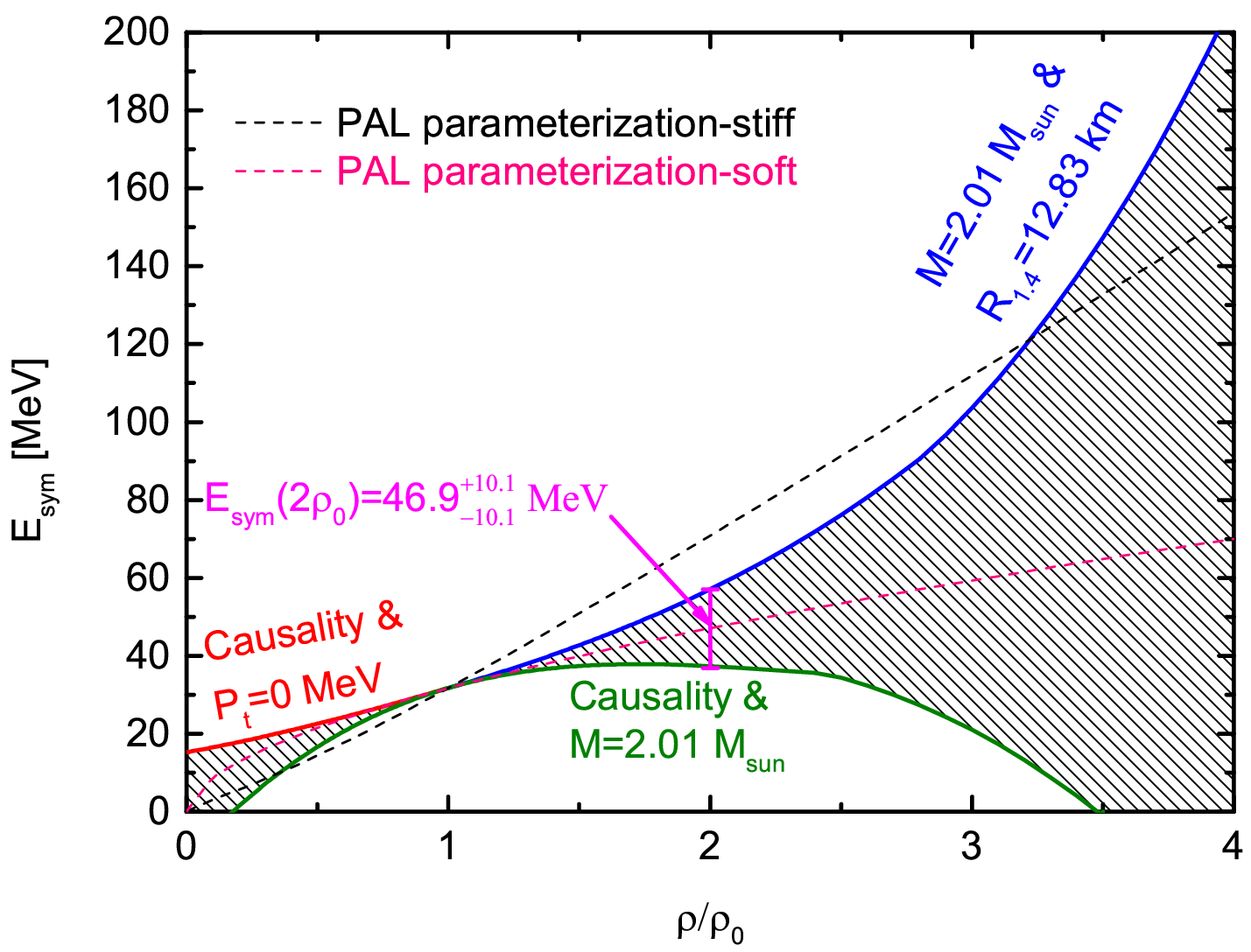}
}
\caption{(Color online) Left: boundaries in the $J_{\rm sym}$ versus $K_{\rm sym}$ plane extracted from the crosslines of the constant surfaces of physical observables and conditions in Fig. \ref{EPJAJ0}. Right: The constrained symmetry energy as a function of baryon density in comparison with the Prakash, Ainsworth and Lattimer parameterizations. The conditions used to infer the boundaries are labeled correspondingly. Taken from ref.~\cite{Zhang2019EPJA}.}
\label{EPJAEsym}
\end{center}
\end{figure*}
\subsubsection{GW170817 implications on the high-density symmetry energy and proton fraction in neutron stars at $\beta$ equilibrium}
Besides the constraints on the EOS discussed above, constraints on the symmetry energy can be extracted from Fig.~\ref{EPJAJ0} in the same way.
Shown in the left window of Fig. \ref{EPJAEsym} are the limiting $K_{\rm{sym}}$ and $J_{\rm{sym}}$ parameters on the constraining boundaries shown in Fig. \ref{EPJAJ0}.
While the $K_{\rm{sym}}$ is limited from left and right by the astrophysical observations, the
$J_{\rm{sym}}$ is still not limited, leading to still large uncertainties at densities above about $2.5\rho_0$.
The corresponding \esym are shown in Fig.~\ref{EPJAEsym}. The conditions used to infer the boundaries are labeled correspondingly.
For comparisons, the Prakash, Ainsworth and Lattimer (PAL) parameterizations for \esym \cite{Prakash1988}
\begin{eqnarray}\label{PALstiff}
  E_{\rm sym}^{\rm stiff}(\rho)&=&12.7(\rho/\rho_0)^{2/3}+38(\rho/\rho_0)^2/(1+\rho/\rho_0),\nonumber\\
  E_{\rm sym}^{\rm soft}(\rho)&=&12.7(\rho/\rho_0)^{2/3}+19(\rho/\rho_0)^{1/2}
\end{eqnarray}
are also shown with the black and magenta dashed lines. As we can see, the PAL stiff symmetry energy is out of the constrained band while the soft one is in it. Thus, the soft symmetry energy is
favored.

Though large uncertainties still exist at high densities, the \esym at densities below $2.5\rho_0$ is well constrained. More quantitatively, the value of \esym at $2\rho_0$ is found to be $E_{\rm sym}(2\rho_0)=46.9\pm10.1$ MeV with an uncertainty of $\sim21\%$. This uncertainty is about twice the uncertainty of symmetry energy at $\rho_0$ $E_{\rm sym}(\rho_0)=31.7\pm3.2$ MeV. Compared to the constrained band of EOS in Fig.~\ref{EPJAEOS}, the band of symmetry energy is much more uncertain. For a comparison, we note that in ref.~\cite{Chen2015}, a value of $E_{\rm sym}(2\rho_0)=40.2\pm12.8$ MeV was suggested by extrapolating the systematics at low densities found earlier from studying terrestrial nuclear laboratory data and predictions of nuclear energy density functionals. Interestingly, the two constraints on $E_{\rm sym}(2\rho_0)$ from analyzing the astrophysical observations and terrestrial laboratory data are in good agreement within their associated uncertainties. It is also interesting to note that a very recent study about the correlations among the neutron star tidal deformability,
the neutron star radius, the root-mean-square radii of neutron drops, and the symmetry energies of nuclear matter at supra-saturation densities within energy density functionals has extracted an upper limit of $E_{\rm sym}(2\rho_0)\leq 53.2$ MeV \cite{PKU-Meng}, consistent with the findings discussed above.

While the constraints on the \esym are not tight at densities above about $2.5\rho_0$, some useful lessons can be learned from the studies discussed here: (1) since the NS radii are known to be most sensitive to the \esym around $2\rho_0$, it is not surprising that the observables related to the radii and physical conditions studied in Fig. \ref{EPJAJ0} are not so restrictive on the \esym at densities above about $2.5\rho_0$. While more precise measurements of NS radii and the tidal deformability in the inspiraling phase of NS mergers may help further narrow down the \esym around
$2\rho_0$, new observables, such as neutrinos from the core of neutron stars, signals from the merging phase of two colliding neutron stars in space or two heavy-nuclei in terrestrial laboratories are needed to probe the \esym at higher densities; (2) both the upper and lower boundaries of the \esym depend on the location of the constant surface of the maximum mass $M_{\rm max}=2.01$ M$_\odot$ in the 3D EOS parameter space, its more precise value will strongly influence the accuracy of determining the \esym above $2.5\rho_0$.

\begin{figure}[h!]
\begin{center}
\resizebox{0.44\textwidth}{!}{
  \includegraphics{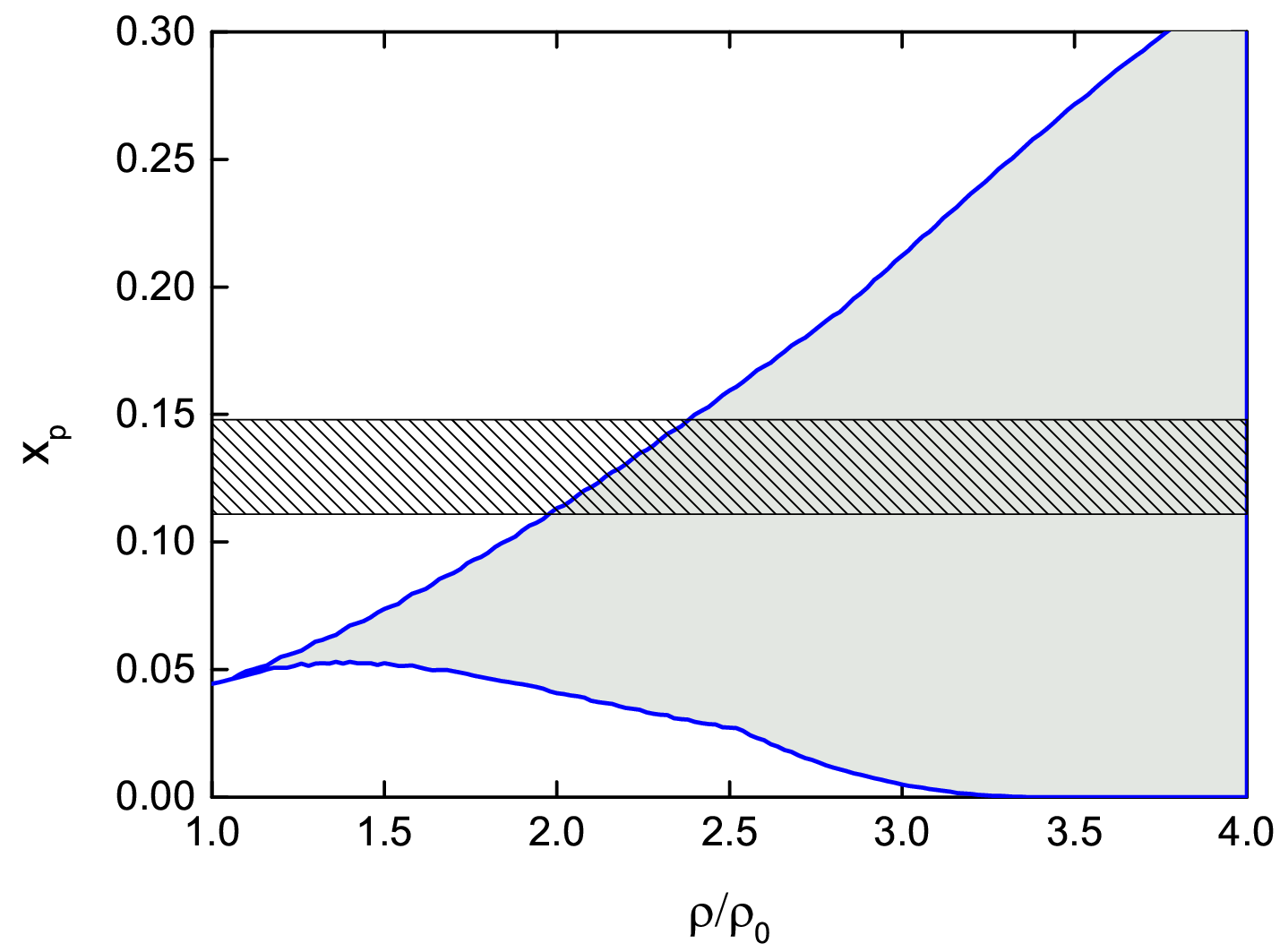}}
  \caption{(color online) Constrained protons fraction $x_p=\rho_p/\rho$ as a function of baryon density in neutron stars at $\beta$ equilibrium based on the parameterized EOS. The upper and lower boundaries correspond to those of the symmetry energy shown in figure~\ref{EPJAEsym}. The horizontal shaded area corresponds to the critical fraction (11.1\% and 14.8\%~\cite{Lattimer1991}) for the direct URCA process to happen in the $npe\mu$ matter. Taken from ref.~\cite{Zhang2019EPJA}.}\label{EPJAxp}
  \end{center}
\end{figure}
\begin{figure*}[h!]
\begin{center}
\resizebox{0.45\textwidth}{!}{
  \includegraphics{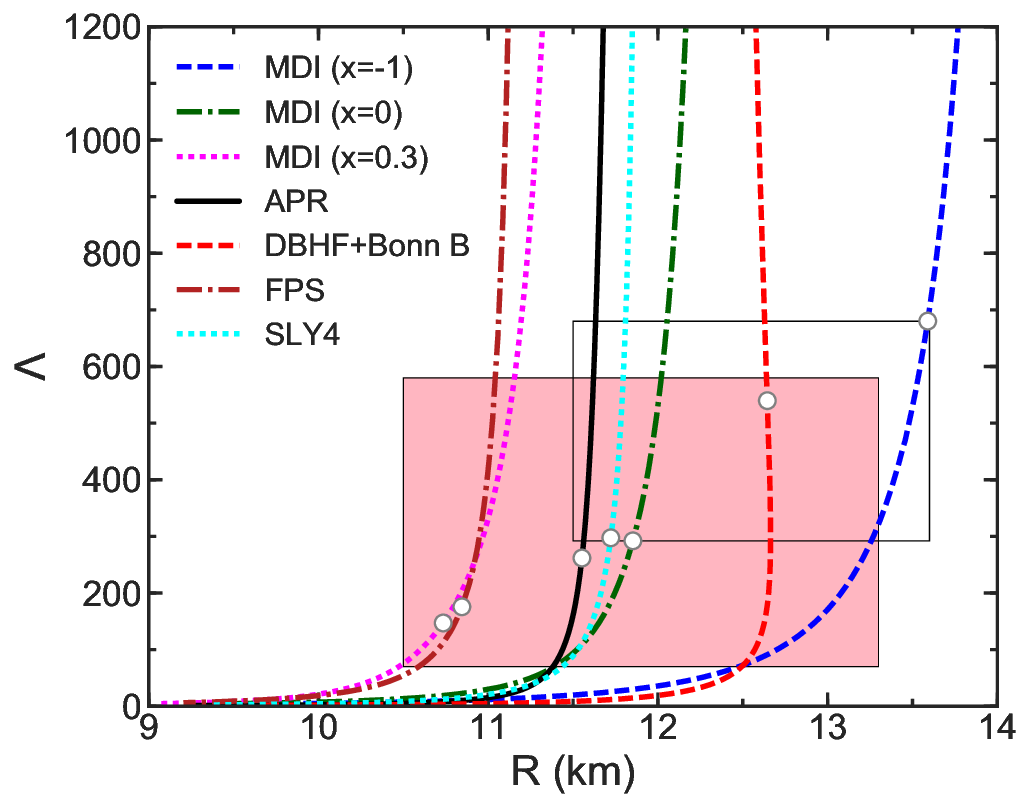}}
 \resizebox{0.45\textwidth}{!}{
  \includegraphics[scale=0.4]{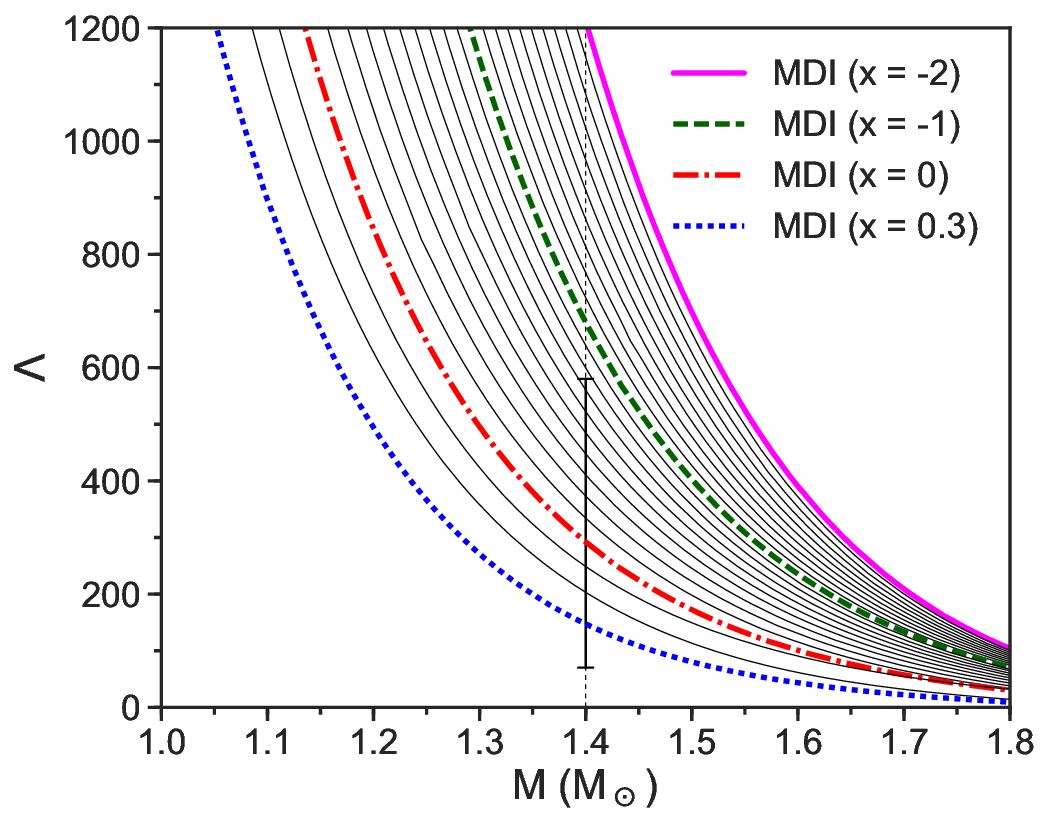}}
  \caption{(color online) Left: Examples of dimensionless tidal deformability as a function of radius. The squared pink region of $70\leq\Lambda\leq580$ and $10.5\leq R\leq13.3$ km corresponds to the constraints of LIGO and Virgo Collaborations in ref.~\cite{LIGO2018}. The squared white region of $292\leq\Lambda\leq680$ and $11.5\leq R\le13.6$ km corresponds to the constraints from heavy-ion collisions in ref.~\cite{Li2006} with the boundaries determined by the MDI EOS with $x = 0$ and $x = -1$. The small while circles indicate $\Lambda_{1.4}$.
Right: Dimensionless tidal deformability $\Lambda$ as a function of stellar mass $M$.
The error bar at $1.4M_{\odot}$ corresponds to the tighter constraints on the tidal deformability
 $\Lambda_{1.4}=[70 - 580]$ based on the refined analysis of GW170817 \cite{LIGO2017}. Results are
 displayed for MDI EOS with values of $x$ between 0.3 and -2 in steps of $\Delta x=0.1$. Taken from Ref.~\cite{Plamen3}.}\label{Lambdar}
  \end{center}
\end{figure*}
The constrained symmetry energy leads directly to some useful information about the composition of neutron stars at $\beta$ equilibrium. For example, the proton fraction $x_p=\rho_p/\rho$ at $\beta$ equilibrium at a given density is uniquely determined by the symmetry energy through the condition $\mu_e=4\delta E_{\rm sym}(\rho)$ for chemical equilibrium. The proton fraction calculated using the constrained \esym is shown as a function of baryon density in Fig.~\ref{EPJAxp}. It is seen that the stiffest symmetry energy can increase the proton fraction up to 30\% before $\rho=4\rho_0$.
An important impact of the proton fraction is on the cooling mechanism of protoneutron stars \cite{Lattimer1991}.  In the $npe\mu$ matter, the threshold proton fraction $x^{DU}_p$ for the fast cooling through the direct URCA process (DU)
\begin{equation}\label{xdu}
x^{DU}_p=1/[1+(1+x_e^{1/3})^3]
\end{equation}
is between 11.1\% to 14.8\% for the electron fraction $x_e\equiv \rho_e/\rho_p$ between 1 and 0.5 \cite{Kla06}. This range is indicated with the horizontal shaded area  in Fig.~\ref{EPJAxp}. It is seen that the minimum critical density enabling the direct URCA process is about $2\rho_0$ with the stiffest symmetry energy. Softer symmetry energies will require larger critical densities.

In the above discussions regarding the extraction of nuclear symmetry energy using the parameterized EOS, the low-order parameters are fixed at their most probable values while the high-order parameters of both the $E_0(\rho)$ and $E_{\rm sym}(\rho)$ are varied. Among the low-order parameters, the $L$ parameter has the highest uncertainty of about $\pm28.1$ MeV. To see its effects without hindering effects of other variables will require the inversion of observables in a four dimensional parameter space. Alternatively, one may select observables that are not sensitive to the high-density EOS of SNM, then one can fix the $J_0$ and focus on the 3D sub-space of \esym parameters $L-K_{\rm sym}-J_{\rm sym}$. Alternative, one may conduct the multivariate Bayesian analyses which have their own challenges related to the prior probability density distributions of some of the parameters.

\subsubsection{GW170817 implications on the isospin-dependence of three-body nuclear force in dense matter}
A precise measurement of the tidal deformability of neutron stars constrain not only the EOS of dense neutron-rich nuclear matter but also the fundamental strong interactions at either the hadronic or quark level underlying the model EOSs. A number of studies have recently examined how the data from GW170817 may help constrain various aspects and/or model parameters of strong interaction.
Unfortunately, mostly because there is so far no firmly determined lower limit of the tidal deformability, the extracted limits on the strong interactions are not so restrictive, indicating again the importance of
firmly establishing the lower limit of $\Lambda_{1.4}$. As we discussed in detail earlier in sect. \ref{3bf}, the $x$ parameter in the MDI EOS controls the competition between the isotriplet and isosinglet channels of the effective three-body nuclear force. It affects significantly the high-density behavior of nuclear symmetry energy. As an example of the impacts of GW170817 on properties of nuclear interactions, we
discuss in the following how the $x$ parameter may be constrained by the reported value of $\Lambda_{1.4}$.

Shown in the left window of Fig.~\ref{Lambdar} are the $\Lambda_{1.4}$ values as functions of radius calculated using several interactions indicated.
The squared pink region of $70\leq\Lambda\leq580$ and $10.5\leq R\leq13.3$ km corresponds to the constraints reported by the LIGO and Virgo Collaborations~\cite{LIGO2018}, while the squared white region of $292\leq\Lambda\leq680$ and $11.5\leq R\le13.6$ km corresponds to the constraints from heavy-ion collisions~\cite{Li2006} with the boundaries determined by the MDI EOSs with $x = 0$ and $x = -1$. The small white circles indicate the results for NSs with $M=1.4$ M$_\odot$. It is seen that the constrained region from heavy-ion collisions overlaps with but is more restrictive than the one from studying the GW170817 event.  Nevertheless, it is interesting to study how the $\Lambda_{1.4}$ itself from GW170817 may help constrain the $x$ parameter of the MDI interaction.
The right window of Fig.~\ref{Lambdar} displays the $\Lambda_{1.4}$ versus stellar mass for the MDI EOS with values of $x$ between -2 and 0.3 in steps of $\Delta x=0.1$. The vertical bar at $1.4M_{\odot}$ indicates the range of $\Lambda_{1.4}$ derived in ref. \cite{LIGO2018}.
It is seen that the upper limit of $\Lambda_{1.4}=580$ is consistent with $\sim x_{gw}^{up}=-0.75$,
which is slightly larger than the upper limit of $x=-1$ derived from nuclear laboratory data. The derived
value $x_{gw}^{up}$ translates directly into an upper limit of the symmetry energy curvature parameter
$L$, i.e., $L_{gw}^{up}\approx 96$ MeV (see Fig. \ref{Plamen18b}). As mentioned earlier, there is currently no firmly established
lower limit of $\Lambda_{1.4}$. Several derived lower limits have been reported but are rather controversial (see, e.g., Ref. \cite{Kiuchi2019} and
references therein). Therefore, at present only an upper limit on $x$ can be derived from the GW170817 results.
\begin{figure}[ht]
\begin{center}
\resizebox{0.78\textwidth}{!}{
  \includegraphics{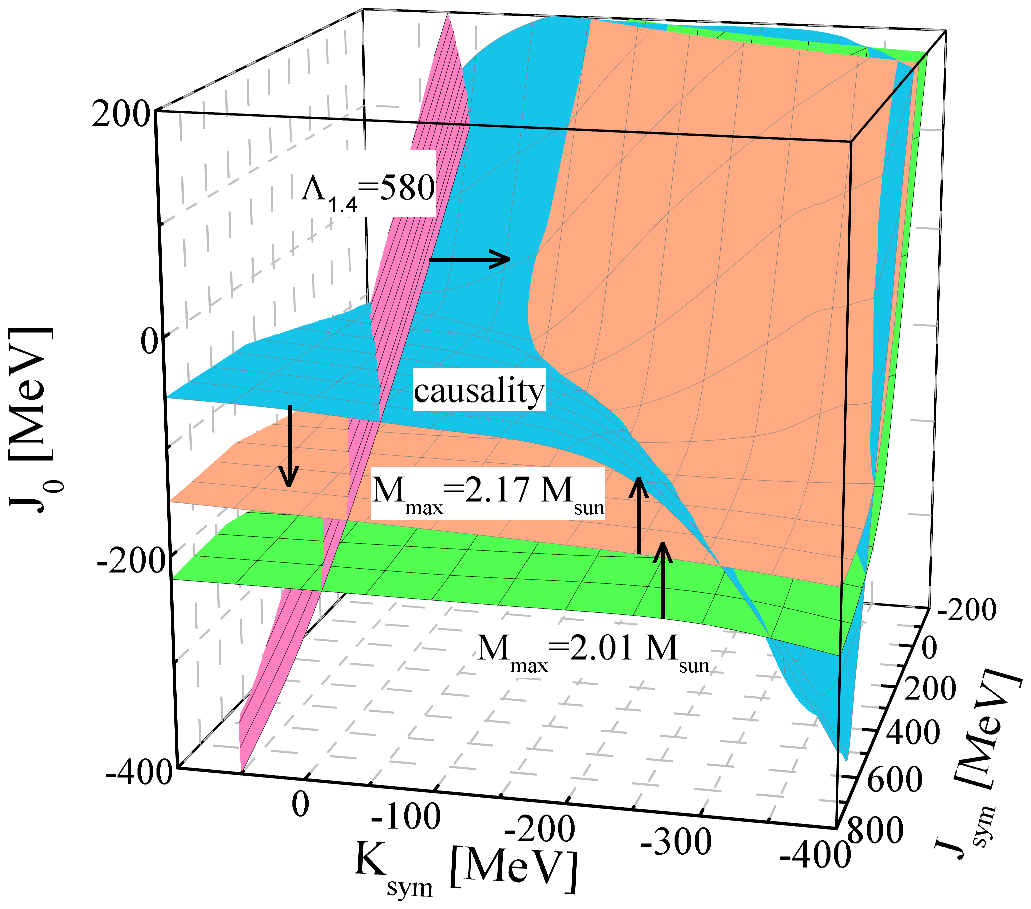}
  }
  \vspace{-2cm}
  \caption{(color online) Constant surfaces of NS maximum mass of $M_{\rm max}=2.01$ M$_\odot$ and $M_{\rm max}=2.17$ M$_\odot$ as well as the maximum tidal deformability
  $\Lambda_{1.4}=580$ (90\% confidence level) for canonical NSs and the causality condition, respectively, in the $J_0-K_{\rm sym}-J_{\rm sym}$ parameter space for high-density neutron-rich nuclear matter. Taken from ref. \cite{ZL217}.
  }\label{mass217}
\end{center}
\end{figure}
\begin{figure*}[ht]
\begin{center}
  \resizebox{0.48\textwidth}{!}{
   \includegraphics[width=17.cm,height=14cm]{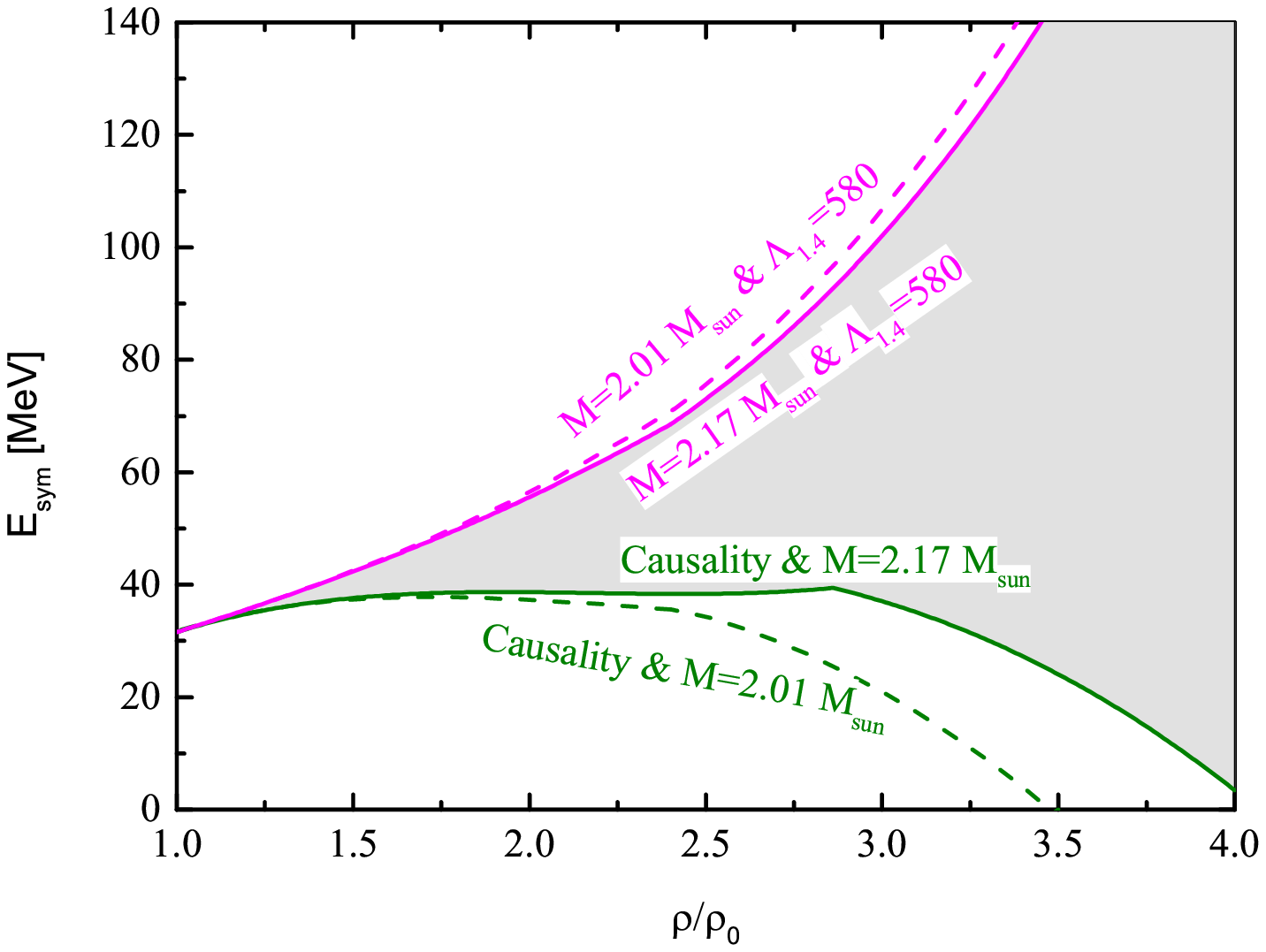}
   }
   \resizebox{0.48\textwidth}{!}{
 \includegraphics[width=17.cm,height=14cm]{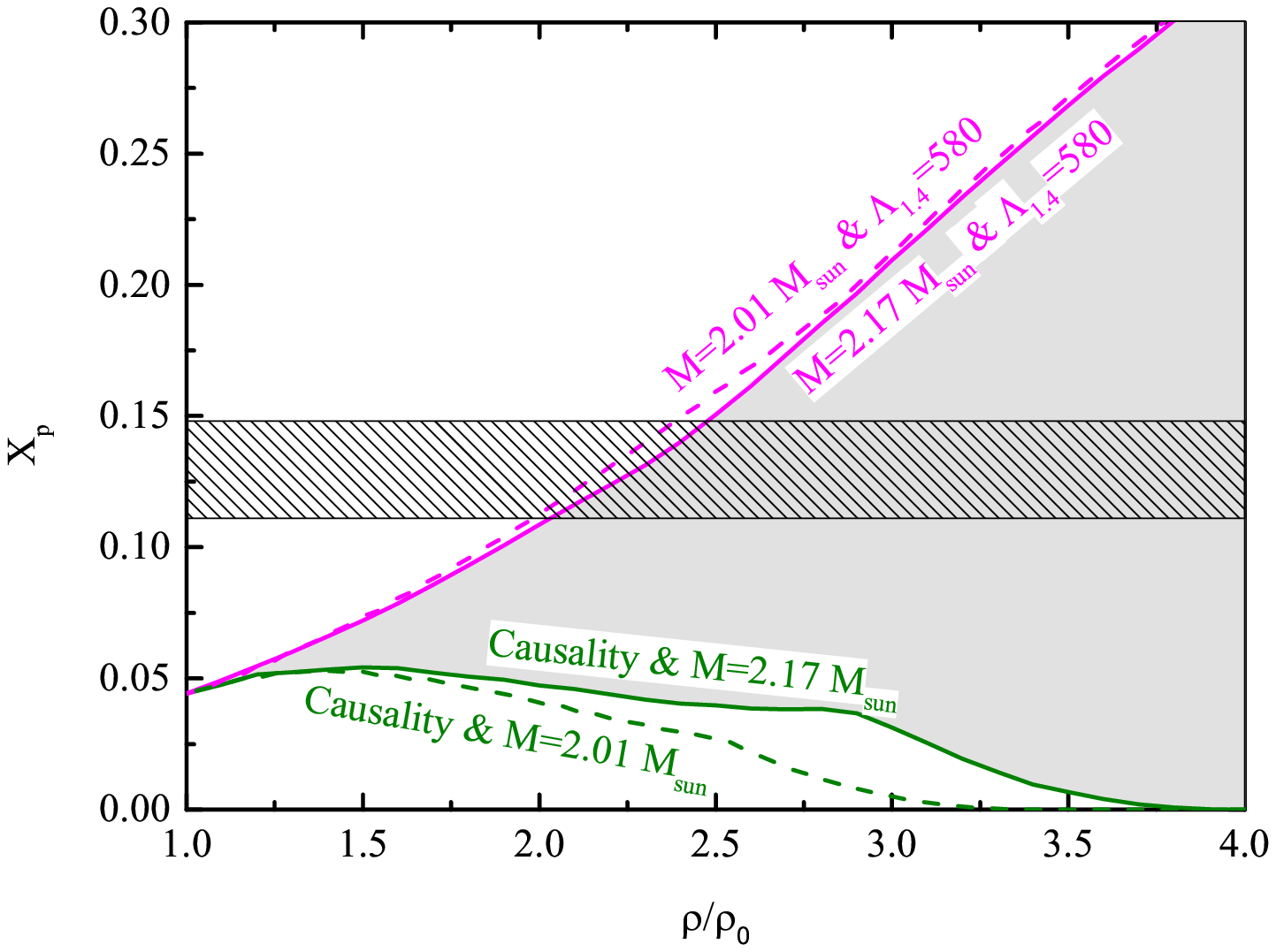}
}
  \caption{(color online) The nuclear symmetry energy in the supra-saturation density region (left) and the corresponding proton fraction in NSs at $\beta$-equilibrium (right). The horizontal band between 11.1\% to 14.8\% is the direct URCA limit for fast cooling of proto-NSs. Taken from ref. \cite{ZL217}.
   }\label{xfraction}
\end{center}
\end{figure*}
\subsection{The absolutely maximum mass of neutron stars in light of GW170817}
Determining the mass boundary between neutron stars and black holes is a longstanding and fundamental question. The absolutely maximum mass of neutron stars were first predicted by Oppenheimer \& Volkoff \cite{Oppenheimer39}. Nowadays, all kinds of improved theories and models have been used to calculate the EOS of dense neutron-rich nuclear matter and predict the corresponding maximum masses of neutron stars, see, i.e., refs. \cite{Dut12,Dutra2014,Chen2009,Colo2014}. The calculated maximum mass is approximately between $1.5$ to $3.0$ M$_\odot$ based on different EOSs \cite{Bombaci1999,Chamel2013,Chamel2013b}. In the following we discuss the latest observations of the NS maximum mass and its implications on the EOS, the theoretically predicted absolutely maximum mass of NSs from causality considerations and the extracted NS maximum mass from analyzing the GW170817-GRB170817A-AT2017gfo events by various groups.

\subsubsection{The observed maximum mass of neutron stars and its implications on the EOS}
Great efforts have been used to measure the mass of neutron stars. In 2012, ref.~\cite{Lattimer2012} summarized $64$ neutron stars with their masses measured in X-ray/optical binaries, double-neutron star binaries, white dwarf-neutron star binaries, and main sequence-neutron star binaries. Though several neutron stars with masses larger than $2.0$ M$_\odot$ have been measured, the large systematic errors have hindered the applications of these observations. For the latest list of NS masses, we refer the reader to ref. \cite{Arizona}. The well accepted maximum observed mass is about $2.0$ M$_\odot$ \cite{Demorest2010,Antoniadis2013}, which puts tight constraints on the EOS especially its isospin-symmetric part. In 2018, the mass of PSRJ2215+5135 was reported to be $2.27^{+0.17}_{-0.15}$ M$_\odot$ \cite{Linares2018}. If this result can be confirmed and its uncertainty narrowed significantly, stricter astrophysical constraints can be put on the EOSs of dense neutron-rich matter.

More recently, in April 2019, the millisecond pulsar J0740+6620 was reported to have a mass of $2.17^{+0.11}_{-0.10}$~M$_\odot$ (68.3\% credibility interval) \cite{M217} based on combined analyses of the relativistic Shapiro delay data taken over 12.5-years at the North American Nanohertz Observatory for Gravitational Waves and the recent orbital-phase-specific observations using the Green Bank Telescope. While the error bars of this mass value are still quite large and several previous instances of revising down the earlier reported Shapiro delay mass measurements may justify some necessary cautions, if this mass stays approximately unchanged, it will certainly help further constrain the EOS of super-dense neutron-rich nuclear matter. 

While the mass of the observed most massive neutron star is likely to keep changing as more accurate measurements and analyses are being carried out, what parts of the EOS are expected to be constrained by the measured maximum mass? Compared to the existing constraints from using $M_{\rm max}=2.01M_\odot$, how much better can the new maximum mass of $M_{\rm max}=2.17M_\odot$ help further constrain the EOS of NS matter?
These questions were recently studied in ref. \cite{ZL217}. It was found that the reported mass $M=2.17^{+0.11}_{-0.10}$~M$_\odot$ of PSR~J0740+6620, if confirmed, not only helps improve quantitatively our knowledge about the EOS of super-dense neutron-rich nuclear matter but also presents some new challenges for nuclear theories. For example, shown in Fig. \ref{mass217} are the constant surfaces of maximum mass $M_{\rm max}=2.01$ M$_\odot$ and $M_{\rm max}=2.17$ M$_\odot$ in the $J_0-K_{\rm sym}-J_{\rm sym}$ parameter space for high-density neutron-rich nuclear matter.
It is seen that the two surfaces are approximately parallel in the whole space. In the front region where the symmetry energy is stiff/high, the lower limit of the skewness $J_0$ increases by approximately 47\% from about $-220$ MeV to $-150$ MeV when the maximum mass increases by about 8\% from 2.01 M$_\odot$ to 2.17 M$_\odot$. Thus, the reported mass of J0740+6620 raises the lower limit of the skewness parameter $J_0$ of SNM significantly. Moreover, the crosslines of the two constant mass surfaces with the constant surfaces of the tidal deformability and causality set the boundaries of the high-density symmetry energy as we discussed earlier. The increase of the maximum mass from 2.01 M$_\odot$ to 2.17 M$_\odot$ will thus also modify the boundaries of the high-density symmetry energy extracted from studying properties of neutron stars. This is illustrated in Fig. \ref{xfraction}. Clearly, the lower boundary of the high-density symmetry energy and the corresponding proton fraction at $\beta$-equilibrium are increased appreciably. More quantitatively, the minimum proton fraction increases from about 0 to 5\% around $3\rho_0$ when the maximum mass of NSs is increased from 2.01 to 2.17 M$_{\odot}$.  However, the upper boundaries from the crosslines of the maximum mass and the tidal deformability are only slightly changed. 

\subsubsection{The predicted absolute maximum mass of neutron stars from causality considerations}
The causality condition should be satisfied by all theoretical models for compact objects, which means that it can set the upper limit on the absolutely maximum mass. Excellent reviews on this topic by Lattimer \& Parakash can be found in refs.~\cite{Lattimer2007,Lattimer2011}. For example, in Lindblom's work and several subsequent studies confirming his results (for a complete list see refs.~\cite{Lattimer2007,Lattimer2011}, they constructed EOSs with a fiducial transition-density $\rho_f$ above which the speed of sound $v_s$ is equal to the speed of light $v_c$. The EOS is normally written as $P(\epsilon)=\epsilon-\varepsilon_f+P_f(\epsilon_f)$ for $\epsilon>\epsilon_f(\rho_f)$. Some empirical or ``realistic'' nuclear EOSs were adopted for $\epsilon<\epsilon_f$. Be design, this kind of EOS satisfies $v_s^2=v_c^2$ when $\rho>\rho_f$. It was found that the redshift only weakly depends on the value of $\rho_f$. For $\rho_f\geq3\times10^{14}$ g cm$^{-3}$, the redshift was found to be $z=(\sqrt{1-2GM/Rc^2})^{-1}\leq0.863$. The latter constant of $z$ can be rewritten as $M/{\rm M}_{\odot}=R/4.16$ km. According to ref. \cite{Lattimer2011}, the above causality condition
is implicitly in the formulation by Rhoades \& Ruffini in 1974~\cite{Rhoades1974}. While the study in ref.~\cite{Glendenning1992} got the same results using a different method. Moreover, the results of ref.~\cite{Rhoades1974} are shown explicitly with the methodology in ref.~\cite{Koranda1997}. We also note that a boundary $M/{\rm M}_{\odot}=R/4.51 {\rm km}$ independent of $\rho_f$ is suggested implicitly in ref.~\cite{Lattimer1990} using a similarly parameterized EOS.

The causality and the allowed maximum mass was studied in an alternative approach in ref. \cite{Zhang2019EPJA} without involving the pre-assumed fiducial density $\rho_f$. In this approach, the condition: $v_s^2=v_c^2$ happens only at the central density of the most massive neutron star calculated: $M_{\rm max}=M(v_s^2=v_c^2)$. This condition defines the causality surface in the 3D EOS parameter space discussed earlier. As shown already with the blue surface in Fig.~\ref{EPJAJ0}, the causality surface constrains the EOS parameter space from the top. Thus, the absolutely maximum mass of neutron stars should be obtained from the causality surface. 

\begin{figure}
\hspace{-1cm}
\resizebox{0.78\textwidth}{!}{
  \includegraphics{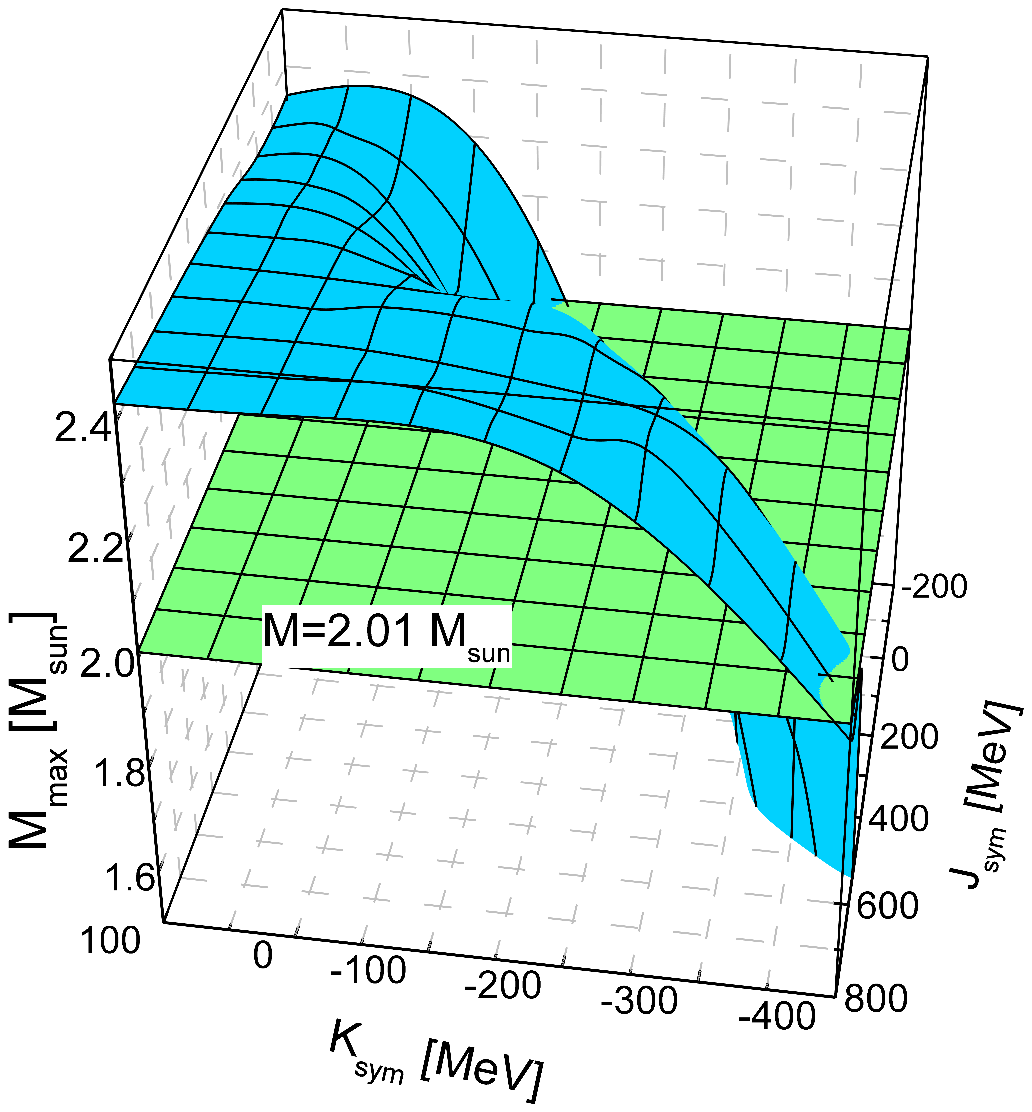}
  }
 \vspace{-1cm}
  \caption{(color online) The mass of the most massive neutron stars on the causality surface as functions of
  $J_{\rm{sym}}$ and $K_{\rm{sym}}$, respectively. Taken from ref. \cite{Zhang2019EPJA}.}
\label{mass1}
\end{figure}

Shown in Fig. \ref{mass1} is the maximum mass of NSs on the causality surface as a function of $J_{\rm{sym}}$ and $K_{\rm{sym}}$. As shown earlier, at each point on the causality surface there is a maximum value of $J_0$ and a corresponding maximum mass of NSs that can be supported. Fig. \ref{mass1} shows clearly that the causality surface sets an absolutely upper limit (the maximum of the maximum masses) for the mass of NSs at M$_{\rm{max}}=2.4$ M$_{\odot}$. However, the actual maximum mass of NSs might be smaller than 2.4 M$_{\odot}$ depending on the high-density behavior of nuclear symmetry energy as we discussed earlier in sect. \ref{MDI-p}. It is seen that the maximum mass decrease quickly when the combinations of the $J_{\rm{sym}}$ and $K_{\rm{sym}}$ make the \esym super-soft, reducing the asymmetric pressure significantly. Thus, the observed maximum mass of NSs not only sets a lower limit for the stiffness of SNM EOS but also
a lower boundary for the high-density symmetry energy as we discussed in the previous subsection. This point is further illustrated by the plane with the confirmed mass M$_{\rm{max}}=2.01$ M$_{\odot}$ of PSR J0348+0432. Since all acceptable EOSs have to be able to support NSs at least as massive as
PSR J0348+0432, the space below the 2.01 M$_{\odot}$ plane is excluded. The crossline of this plane with the causality surface sets a boundary in the $J_{\rm{sym}}$ versus $K_{\rm{sym}}$ plane, thus a limit on the high-density behavior of nuclear symmetry energy. As shown in Fig. \ref{xfraction}, when the observed maximum mass increases, the corresponding lower boundary for the high-density \esym also moves up accordingly.

\begin{figure}
\begin{center}
\resizebox{0.47\textwidth}{!}{
 \includegraphics{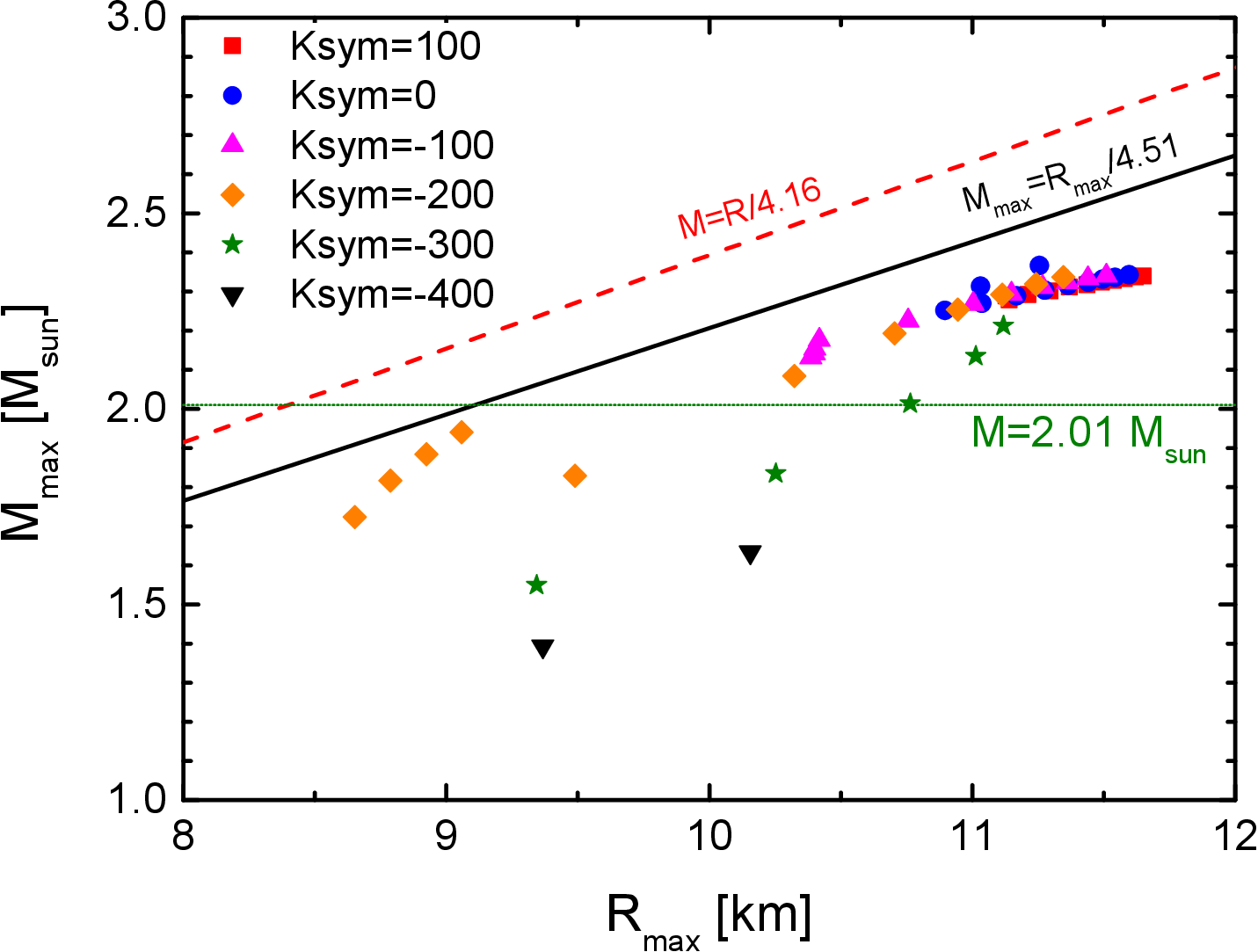}}
 \vspace{0.6cm}
\caption{(Color online) The maximum mass and corresponding radius on the causality surface in figure~\ref{EPJAJ0} in comparison with causality constraints on the maximum mass and corresponding radius suggested in ref.~\cite{Lindblom1984} (red dashed line) and ref.~\cite{Lattimer1990} (black solid line). The maximum observed mass of $2.01$ M$_\odot$ is shown as a reference. Taken from ref.~\cite{Zhang2019EPJA}.}
\label{mass2}
\end{center}
\end{figure}

The maximum mass and the corresponding radius of NSs on the causality surface are shown in Fig.~\ref{mass2} in comparison with the causality constraints derived in ref.~\cite{Lindblom1984} (red dashed line) and ref.~\cite{Lattimer1990} (black solid line). The maximum observed mass of $2.01$ M$_\odot$ is shown as a reference. While the relation between the mass and radius of the most massive neutron stars varies with the EOS parameters, the maximum mass reaches a limit of about $2.4$ M$_\odot$ independent of the EOS used. While the radius of the most massive neutron stars is unlikely to be measured anytime soon, it is about 11.5 km.
The absolutely maximum mass of $2.4$ M$_\odot$ is lower than the maximum masses from the two suggested scalings from refs.  \cite{Lindblom1984,Lattimer1990}. This is because the causality limit $v_s^2=v_c^2$ is allowed to be reached in the study of ref. \cite{Zhang2019EPJA} only at the central density of the most massive neutron stars instead of at all densities above the assumed fiducial densities in refs. \cite{Lindblom1984,Lattimer1990}. Very interestingly, the latest and improved version of the zero temperature quark-hadron crossover EOS, QHC19, of Baym et al. \cite{Baym19}, gives an absolutely maximum mass of 2.35 M$_\odot$ at the casual limit consistent with that found in ref.  \cite{Zhang2019EPJA}.
\begin{figure}
\begin{center}
\resizebox{0.43\textwidth}{!}{
 \includegraphics{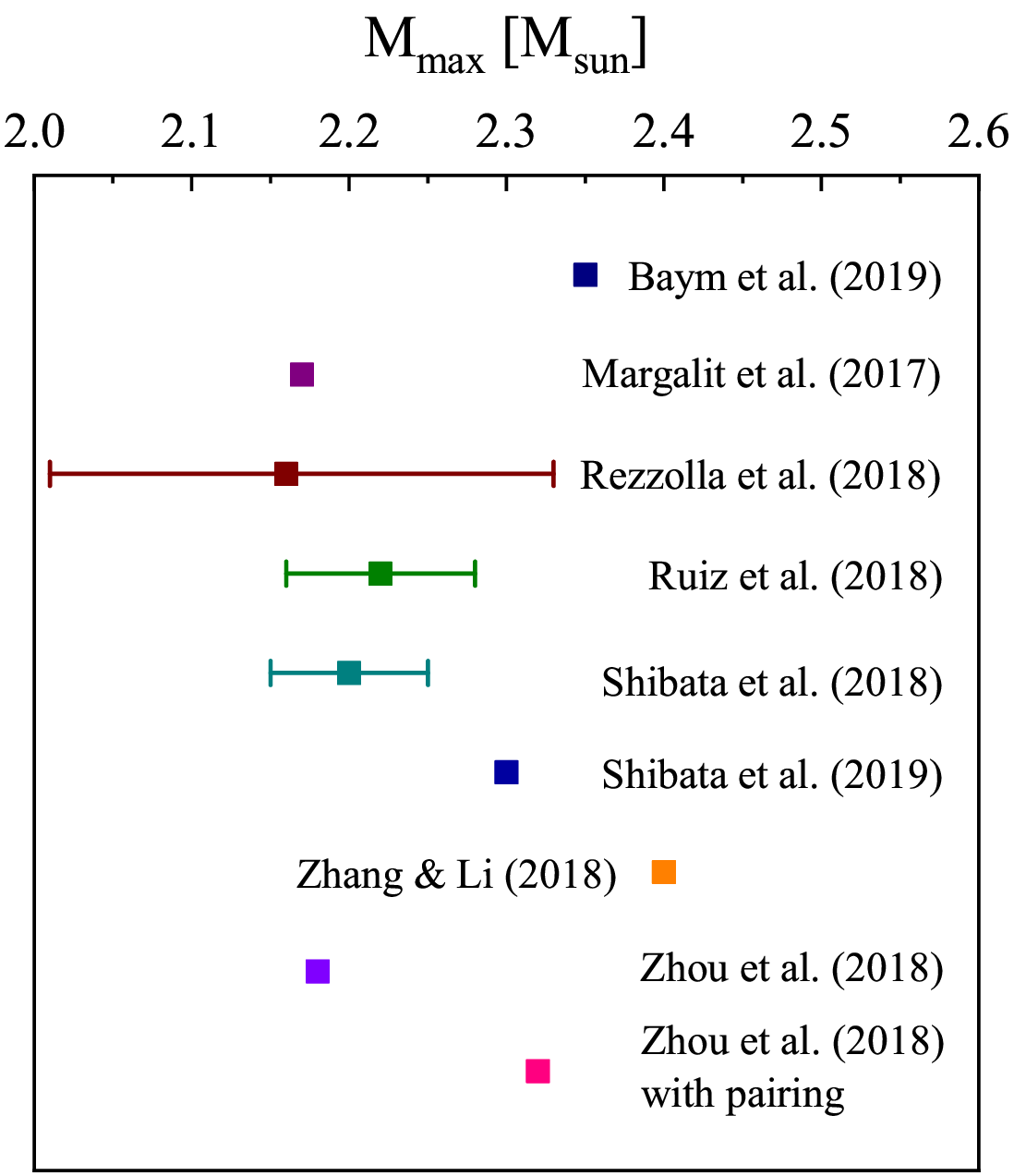}}
 \vspace{0.6cm}
\caption{(Color online) Constraints on the maximum mass of non-rotating neutron stars $M_{\rm max}$ after the detection of GW170817~\cite{Zhang2019EPJA,Baym19,Margalit2017,Shibata2017,Rezzolla2018,Ruiz2018,Zhou2018,Shibata2019}. }
\label{mass}
\end{center}
\end{figure}
 \subsubsection{The extracted maximum mass of neutron stars from analyzing the GW170817-GRB170817A-AT2017gfo events}
 While the fate of the remanent formed in the aftermath of GW170817 is still unclear, many interesting studies have been carried out to estimate the maximum mass of neutron stars using signals from GW170817. Besides emitting gravitational waves, GW170817 also emitted two strong electromagnetic signals: a short gamma-ray burst GRB170817A with a time delay of $\sim1.7$ s and a kilonova AT2017gfo powered by the radioactive decay of $r$-process nuclei synthesized in the ejecta a few days after the merger \cite{LIGO2017,LIGO2017b}. By analyzing these observations, the maximum mass of neutron stars has been suggested to be: $2.17$ M$_\odot$ at 90\% confidence~\cite{Margalit2017}, $2.16^{+0.17}_{-0.15}$ M$_\odot$ at 90\% confidence~\cite{Rezzolla2018}, $2.16-2.28$ M$_\odot$ when the ratio of the maximum mass of a uniformly rotating neutron star (the supramassive limit) over the maximum mass of a nonrotating star is within $1.2\leq\beta \leq1.27$~\cite{Ruiz2018}, $2.15-2.25$ M$_\odot$ after reducing effects of gravitational-wave emission, long-term neutrino emission, ejected mass, and rotation from the total mass of GW170817 $2.73\sim2.78$ M$_\odot$~\cite{Shibata2017}, $2.18$ M$_\odot$ (2.32 M$_\odot$ when pairing is considered) using the upper limit of $\Lambda_{1.4}=800$ (90\% confidence)~\cite{Zhou2018} and $2.3$ M$_\odot$ considering the conservation laws of energy and angular momentum self-consistently~\cite{Shibata2019}. Applications of these constrained $M_{\rm max}$ have been discussed in many recent studies~\cite{Raithel2018,Annala2018,Lim2018,Bauswein2017,Ellis2018,Han2018,Koliogiannis2018,Cowan2019}. The constraints on the maximum mass of neutron stars $M_{\rm max}$~\cite{Zhang2019EPJA,Radice2018,Baym19,Margalit2017,Shibata2017,Rezzolla2018,Ruiz2018,Zhou2018,Shibata2019} are summarized in Fig. \ref{mass}. While the collection might be incomplete, it is seen that all constrained maximum masses reported so far are less than the predicted absolutely maximum mass of $2.4$ M$_{\odot}$.

In summary of this section, the study of \esym effects on properties of non-rotating neutron stars has a long and fruitful history. This study received a strong boost from the first detection of a binary neutron star merger event. Pre-GW170817, it was known that the radii of neutron stars are sensitive to the \esym around $2\rho_0$. In fact, using the \esym constrained by terrestrial nuclear experiments, the radii of canonical neutron stars were predicted to be in the range of 11.5 km $\leq R_{1.4}\leq13.6$ km. The majority of analyses of both the X-ray and gravitational wave data found values of $R_{1.4}$ consistent with the earlier prediction within about 1km. In particular, the post-GW170817 analyses of the tidal deformability using many different approaches found a fiducial value of $R_{1.4}=12.42$ km with the lower and upper limit of $10.95$ km and $13.23$ km, respectively.
The extracted radii and tidal deformability as well as the observed maximum mass of neutron stars have been used together with the causality condition to constrain the \esym at supra-saturation densities by numerically solving the NS inverse-structure problem in a multi-dimensional EOS parameter space.

While the \esym around twice the saturation density has been constrained to $E_{\rm sym}(2\rho_0)=46.9\pm10.1$ MeV, it remains very uncertain at higher densities. While more precise measurements of neutron star radii and the tidal deformability in the inspiraling phase of neutron star mergers are expected to help further narrow down the $E_{\rm sym}(2\rho_0)$ around twice the saturation density,
new observables are needed to probe the \esym at higher densities. These new observables should carry information about the merging phase of two colliding neutron stars in space or two heavy-nuclei in terrestrial laboratories where super-dense neutron-rich matter are expected to be formed. Moreover, an absolutely maximum mass of $2.4$ M$_\odot$ independent of the EOS was predicted. Other post-GW170817 analyses using various approaches extracted a fiducial maximum masses between about $2.17$ M$_{\odot}$ to $2.30$ M$_{\odot}$ for the remanent of GW170817.  It remains an interesting question and outstanding challenge to pin down the mass boundary between massive neutron stars and black holes. Future observational determination of the fate of NS merger remanent will be very useful.

\section{Symmetry energy effects on properties and gravitational wave emissions of rotating neutron stars}
All neutron stars were born rotating. Compared to modeling non-rotating neutron stars and
understanding their properties, rotating neutron stars are more complex to study but provide more observables and thus new physics opportunities.
In this section, we examine effects of nuclear symmetry energy on several properties of both slowly and fast rotating neutron stars.

\subsection{Symmetry energy effects on the moment of inertia of slowly rotating neutron stars}\label{sec6:inertia}
\begin{figure*}[t!]
\centering
\includegraphics[scale=0.3]{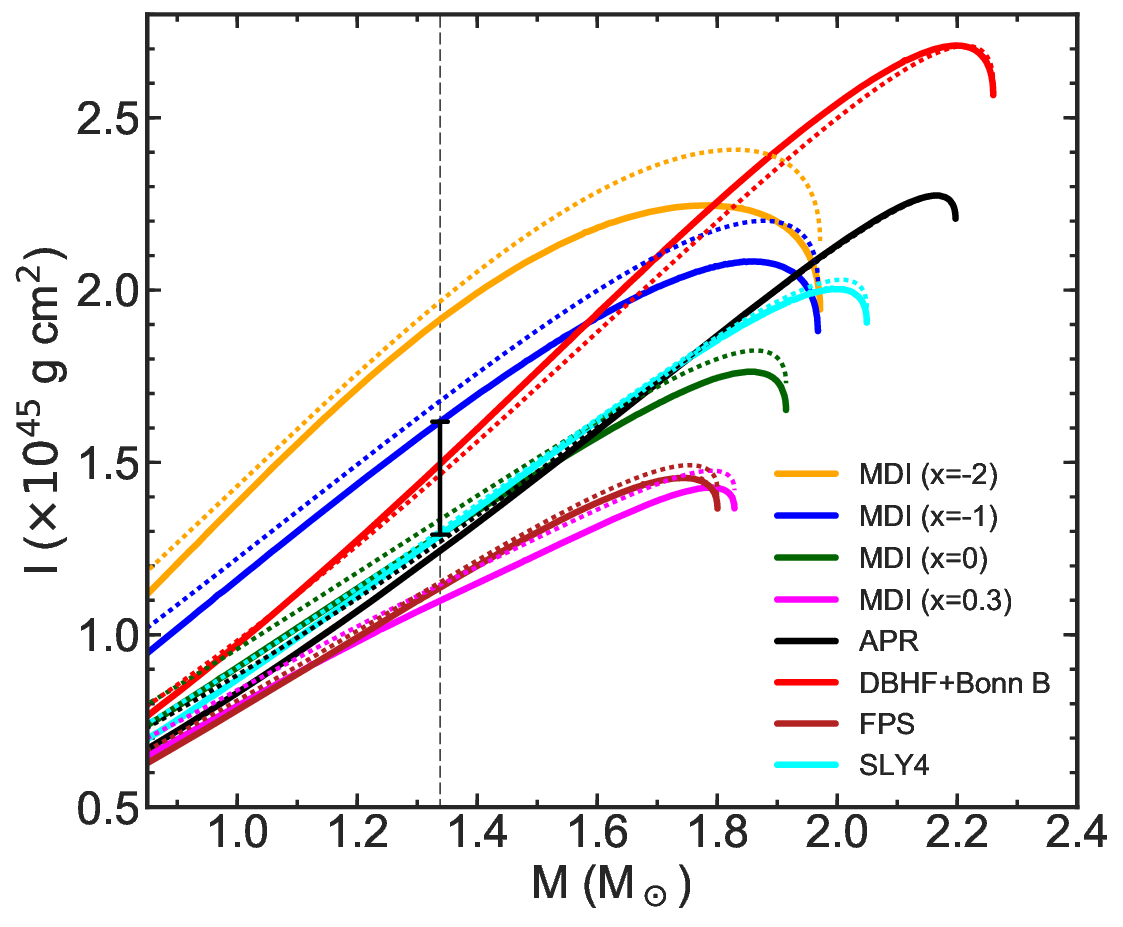}
\includegraphics[scale=0.3]{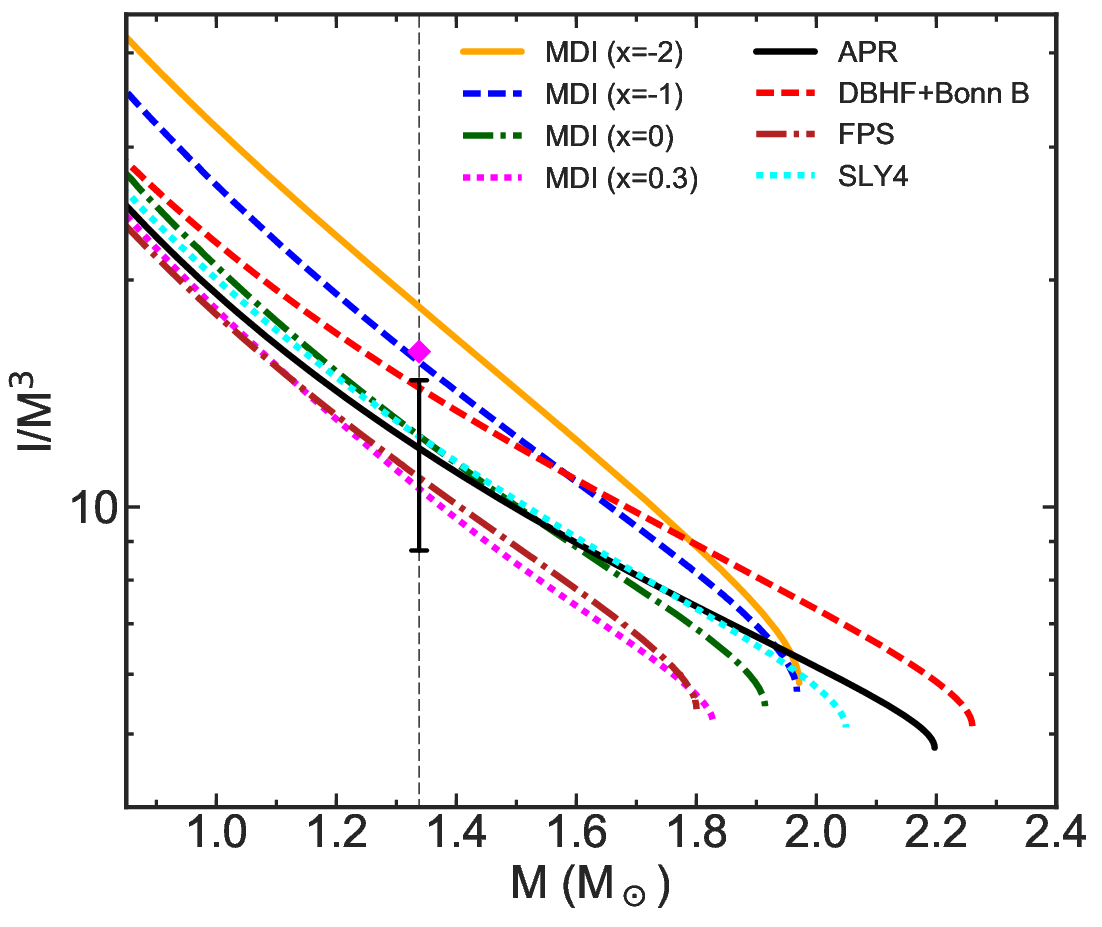}
\includegraphics[scale=0.3]{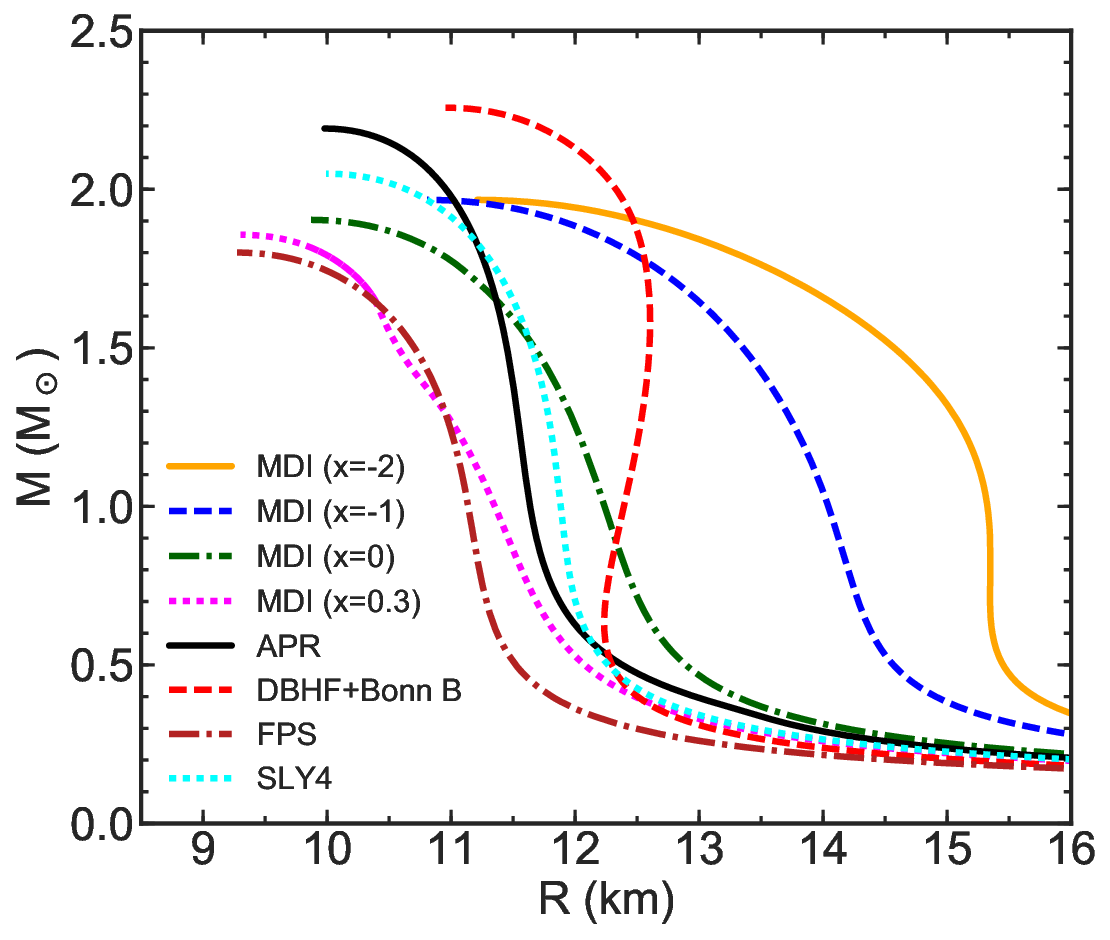}
\caption{(Color online) Total moment of inertia $I$ (left panel) and dimensionless moment of inertia $\bar{I}=I/M^3$
(middle panel) as a function of neutron-star mass $M$, and mass-radius relation (right panel). The solid curves in the
left window denote the exact calculation of $I$ and the dotted curves represent $I$ as estimated with the approximate
relation given by Eq.~(\ref{eq8_sec6}). Results with the MDI EOS are shown for several representative values of the
parameter $x$: -2, -1, 0 and 0.3. In addition, results with a number of other EOSs frequently used in astrophysical applications
are also displayed: APR~\cite{Akmal:1998cf}, DBHF+Bonn B~\cite{Alonso:2003aq,Krastev:2006ii}, FPS~\cite{P&R:1989}
and SLY4~\cite{SLy}. The dashed vertical lines at $M=1.338M_{\odot}$ represent the precisely measured mass of
PSR J0737-3039A. In the left window the error bar between the $x=-1$ and $x=0$ curves indicates the constraint on $I$ of
pulsar A based on terrestrial nuclear laboratory data from heavy-ion collisions. In the middle window the error bar denotes
the constraints on $I$ deduced by Landry and Kumar \cite{Landry:2018jyg} based the refined analysis of GW170817
\cite{LIGO2017}. The upper bound from the minimal-assumption analysis of GW170817 \cite{LIGO2017}
is also shown with a ``diamond" \cite{Landry:2018jyg}.} \label{fig3_sec6}
\end{figure*}

For rotational frequencies much lower than the Kepler frequency (the highest possible rotational frequency supported by
a given EOS), i.e. $\nu/\nu_k<<1$ ($\nu=\Omega/(2\pi)$), the deviations from spherical symmetry are very small and the
moment of inertia can be approximated from spherical stellar models. Below we first recall briefly this slow-rotation approximation,
see e.g. Ref.~\cite{Hartle:1967he}. In the slow-rotation limit the metric can be written in spherical polar coordinates
as
\begin{eqnarray}\label{eq1_sec6}
ds^2=&-&e^{2\Phi(r)}dt^2+\left(1-\frac{2m}{r}\right)^{-1}dr^2  \nonumber \\
 &-&2\omega r^2\sin^2\theta dt d\phi+r^2(d\theta^2+\sin^2\theta d\phi^2).
\end{eqnarray}
The neutron star moment of inertia is calculated in this case by solving the conventional TOV equations
together with an equation for the rotational frequency. For a slowly-rotating neutron star the moment of inertia can be
written as
\begin{equation}\label{eq2_sec6}
I = \frac{8\pi}{3}\int^R_0(\epsilon+p)e^{-\Phi}\left(1-\frac{2m}{r}\right)^{-1}
\frac{\bar{\omega}}{\Omega}r^4dr,
\end{equation}
where the metric potential $\Phi(r)$ in the stellar interior is defined by
\begin{equation}\label{eq3_sec6}
\frac{d\Phi}{dr}= ( m + 4\pi r^3 p)\left(1-\frac{2m}{r}\right)^{-1}\quad
(r<R_{star}),
\end{equation}
$\Omega$ is the angular velocity of a uniformly rotating neutron star, and $\bar{\omega}\equiv \Omega-\omega$
is the dragging rotational velocity (the angular velocity of the star relative to a local
inertial frame rotating at $\omega$), with $\omega(r)\equiv(d\phi/dt)_{ZAMO}$ the Lense - Thirring angular
velocity of a zero - angular - momentum observer (ZAMO). Inside the star $\bar{\omega}$ satisfies the equation
\begin{equation}\label{eq4_sec6}
\frac{1}{r^3}\frac{d}{dr}\left(r^4j\frac{d\bar{\omega}}{dr}\right)+4\frac{dj}{dr}\bar{\omega}=0\quad
(r<R_{star}),
\end{equation}
where
\begin{equation}\label{eq5_sec6}
j\equiv\left(1-\frac{2m}{r}\right)^{1/2}e^{-\Phi}.
\end{equation}
The second-order differential equation (\ref{eq4_sec6}) can be transformed into a first-order differential equation
by introducing $\xi \equiv d\ln \bar{\omega}/d \ln r$. Then
\begin{equation}\label{eq6_sec6}
\frac{d\xi}{dr}=\frac{4\pi r^2(\epsilon+p)(\xi+4)}{r-2m} -\frac{\xi(\xi-3)}{r}\quad
(r<R_{star}),
\end{equation}
with the boundary condition $\xi(r=0)=0$. The total moment of inertia of a slowly rotating neutron star is then
given by
\begin{equation}\label{eq7_sec6}
I = \frac{R^3 \xi_R}{6+2\xi_R},
\end{equation}
where $\xi_R=\xi(r=R)$.

With a given EOS the TOV equations and Eq. (\ref{eq6_sec6}) are integrated simultaneously from the center
of the star, where the central density $\rho_c=\rho(0)$ must be specified, to its surface. By sampling
$\rho_c$ values up to $\rho_{\rm c,max}$, the central density for which the neutron star mass reaches a maximum
$M_{\rm max}$, one generates a sequence of stable neutron-star configurations. Beyond $\rho_{\rm max}$, the stars
become unstable to radial perturbations~\cite{Harrison:1965}. For neutron stars with masses greater than
$1M_{\odot}$ Lattimer and Schutz~\cite{Lattimer:2005} found that the moment of inertia computed through
the above formalism can be very well approximated by the following empirical relation:
\begin{equation}\label{eq8_sec6}
I\simeq (0.237\pm 0.008)MR^2\left[1+4.2\frac{Mkm}{M_{\odot}R}+90\left(\frac{Mkm}{M_{\odot}R}\right)^4\right]
\end{equation}
The above equation is shown \cite{Lattimer:2005} to hold for a wide class of EOSs except for ones with
appreciable degree of softening, usually indicated by achieving a maximum mass of $\sim 1.6M_{\odot}$
or less.

For rotational frequencies much smaller than the Kepler frequency the deviations from spherical symmetry
are negligible and the moment of inertia can be computed applying the slow-rotation approximation discussed
briefly above. In the left window of Fig.~\ref{fig3_sec6} moment of inertia is shown as a function of stellar
mass. The solid curves represent the exact calculation and the dotted curves denote $I$ as computed with the
empirical relation Eq. (\ref{eq8_sec6}). As seen in the right window of Fig.~\ref{fig3_sec6}, above
$\sim 1.0M_{\odot}$ the neutron star radius  remains approximately constant before reaching the maximum mass
supported by a given EOS. Accordingly, the moment of inertia ($I\sim MR^2$) increases almost linearly with stellar mass
for all models. Right before reaching the maximum mass, the neutron star radius starts decreasing, which causes
the sharp drop in the moment of inertia. Because $I$ is proportional to $M$ and the square of $R$, it is more
sensitive to the density dependence of the nuclear symmetry energy, which determines the neutron star radius. Here
we recall that EOSs with $x=-2$ and $x=-1$ have stiffer symmetry energy, with respect to the rest of the MDI EOSs
employed here (see Fig.~\ref{Plamen18a}), which result in neutron star models with larger $R$ and, in turn, $I$.
In Fig.~\ref{fig3_sec6} it is seen that for fixed $M$ the moment of inertia exhibits a considerable
variation as $x$ changes from 0.3 to -2, together with a corresponding variation in $R$ (right window of Fig.
\ref{fig3_sec6}), which clearly signifies the dependence on $E_{\rm sym}$.

The results in the left window of Fig.~\ref{fig3_sec6} also illustrate that Eq.~(\ref{eq8_sec6}) is a very
good approximation for the moment of inertia of slowly-rotating neutron stars. This approximate relation is also
frequently written in terms of the compactness $\beta=M/R$ as, see e.g., Ref.~\cite{Lattimer:2015nhk}
\begin{equation}\label{eq9_sec6}
I\simeq (0.237\pm 0.008)MR^2[1+2.844\beta+18.91\beta^4].
\end{equation}
An improved relation, limited to $\beta\geq 0.1$ and based on a set of piecewise polytropic EOSs supporting
neutron-star models with $M_{\rm max}\geq 1.97M_{\odot}$ is derived in Ref.~\cite{Steiner16}
\begin{eqnarray}\label{eq10_sec6}
I\simeq MR^2[0.247\pm 0.002 &&+ (0.642\pm 0.012)\beta  \nonumber  \\
&&+ (0.466\pm 0.096)\beta^2].
\end{eqnarray}

The discovery of the extremely relativistic binary pulsar PSR J0737-3039A,B provides an unprecedented
opportunity to test General Relativity and the physics of pulsars~\cite{Burgay:2003jj}. Lattimer and
Schutz~\cite{Lattimer:2005} suggested that a measurement of the moment of inertia of component A of
the system accurate to 10\% is sufficient to place a firm constraint on the neutron-star EOS. Although
such a measurement is perhaps several years away~\cite{Lattimer:2015nhk}, given that the masses of both stars
are already accurately determined by observations, a measurement of the moment of inertia of even one neutron
star could have enormous importance for the neutron star physics and the EOS of dense neutron-rich matter
\cite{Lattimer:2005}. A 10\% error in the measurement of $I$ will dominate the uncertainty ($\sim 6-7\%$) of
a radius measurement \cite{Lattimer:2015nhk}. This would in turn place a strong constraint on $R$, complimentary
to the gravitational wave observations~\cite{LIGO2017} and X-ray pulsar timing measurements of $R$
\cite{Watts:2016uzu}. Independent neutron-star radius constraints have been recently established by various
research groups \cite{LIGO2017,Nandi2018,Lourenco2019,Rai,Fattoyev2018,Malik2018,Zhou2019,
Most,Annala2018,Lim2018,Radice2018,Tews2018,De2018,Bauswein2017}
based on the gravitational-wave measurement of neutron-star tidal deformability $\Lambda$ in GW170817,
using both the original~\cite{LIGO2017} and refined event analyses~\cite{LIGO2017}.
On the other hand, it has already been pointed out in the literature and discussed in detail in Sect. \ref{GW-LR}
that translating $\Lambda$ measurements
directly into $R$ constraints has to be taken with caution \cite{Zhu:2018ona}, and that both $\Lambda$ and $R$ must be
measured independently \cite{Plamen3,Zhang2018JPG} to extract meaningful constraints on the EOS of dense
neutron-rich matter. Thus, theoretical predictions of the neutron-star moment of inertia are very timely and important
in the ongoing efforts to determine the exact details of the EOS.

\begin{figure}[t!]
\centering
\includegraphics[scale=0.4]{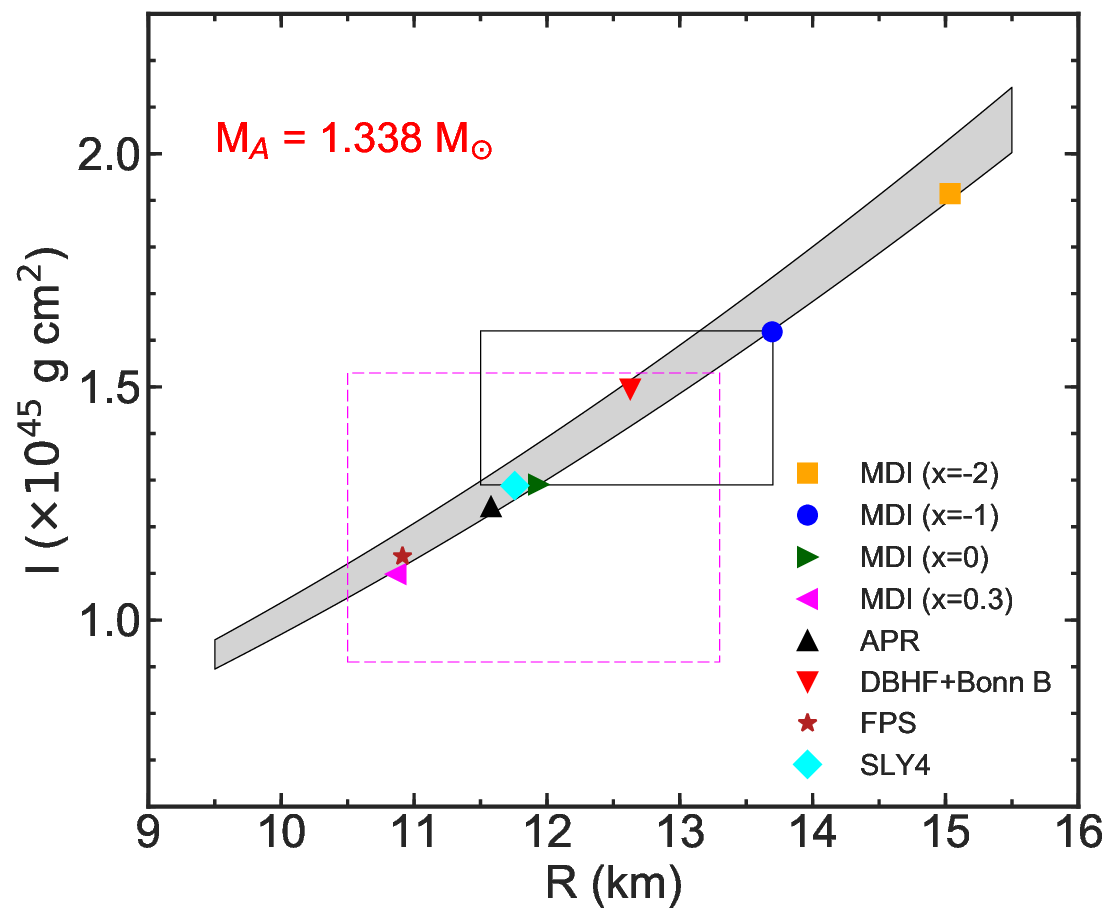}
\caption{(Color online) Radii and moments of inertia predicted for a neutron star of mass
$M_A=1.338M_{\odot}$ (PSR J0737-3039A). The light-grey shaded region indicates the empirical
relation Eq.~(\ref{eq8_sec6}). It is seen that it is an excellent approximation for the neutron
star models considered in this work. The larger rectangular region (dashed magenta lines) represents
the tighter constraints from the GW170817 observation \cite{LIGO2017} deduced by Ref.~\cite{Landry:2018jyg}.
The rectangular region (solid black lines) of $I_{\star}=[1.29 - 1.62]\times 10^{45}$~g~cm$^2$ and
$R=[11.5 - 13.6]$ km represents the constraints on $I$ and $R$ from heavy ion collision data.} \label{fig4_sec6}
\end{figure}

Previously, calculations of the moment of inertia of PSR J0737-3039A ($M_A=1.338M_{\odot}$, $\nu_A=44.05Hz$) have
been reported by Morrison et al.~\cite{Morrison:2004} and Bejger et al.~\cite{BBH2005}. More recently, Landry and
Kumar \cite{Landry:2018jyg} derived constraints on $I$ of pulsar A based on the $\Lambda$ measurement in GW170817
and universal relations among neutron-star observables. They quote $I_{\star}=1.15^{+0.38}_{-0.24}\times 10^{45}$ g cm$^2$ for
component A of the system. In Ref. \cite{Lim:2018xne} the authors applied Bayesian posterior probability distributions of the
nuclear EOS that incorporates information from microscopic many-body theory and empirical data of finite nuclei to
compute the moment of inertia of PSR J0737-3039A with a most probable value of $I_{\star}=1.35\times 10^{45}$ g cm$^2$
and a range of $I_{\star}=[0.98 - 1.48]\times 10^{45}$ g cm$^2$. Table \ref{tab1_sec6} shows the moment of inertia
(and other selected quantities) of PSR J0737-3039A. The results with the APR EOS are in very good agreement with those by
\cite{Morrison:2004} ($I^{APR}=1.24\times 10^{45}$ g cm$^2$) and \cite{BBH2005} ($I^{APR}=1.23\times 10^{45}$ g cm$^2$).
The last column of Table \ref{tab1_sec6} also includes results computed with the empirical relation (Eq.~(\ref{eq8_sec6})).
From a comparison with the results from the exact numerical calculation it is seen that Eq.~(\ref{eq8_sec6}) is
an excellent approximation for the moment of inertia of slowly-rotating neutron stars. (The average uncertainty of
Eq.~(\ref{eq8_sec6}) is $\sim 2\%$.) Our results for the MDI EOS with $x=0$ and $x=-1$ allowed us to constrain the
moment of inertia of PSR J0737-3039A  to be in the range $I_{\star}=[1.29 - 1.62]\times 10^{45}$~g~cm$^2$. These
limits are indicated by the error bar in the left window of Fig. \ref{fig3_sec6}, and overlap considerably with the
very recent constraints by Landry and Kumar \cite{Landry:2018jyg}, $I_{\star}=[0.91 - 1.53]\times 10^{45}$~g~cm$^2$,
based on the GW170817 observation. In the middle window of Fig.~\ref{fig3_sec6} we take another view of $I$,
where we display the dimensionless moment of inertia $\bar{I}=I/M^3$ as a function of the neutron-star mass $M$.
The error bar in the figure denotes the constraints on $I$ of pulsar A obtained in Ref.~\cite{Landry:2018jyg} from the
refined analysis of the gravitational wave data~\cite{LIGO2017}. The upper bound from the minimal-assumption
analysis of GW170817 \cite{LIGO2017} is also shown with a "diamond" \cite{Landry:2018jyg}. It is seen
that predictions with the MDI EOS with $x=-1$ and $x=0$ are within the gravitational-wave based (including the upper
bound) constraints.

\begin{table}[t!]
\caption{Numerical results for PSR J0737-3039A ($M_A=1.338M_{\odot}$, $\nu_A=44.05Hz$). The first column identifies
the equation of state. The remaining columns exhibit the following quantities: compactness $M/R$ (dimensionless),
radius (km), total moment of inertia $I_{\star}$ ($\times 10^{45}$ g cm$^2$), total moment of inertia $I^{LS}$
($\times 10^{45}$ g cm$^)$)  as computed with Eq.~(\ref{eq8_sec6}).}
\begin{center}
\begin{tabular}{lcccc}\hline \hline \label{tab1_sec6}
EOS &  $\beta$  & $R$ & $I_{\star}$ & $I^{LS}$      \\
\hline\hline
      MDI (x=-2)  &   0.13 &  15.03 &  1.91 &  1.97 \\
      MDI (x=-1)  &   0.14 &  13.70 &  1.62 &  1.68 \\
      MDI (x=0)   &   0.17 &  11.94 &  1.29 &  1.33 \\
      MDI (x=0.3) &   0.18 &  10.86 &  1.10 &  1.14 \\
      APR         &   0.17 &  11.58 &  1.24 &  1.27 \\
      DBHF+Bonn B &   0.16 &  12.63 &  1.50 &  1.46 \\
      FPS         &   0.18 &  10.91 &  1.14 &  1.15 \\
      SLY4        &   0.17 &  11.76 &  1.29 &  1.30 \\
\hline
\end{tabular}
\end{center}
\end{table}

This is best illustrated in Fig.~\ref{fig4_sec6} where the moment of inertia of PSR J0737-3039A is shown as a function
of stellar radius $R$. This allows for a direct comparison with the gravitational-wave constraints based on the
GW170817 event. The larger rectangular region (dashed magenta lines) represents the tighter constraints from the GW170817
observation \cite{LIGO2017} deduced by Ref.~\cite{Landry:2018jyg}. The rectangular region (solid black lines)
of $I_{\star}=[1.29 - 1.62]\times 10^{45}$~g~cm$^2$ and $R=[11.5 - 13.6]$ km represents the constraints on
$I$  and $R$ from analyzing heavy ion collision data \cite{Chen05a,LiChen05,Li:2005sr,Li2006}.
As explained in Ref.~\cite{Li2006} the minimum radius is extended to
11.5 km (from 11.9 km as obtained with the MDI ($x=0$) EOS) to account for the remaining uncertainty in the symmetric
part of the EOS. While both regions, based on the GW170817 analysis and heavy-ion collision data, reasonably overlap,
the constraints from nuclear laboratory data appear to be more restrictive. The light-grey shaded region in Fig.
\ref{fig4_sec6} denotes the empirical relation Eq. (\ref{eq8_sec6}). It is seen that all neutron-star models
considered here are within this empirical band.

\subsection{Symmetry energy effects on the ellipticity and GW emissions of slowly rotating
deformed pulsars}\label{sec6:slow_rotation}

\begin{figure*}[t!]
\centering
\includegraphics[scale=0.4]{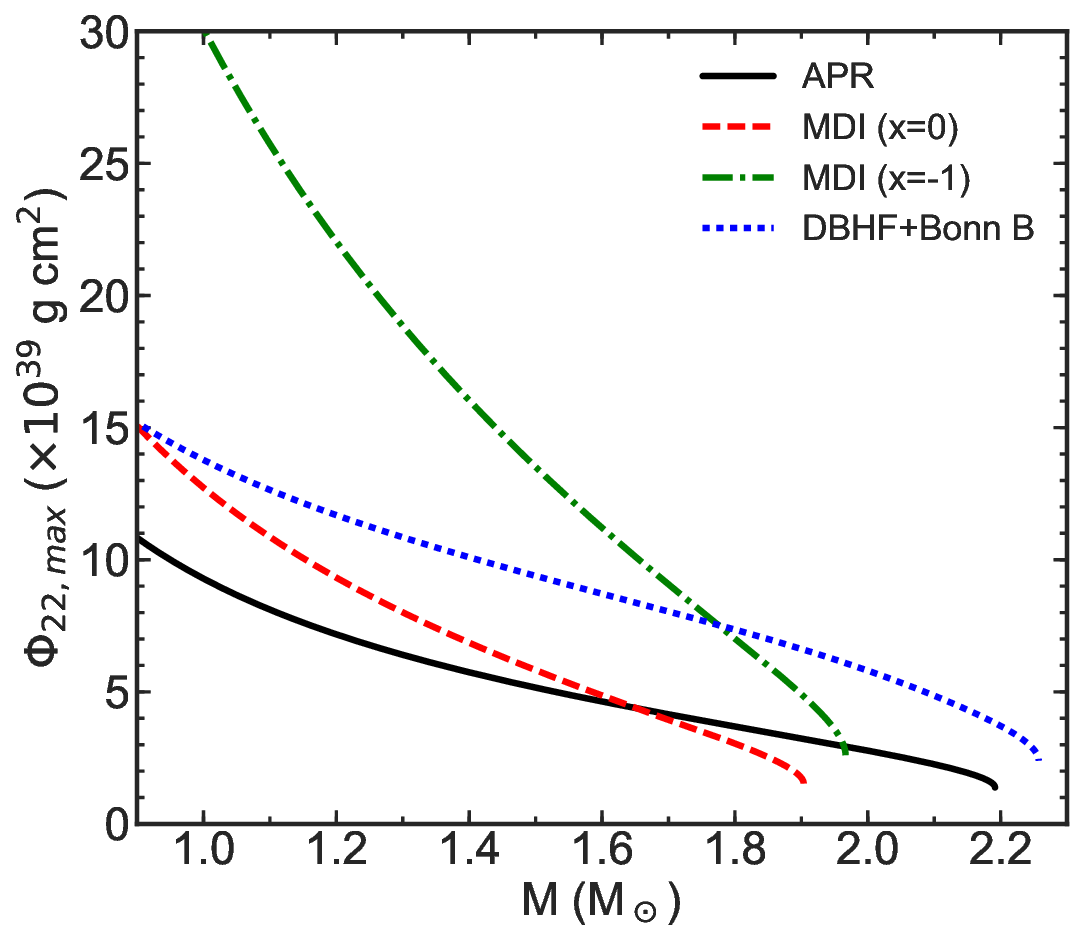}
\includegraphics[scale=0.4]{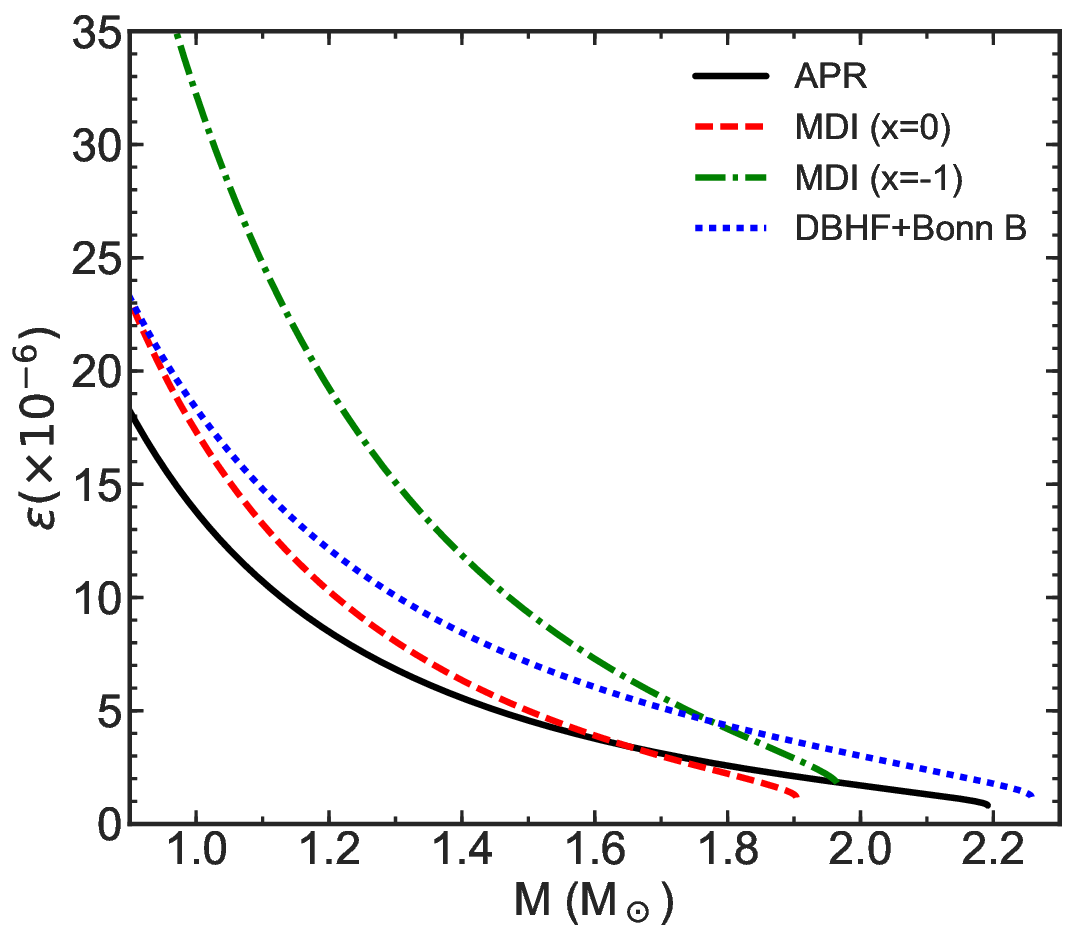}
\caption{(Color online) Neutron star quadrupole moment (left panel) and ellipticity (right panel).
Adapted from Refs.~\cite{Krastev:2008PLB,Krastev:2017NOVA}.} \label{fig5_sec6}
\end{figure*}

In the following we review the effects of $E_{\rm sym}$ on the gravitational wave emission to be expected from slowly
rotating deformed pulsars. The discussion is based mainly on Refs. \cite{Krastev:2008PLB,Krastev:2017NOVA}.
(Rapidly) rotating neutron stars are among the major candidates for sources of continuous GWs potentially detectable by
the LIGO~\cite{Abbott:2004ig} and VIRGO (e.g. Ref.~\cite{Acernese:2007zzb}) observatories. In order to generate GWs over
an extended time period, a neutron star must have some kind of a long-living axial asymmetry, e.g., a ``mountain" on its
surface \cite{Jaranowski:1998qm}. Various mechanisms that cause such asymmetries have been discussed in the literature:
\begin{enumerate}[label=(\roman*)]
  \item{Anisotropic stress built up during the crystallization period of the neutron star crust may be able
  to support long term asymmetries, such as static ``mountains" on the neutron star surface~\cite{PPS:1976ApJ}.}
  \item{Additionally, because of its violent formation in supernova the rotational axis of a neutron star may not
  necessarily be aligned with its principal moment of inertia axis, which results in a neutron star
  precession~\cite{ZS:1979PRD}. Even if the pulsar remains symmetric with respect to its rotational
  axis, because of this it generates GWs \cite{ZS:1979PRD,Z:1980PRD}.}
  \item{Also, neutron stars have extremely strong magnetic fields, which could create magnetic pressure and,
  in turn, deform the pulsar, if the magnetic and rotational axes do not coincide~\cite{BG:1996AA}.}
\end{enumerate}
The listed mechanisms generally lead to a triaxial pulsar configuration. GWs are characterized by a very small
dimensionless strain amplitude, $h_0$. The magnitude of $h_0$ depends on how much the pulsar is distorted from
axial symmetry which depends on details of the EOS of dense  neutron-rich matter. In Refs. \cite{Krastev:2008PLB,Krastev:2017NOVA}
it was demonstrated that $h_0$ depends on $E_{\rm sym}(\rho)$, and calculations with the MDI EOS with $x=0$ and $x=-1$
set the first nuclear constraints on the GW strength to be expected from selected {\it slowly} rotating pulsars.

To facilitate the following discussions, the formalism used to compute the GW strain amplitude is briefly recalled here
following closely the discussion of Ref.~\cite{Krastev:2008PLB}. A rotating neutron star generates GWs if it has some
long-living axial asymmetry. As already discussed above, there are several mechanisms that could cause stellar
deformations, and in turn GW emission. Generally, such processes result in triaxial neutron star configuration,
which in the quadrupole approximation, would generate GWs at {\it twice} the rotational frequency of the star,
$f=2\nu$ \cite{Abbott:2004ig}. These waves are characterized by a strain amplitude at the Earth's vicinity
given by \cite{HAJS:2007PRL}
\begin{equation}\label{eq11_sec6}
h_0=\frac{16\pi^2G}{c^4}\frac{\epsilon I_{zz}\nu^2}{r}.
\end{equation}
In the above equation $G$ is Newton's constant, $c$ is the speed of light, $\nu$ is the neutron star rotational
frequency, $I_{zz}$ is the stellar principal moment of inertia, $\epsilon=(I_{xx}-I_{yy})/I_{zz} $ is its
equatorial ellipticity, and $r$ is the distance to Earth. The ellipticity is related to the maximum quadrupole
moment of the star via~\cite{Owen:2005PRL}
\begin{equation}\label{eq12_sec6}
\epsilon = \sqrt{\frac{8\pi}{15}}\frac{\Phi_{22}}{I_{zz}},
\end{equation}
where for {\it slowly} rotating (and static) neutron stars
$\Phi_{22}$ can be written as~\cite{Owen:2005PRL}
\begin{eqnarray}\label{eq13_sec6}
\Phi_{\rm 22,max} = 2.4 &\times& 10^{38}g\hspace{1mm}cm^2 \left(\frac{{\sigma}_{\rm max}}{10^{-2}}\right) \nonumber \\
&\times& \left(\frac{R}{10km}\right)^{6.26} \left(\frac{1.4M_{\odot}}{M}\right)^{1.2}.
\end{eqnarray}
In this expression ${\sigma}_{\rm max}$ is the breaking strain of the neutron star crust which is rather uncertain at present time.
Although earlier studies estimated the value of the breaking strain to be in the range of ${\sigma}_{\rm max}=[10^{-5}-10^{-2}]$
\cite{HAJS:2007PRL}, more recent investigations using molecular dynamics suggested that the breaking strain could be as large as
${\sigma}_{\rm max}=0.1$~\cite{Horowitz:2009ya,Caplan:2016uvu}. This could support significant ``mountains" with large asymmetries
and ellipticity $\epsilon$ as large as $10^{-6}$ to $10^{-5}$ \cite{Caplan:2016uvu}. More recently, Baiko and Chugunov
\cite{Baiko:2018jax} argued that the maximum strain for the stretch deformation sustainable elastically is 0.04. As already
mentioned earlier in this review, the physics of the neutron-star crust is a very active area of research and the exact value
of $\sigma_{\rm max}$ is still to be determined. In a previous work \cite{Krastev:2008PLB} a rather conservative value of
${\sigma}_{\rm max}=0.01$ was used to compute $h_0$ and to set the first nuclear constraint on the GW signals to be
expected from several pulsars close to Earth. In this review, to estimate the maximum value of $h_0$, ${\sigma}_{\rm max}$ is taken
as 0.1 (as done in Ref. \cite{Krastev:2017NOVA}) to revisit some key questions. In particular, observations of continuous GWs
and/or theoretical upper limits on $h_0$ can be used to establish observational limits on $\epsilon$ \cite{Caplan:2016uvu}.
Extensive searches of continuous GWs have already been performed \cite{Aasi:2013sia,Abbott:2017ylp}, and new ones are under
way. No source of continuous GWs has been reported so far, but tighter upper limits on $\epsilon$ were established,
with $\epsilon < 10^{-8}$ obtained in the most sensitive case. According to Ref.~\cite{Caplan:2016uvu}, neutron star crust
could support large mountains leading to ellipticity larger than $10^{-8}$. However current observations suggest that, at
least for the target pulsars included in these searches, such large asymmetries did not form. Processes that are
responsible for creating mountains on neutron stars are currently largely unknown. One possibility that could cause
mountains is matter accretion where material may aggregate asymmetrically due to temperature gradients or strong magnetic
fields \cite{Caplan:2016uvu}.

Eqs.~(\ref{eq11_sec6}) and (\ref{eq12_sec6}) show that $h_0$ does not depend on $I_{zz}$ and the total dependence
upon the underlying EOS is carried by $\Phi_{22}$. Eq. (\ref{eq11_sec6}) can be therefore rewritten as
\begin{equation}\label{eq14_sec6}
h_0=\chi\frac{\Phi_{22,max}\nu^2}{r},
\end{equation}
with $\chi=(136.53\pi^5)^{1/2}G c^{-4}$.

For the purpose of this review, for slowly rotating neutron stars the moment of inertia is computed via
the empirical relation, Eq.~(\ref{eq8_sec6}). The quadrupole moment and ellipticity are calculated through
Eqs. (\ref{eq13_sec6}) and (\ref{eq12_sec6}) respectively. For rotational frequencies up to $\sim 300Hz$, global
properties of rotating neutron stars remain approximately constant \cite{Plamen2}. Therefore for slowly
rotating stars, if one knows the pulsar's rotational frequency and its distance from the Earth, the above
formalism can be readily applied to estimate the GW strain amplitude.

The focus is specifically on the effect of $E_{\rm sym}$ on the neutron-star quadrupole moment and ellipticity,
and strain amplitude of GWS from slowly rotating deformed pulsars. The neutron star quadrupole moment $\Phi_{22}$
and ellipticity $\epsilon$ are shown in the left and right windows of Fig. \ref{fig5_sec6} respectively. The
quadrupole moment decreases with increasing $M$ for all EOSs. As noted previously in Ref \cite{Krastev:2008PLB},
this decrease is dependent upon the EOS and is most pronounced for the MDI EOS with $x=-1$. This trend is explained
in terms of increasing the central density with the neutron star mass. Heavier pulsars have higher central density
and because $\Phi_{22}$ measures the pulsar's degree of distortion (Fig.~\ref{fig5_sec6}, left panel), they also
exhibit smaller deformations compared to stars with lower central densities. These findings are consistent
with previous investigations suggesting that more centrally condensed stellar configurations are less deformed by
rapid rotation \cite{1984Natur.312..255F}. As concluded in Ref.~\cite{Krastev:2008PLB}, such neutron star models are
also to be expected to be more ``resistant'' to any kind of distortion.

Since $\epsilon$ is proportional to $\Phi_{22}$, it also decreases with the increase of the neutron star
mass (see the right window of Fig~\ref{fig5_sec6}). It is observed that models with stiffer $E_{\rm sym}$, such
as the MDI EOS with $x=-1$, favor larger crust  ``mountains''. These results are consistent with recent investigations
\cite{Fattoyev:2013rga}, involving one of us, where it was demonstrated that gravitational wave signals could
provide critical information about the high-density behavior of nuclear symmetry energy.

\begin{figure}[t!]
\centering
\includegraphics[scale=0.4]{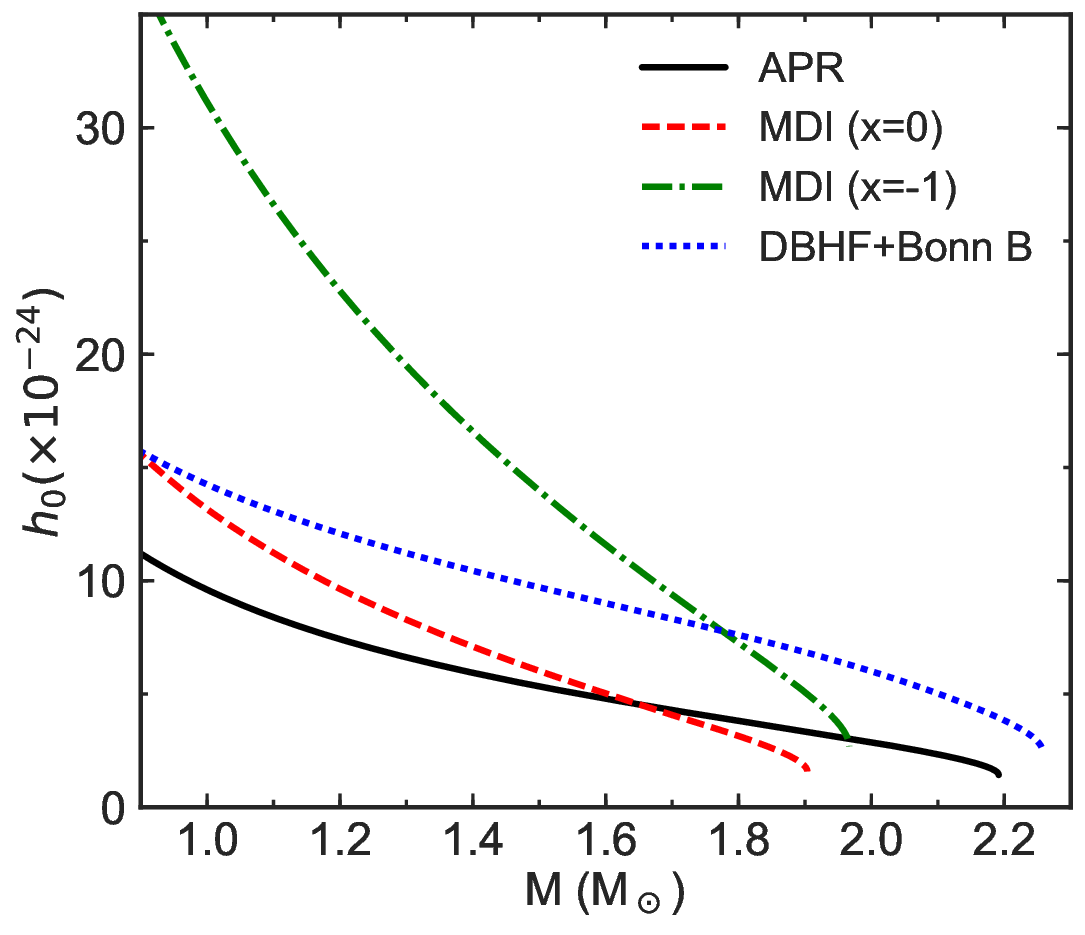}
\caption{(Color online) Gravitational-wave strain amplitude $h_0$ as a function of the neutron-star mass $M$. Results are
shown for PSR J0437-4715. Adapted from Ref.~\cite{Krastev:2017NOVA}.} \label{fig6_sec6}
\end{figure}

The general behavior of the GW strain amplitude is shown in Fig.~\ref{fig6_sec6}, where the specific case
is for PSR J0437-4715 and illustrates the main features of $h_0$ as a function of $M$. Since $h_0$ is proportional
to $\Phi_{22}$ it follows closely the trend of $\Phi_{22}$. Fig. \ref{fig6_sec6} demonstrates that $h_0$
depends on the details of the EOS and $E_{\rm sym}(\rho)$. This dependence is stronger for EOSs with
stiffer symmetry energy, e.g., the MDI EOS with $x=-1$. The dependence on $E_{\rm sym}$ is also stronger for
lighter neutron star models. As already discussed, such pulsars have lower central densities and
are therefore less bound by gravity. This makes them more easily distorted by various mechanisms,
and as a result, greater prospects for stronger continuous GWs. The GW strain amplitude computed
with the MDI EOS with $x=0$ and $x=-1$ serves as a limit on the maximum $h_0$. These estimates
ignore the distance measurement uncertainties, and they also should be considered as {\it upper}
limits of $h_0$ because $\Phi_{22}$ has been evaluated with $\sigma=0.1$, and it may go as low as
$10^{-5}$ as mentioned previously. Ref.~\cite{Krastev:2008PLB} used a conservative value of
$\sigma_{\rm max}=0.01$ to set the first direct nuclear constraint on $h_0$ of GWs to be expected for
several pulsars close to Earth. Depending on the derails of the EOS, the maximal $h_0$ was found
to be in the range of $\sim[0.5-1.5]\times 10^{-24}$. More recently, Ref.~\cite{Krastev:2017NOVA}
used $\sigma_{\rm max}=0.1$ to revisit this result and estimate $h_0$ for a larger sample of selected
pulsars deducing a wider limit on $h_0$ of $\sim[0.2-31.1]\times 10^{-24}$.

The values of $\epsilon$ shown in Fig.~\ref{fig5_sec6} (right panel) are larger than the current
upper limits ($\epsilon<10^{-8}$) deduced by the recent searches of GWs from known pulsars \cite{Abbott:2017ylp}.
Because no sources were detected, as pointed out in Ref.~\cite{Caplan:2016uvu}, this may signify
that large mountains do not form on the neutron star surface. On the other hand, neutron stars
are small and it is difficult to measure precisely their radii, and extensive all sky GW searches are
very computationally demanding \cite{Caplan:2016uvu,Bejger:2017kuu}. Observational limits on $\epsilon$ and $h_0$ are
expected to be improved in the next few years as the GW detectors continue to improve their sensitivity.
Together with advanced computational techniques, this will bring us closer to detection of GWs
from deformed pulsars.

\subsection{Symmetry energy effects on the mass-radius relation, moment of inertia and GW emissions of rapidly rotating neutron stars}
Equations of stellar structure of rapidly rotating neutron stars are considerably more
complex than those of spherically symmetric stars~\cite{Weber:1999a}. These complications
arise due to the rotational deformations in rotating stars (i.e., flattening at the poles
and bulging at the equator), which lead to a dependence of the star's metric on the polar
coordinate $\theta$. In addition, rotation stabilizes the star against gravitational collapse
and therefore rotating neuron stars are more massive than static ones. A larger mass, however,
causes greater curvature of space-time. This renders the metric functions frequency-dependent.
Finally, the general relativistic effect of dragging the local inertial frames implies the
occurrence of an additional non-diagonal term, $g^{t\phi}$, in the metric tensor $g^{\mu\nu}$.
This term imposes a self-consistency condition on the stellar structure equations, since the
degree at which the local inertial frames are dragged along by the star, is determined by the
initially unknown stellar properties, such as the mass and rotational frequency~\cite{Weber:1999a}.

Rapidly rotating neutron stars have been studied extensively in the literature and the interest
in these exotic objects have been greatly renewed after the direct detection of GWs \cite{Abbott:2016blz},
as such stars are among the major candidates for sources of continuous gravitational radiation, e.g., see
Ref. \cite{Paschalidis:2016vmz} for a recent review and references therein. Several
open-source computer codes, such as RNS \cite{Stergioulas:1994ea,Stergioulas:2003yp} and LORENE
\cite{LORENE}, solving the structure equations of rapidly rotating neutron stars, exist as public
domains and have been widely used in various studies.

\begin{figure}[t!]
\centering
\includegraphics[scale=0.4]{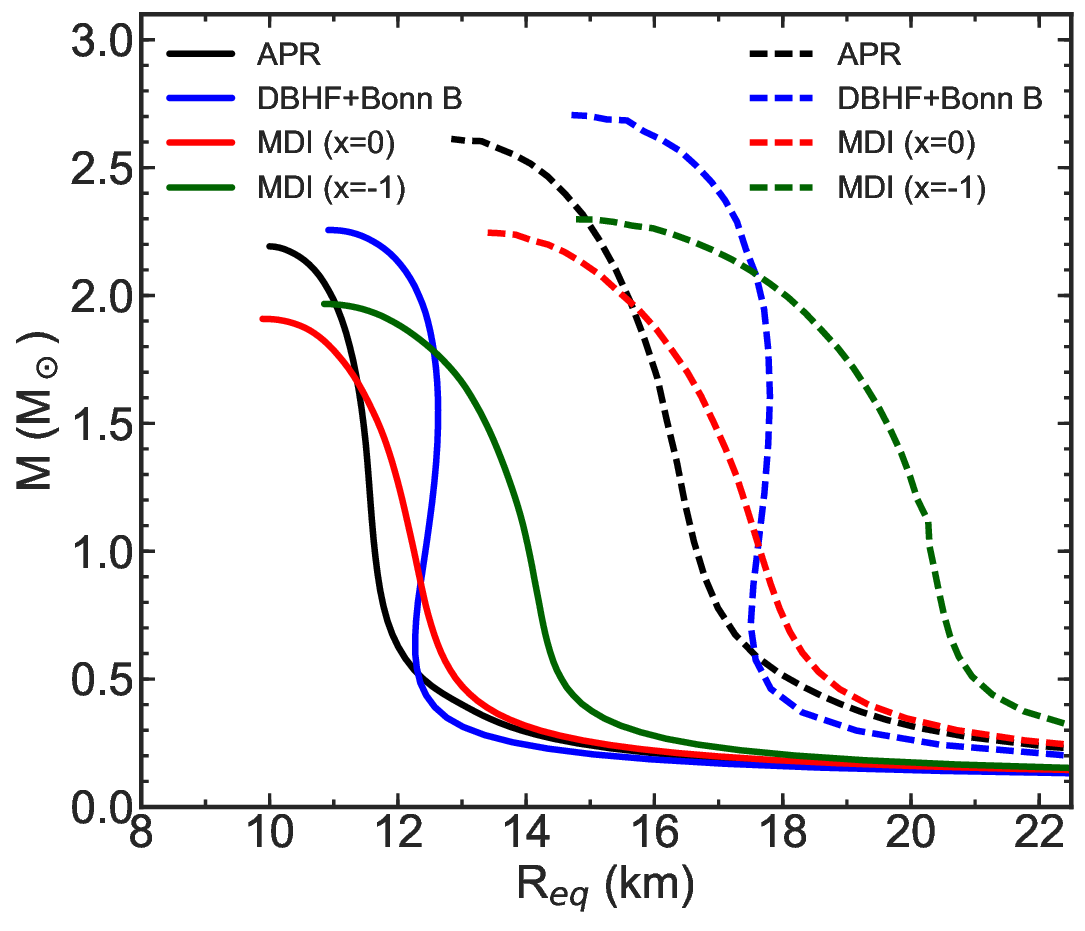}
\caption{(Color online) Mass as a function of equatorial radius. Both static (solid curves) and
Keplerian (dashed curves) sequences are shown. Adapted from Ref. \cite{Plamen1}.} \label{fig7_sec6}
\end{figure}

Ref. \cite{Plamen1} used the RNS code to construct one-parameter 2-D stationary configurations of rapidly
rotating neutron stars to investigate the effects of $E_{\rm sym}$ on various neutron star properties.
These studies employed the APR, DBHF+Bonn B and the MDI with $x=0$ and $x=-1$ EOSs. As explained earlier
in this review, the two values of $x$ were chosen as they are consistent with the isospin
diffusion data of heavy-ion collisions. The Keplerian (and static) sequences computed with RNS and these
EOSs are shown in Fig. \ref{fig7_sec6}. The sequences terminate at the ``maximum mass" point for each EOS.
It is clearly seen that rapid rotation at $\nu=\nu_k$ increases considerably the mass that can be
supported against collapse while also increasing the equatorial radius $R_{\rm eq}$. It was found in Ref.
\cite{Plamen1} that rotation at the Kepler frequency increases the NS mass up to $\sim 17\%$ with respect to static
configurations, depending on the EOS. The equatorial radius increases by several kilometers while the polar
radius decreases by several kilometers leading to an overall oblate shape of the star \cite{Plamen1}. For
each EOS the upper mass-limit is obtained for a configuration at the mass-shedding limit where $\nu=\nu_k$,
with central density $\sim 15\%$ below that of the static model with the largest mass. It was also found that
the MDI ($x=0$) EOS permits larger rotational frequency than that of the MDI ($x=-1$) EOS \cite{Plamen1}. This
is because the MDI EOS with $x=0$ has softer $E_{\rm sym}$ and results in more centrally condensed stellar
models. Such configurations can withstand larger rotational frequencies as they are bounded by stronger gravity
\cite{Bejger:2006hn}.

\begin{figure}[t!]
\centering
\includegraphics[scale=0.4]{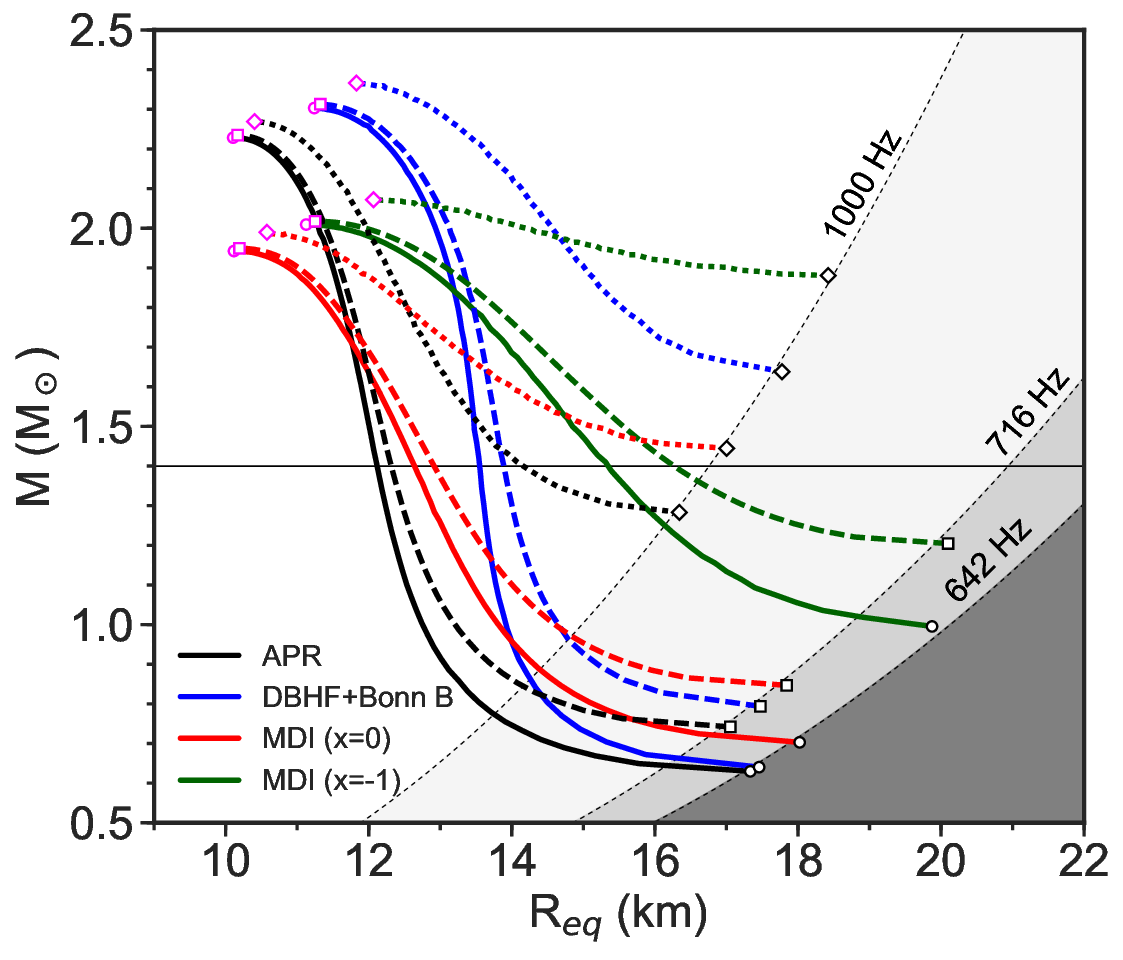}
\caption{(Color online) Gravitational mass versus circumferential radius for neutron stars
rotating at $642Hz$ (solid curves), $716Hz$ (dashed curves) and $1000Hz$ (dotted curves). Data taken from
Ref. \cite{Plamen1}. See text for details.} \label{fig8_sec6}
\end{figure}

The effects of the symmetry energy are further illustrated in Fig. \ref{fig8_sec6}, which shows neutron-star
models rotating at various fixed frequencies. Namely, sequences spinning at $642$ Hz (solid lines), $716$ Hz
(dashed lines) and $1000$ Hz (dotted lines) are displayed. (Rotational frequencies $642$ Hz and $716$ Hz are
the spinning rates of the two fastest pulsars PSR1937+214 \cite{Backer:1982} and J1748-2446ad
\cite{Hessels:2006ze}.) It is clearly seen in the figure that the range of the allowed masses by a given
EOS for rapidly rotating neutron stars becomes narrower than that of static models. This effect becomes stronger
with increasing rotational frequency and is EOS dependent. In particular, it depends on the nuclear
symmetry energy -- configurations with stiffer $E_{\rm sym}$, such as those with the MDI EOS with $x=-1$,
exhibit greater reduction of the allowed gravitational masses with increasing $\nu$. For instance, for models
rotating at $1000$ Hz for the MDI ($x=-1$) EOS the allowed mass range is $\sim 0.2M_{\odot}$. As
already discussed in the literature \cite{Bejger:2006hn}, this observation could explain why such rapidly
rotating neutron stars are so rare -- their allowed masses fall within a very narrow range. This restriction follows
from the stability with respect to mass-shedding from the star's equator requirement, which implies that at
a given gravitational mass the equatorial radius $R_{\rm eq}$ must be smaller than the $R_{\rm eq}^{\rm max}$ corresponding
to the Keplerian limit. As reported by Bejger et al. \cite{Bejger:2006hn} $R_{\rm eq}^{\rm max}$ is very well
approximated by the expression for the orbital frequency for a test particle at $r=R{\rm eq}$ in the Schwarzschild
space-time created by a spherical mass. The equation satisfying $\nu_{\rm orb}^{\rm Schw.}$, represented by
the dotted contours separating the shaded regions in Fig. \ref{fig8_sec6} is given by
\begin{equation}\label{eq15_sec6}
R_{\rm max}=\chi\left(\frac{M}{1.4M_{\odot}}\right)^{1/3}km,
\end{equation}
with $\chi=22.52$ for rotational frequency $\nu=642$ Hz, $\chi=20.94$ for $\nu=716$ Hz and
$\chi=16.76$ for $\nu=1000$ Hz, respectively.

\begin{figure}[t!]
\centering
\includegraphics[scale=0.7]{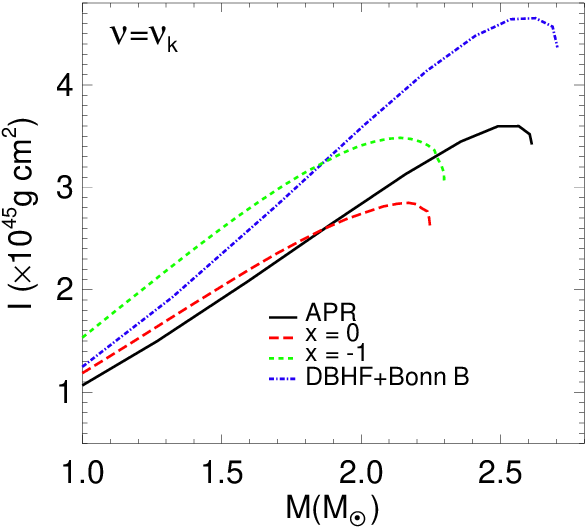}
\caption{(Color online) Total moment of inertia for Keplerian models. The neutron star sequences are computed
with the RNS code. Taken from Ref. \cite{Plamen2}.} \label{fig9_sec6}
\end{figure}

\begin{figure}[t!]
\centering
\includegraphics[scale=0.55]{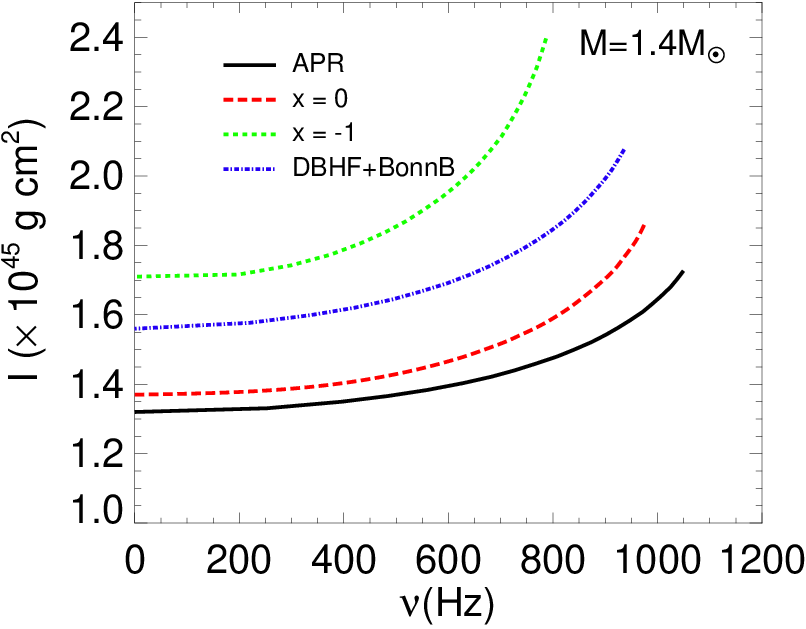}
\caption{(Color online) Total moment of inertia as a function of rotational frequency for
stellar models with mass $M=1.4M_{\odot}$. Taken from Ref. \cite{Plamen2}.} \label{fig10_sec6}
\end{figure}

Rapid rotation also affects significantly the neutron star moment of inertia where the magnitude
of the effect depends on the details of the EOS, and in particular on the density dependence of
$E_{\rm sym}$. In Ref. \cite{Plamen2} the moment of inertia was calculated numerically with the
RNS code, and the imprints of the nuclear symmetry energy were examined employing the MDI EOS.
The interested reader is referred to the original publications by Nikolaos Stergioulas, e.g., Refs.
\cite{Stergioulas:1994ea,Stergioulas:1997ja,Stergioulas:2003yp}, for details on the structure
equations of rapidly rotating neutron stars and implemented numerical scheme. The moment of inertia
versus stellar mass is shown in Fig. \ref{fig9_sec6} for neutron star models rotating at the mass-shedding
(Kepler) frequency. It is seen that the moments of inertia of rapidly rotating neutron stars are significantly
larger than those of slowly rotating models (for a fixed mass). This is easily understood in terms of the
increased (equatorial) radius (Fig. \ref{eq7_sec6}). In addition, $I$ increases with rotational
frequency $\nu$ at a rate dependent upon the details of the EOS. This effects is best seen in Fig. \ref{fig10_sec6}
where the moment of inertia is displayed as a function of rotational frequency for stellar models with
a fixed mass $M=1.4M_{\odot}$. The neutron star sequences shown in Fig. \ref{fig10_sec6} terminate at
the Kepler frequency. It is seen that at the lowest frequencies $I$ remains approximately constant
for all EOSs (which justifies the use of the slow-rotation approximation Eq. (\ref{eq8_sec6})).
As the stellar models approach the Kepler frequency, $I$ exhibits increases sharply. This is because of
the large increase of the circumferential radius as the star approaches the ``mass-shedding point''.
As pointed out previously by Friedman et al. \cite{1984Natur.312..255F}, properties of rapidly
rotating neutron stars display greater deviations from those of spherically symmetric (static) stars
for configurations computed with stiffer EOSs. This explains why the momenta of inertia of rapidly rotating
neuron star configurations with the MDI ($x =-1$) EOS show the greatest deviation from those of static
models. As already discussed in this review, since the ``stiffness'' of the MDI EOS is mainly controlled
by the density dependence of $E_{\rm sym}$, the degree of deviation of $I$ of rapidly rotating neutron stars
from those of static models is an effect clearly due to the nuclear symmetry energy.

Rapidly rotating neutron stars are among the major candidates for sources of continuous gravitational waves
and the LIGO and Virgo Collaboration reported their latest results from searching GWs from known
pulsars \cite{Abbott:2019bed}. Given a measure of the pulsar rotational frequency $\nu$, magnitude of
its time derivative $|\dot{\nu}|$ and distance $r$, the GW signal can be constrained assuming that
all of the star's rotational energy is lost via gravitational radiation \cite{Abbott:2019bed}. This
theoretical value, called {\it spin-down limit}, is given by
\begin{equation}\label{eq16_sec6}
h_0^{sd}=8.06\times 10^{-19}I_{38}^{1/2}\left[\frac{1kpc}{r}\right]\left[\frac{\dot{\nu}}{Hz/s}\right]^{1/2}
\left[\frac{Hz}{\nu}\right]^{1/2},
\end{equation}
where $I_{38}$ is the neutron star moment of inertia in units of $10^{38}$ kg m$^2$. Usually in these searches
the canonical value of $I\approx 10^{38}$ kg m$^2$ (or 10$^{39}$ g cm$^2$) is assumed \cite{Abbott:2019bed}.
Then, the corresponding spin-down limit on the pulsar's fiducial ellipticity can be obtained from
Eq. (\ref{eq11_sec6})
\begin{equation}\label{eq17_sec6}
\epsilon^{sd}=0.237 I_{38}^{-1}\left[\frac{h_0^{sd}}{10^{-24}}\right]\left[\frac{Hz}{\nu}\right]^2
\left[\frac{r}{1kpc}\right],
\end{equation}
which does not depend on the star's distance \cite{Abbott:2019bed}.

PSR B1937+21 is a particularly interesting example of a rapidly rotating neutron star rotating at
642 Hz~\cite{Backer:1982}. Since its first observation in 1982, this pulsar has been studied extensively
and an observed spin-down rate has been measured. Using the spin-down rate, an upper limit on the
gravitational wave strain amplitude was obtained. The spin-down rate corresponds to a loss in kinetic
energy at a rate of $\dot{E}$ = $4\pi^2I_{zz}\nu|\dot{\nu}|\sim$[0.6 -- 3.1] $\times$10$^{36}erg/s$,
depending on the EOS. Assuming that the energy loss is completely due to the gravitational radiation,
the gravitational wave strain amplitude can be calculated through Eq. (\ref{eq16_sec6}). Similar calculations
for this pulsar and others with an observed spin-down rate have been performed in the past
\cite{Abbott:2004ig,Abbott:PRD2007}. These calculations have provided estimates for the GW strain
amplitude of selected pulsars for which the spin-down rates are available, and also upper bounds for
their ellipticities using the quadrupole model \cite{Abbott:2004ig}. However, such calculations simply assumed
the ``fiducial'' value of $10^{45}$$g$ $cm^2$ for the moment of inertia $I_{zz}$ in all estimates. On the other
hand, the neutron star moment of inertia is sensitive to the details of the EOS of stellar matter, and especially
to the density dependence of the nuclear symmetry energy~\cite{Plamen2}. Moreover, $I_{zz}$ increases with
increasing rotational frequency (see Fig.~\ref{fig10_sec6}) and the differences with the static values of the
moment of inertia  could be significant, particularly for rapidly rotating neutron stars.

\begin{figure*}[t!]
\centering
\includegraphics[scale=0.6]{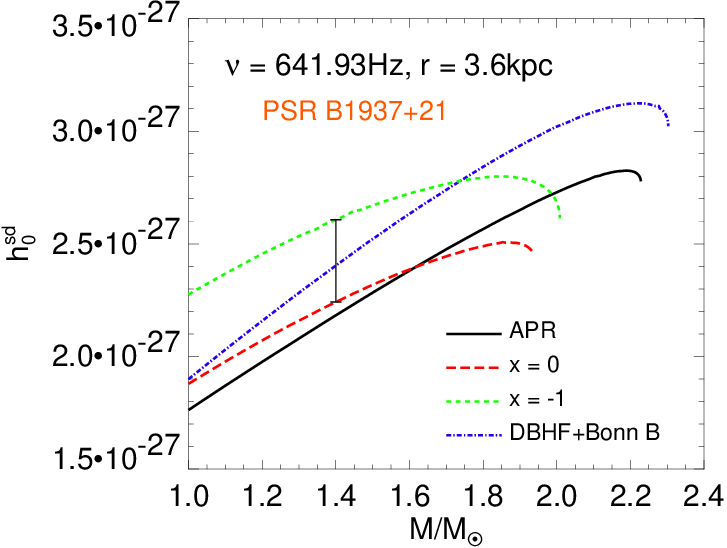}
\includegraphics[scale=0.6]{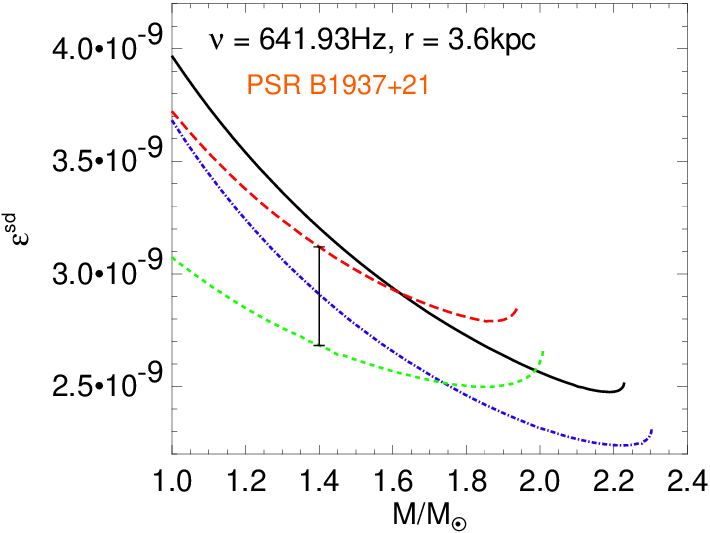}
\caption{(Color online) Gravitational wave strain amplitude $h_0^{sd}$ (left panel) and ellipticity $\epsilon^{sd}$
(right panel), deduced from the spin-down rate of PSR B1937+21. Taken from Ref. \cite{Krastev:2010NOVA}.} \label{fig11_sec6}
\end{figure*}

The GW strain amplitude of PSR B1937+21 was calculated in Ref. \cite{Krastev:2010NOVA} and  is shown in
Fig.~\ref{fig11_sec6} (left panel). Because the MDI EOS is constrained by available nuclear laboratory data,
the results with the $x=0$ and $x=-1$ EOSs place a rather conservative {\it upper} limit on the gravitational
waves to be expected from this  pulsar, provided the {\it only} mechanism accounting for its spin-down rate is
gravitational radiation. Under these circumstances, the upper limit of the strain amplitude, $h_0^{sd}$,
for neutron star models of $1.4M_{\odot}$ was found to be in the range of $h_0^{sd}=[2.24-2.61]\times 10^{-27}$
\cite{Krastev:2010NOVA}. Similarly, the upper limit of the ellipticity of PSR B1937+21 was found to be in the
range of $\epsilon^{sd}=[2.68-3.12]\times 10^{-9}$ \cite{Krastev:2010NOVA}, and is shown in Fig. \ref{fig11_sec6}
(right panel). So far the searches for GWs from isolated pulsars have found no evidence for a continuous GW
signal but derived very restrictive upper spin-down limits for several pulsars (see Ref. \cite{Abbott:2019bed}
and discussions therein). Further improvements to the sensitivity of the GW detectors are expected to surpass
the current spin-down limits for a much wider class of pulsars and bring us closer to detecting continuous GWs.

In summary of this section, while the effect of rotation on the maximum mass is only about 20\% even at the Kepler frequency,
its effect on the radius can be several kilometers depending on
the rotation frequency. Most of the properties of rotating neutron stars, such as their moments of inertia, quadrupole
deformations/ellipticities and equatorial radii depend on the symmetry energy to various degrees. Effects of the \esym on
these properties are clearly reflected in the strain amplitudes of gravitational waves emitted from isolated pulsars mostly
through the moment of inertia. One major uncertain is the breaking strain of neutron star crust. Very rich and interesting
physics regrading nuclear symmetry energy may be extracted from future measurements of either the moment of inertia directly
and/or indirectly from the continuous gravitational waves emitted by isolated pulsars.

\section{Symmetry energy effects on oscillations of neutron stars}
It has long been known that oscillations of neutron stars carry important information about their internal structures and the underlying EOS \cite{Thorne1967,Thorne1969a,Thorne1969b,Thorne1970}, see, e.g., ref. \cite{Cox-book} for an earlier review. Based on the pioneering work on the quasi-normal oscillations (QNO) of static and spherical celestial bodies \cite{Thorne1967,Thorne1969a,Thorne1969b,Thorne1970}, the operable-numerical-calculation formalism for the QNO of neutron stars was well established \cite{Lindblom83,Detweiler85,Chandrasekhar1991a,Chandrasekhar1991b}. In this formalism, by expanding the perturbed metric tensor for the QNO in spherical harmonic functions, the oscillation modes were divided into two categories \cite{Benhar1999}. They are the axial and polar modes in which the spherical harmonics transform under a parity operation as $(-1)^{l+1}$ and $(-1)^{l}$, respectively.

Many pulsation modes of neutron stars can be excited. For example, the following modes may be excited for non-rotating neutron stars \cite{Chandrasekhar1991a,Chandrasekhar1991b,Benhar1999,Kokkotas1992,Andersson01}:
(1) The high frequency pressure p-modes, where the fundamental p-mode is called f-mode. The
latter is one of the most important and widely studied modes. Its frequency is in the range of about 2-4 kHz while the first overtone of  p-mode is over
4 kHz. (2) The low frequency gravity g-modes are associated with the fluid buoyancy. They are related to the gradients of internal composition and temperature
of neutron stars. Their typically frequencies are about a few hundred Hz. (3) The w-modes are purely due to general relativistic effects. They are thus only associated
with the space-time curvature. They have relatively higher frequencies (above 7 kHz) and shorter damp times
(tens of milliseconds). Many interesting researches have been done on this topic in recent years, see, e.g. refs.  \cite{Chandrasekhar1991a,Chandrasekhar1991b,Kokkotas1992,Benhar1999,Andersson01,Wen2009,Wen2012,Sotani2017,Wen2019}.
One outstanding feature is the scaling (i.e., relations independent of the EOSs) of some combinations of the frequency, mass and radius of the f, p and w modes. Various proposals have been made to make good use of these scalings.

For rotating neutron stars,  the low-frequency r-mode has received much attention, see, e.g., ref. \cite{Andersson01} for an earlier review. It belongs to a class of purely axial inertial modes with the Coriolis force as the restoring force. Because of the Chandrasekhar-Friedmann-Schutz (CFS) instability \cite{Chandrasekhar70,Friedmann78}, detectable gravitational waves can be generated by the
r-mode. Meanwhile, it can also be used to explain why the young pulsars have relatively slower spin periods \cite{Andersson01,Lindblom98,Owen98}.
In addition, crustal oscillations, i.e., the torsional modes may also be excited \cite{Schumaker1983,Samuelsson2007,Sotani2007,Steiner2009}. The restoring force for the torsional oscillations are believed to come from the shear modulus of the solid crust.

Gravitational radiations from oscillating neutron stars are expected to carry invaluable information about stellar properties and the EOSs of dense neutron-rich matter. This longstanding hope has
received a strong boost recently from the GW170817 event. In fact, some interesting observations and predictions have been made about the various oscillation modes and their relations with the EOS.
In this section, we discuss a few effects of nuclear symmetry energy on some features of these oscillation modes. Hopefully, some of the effects will be detected using advanced gravitational wave detectors \cite{Mot18,Blazquez2014}, such as the  Einstein Telescope \cite{Pitkin2011,sensitivities}.

\subsection{Symmetry energy effects on the Brunt-V\"ais\"al\"a frequency and the core g-modes}
For ease of the following discussions, lets first recall here the definition of the Brunt-V\"ais\"al\"a frequency \cite{Brunt}.
Considering a fluid element at mass density $\rho_e$ in a fluid with density stratification in the vertical direction $z$, if it is displaced by $\Delta z$ from its original position it then experiences a vertical force proportional to $-g\partial \rho(z)/\partial z$ where $g=|{\bf g}|$ is the local gravitational acceleration.
The $\Delta z$ then satisfies a second-order differential equation after applying Newton's second law to the motion of the fluid element. Its solution has the form
$\Delta z= \Delta z_0\cdot e^{\sqrt{-N^2}t}$ where
\begin{equation}
N=\sqrt{-\frac{g}{\rho_e}\frac{\partial \rho(z)}{\partial z}}
\end{equation}
is the so-called Brunt-V\"ais\"al\"a frequency.
If the force is back towards the initial position (restoring), the stratification is said to be stable and the fluid parcel oscillates vertically. In this case, $N^2 > 0$ and $N$ is the angular frequency of oscillation. While $N^2 < 0$ indicates that the stratification is unstable, leading to overturning or convection.

Detailed discussions of the Brunt-V\"ais\"al\"a frequency and its effects on the g and f modes of neutron star oscillations can be found in Sections 17.2 and 17.8 of Cox' book \cite{Cox-book}. Various kinds of g-modes may be triggered, such as those associated with the thermally induced buoyancy in warm neutron stars \cite{Mc1,Mc2}, crustal g-modes due to the composition discontinuities in the outer envelope of cold neutron stars \cite{Fin87}, and the g-modes due to the buoyancy induced by proton-neutron composition gradient in the cores of neutron stars \cite{RG1,DongLai}. While all of these oscillations strongly depend on the EOS, it is particularly interesting to note that the core g-mode is directly related to the density dependence of nuclear symmetry energy \cite{RG1,DongLai}. In cold neutron stars, it is known that \cite{Cox-book,RG1,DongLai}
\begin{equation}
N^2=g^2\left({1\over c_e^2}-{1\over c_s^2}\right)
\end{equation}
where $c_s$ is the adiabatic sound speed given by
$
c_s^2=\left({\partial P\over\partial\rho}\right)_{x}
$
taken at a constant proton fraction $x$ and $c_e$ is given by
$
c_e^2={dP\over d\rho}
$
taken at $\beta$ equilibrium.  As discussed in refs. \cite{RG1,DongLai}, the fluid element considered is always in pressure equilibrium with the
surroundings, but its composition stays a constant during the adiabatic process because the timescale for the weak interaction leading to $\beta$ equilibrium is much longer than the dynamical timescale.
Thus, the  ``adiabatic'' condition implies a constant composition characterized by a fixed proton fraction $x$. Therefore, the convective stability or condition for stable g-mode $N^2>0$ is
equivalent to requiring $c_s^2-c_e^2>0.$ Dong Lai has further expressed the Brunt-V\"ais\"al\"a frequency in terms of the density dependence of nuclear symmetry energy and investigated its effects on the g-mode frequency \cite{DongLai}.
\begin{figure}
  \centering
   \resizebox{0.8\textwidth}{!}{
  \includegraphics{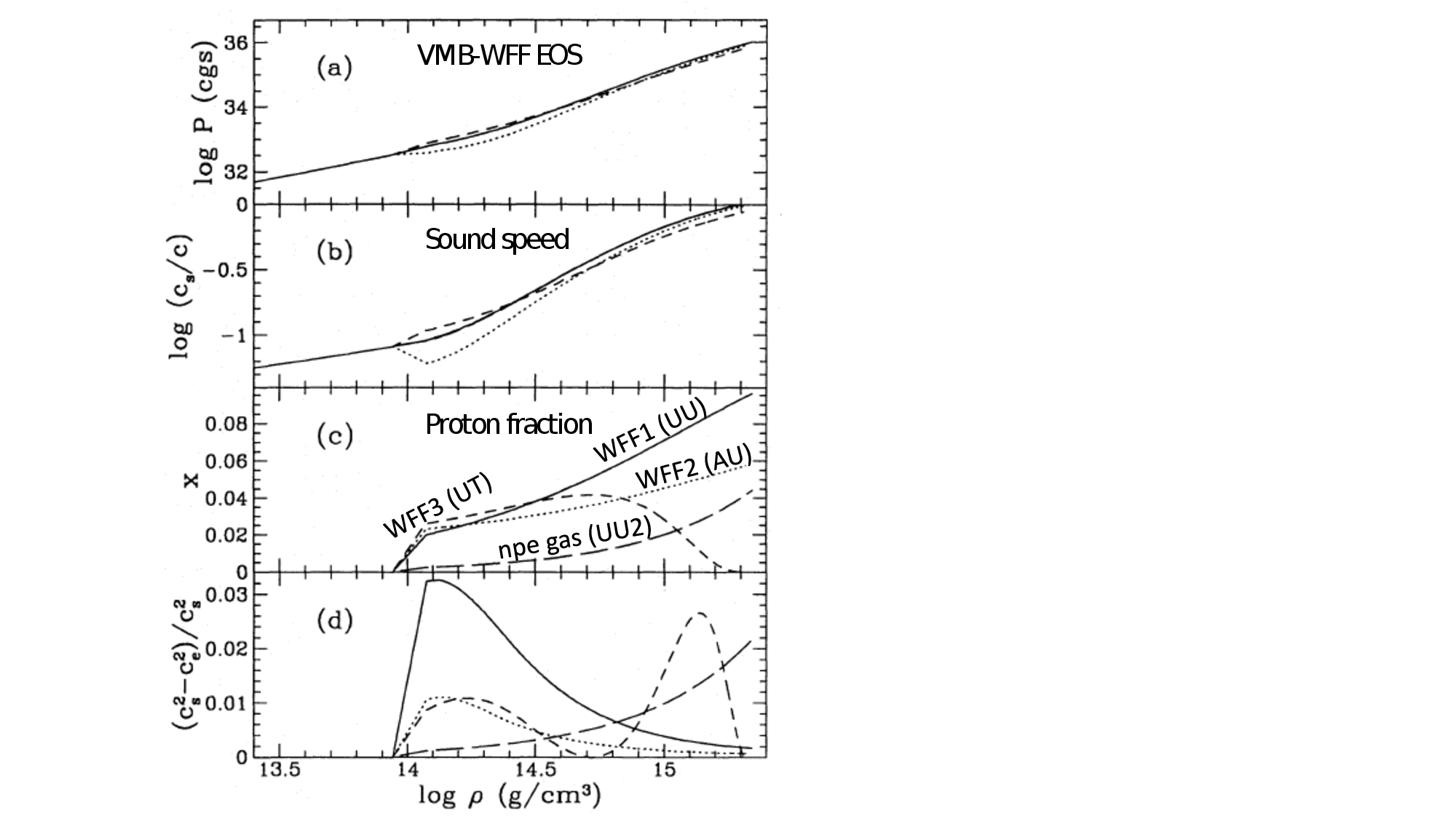}
  }
  \vspace{0.3cm}
  \caption{
(a) pressure $P$; (b) adiabatic sound speed $c_s$; (c) the proton fractions $x$ corresponding to the symmetry energy functions labeled as WFF1 (UU), WFF2 (AU) and WFF3 (UT) in Fig. \ref{examples} as well as with only the kinetic symmetry energy for the $npe$ gas model (UU2, long dashed line); (d) the fractional difference between $c_s^2$ and $c_e^2$ used in calculating the Brunt-V\"ais\"al\"a frequency by Dong Lai \cite{DongLai} using four EOSs based on the Variational Many-Body (VMB) theory of Wiringa, Fiks and Fabrocini (WFF) \cite{Wiringa1988}. Taken from ref. \cite{DongLai}.}
\label{DongFig1}
\end{figure}
Assuming neutron stars are made of $npe$ matter with an electron fraction $x_e$ and using Dong Lai's notations with $n$ as the baryon number density, $\mu_n,\mu_p$ and $\mu_e$ as the chemical potentials of neutrons, protons and electrons, respectively, the difference between $c_s^2$ and $c_e^2$ can be written as \cite{DongLai}
\begin{eqnarray}\label{Dong}
c_s^2-c_e^2&=&{n\over\rho+P/c^2}\left({\partial P\over\partial n}
-{dP\over dn}\right)=-{n\over\rho+P/c^2}
\left({\partial P\over\partial x}\right){dx\over dn} \nonumber\\
&=&-{n^3\over\rho+P/c^2}\left[{\partial\over \partial n}(\mu_e+\mu_p-\mu_n)
\right]{dx\over dn}
\end{eqnarray}
where the partial derivative is taken with respect to $n$ or $x$.  At $\beta$ equilibrium, the relation $x(n)$ is determined by $\mu_n-\mu_p=\mu_e$ and the charge neutrality condition $x_e=x$.
The ${dx\over dn}$ can then be written as \cite{DongLai}
\begin{equation}
{dx\over dn}=-\left[{\partial\over \partial n}(\mu_e+\mu_p-\mu_n)
\right]
\cdot\left[{\partial\over \partial x}(\mu_e+\mu_p-\mu_n)\right]^{-1}.
\end{equation}
The Eq. (\ref{Dong}) then becomes
\begin{eqnarray}
c_s^2-c_e^2&=&{n^3\over\rho+P/c^2}
\left[{\partial\over \partial n}(\mu_e+\mu_p-\mu_n)\right]^2\nonumber\\
&\cdot&\left[{\partial\over \partial x}(\mu_e+\mu_p-\mu_n)\right]^{-1}.
\end{eqnarray}
The convective stability $c_s^2-c_e^2>0$ thus requires
\begin{equation}\label{c-condition}
\frac{\partial}{\partial x} (\mu_e+\mu_p-\mu_n)>0
\end{equation}
where the electron chemical potential $\mu_e=\hbar c(3\pi^2n)^{1/3}x^{1/3}$
while the $\mu_n-\mu_p$ can be obtained from
\begin{equation}\label{Cnp}
\mu_n-\mu_p=-{\partial E_n\over \partial x}.
\end{equation}
 As discussed early, often one uses the parabolic approximation of the EOS, i.e., $E(n,x)=E_0(n)+E_{\rm sym}(n)(1-2x)^2$. Within this approximation,
 the Eq. (\ref{Cnp}) is reduced to
\begin{equation}
\mu_n-\mu_p=4E_{\rm sym}(n)(1-2x).
\end{equation}
Moreover, since $\partial \mu_e/\partial x=\mu_e/3x$ and $\partial (\mu_n-\mu_p)/\partial x=-8E_{\rm sym}(n)$, the convective stability condition Eq. (\ref{c-condition}) can be further reduced to
\begin{equation}
\frac{\mu_e}{3x}+8E_{\rm sym}(n)>0.
\end{equation}
Clearly, the density dependence of nuclear symmetry energy $E_{\rm sym}(n)$ plays a key role in determining the density range where the g-mode is stable.
Sometimes, one parameterizes the EOS by explicitly separating out the kinetic and potential contributions, namely,
$
E_n(n,x)=T_n(n,x)+V_0(n)+V_2(n)(1-2x)^2
$
where
\begin{equation}
T_n(n,x)={3\over 5}{\hbar^2\over 2m_n}(3\pi^2n)^{2/3}[x^{5/3}
+(1-x)^{5/3}]
\end{equation}
is the kinetic energy of a free Fermi gas of neutrons and protons, $V_0(n)$ and $V_2(n)$ are the isospin-independent and isospin-dependent parts of nucleon potential energy, respectively.
In this case,
\begin{equation}
\mu_n-\mu_p=4(1-2x)V_2+{\hbar^2\over
2m_n}(3\pi^2n)^{2/3}\left[(1-x)^{2/3}-x^{2/3}\right]
\end{equation}
and the convective stability condition Eq. (\ref{c-condition}) can be written as \cite{DongLai}
\begin{equation}
{1\over 3x}\mu_e+8V_2+{2\over 3}{\hbar^2\over 2m_n}(3\pi^2n)^{2/3}
\left[(1-x)^{-1/3}+x^{-1/3}\right]>0.
\end{equation}
Now, the individual roles of the kinetic and potential parts of the symmetry energy are clearly revealed. As it was pointed out by Dong Lai \cite{DongLai}, unless the potential symmetry energy $V_2$ is extremely negative, the core g-modes always exist.

\begin{figure}
  \centering
    \vspace{-0.5cm}
   \resizebox{0.9\textwidth}{!}{
  \includegraphics{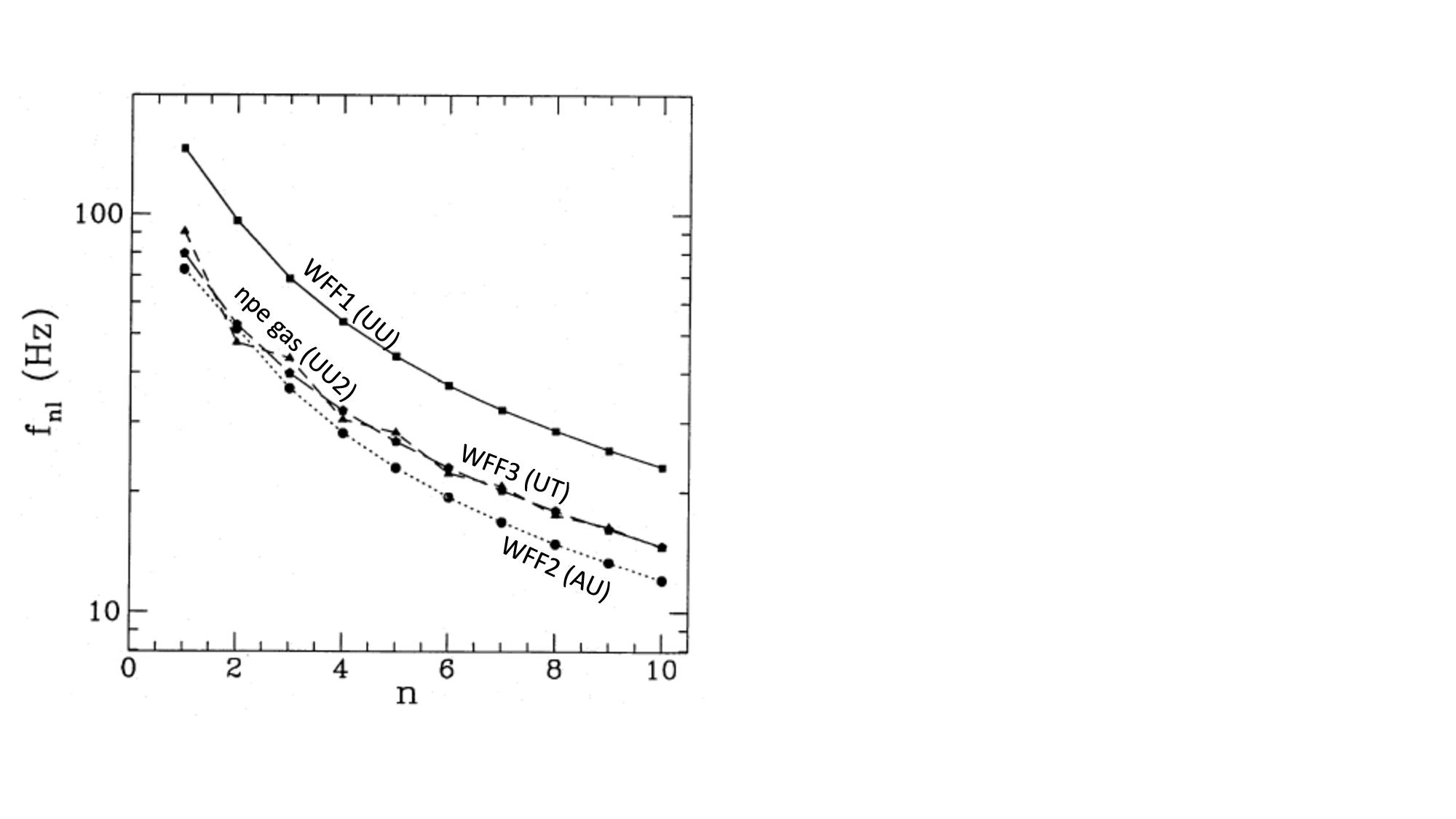}
  }
  \vspace{-1.cm}
  \caption{ The frequencies of the first ten quadrupole ($l=2$) g-modes based
on four WFF EOS models. Here $n$ specifies the radial
order of the g-mode. The solid line and squares for model WFF1 (UU), the dotted line and round circles are for model WFF2 (AU),
the short-dashed line and triangles for model WFF3 (UT), and the long-dashed line labeled as UU2 is with the \esym for free nucleon gas. Taken from ref. \cite{DongLai}.}
\label{DongFig2}
\end{figure}

Using the EOSs and related \esym predicted by the Variational Many-Body (VMB) theory of Wiringa, Fiks and Fabrocini (WFF) \cite{Wiringa1988}, Dong Lai \cite{DongLai} investigated effects of \esym on
the Brunt-V\"ais\"al\"a frequency and the core g-modes. The three typical density dependences of nuclear symmetry energy predicted by WFF are shown in Fig. \ref{examples}. The results obtained using the nuclear potential UV14+UVII are labeled as WFF1 (in Fig. \ref{examples} or UU (in ref.\cite{DongLai}), those obtained by using the AV14+UVII interaction as WFF2 or AU, while those from using the
interaction UV14+TNI as WFF3 or UT. For comparisons, Dong Lai also constructed a fourth model labeled as UU2 having the same $P(n)$, $\rho(n)$,
and $c_e(n)$ as UU. However, the proton fraction $x$ and the adiabatic speed of sound $c_s$ in UU2 were obtained by using the \esym of a free nucleon gas. Thus a comparison of results using the UU and UU2 can reveal effects of the symmetry energy. Shown in Fig. \ref{DongFig1} are the pressure $P$, adiabatic sound speed $c_s$, the proton fractions $x$ and the fractional difference between $c_s^2$ and $c_e^2$ using the four EOSs. While the pressure  $P(n)$ and adiabatic sound speed $c_s$ are not so different as one expects, the proton fraction $x$ and the fractional difference
$(c_s^2-c_e^2)/c_s^2$ are rather different, indicating strong effects of nuclear symmetry energy. Moreover, the UU and UU2 lead to dramatically different values of $x$ and $(c_s^2-c_e^2)/c_s^2$ as functions of density. The resulting core g-mode frequencies obtained within Newtonian dynamics are shown in Fig. \ref{DongFig2}. Interestingly,  the frequency of core g-mode oscillations of neutron stars
was found to depend not only on the pressure-density relation, but also sensitively on the density dependence of nuclear symmetry energy.

In short, the Brunt-V\"ais\"al\"a  and the core g-mode frequencies are sensitive to the density dependence of nuclear symmetry energy. Earlier explorations cited above have used predictions by the well-known WFF EOSs and some approximations. As discussed earlier, significant constraints on the \esym especially around and below the saturation densities of nuclear matter have been obtained over the last few years. Since resonant excitations of the g-modes play an important role in tidal heating of binary neutron stars \cite{DongLai}, in light of the GW170817 event timely new investigations of the Brunt-V\"ais\"al\"a  as well as the crustal and core g-mode frequencies using the latest knowledge about the \esym would be particularly interesting \cite{Lai2019}.

\begin{figure}
\includegraphics[scale=1.2]{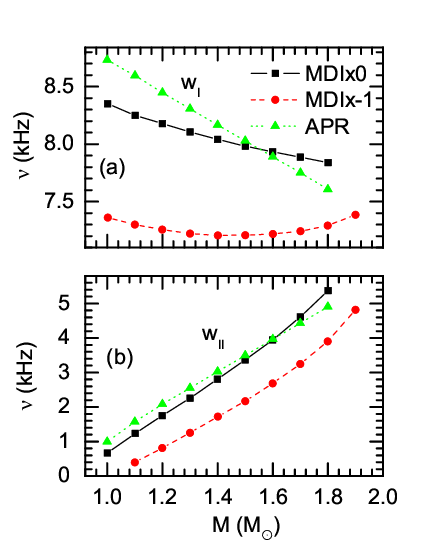}
\caption{\label{fig3-fM}(Color online) The frequency $\nu$ of $w_{I}$-mode (panel a)
and $w_{II}$-mode vs the stellar mass $M$ ( panel b). Taken from Ref. \cite{Wen2009}.}
\end{figure}
\begin{figure}
\includegraphics[scale=1.2]{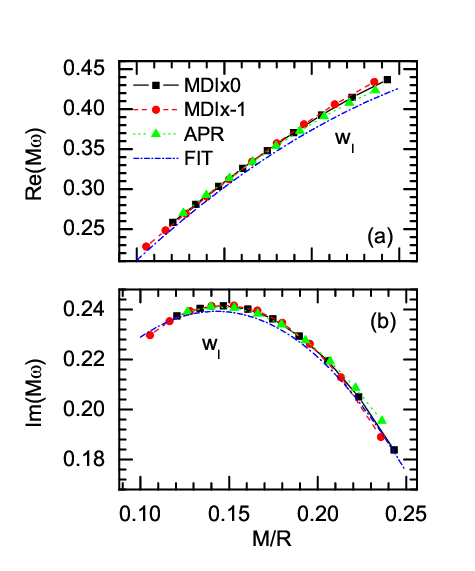}
\caption{\label{fig5-wI-MR}(Color online) The universal relation between the scaled Eigen-frequency, $\omega$, of the $w_{I}$-mode and the
compactness $M/R$. The FIT is a universal relation found by Tsui et al. \cite{Tsui2005a}. Taken from Ref. \cite{Wen2009}.}
\end{figure}
\subsection{Symmetry energy effects on frequencies of w-modes}
The w-modes can be divided into several categories \cite{Benhar1999}: (1) The first standard axial w-mode  $w_{I}$;  (2) The additional axial w-mode $w_{II}$; (3) The second axial w-mode $w_{I2}$ and the third axial w-mode $w_{I3}$, etc. The salient feature of the w-mode is its high frequency and rapid damping.
Chandrasekhar \& Ferrari \cite{Chandrasekhar1991a,Chandrasekhar1991b} showed that the axial w-mode can be described by the following second-order differential equation for a function $z(r)$ constructed from the radial part
of the perturbed axial metric functions
\begin{equation}\label{RW}
\frac{d^{2}z}{dr_{*}^{2}}+[\omega^{2}-V(r)]z=0,
\end{equation}
where $\omega (=\omega_{0}+i\omega_{i})$ is defined as the complex
eigen-frequency of the axial w-mode.
The tortoise coordinate $r_{*}$ and a potential function $V$ are defined inside neutron stars as
  \begin{equation}\label{rstar}
 r_{*}=\int_{0}^{r}e^{\lambda-\nu}dr \  (or \
 \frac{d}{dr_{*}}=e^{\lambda-\nu}\frac{d}{dr})
 \end{equation}
and
  \begin{equation}\label{VP}
V=\frac{e^{2\nu}}{r^{3}}[l(l+1)r+4\pi r^{3}(\rho-p)-6m]
 \end{equation}
where  $l$ is the spherical harmonics index (often only the case $l=2$ is considered), $\rho$
and $p$ are the density and pressure, and $m$ is the mass inside radius $r$, respectively.
The metric functions $e^{\nu}$ and $e^{\lambda}$ are
obtained from the line element
 \begin{equation}
-ds^{2}=-e^{2\nu}dt^{2}+e^{2\lambda}dr^{2}+r^{2}(d\theta^{2}+\textrm{sin}^{2}\theta
d\phi^{2}).
 \end{equation}
While outside the star, the Eqs.\ \ref{rstar} and \ref{VP} are reduced to
 \begin{equation}
 r_{*}=2M \textrm{ln}(r-2M) \  (or \
 \frac{d}{dr_{*}}=\frac{r-2M}{r}\frac{d}{dr})
 \end{equation}
and
  \begin{equation}
V=\frac{r-2M}{r^{4}}[l(l+1)r-6M]
 \end{equation}
 with $M$ being the total gravitational mass of the neutron star. Boundary conditions and technical details of solving these equations can be found in refs. \cite{Benhar1999,Wen2009}.

Effects of nuclear symmetry energy \esym on the w mode were studied in ref. \cite{Wen2009} using the APR and the MDI EOSs with $x$=-1 and 0. The MDI symmetry energy functions with $x$=-1 and 0 are shown in Fig. \ref{Esym-Xu2} while the corresponding mass-radius relations of neutron stars are shown in Fig. \ref{LiSteiner}. Again, these two $x$-parameters were selected by fitting the isospin diffusion data of heavy-ion reactions \cite{Chen05a,Chen2005} as we discussed earlier in detail in Sect. \ref{MDI-p}. The predicted frequencies of $w_{I}$-mode  and $w_{II}$-mode are shown in Fig. \ref{fig3-fM}. Obviously, the symmetry energy has a clear imprint on the w-mode frequencies.
\begin{figure*}
\resizebox{0.49\textwidth}{!}{
 \includegraphics{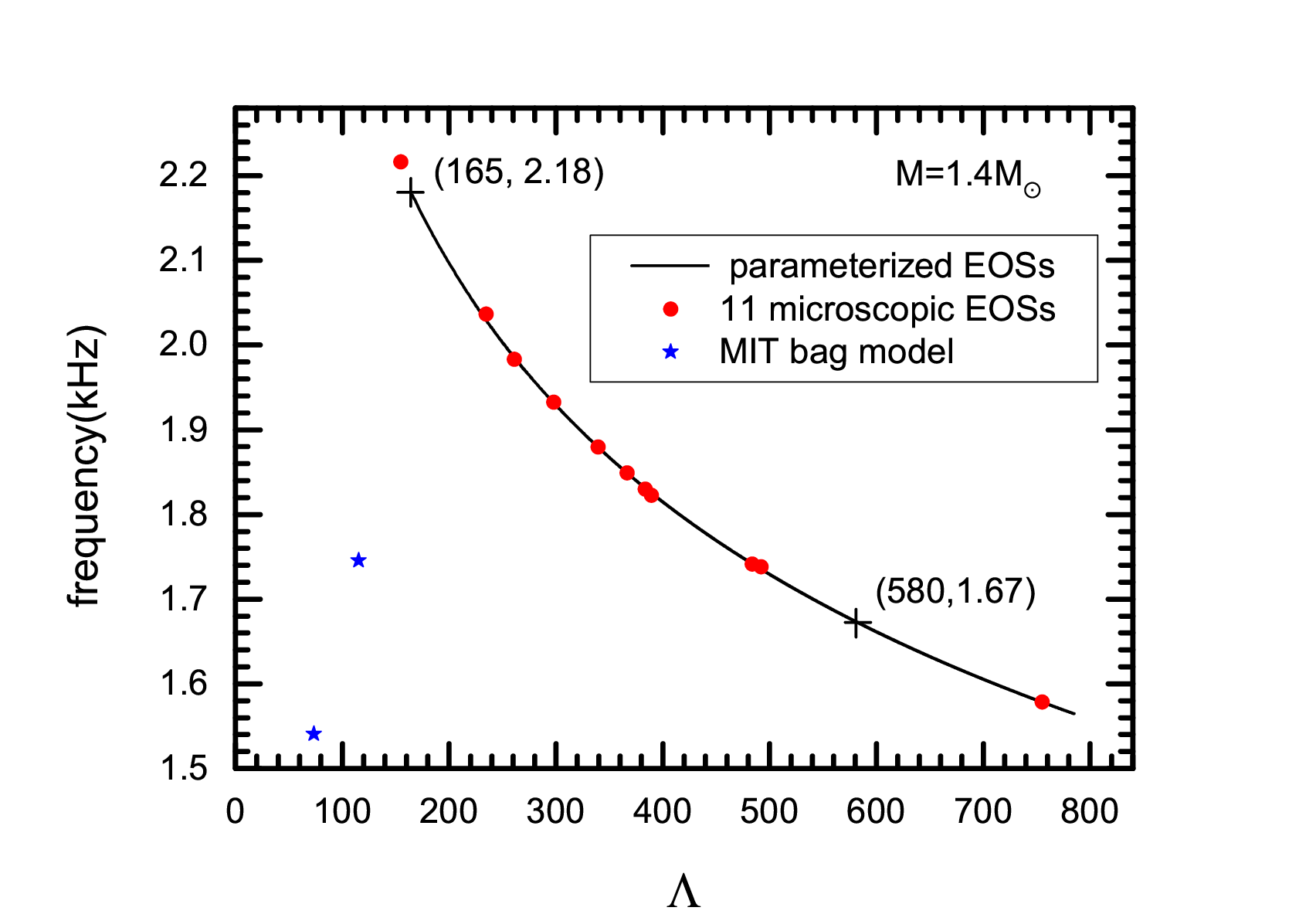}
}
\resizebox{0.49\textwidth}{!}{
 \includegraphics{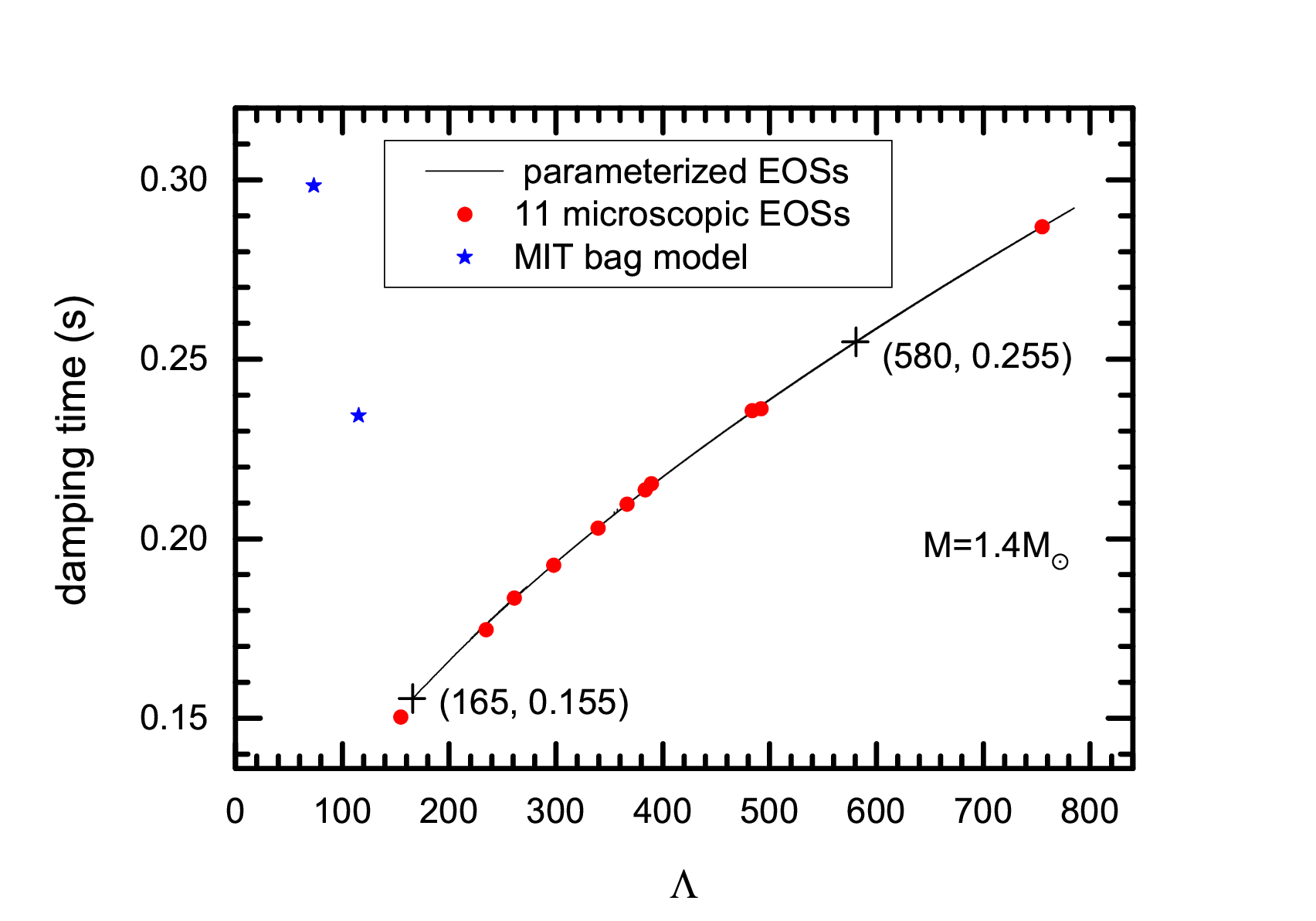}
}
\caption{ \label{fig1-fnamda} The f-mode frequency (left) and damping time (right) as functions of the tidal deformability of canonical neutron stars.
 The correlation between the f-mode frequency and the tidal deformability of canonical neutron stars using 23,000 phenomenological EOSs (solid black line), 11 microscopic EOSs (red dots) and 2 EOSs for quark stars within the MIT bag model (blue stars).  The cross at (165,2.18) corresponds to the lower limit of the tidal deformability predicted using the parameterized EOS satisfying all known constraints from terrestrial nuclear experiments, while the cross at (580, 1.67) corresponds to the upper limit of the tidal deformability of neutron stars extracted from the GW170817 event by LIGO and VIRGO Collaborations. Figures adapted from Ref. \cite{Wen2019}.}
\end{figure*}

It has been well known that a universal scaling exists between the scaled eigen-frequencies of w-modes and some global properties of neutron stars \cite{Benhar1999,Tsui2005a,Tsui2005b,Andersson1998}.
For example, shown in Fig. \ref{fig5-wI-MR} are the mass scaled real and imaginary frequencies versus the compactness $M/R$. They show very nice universality independent of the EOS.
The curves (Fit) best fit the scaling relations found earlier by Tsui et al. \cite{Tsui2005a}. It was hoped that if the masses and radii of neutron
stars are determined accurately, the universal scaling behaviors provide a way to infer the frequency and damping time of gravitational waves from the w-mode.
Of course, if both the frequency and damping time of the w-modes are observed precisely, they will enable the extraction of the masses and radii of neutron stars, leading to
further constraints on the EOS of dense neutron-rich nuclear matter \cite{Sotani2017,Fernandez2019}.

In short, the density dependence of nuclear symmetry energy $E_{\rm sym}(\rho)$ has a
clear imprint on both the frequency and the damping time of the axial w-modes. However, compared to the core g-mode where the frequency is directly related to the gradient of the composition uniquely determined by the symmetry energy, the \esym affects the w-mode's real and imaginary frequencies through the pressure in the same way as the EOS of symmetric nuclear matter (SNM). Namely, effects of the \esym may be mimicked by readjusting the stiffness of the SNM EOS.

\subsection{Symmetry energy effects and GW170817 implications on the frequency and damping time of the f-mode }
Since the f-mode has a relatively lower frequency (1$\sim$ 3 kHz), it is easier to be excited compared to the p-mode and w-mode \cite{Lau2010,Kokkotas2001}.
To our best knowledge, similar to the w-mode, all features of the f-mode are determined by the total pressure as a function of energy density.
The \esym affects the f-mode through its contribution to the total pressure. However, unlike the g-mode, the f-mode has no unique and explicit dependence on the symmetry energy.

The formalism for calculating the frequency and damping time of the f-mode can be found in refs. \cite{Lindblom83,Detweiler85}. The f-mode has been studied extensively in the literature.
The frequency and damping time of f-mode were found to scale universally with the compactness or its variations \cite{Wen2009,Blazquez2014,Andersson1998,Lau2010,Andersson1996,Benhar2004,Tsui2005,Blazquez2013,Chirenti2015,Stergioulas2018}. Interestingly, Sham et al. further investigated the universal scalings in the Eddington-inspired Born-Infeld (EiBI) gravity, the universal relations obtained in general relativity were found valid also in the EiBI gravity \cite{Sham2014}. Thus, much interest has been focused on verifying and using the scalings to infer the mass and radius of neutron stars and hopefully further infer the underlying EOS.

To our best knowledge, so far there is still no observational evidence of the f-mode. It is thus very interesting that it was found very recently that the f-mode frequency and damping time scale with the tidal deformability with all EOSs except those involving phase transitions to quark stars \cite{Wen2019}.  Shown in Fig. \ref{fig1-fnamda} are the f-mode frequency and damping time as functions of
the tidal deformability for canonical neutron stars. These results were obtained from using 40,000 hadronic EOSs randomly generated using Eqs. (\ref{E0para} and \ref{Esympara}) with parameters within the currently known constraints (labeled as parameterized EOSs) \cite{Zhang2018}, 11 EOSs from microscopic nuclear many-body theories for normal neutron stars (marked as 11 microscopic EOSs) and 2 EOSs from the MIT bag model for quark stars \cite{Lattimer2012,Wiringa1988,Zhang2018,ALF2,AP34,ENG,Muther1987,SLy,Zhu2018,Wen2019,Chodos1974,Glendenning1997}. Except for quark stars, all of the widely different hadronic EOSs predict approximately the same correlations  between the f-mode frequency (damping time) and the tidal deformability. Constraints on the latter from the GW170817 event thus limit the f-mode frequency and damping time. More quantitatively, it was found that the range of tidal deformability $\Lambda_{1.4}=190^{+390}_{-120}$ from GW170817 \cite{LIGO2018} implies that the f-mode frequency and damping time of a  canonical neutron star are in the range of 1.67 kHz $\leq f\leq $ 2.18 kHz and 0.155 s $\leq \tau\leq $ 0.255 s, respectively \cite{Wen2019}. Of course, this prediction awaits observational confirmations. Nevertheless, the GW170817 implications on the f-mode frequency and damping time are useful for designing detectors to search for signatures of the f-mode oscillations of neutron stars.

\begin{figure*}[h!]
\resizebox{0.49\textwidth}{!}{
\includegraphics{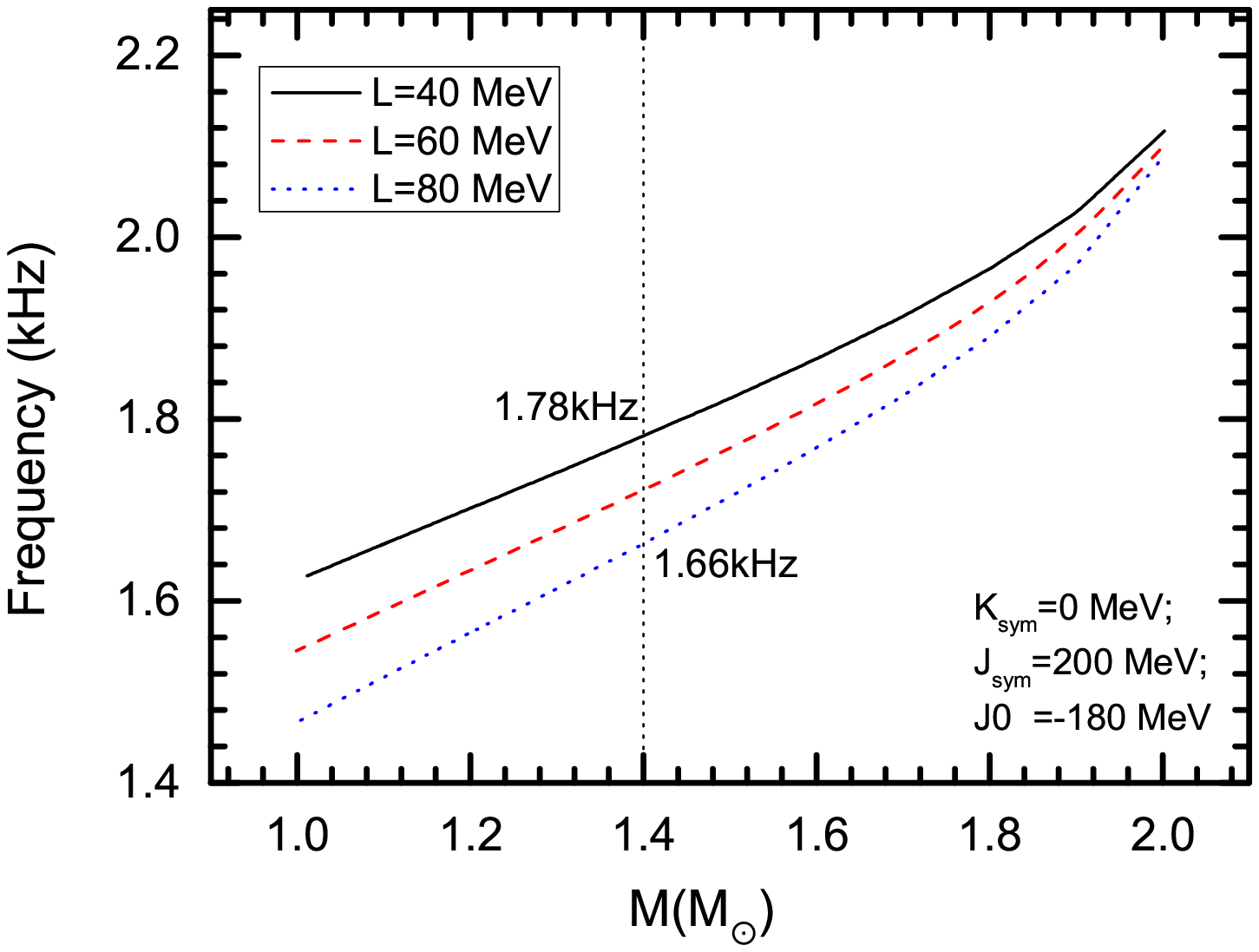}}
\resizebox{0.49\textwidth}{!}{
\includegraphics{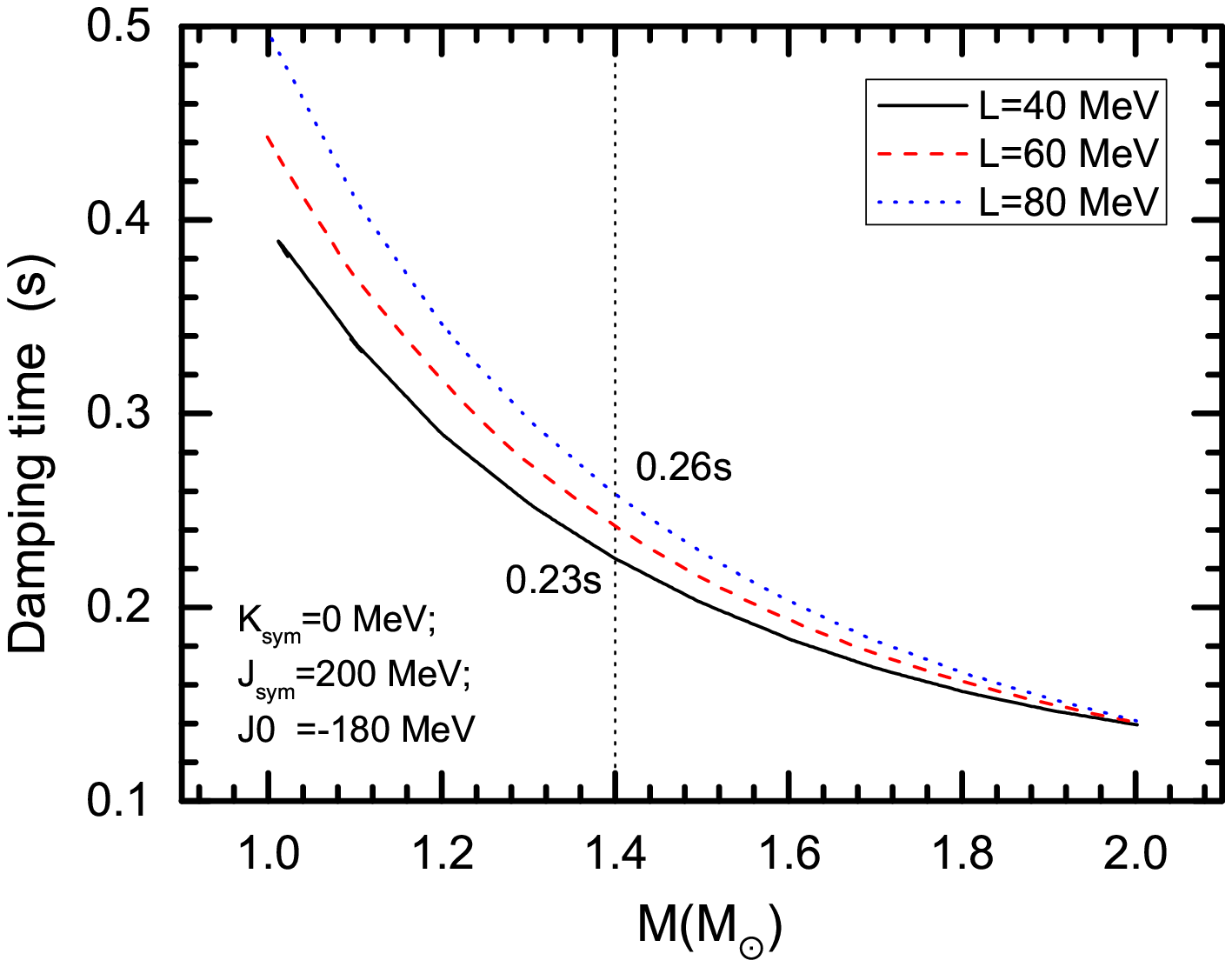}}
\caption{ \label{ft-esym} Effects of the symmetry energy slope parameter $L$ on the frequency (left) and damping time (right) of f-mode oscillations of neutron stars.}
\end{figure*}

In discussing potential applications of the various scalings of the f-mode frequency and damping time, one often optimistically assumes that both the frequency and damping time can be measured so precisely such that the scalings can be used to infer subsequently both the masses and radii of neutron stars. However, it was already pointed out that while the f-mode frequency could be detected very precisely, it is very difficult to extract accurately the damping time from observations \cite{Kokkotas2001}. Thus, the observed correlations of the $f$ and $\tau$ with the tidal deformability $\Lambda$ are very useful for accurately determining properties of the f-mode oscillations. As the accuracy of measuring the tidal deformability further improves with more neutron star mergers events, more knowledge about the f-mode properties may be indirectly inferred.

Effects of nuclear symmetry energy slope $L$ on the f-mode frequency and damping time as functions of mass are shown in Figs. \ref{ft-esym}. For this demonstration, the parametric EOS of Eqs. (\ref{E0para}) and (\ref{Esympara}) is used.  Three values of the slope parameter $L=$ 40 MeV, 60 MeV and 80 MeV are used while the parameters $K_{\rm sym}$, $J_{\rm sym}$ and $J_{0}$ are fixed at 0 MeV, 200 MeV and -180 MeV, respectively. It is seen that the $L$ parameter of the symmetry energy has appreciable effects in the range of 7-13\% on both the f-mode frequency and damping time. Similarly, effects of the $K_{\rm sym}$, $J_{\rm sym}$ can be studied. Overall, however, effects of the symmetry energy on the f-mode properties are small.

In short, the f-mode frequency (and damping time) scales with the tidal deformability of canonical neutron stars with hadronic nuclear EOSs.
Quark stars show clear deviations from the scaling. When examined as functions of neutron star masses,  both the f-mode frequency and damping time show appreciable dependence on the symmetry energy.  More precise measurements of the tidal deformability with more NS merger events will allow us to better understand properties of the f-mode oscillations which by themselves are hard to be detected.

\begin{figure*}
\centering
\resizebox{0.49\textwidth}{!}{
\includegraphics{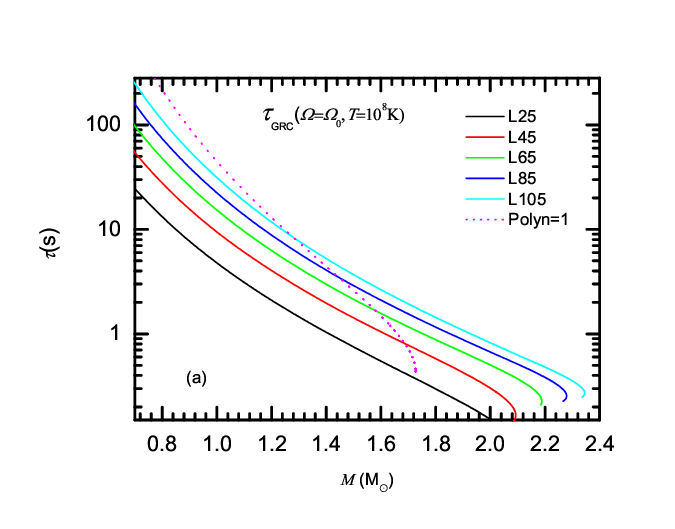}
}
\resizebox{0.49\textwidth}{!}{
\includegraphics{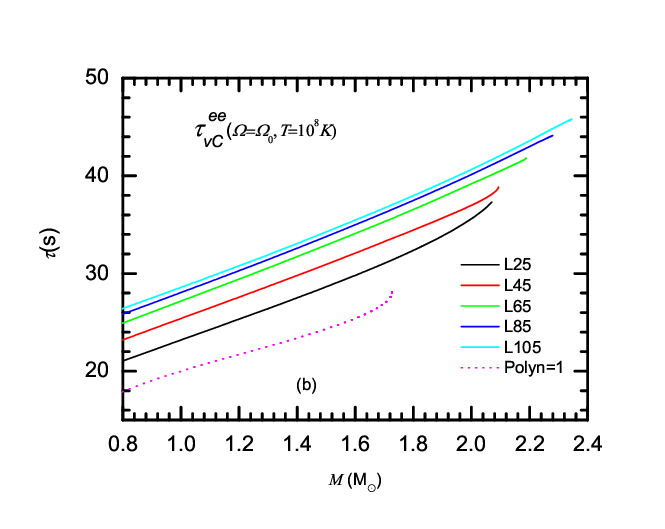}
}
\caption{\label{Wfigr1}(Color online) (a) The time scales for the gravitational radiation driven r-mode instability for a neutron star with rigid crust as a function of the stellar masses, where $\Omega=\Omega_o=\sqrt{{3 G M}/({4R^{3}}) }$ and $T=10^{8}$ K. (b) The corresponding viscous dissipation time scales due to the electron-electron scattering at the core-crust boundary layer. Taken from ref. \cite{Wen2012}.}
\end{figure*}

\subsection{Symmetry energy effects on the r-mode instability boundary}
The r-mode of oscillating neutron stars triggered by the Chandrasekhar-Friedmann-Schutz (CFS) instability \cite{Chandrasekhar70,Friedmann78} has long been recognized as a potentially useful probe of the inner structure and EOS of neutron stars \cite{Andersson01,Lindblom98,Owen98,Andersson11,Kokkotas11}. As a purely axial inertial mode with the Coriolis force as its restoring agent, the r-mode is one way of generating gravitational waves. The r-mode instability boundary has been used to explain the limited spin-up of accreting millisecond pulsars in the low mass X-ray binaries (LMXBs) \cite{Bildsten1998,Andersson1999}. The r-mode instability window often shown in the plane of frequency versus temperature, defined as the frequency above which the CFS instability of r-mode will be triggered, is  expected to be below the Kepler frequency but higher than the highest frequency of 716 Hz observed so far \cite{Hessels:2006ze}. Much efforts have been devoted to understanding the physics of r-mode and locating its instability window \cite{Andersson01,Lindblom98,Owen98,Idrisy2015,Kantor2017,Mukhopadhyay2018}. The nuclear symmetry energy \esym has been found to play a significant role in determining the r-mode instability window \cite{Wen2012,Vid2012}. An extensive review about \esym effects on the oscillation modes by W.G. Newton {\it et al.} can be found in ref. \cite{Newton14}.
For completeness of the present review, we summarize in the following the most important findings about \esym effects on the r-mode instability window.

The \esym comes into the physics of r-mode instability mostly through its effects on the core radius, crust thickness as well as the core-crust transition density and pressure \cite{Lattimer04,Cen09,Leh09}. The solid crust of neutron stars plays a key role in determine the r-mode instability window \cite{Bildsten00,Andersson00,Lindblom00,Rieutord01,Peralta06,Glampedakis06}.  Earlier studies have shown that the dissipation in the viscous boundary layer (between the crust and core) is sensitive to the crust thickness, and thus will influence the instability window \cite{Lindblom00,Levin2001}. While a real crust is  probably elastic, enabling the core oscillation to penetrate into the crust \cite{Andersson11,Peralta06,Glampedakis06}, the crust is often assumed to be perfectly rigid \cite{Bildsten00,Andersson00,Lindblom00}, thus providing an upper limit on the instability window.
Lindblom and Owen {\it et al.} have shown that the growth timescale of the r-mode instability in a neutron star with rigid crust can be written as \cite{Lindblom98,Owen98,Lindblom00}
\begin{eqnarray}\label{TGR}
 {1\over\tau_{GR}}& =&  {32\pi G \Omega^{2l+2}\over c^{2l+3}}
{(l-1)^{2l}\over [(2l+1)!!]^2} \nonumber \\
 & & \times\left({l+2\over
l+1}\right)^{2l+2} \int_0^{R_c}\rho r^{2l+2} dr
\end{eqnarray}
where $\Omega$ is the angular frequency and $R_c$ is the radius of the core. The timescale for the viscous damping of the r-mode at the boundary layer with the rigid crust and fluid core can be
written as \cite{Lindblom00}
\begin{equation}\label{TV}
 \tau_v  = \frac{1}{2\Omega} \frac{{2^{l+3/2}(l+1)!}}{l(2l+1)!!{\cal
 I}_l}
 \times \sqrt{2\Omega
R_c^2\rho_c\over\eta_c} \int_0^{R_c}
{\rho\over\rho_c}\left({r\over R_c}\right)^{2l+2} {dr\over R_c},
\end{equation}
where  $\rho_c$ and $\eta_c$ are the density and fluid viscosity at the core-crust interface, respectively.
Often, only the case $l=2$ ,  ${\cal I}_2 =0.80411$~\cite{Lindblom00,Rieutord01} are considered.
The shear viscosity of neutron stars are temperature dependent. It is dominated by neutron-neutron and electron-electron scatterings above and below $10^9$~K, respectively.
For example, the shear viscosity due to electron-electron scattering depends on the  density
and temperature \cite{Flowers79,Cutler87} according to
$
 \eta_{ee}=6.0\times 10^6 \rho^2 T^{-2}~ (g\cdot cm^{-1}\cdot
 s^{-1}).
$
As we have discussed earlier in this review, the \esym affects significantly the core radius as well as the core-crust transition properties. Indeed, as shown in Fig.\ \ref{Wfigr1} the
$L$ parameter of \esym affect both the r-mode growth timescale and its damping time scale. In this example, the neutron star is modeled with a rigid crust
for the five adopted EOSs with $L$=25-105 MeV (labeled as $L25, L45$ etc). Interestingly, both time scales show obvious mass dependences. While the time scales of the gravitational radiation decreases for more massive neutron stars, the time scales of the shear viscosity have the opposite trend. The obvious $L$ dependences of both time scales were understood from its opposite effects on the core radius and the core-crust transition density \cite{Wen2012}.

\begin{figure}
\resizebox{0.5\textwidth}{!}{
\includegraphics {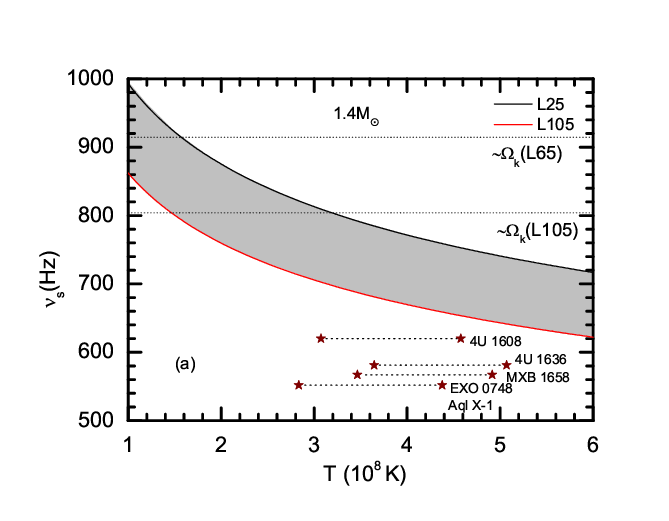}}
\caption{\label{Wfigr2} (Color online) The lower boundary of r-mode instability window for  canonical ($1.4 M_{\odot}$) neutron stars with the slope of the symmetry energy $L$ between 25 and 105 MeV.
Kepler frequencies with two typical $L$ values are indicated by the dotted lines. Locations of several observed LMXBs with short recurrence times \cite{Watts08,Keek10} are also shown. Taken from Ref. \cite{Wen2012}. }
\end{figure}
For a given temperature, the critical frequency above which stars become unstable is determined by equating the gravitational radiation and viscosity timescales. Naturally, the r-mode instability boundary in the plane of frequency versus temperature is $L$ dependent. For example, shown in Fig.\ \ref{Wfigr2} are the r-mode instability windows for canonical neutron stars with $L$ in the range of 25 to 105 MeV. Effects of the $L$ parameter are obvious. If both the internal temperatures and masses of the observed fast pulsars are known, they can limit the r-mode instability and subsequently the $L$ parameter. Several such candidates, such as the 4U 1608-522 at 620Hz, EXO 0748-676 at 552Hz, 4U 1636-536 at 581Hz,
Apl X-1 at 549Hz and MXB 1658-298 at 567Hz \cite{Watts08,Keek10} were considered in ref. \cite{Wen2012}. Unfortunately, their masses are not known. It is seen from Fig.\ \ref{Wfigr2} that they fall below the instability boundary for canonical neutron stars. However, some of them do enter into the uncertainty range of the instability boundary between the $L=25$ and $L=105$ MeV curves \cite{Wen2012}. It is worth noting that the range of $L$ considered in ref. \cite{Wen2012} is significantly broader than the currently known range.  In fact, none of the NSs mentioned above enters into the
narrowed uncertain region of the r-mode instability window.

In short, the slope of the symmetry energy $L$ affects the time scales of both the r-mode gravitational radiation and its viscous damping. Consequently, the r-mode instability window in the plane of frequency and temperature depend appreciably on the nuclear symmetry energy. Constraining the symmetry energy from studying the r-mode instability window, however, has the prerequisite of knowing mechanical properties of the crust as well as the thermal and transport properties of the core-crust transition region.

\subsection{Symmetry energy effects on the crust torsional modes}
Boosted by the observations of quasiperiodic oscillations (QPOs) following giant flares in soft gamma-ray repeaters (SGRs \cite{Strohmayer2006,Samuelsson2007,Israel2005,Watts2006,Strohmayer2005}, extensive studies on the torsional crust oscillations have been carried out, see, e.g., refs.
\cite{Schumaker1983,Samuelsson2007,HC1980,McDermott1988,Carroll1986,Strohmayer1991a,Lee2006,Leins1994,Messios2001}.
The shear modulus of the crust are thought to be the restoring force of torsional modes, see, e.g., discussions in refs. \cite{Schumaker1983,Samuelsson2007,Sotani2007,Steiner2009}.
To our best knowledge, effects of nuclear symmetry energy on the frequency of torsional oscillation was first studied  by Steiner and Watts  \cite{Steiner2009}. Later, systematic studies have been carried out by Newton {\it et al.} \cite{Newton14,Gearheart2011} and  Sotani  {\it et al.} \cite{Sotani2018,Sotani2007,Sotani2011,Sotani2012,Sotani2013a,Sotani2013b,Sotani2014a,Sotani2014b,Sotani2015,Sotani2016}.
For reviews of the work by these two groups, we refer the readers to refs. \cite{Sot19,Newton14}.

In the literature, various approximations have been used within both Newtonian gravity and general relativity to estimate the frequencies of torsional modes. While often quantitatively different, they share the common features that the torsional crust oscillation frequency is closely related to the radius of the core and the thickness of the crust as well as the shear waves speed $v_s$. For example,
using the relativistic Cowling approximation Samuelsson and Andersson \cite{Samuelsson2007} derived the following approximate formulae for the fundamental modes $_{\rm 0}t_{\rm l}$, $\omega_0$, and of the overtones $_{\rm n \neq 0}t_{\rm l}$,  $\omega_n$
\begin{eqnarray}\label{tfrequency}
	&&\displaystyle\omega^2_0 \approx \frac{e^{2\nu}v_s^2(l-1)(l+2)}{2RR_c}, \notag \\
	&& \omega^2_n \approx e^{\nu - \lambda} {n \pi v_s \over \Delta} \bigg[ 1 + e^{2\lambda} {(l-1)(l+2)\over 2 \pi^2} {\Delta^2 \over R R_c} {1 \over n^2}\bigg]
\end{eqnarray}
where $n,l$ are the number of radial and angular nodes the mode has respectively. The $M, R, R_{\rm c}$ and $\Delta$ are the stellar mass and radius, the radius out to the core-crust boundary and the thickness of the crust, respectively. These formulae have been used in several studies incorporating different microscopic physics for the crust and sometimes using different models for the core. Often the
core and the crust are not calculated self-consistent within the same model framework. Nevertheless, some interesting information about the \esym has been extracted, albeit sometimes quantitatively different.

\begin{figure}
\begin{center}
\includegraphics[scale=0.6]{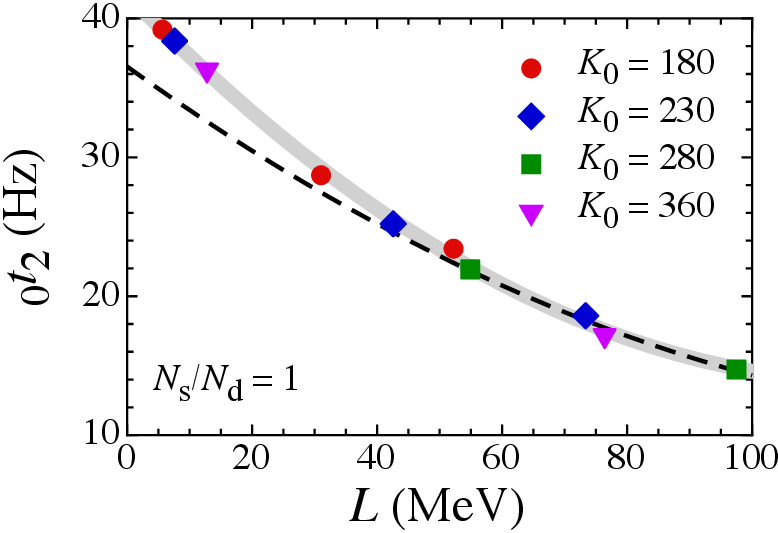}
\end{center}
\caption{(Color online) The fundamental frequencies ${}_0t_2$ of the $\ell=2$ torsional oscillations globally excited in the phases of spherical (solid line) and cylindrical (dashed line) nuclei calculated for various stellar EOS models with $M=1.4M_\odot$, $R=12$ km and the ratio of superfluid neutrons to the dripped neutrons $N_s/N_d=1$. Taken from ref. \cite{Sot19}.}\label{SotaniL}
\end{figure}

As an interesting example, Sotani {\it et al.} investigated effects of the crust and the pasta phase by fixing the core radius and mass. Thus, effects of the \esym on the stellar bulk properties are not fully taken into account in examining the dependence of the torsional crust oscillation frequencies on the symmetry energy.  As shown in Fig. \ref{SotaniL}, they found that the fundamental frequencies are almost independent of the incompressibility $K_0$ of symmetric nuclear matter, but strongly depend on the slope parameter $L$. This is true in both cases considering only spherical nuclei and including also cylindrical nuclei in the pasta. The latter was found to play an appreciable role when the $L$ is low. The strong $L$ dependence shown in Fig. \ref{SotaniL} indicates that
one may use it to further constrain the value of $L$ by identifying the observed QPOs as manifestations of various crustal torsional oscillations. Indeed, this has been tested in several studies.
However, the conclusions depend on the detailed microphysics considered for the crust. For example, in one scenario considering only spherical nuclei in the crust, they can explain all the observed low frequency QPOs in terms of the crustal torsional oscillations if they use $L$ in the range of 100 and 130 MeV in tension with most of the available terrestrial laboratory constraints shown in Fig. \ref{Esym0-Li}. Interestingly, they have shown more recently that considering the pasta phase containing cylindrical nuclei and constraints on $K_0$, frequencies of the observed QPOs can now be explained satisfactorily with $L$ in the range of 58 and 73 MeV in good agreement with the constraints from other analyses of the QPOs and the terrestrial laboratory data \cite{Sot19}.

\begin{figure}
\resizebox{0.45\textwidth}{!}{
\includegraphics{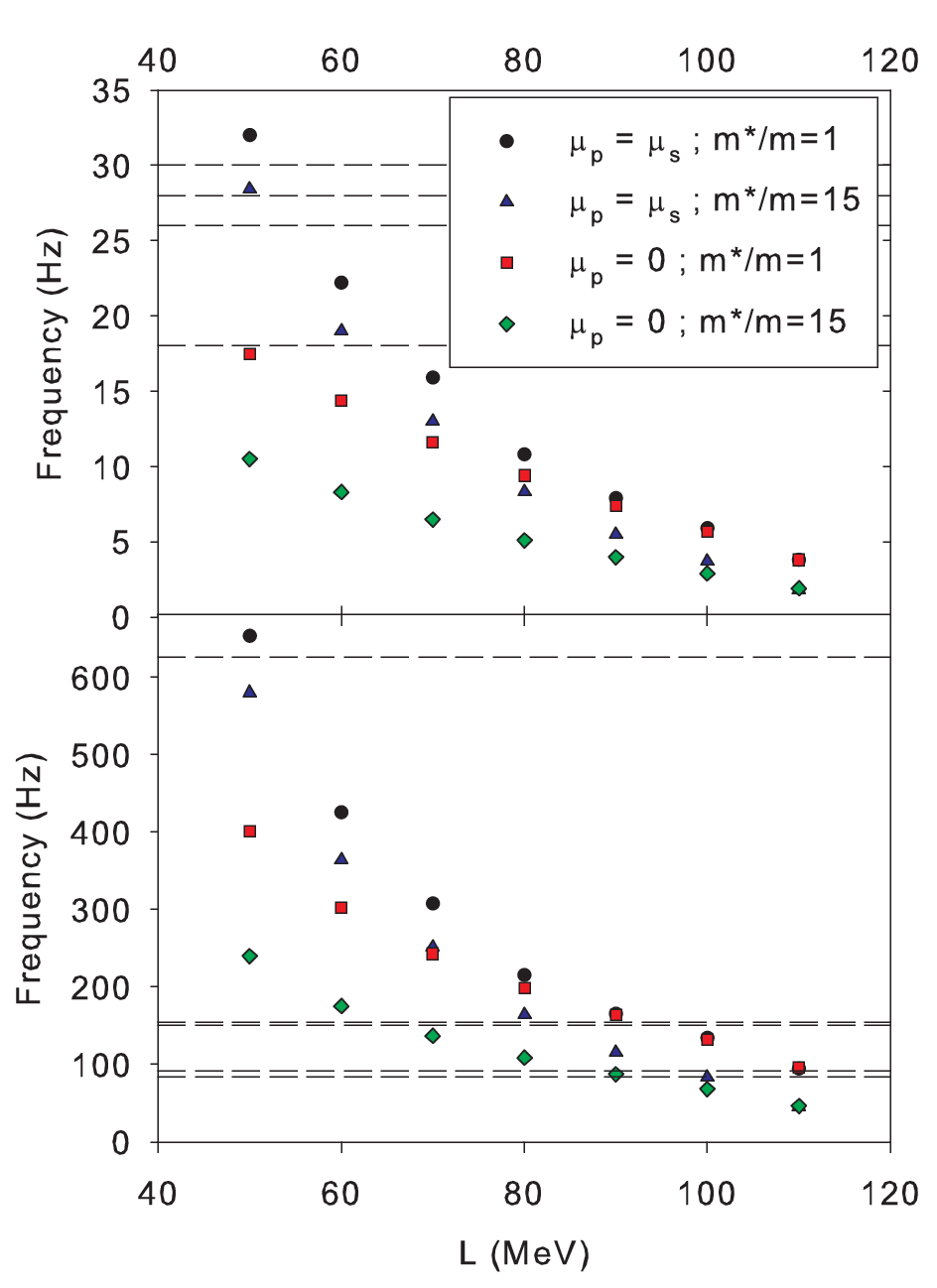}
}
\caption{(Color online) The frequency of the fundamental mode of torsional oscillation (top) and its first overtone (bottom) for a 1.4$M_{\odot}$ neutron star vs the slope parameter $L$.  For comparison with observations, the possible candidate frequencies: 18, 26, 28, 30Hz for the fundamental modes and 84, 92, 150, 155, 625Hz for the first overtone are marked by the dashed lines. Taken from ref. \cite{Gearheart2011}.}\label{tmode-mike}
\end{figure}

The speed of shear waves $v_{\rm s}$ in the expressions of $_{\rm n}t_{\rm l}$ is related to the shear modulus $\mu$ and the mass density $\rho$ at the base of the crust by $v_{\rm s}^2 = \mu/\rho$.
Based on molecular dynamics simulations in refs. \cite{Ogata1990,Strohmayer1991,Hughto2008} and assuming the crust is made of nuclei on the Coulomb lattice, the value of $\mu$ has been parameterized as
\begin{equation}\label{eq:shear_mod}
	\mu_{\rm s} = 0.1106 \left(\frac{4\pi}{3}\right)^{1/3} A^{-4/3} n_{\rm b}^{4/3} (1-X_{\rm n})^{4/3} (Ze)^2
\end{equation}
where $A$ and $Z$ are the mass and charge numbers of the nuclei at the base of the crust. The $n_{\rm b}$ is the baryon number density and $X_{\rm n}$ is the density fraction of free neutrons at the base of the crust. Therefore, the torsional mode frequency spectrum also depends on the microscopic structure and composition of the crust through the shear modulus \cite{Newton12,Gearheart2011}.
Gearheart {\it et al.} \cite{Gearheart2011} mimicked the limiting effects of the pasta by setting its $\mu$ either at zero as in the liquid core or $\mu_{\rm s}$ as in the solid crust with the corresponding transition densities properly evaluated using the compressible liquid drop model discussed in Sect. \ref{cc-tran}.  They also examined effects of entrainment due to
scatterings of superfluid neutrons off the crustal lattice by multiplying the frequencies by a mesoscopic effective mass term $\epsilon_{\star} = { (1 - X_{\rm n}) /[1 - X_{\rm n} (m_{\rm n} / m_{\rm n}^*)]}$ where $m_{\rm n}^*$ is the mesoscopic effective neutron mass. The maximum effect of entrainment using $m_{\rm n}^* / m_{\rm n}$ = 1 (no entrainment) and $m_{\rm n}^* / m_{\rm n}$ = 15 (maximum entrainment) were compared with that of the pasta phase using $\mu$=0 or $\mu=\mu_s$ in Fig. \ref{tmode-mike}. Several interesting observations can be made by examining their results:
(1) the frequency decreases as $L$ increases mostly because the radius $R$, in the denominator of Eq. (\ref{tfrequency}), increases with $L$, (2) The shear modulus of the pasta plays a very significant role. The frequencies may be reduced up to a factor of $\approx 2$; (3) The entrainment plays a relatively smaller effects by reducing the frequencies by about 20-30\% at most. The observed level of dependences on the very uncertain properties of the pasta makes drawing conclusions from comparing the calculated and observed frequencies rather conditional. In Fig. \ref{tmode-mike},
the horizontally dashed lines indicate the measured QPO frequencies from SGRs. Requiring only the calculated fundamental frequencies to fall somewhere in the range of 18 - 30Hz limits the value of L
to be approximately less than 60 MeV. As discussed in detail in refs. {\cite{Gearheart2011}, if additionally the observed 625Hz mode is to be matched to the 1st overtone, then one may conclude that $L \lesssim 60$ MeV \emph{and} the pasta phases should have mechanical properties approaching that of an elastic solid.

In short, the frequencies of the torsional crust oscillations are very sensitive to the \esym through several stellar bulk properties, the core-crust transition density as well as the microphysics about the crust and the possible pasta at its base. They are one of the only few quantities that can be actually compared with observations assuming the QPOs from SGRs are indeed due to the torsional oscillations.
While several analyses of the torsional oscillations have generally indicated consistently that $L$ should be around 60 MeV, many interesting but poorly known microphysics and related quantities need to be further studied before a firm conclusion about the \esym can be drown from comparing the calculated frequencies for the torsional crust oscillations with the observed frequencies of the QPOs.

\section{Symmetry energy effects on the EOS-gravity degeneracy of massive neutron stars}
While gravity is the first force discovered in nature, the quest to unify it with other fundamental forces
remains elusive because of its apparent weakness at
short-distance \cite{Ark98,Pea01,Hoy03,Long03,Jean03,Boehm04a,Boe04,Dec05}. In fact, the nature of gravity has been identified
as one of the eleven greatest unanswered questions of physics \cite{11questions}. Since Einstein's general relativity (GR) theory for gravity has passed successfully all tests
in the solar system but not fully tested yet in the strong-field domain \cite{Yun16}, searches for evidence of possible deviation from GR is at
the forefront of natural sciences \cite{11questions,Ark98,Pea01,Hoy03,Long03,Fujii71,Adel09,Aza08,Kap07,Psa08,Nes08,Kam08,Ger10,Luc10}.
It is fundamentally important to test whether the GR will break down at the strongest possible gravitational field reachable using various probes by human beings.
Moreover, to overcome known problems associated with GR's predictions of singularities in the Big Bang and inside black holes has also been stimulating active searches for alternative theories of gravity.

To our best knowledge, the strongest gravity presents in events involving black holes and so far all observations are in good agreement with GR predictions. Excitingly and fortunately, we have all witnessed very recently great breakthroughs in exploring gravity in its most extreme limit and on a mass scale that was not accessible before.
Perhaps well known to everybody at least in the physics community, the two historically most defining events directly verifying GR predictions 
are so far (1) the first detection of gravitational waves from the inward spiral and merger of a pair of black holes of around 36 M$_{\odot}$ and 29
M$_{\odot}$ and the subsequent ``ringdown" of the single resulting black hole by LIGO and VIRGO Collaborations \cite{LIGO1} and (2) the event-horizon-scale images of the supermassive black hole of mass M = $(6.5\pm 0.7)\times 10^9$ M$_{\odot}$ in the center of the giant galaxy M87 by the Event Horizon Collaboration \cite{M87}. The observed image of this black hole was found consistent with expectations for the shadow of a Kerr black hole as predicted by GR \cite{M87}.

While neutron stars are not as massive as black holes, they are still among the densest objects with the strongest gravity in the Universe. They thus are also ideal places to test gravity theories. However, it is well known that there is a degeneracy between the EOS of dense neutron-rich matter and the gravity theory in determining properties of neutron stars. Basically, this is because in applying the variational principle to the total action to describe properties of neutron stars one can modify the gravity, matter content and/or their couplings. It has been pointed out that this degeneracy is tied to the fundamental Strong Equivalence Principle \citep{Yun10}. For a very recent review on alternative gravity theories for strong fields and their applications to neutron stars, we refer the reader to ref. \cite{LJShao}.

Interestingly, some alternative gravity theories that have all passed low-field tests but are different from GR in the strong-field regime do sometimes predict neutron stars to have significantly different properties compared to their GR counterparts \citep{Ded03}. To break the EOS-gravity degeneracy requires simultaneously measuring at least two independent observables. Due to the strong diversity of  alternative gravity theories, results of ongoing studies are strongly model dependent, see, e.g., refs. \cite{wen,Aza08,Ger01,Wis02,Kri09}.  Since the \esym is the most uncertain part of the EOS of dense neutron-rich nucleonic matter, better knowledge about the \esym may thus play significant roles in breaking the EOS-gravity degeneracy. Using a few examples from the literature, we illustrate in the following some effects of the \esym in determining the strength of gravity and/or properties of neutron stars within GR and several alternative gravity theories.

\begin{figure}
  \centering
  \resizebox{0.48\textwidth}{!}{\includegraphics{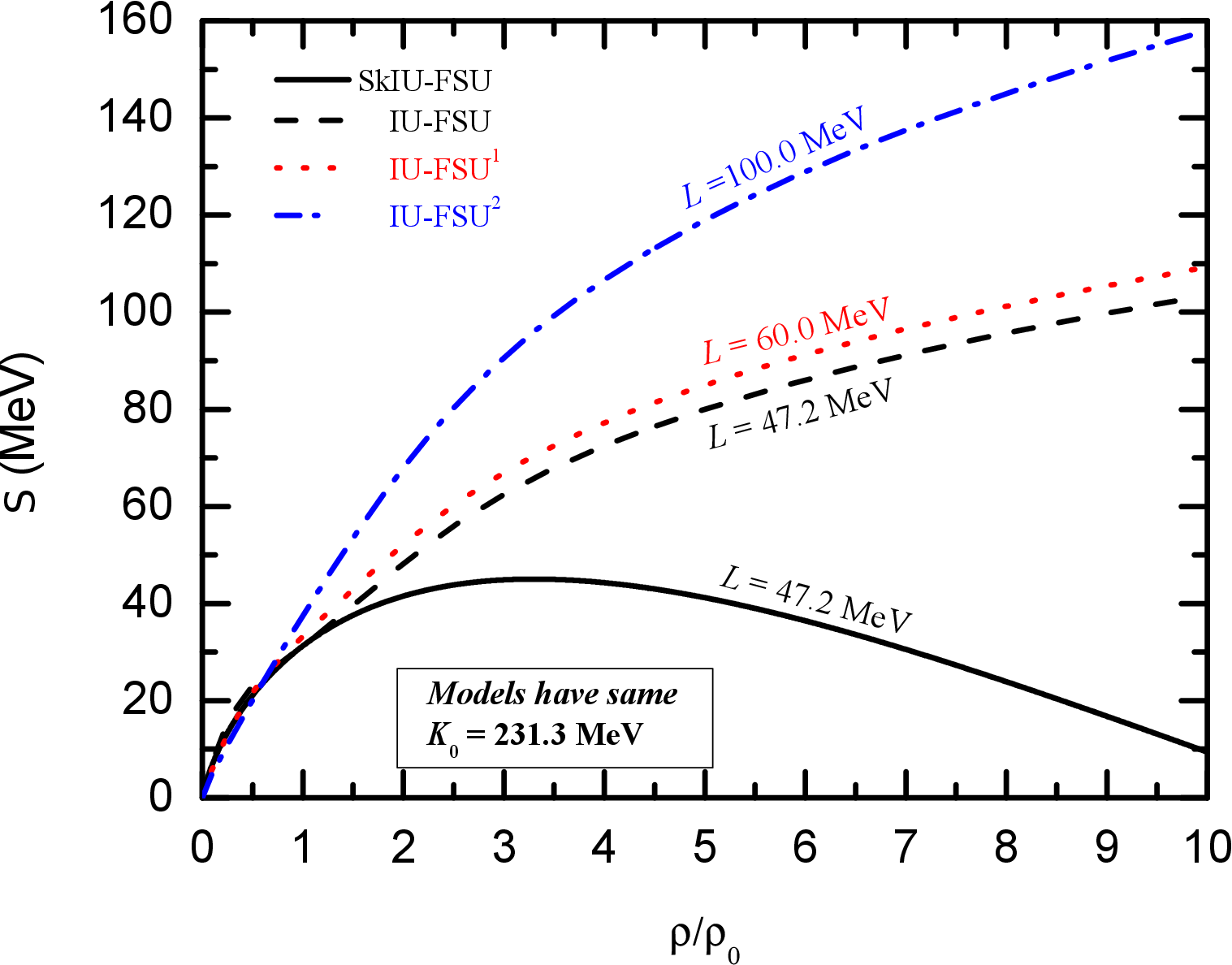}}
  \caption{Examples of nuclear symmetry energy as a function of density within the relativistic mean-field (RMF) model and the Skyrme-Hartree-Fork (SHF). By construction, the corresponding EOSs all have the same incompressibility $K_0 = 231.3$ MeV and almost the same \esym at sub-saturation densities while some of them have the same $L$ but different high density symmetry energies.
 Taken from ref. \cite{XTHE}.}
\label{HeFig1}
\end{figure}
\subsection{Example-1: \esym effects on the binding energy, surface curvature and red shift of massive neutron stars in Einstein's General Relativity (GR) theory of gravity}
\begin{figure}
 \centering
  \resizebox{0.48\textwidth}{!}{  \includegraphics{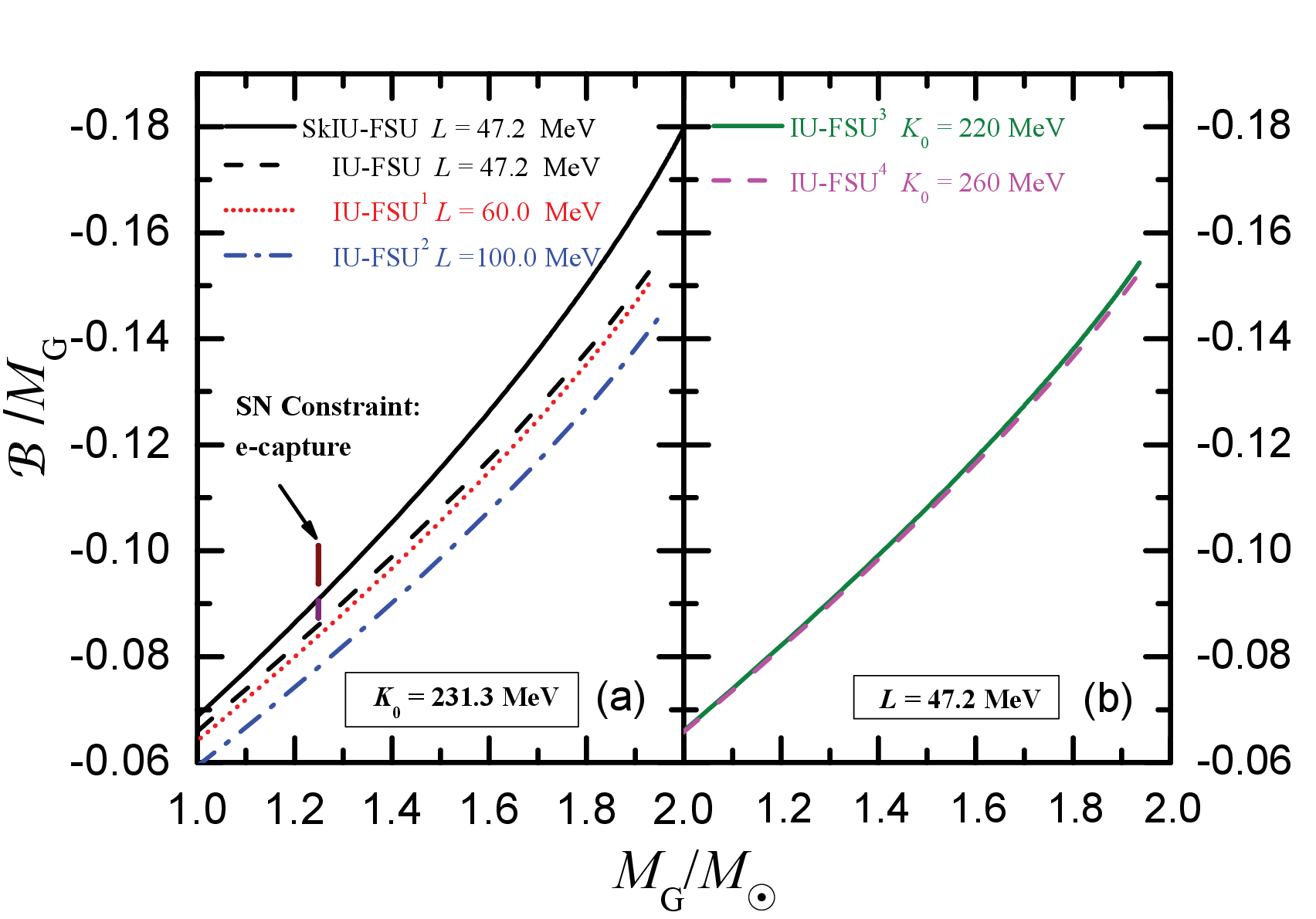}  }
 \resizebox{0.48\textwidth}{!}{  \includegraphics{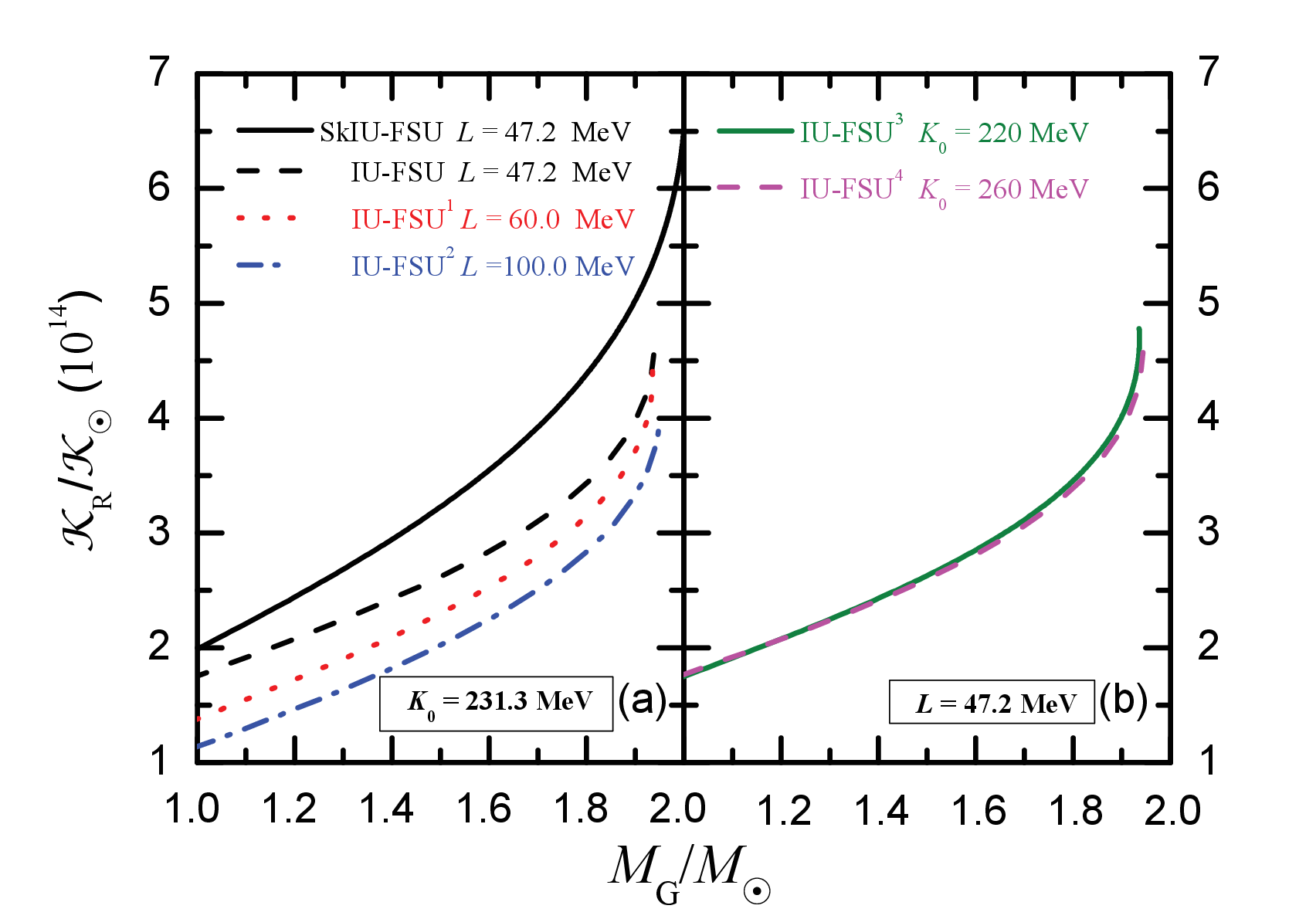}}
 \resizebox{0.48\textwidth}{!}{  \includegraphics{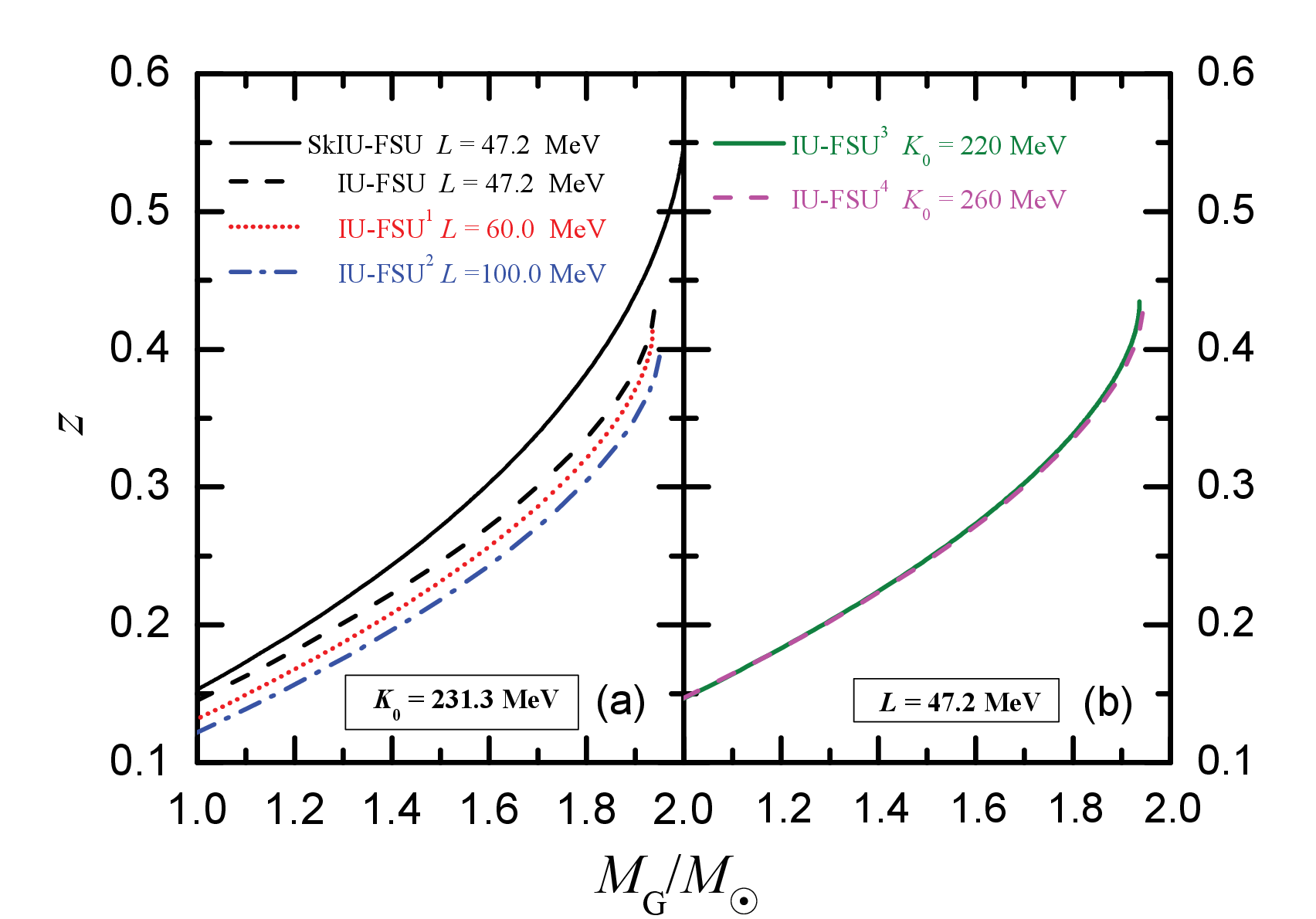}}
 \vspace{1.4cm}
  \caption{(Color online) Gravitational binding energy (upper window), surface curvature (middle window) and surface redshift (lower window) as functions of neutron star mass within GR using EOSs with different symmetry energies shown in Fig. \ref{HeFig1}. Taken from ref. \cite{XTHE}.}
  \label{HeFig2}
\end{figure}

How strong is the gravity of neutron stars? Several quantities including the binding energy and space-time curvature have been used to measure the strength of gravity. The gravitational binding energy is defined as $\mathcal{B}\equiv M_{\rm G}-M_{\rm
B}$~\cite{Wei72} where $M_{\rm G}$ is the gravitational mass and $M_{\rm B}\equiv Nm_{\rm B}$ is the baryon mass.
The total number $N$ of baryons can be found by a volume
integration of the baryon density $\rho_{\rm B}(r)$ via
\begin{eqnarray}
N=\int_{0}^{R}4\pi r^{2} \rho_{\rm B}
(r)\left[1-\frac{2GM(r)}{c^{2}r}\right]^{-1/2}dr.
\end{eqnarray}
The binding energy of a neutron star relative to its gravitational mass $M_{\rm G}$ is often used to measure the strength of gravity. Another frequently used measure of gravity is the
Kretschmann invariant~\cite{Eks14}
\begin{eqnarray}
&&\mathcal{K}^{2} =
\kappa^{2}\bigg[3\bigg(\mathcal{E}(r)+P(r)\bigg)^2-4\mathcal{E}(r)P(r)\bigg]\nonumber\\
&&-\kappa\mathcal{E}(r)\frac{16GM(r)}{c^2r^3}+\frac{48G^{2}M^{2}(r)}{c^4r^6}
\label{EqR}
\end{eqnarray}
with $\kappa \equiv 8\pi G/c^{4}$. Of course, the strength of gravity depends on the position. Often, for comparisons one may use the surface gravity. The latter can also be measured by using the
surface gravitational redshift
\begin{equation}
z = \frac{1}{\sqrt{1-2GM/c^2R}}  - 1.
\end{equation}

\begin{figure*}
\begin{center}
 \resizebox{0.53\textwidth}{!}{\includegraphics{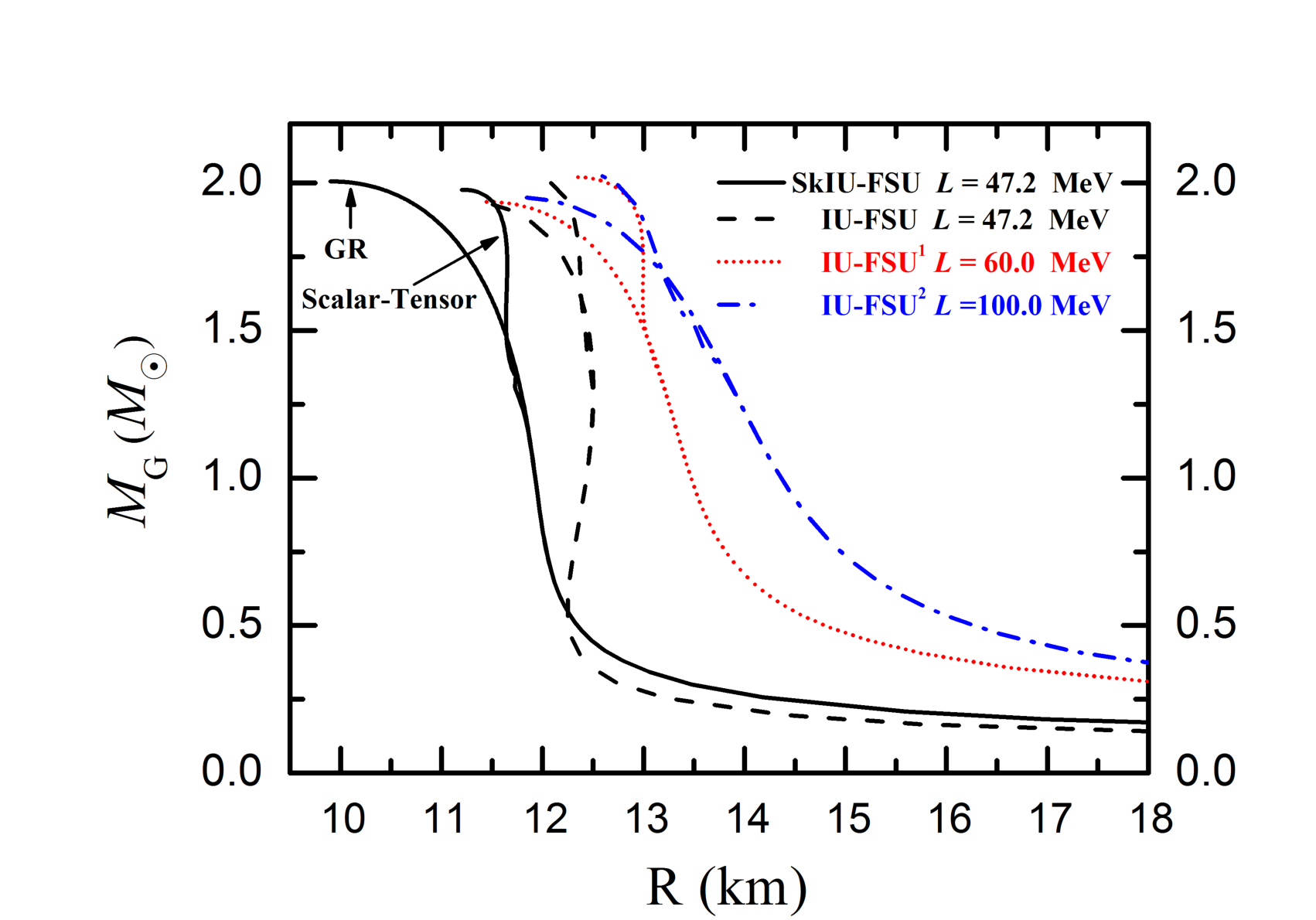}}
  \resizebox{0.41\textwidth}{!}{\includegraphics{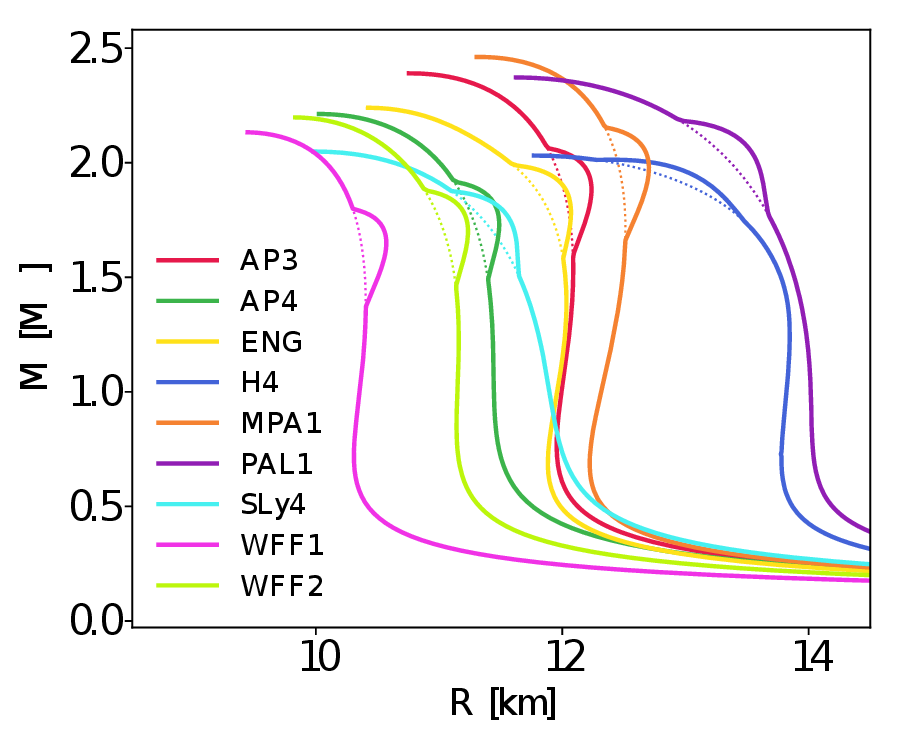}}
\end{center}
  \caption{(Color online) The mass versus radius relation within GR the Damour \& Esposito-Far\`ese (DEF) theory of gravity. Left: using the same four EOSs shown in Fig. \ref{HeFig1} and the gravity parameters of $\{\alpha_0,\beta_0\} =\{\sqrt{2.0 \times 10^{-5}}, -5.0\}$, taken from ref. \cite{XTHE}. The left branch is from GR calculations while the right one is from the DEF theory. Right: predictions from the GR (dotted lines) and DEF theory with $\left| \alpha_0 \right| = 10^{-5}$ and $\beta_0 = -4.5$ (solid lines) using nine popular EOSs for neutron star matter, taken from ref. \cite{LJShao}.}
  \label{He-Shao}
\end{figure*}

In a recent study \cite{XTHE}, He {\it et al.} studied effects of the \esym on the above three measures of gravity within Einstein's GR. To reveal effects of the  \esym at different densities, they constructed
several EOSs for neutron-rich nucleonic matter using the IU-FSU RMF model and the SkIU-FSU SHF approach. For example, shown in Fig.\ \ref{HeFig1} are the resulting nuclear symmetry energy as a function of density. By construction, the corresponding EOSs all have the same incompressibility $K_0 = 231.3$ MeV and almost the same \esym at sub-saturation densities while some of them have the same $L$ but different behaviors at high densities. In particular, the SkIU-FSU and IU-FSU models both have the magnitude $E_{\rm sym}(\rho_0)= 31.3$ MeV and the slope $L=47.2$ MeV for the symmetry energy. For comparisons, by properly adjusting all parameters of the RMF model, they also created two additional RMF models with differing incompressibility coefficients of $K_0 =220$ MeV and $K_0 = 260$ MeV but the same $L=47.2$ MeV. Results of their studies on the strength of gravity on the surface of neutron stars are shown in Fig.\ \ref{HeFig2}. Several interesting observations were made: (1) all three measures are sensitive to the \esym but are almost independent of the compressibility of symmetric nuclear matter, (2) more massive neutron stars are more sensitive to the variation of symmetry energy especially at supra-saturation densities as they are probing more denser matter than the light ones,
(3) all three measures reveal rather consistent information about effects of the EOSs on the strength of gravity. Moreover, the surface curvature appears
to be most sensitive to the variation of symmetry energy.

In the context of this subsection, it is interesting to note here  that the test of strong-field gravity
using the binding energy of the most massive neutron star observed so far, the PSR J0348+0432 of mass $2.01\pm0.04M_{\odot}$,
found no evidence of any deviation from the GR prediction \cite{Antoniadis2013}. The analysis used a very stiff EOS with an incompressibility of $K_0=546$ MeV
at normal nuclear matter density \cite{Ha81,Serot79}. While such a super-stiff incompressibility is more than twice the value of $K_0=240 \pm 20$ MeV extracted from analyzing various experiments
in terrestrial nuclear laboratories during the last 30 years, the finding by Newton {et al.} \cite{Newton} and later by He {et al.} \cite{XTHE} that the binding energy is approximately independent of the $K_0$ makes the conclusion of ref. \cite{Antoniadis2013} more reliable despite of the unreasonably stiff incompressibility used.

\subsection{Example-2:  \esym effects on the mass-radius relation of massive neutron stars in the Damour \& Esposito-Far\`ese (DEF) scalar-tensor theory of gravity}
To illustrate EOS effects on properties of neutron stars in the non-perturbative strong-gravity regime, here we cite two comparisons of the mass-radius relations calculated within the Damour \& Esposito-Far\`ese (DEF) model of the scalar-tensor gravity theory~\cite{DEF} and the GR, respectively.
In the DEF theory, the action has the form
\begin{eqnarray}
S &=& \frac{c^4}{16 \pi G} \int d^4x \sqrt{-g^{\ast}}\left[R^{\ast} -2
g^{\ast \mu \nu} \partial_{\mu} \varphi \partial_{\nu} \varphi
-V(\varphi)\right]  \nonumber\\
&+& S_{\rm matter}\left(\psi_{\rm matter};
A^2(\varphi)g^{\ast}_{\mu\nu} \right)
\end{eqnarray}
where $R^{\ast}$ is the Ricci scalar curvature with respect to the
so-called {\sl Einstein frame} metric $g^{\ast}_{\mu\nu}$ and
$V(\varphi)$ is the scalar field potential. The GR is
automatically recovered in the absence of the scalar field. The stellar structure is determined by
a set of first-order differential equations that can be solved once the EOS and boundary conditions are specified \cite{DEF}.
The gravity-matter coupling function has the form
\begin{equation}
A(\varphi) = \exp\left(\alpha_0\varphi + \frac{1}{2}\beta_0
\varphi^2\right).
\end{equation}

Various studies using different $\alpha_0$ and $\beta_0$ parameters have been reported in the literature. For example, shown in the left window of Fig. \ \ref{He-Shao} are predictions by He {\it et al.} \cite{XTHE} within the DEF theory using the same four EOSs shown in Fig. \ref{HeFig1} and the gravity parameters of $\{\alpha_0,\beta_0\} =\{\sqrt{2.0 \times 10^{-5}}, -5.0\}$. It is seen that predictions using the two theories split above about 1.5M$_{\odot}$. For each EOS, the left branch is from GR calculations while the right one is from the DEF theory. The right window shows predictions by Shao using nine popular EOSs for neutron star matter within the DEF theory with $\left| \alpha_0 \right| = 10^{-5}$ and $\beta_0 = -4.5$ (solid lines) and the GR theory (dashed lines) \cite{LJShao}. Despite of the different parameters used, two major features are shared by both predictions. Firstly, regardless of the EOSs used, the GR and DEF theories predict almost the same mass-radius relations for neutron stars lighter than about 1.5M$_{\odot}$. Secondly, the DEF predicts ``bumps'', namely, some more massive neutron stars having significantly larger radii compared to the GR predictions.
While one can numerically show that the sizes of the bumps depend on the value of $\beta_0$ in the matter-gravity coupling function, astrophysical observations \cite{beta-c} indicate that $\beta_0\geq -5$.  Moreover, as shown in the right panel of Fig. \ \ref{He-Shao}, the bumps are almost invisible with $\beta_0= -4.5$. Thus, within the small parameter range of $-4.5\leq \beta_0\geq -5$, the differences between GR and DEF predictions are small. Going beyond this limit or changing the sign of  $\beta_0$, one can show purely numerically that the DEF may give a maximum mass less than that predicted by GR \cite{Farrooh19}.

It is interesting to note that the bumps can not be mimicked by varying the EOS in GR. Shao pointed out that this bumpy structure is a very distinct feature for spontaneously scalarized neutron stars \cite{LJShao}. However, given the current and expect precisions of measuring the radii of neutron stars in the near future, observational signatures of the bumps are hard to be obtained. On the other hand, comparing the predicted mass-radius relations with the EOSs used by He {\it et al.}, it is seen that in both gravity theories, the major source of uncertainties for the radii is still the density dependence of nuclear symmetry energy.
\begin{figure*}
 \centering
  \resizebox{0.4\textwidth}{!}{
  \includegraphics{HKFig.eps}
  }
  \hspace{0.5cm}
  \resizebox{0.49\textwidth}{!}{
  \includegraphics{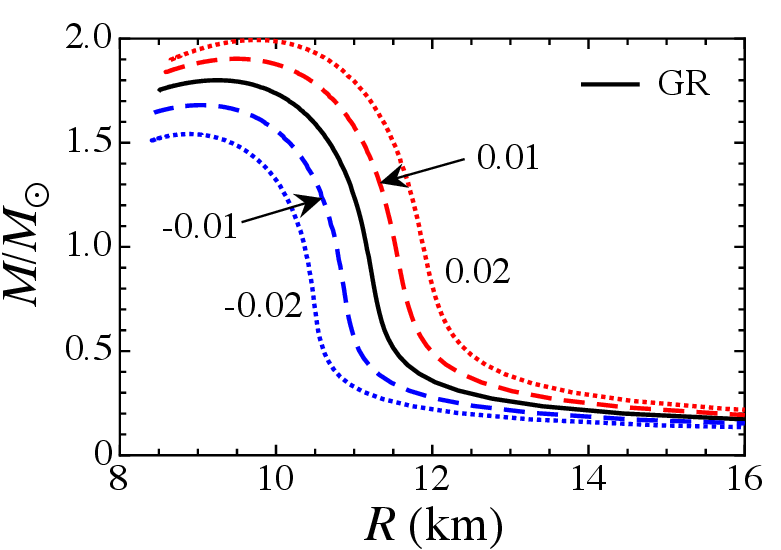}
}
  \caption{(Color online) Left: Gravitational mass $M$ plotted against the central density
$\epsilon_c$ for the APR EOS. The values of the coupling parameter
$8\pi\kappa\epsilon_0$ (with $\epsilon_0=10^{15} {\rm g\ cm}^{-3}$) are
shown in parentheses. Note that $\kappa = 0$ corresponds to the GR limit (solid). Taken from ref. \cite{Sham2014}.
Right: Neutron star mass-radius relations in the EiBI gravity theory constructed from FPS EOS. The labels on lines denote the values of $8\pi\epsilon_0\kappa$ while the solid line is from using the GR for gravity corresponding to $\kappa=0$. Taken from ref. \cite{Sota14}.}
  \label{MR-EiBI}
\end{figure*}

\begin{figure*}[t]
\begin{center}
  \resizebox{0.38\textwidth}{!}{
  \includegraphics{Chuck.eps}
  }
  \hspace{0.3cm}
  \resizebox{0.54\textwidth}{!}{
  \includegraphics{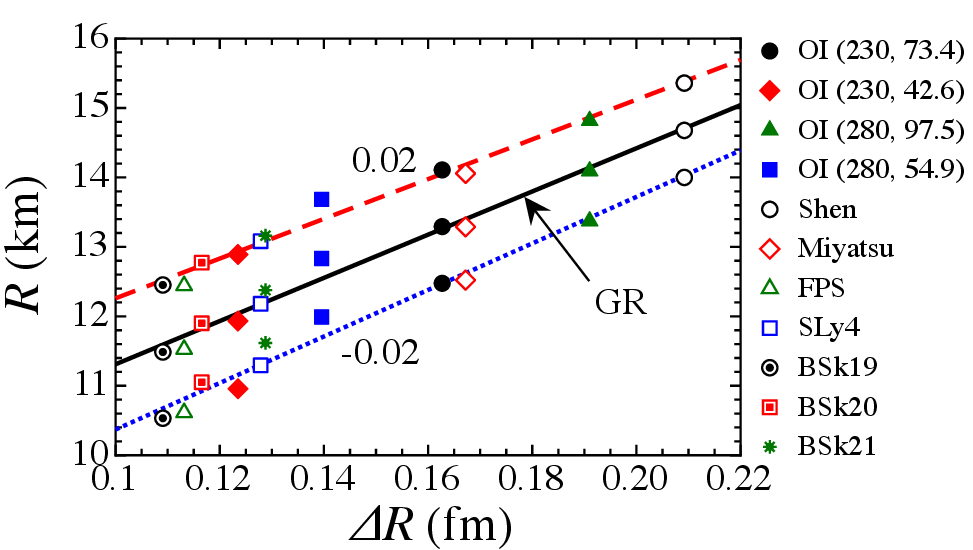}
}
\end{center}
\caption{(Color online) Radii of neutron stars with $0.5 M_\odot$, $R_{05}$ as a function of neutron skin thickness $\Delta R$ of $^{208}$Pb using the indicated nuclear EOSs.
Left: calculations within GR using several EOSs from relativistic mean-field models, taken from ref. \cite{Chuck03}.
Right: calculations for neutron star radii in both GR (solid line) and EiBI (broken and dotted lines) gravity theory with $8\pi\epsilon_0\kappa=-0.02$, 0, and 0.02, respectively.
The sizes of neutron-skin in $^{208}$Pb are calculated by using (1) the extended Thomas-Fermi theory using the EOSs indicated by OI($K_0$,$L$), (2) several RMF and (3) several SHF EOSs. Taken from ref. \cite{Sota14}.}
\label{Chuck-Sotani}
\end{figure*}

\subsection{Example-3: \esym effects on the mass-radius relation in the Eddington-inspired Born-Infeld (EiBI) theory of gravity}
The Eddington-inspired Born-Infeld (EiBI) gravity theory proposed in ref. \cite{Banados10} has the appealing feature that it reduces to GR in vacuum and can avoid
the Big Bang singularity.  While it has some pathologies, such as surface singularities
and anomalies associated with phase transitions in compact stars, applications of the EiBI gravity in studying properties of neutron stars have been found useful, see, e.g., refs. \cite{Sham2014,Sota14}.
The EiBI theory is based on the action \cite{Banados10}
\begin{eqnarray}
S &=& {1 \over 16 \pi} {2 \over \kappa} \int d^4x \left( \sqrt{
\left| g_{\mu\nu} + \kappa R_{\mu\nu} \right| } - \lambda \sqrt{-
g} \right) \nonumber  \\
\cr
&& + S_M \left[ g, \Psi_M \right]
\label{eq:EiBI_action}
\end{eqnarray}
where $R_{\mu\nu}$ is the symmetric part of the Ricci tensor and
$S_M \left[ g, \Psi_M \right]$ is the matter action.
The dimensionless constant $\lambda$ is related to the cosmological constant by $\Lambda=(\lambda - 1)/\kappa$.
The $\kappa$ and $\lambda$ are two model parameters. Different values are often taken in various studies.
For example, in the studies of both ref. \cite{Sham2014} and ref. \cite{Sota14}, $\lambda$ was set to $1$ while the $\kappa$ is varied under the condition $8\pi\kappa\epsilon_0 < 0.1$,
where $\epsilon_0 = 10^{15} {\ \rm g\ cm}^{-3}$.  Interestingly, effects on the structure of neutron stars by varying the $\kappa$ parameter were found to depend strongly on the EOS especially its \esym term. As examples of neutron star models in the EiBI theory, shown in Fig. \ref{MR-EiBI} are the mass-central density (left) and mass-radius (right) relations constructed from using different EOSs and $\kappa$ parameters. Shown in the left window are the gravitational mass $M$ plotted against the central density
$\epsilon_c$ for the APR EOS \cite{APR}. The values of the coupling parameter
$8\pi\kappa\epsilon_0$ (with $\epsilon_0=10^{15} {\rm g\ cm}^{-3}$) are
shown in parentheses. Note that the GR limit (solid) corresponds to $\kappa = 0$. Clearly, the value of $\kappa$ parameter affects significantly the maximum mass reached.
Similarly, shown in the right window are the mass-radius correlation with the FPS EOS \cite{FPS} by varying the $\kappa$ parameter. Comparing the results of the two windows, the EOS-gravity degeneracy within one class of gravity theory is clearly demonstrated. The same mass-radius correlation can be obtained by varying either the gravity coupling constant $\kappa$ in the EiBI theory or the EOS of neutron star matter. Thus, without the precise knowledge about the EOS, it is impossible to distinguish the EiBI from the GR theory for gravity. Comparing to the mass-radius correlation from the
DEF scalar-tensor theory discussed in the previous subsection, it is interesting to note that the EiBI and GR predictions are different in the whole range of mass while the predictions from DEF and GR are different only for massive neutron stars. This feature may allow one to distinguish the EiBI from DEF if one can measure accurately the radii of very light neutron stars as we shall discuss next.

Efforts have been made to break the EOS-gravity degeneracy, see, e.g., ref. \cite{Psa08} for a review. As an example relevant to this review, we discuss next a proposal made by Sotani \cite{Sota14}.
Earlier, it was first proposed by Carriere {\it et al.} \cite{Chuck03} that a precise measurement of neutron skin $\Delta R$ in $^{208}$Pb may help constrain the radii $R_{05}$ of low mass ($\simeq\!0.5 M_\odot$) neutron stars. It is based on their findings within the RMF model for nuclear EOS and the GR for gravity that there is a strong correlation between $R_{05}$ and $\Delta R$ depending on the EOSs used only weakly as similar densities are reached in both objects, as shown in the left window of Fig. \ref{Chuck-Sotani}. On the other hand, as shown in the right window of Fig. \ref{Chuck-Sotani}, the $R_{05}$-$\Delta R$ correlation depends significantly on the $\kappa$ parameter in the EiBI theory. The EOSs used in this study have even less effects on the $R_{05}$-$\Delta R$ correlation compared to the findings of ref. \cite{Chuck03}. However, it is worth noting the outstanding role of the \esym in this correlation. Calculations using a phenomenological EOS proposed by Oyamatsu and Iida \cite{Oya2003}, labeled by the OI($K_0$,$L$) in the right window, show clearly that the $L$ makes both the $R_{05}$ and $\Delta R$ increase independent of the gravity theory used. While the $R_{05}$ depends also on the gravity theory and its strength as one expects.

Since it is observationally extremely hard to measure the radii of light neutron stars as they are hard to be formed in the first place, if the $\Delta R$ can indeed be measured precisely in terrestrial laboratories, its strong correlation with $R_{05}$ may help constrain the $\kappa$ parameter in the EiBI gravity theory. A none-zero value of $\kappa$ will be a clear indication of deviations from the GR.
Most of other gravity theories, such as the DFE scalar-tensor theory, do not expect GR to be broken in the low-gravity field reached in such low-mass neutron stars. The proposed study about the radii of light neutron stars hopefully can be done in the near future to help distinguish the EiBI theory from other theories for gravity.

\subsection{Example-4: \esym effects on the mass-radius relation in a non-Newtonian theory of gravity with Yukawa correction}
\begin{figure*}
\begin{center}
 \resizebox{0.48\textwidth}{!}{
\includegraphics[width=8cm,height=6.5cm]{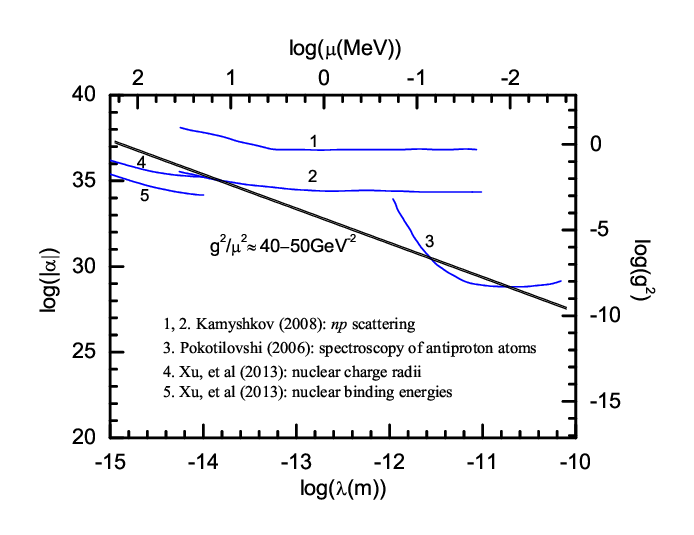}
}
  \resizebox{0.48\textwidth}{!}{
\includegraphics[width=8cm,height=6.5cm]{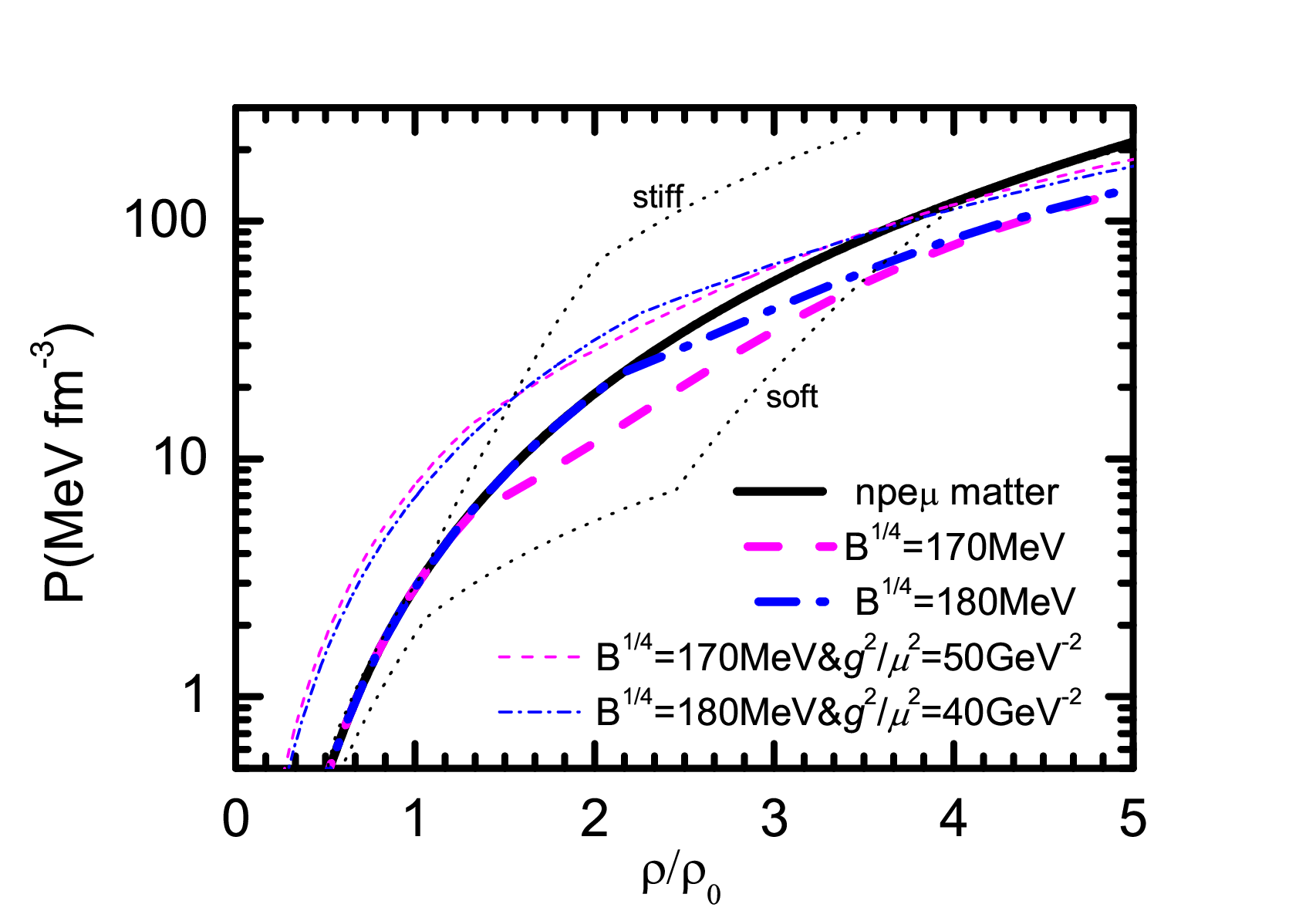}
}
\caption{\label{WFig1} (Color online) Left: Constraints on the strength and range of the Yukawa term from terrestrial nuclear
experiments in comparison with a fit with $g^{2}/\mu^{2}\approx 40-50$
GeV$^{-2}$; Right: Model EOSs for hybrid stars with and without using the
Yukawa term with a MIT bag constant $B^{1/4}$=170 MeV and
$B^{1/4}$=180 MeV, respectively. The soft and stiff EOSs for $npe$ matter are taken from refs. \cite{Heb13}. Taken from ref. \cite{Wlin14}.}
\end{center}
\end{figure*}
In the weak-field limit, some modified gravity theories, e.g., f(R) \cite{fr1,fr2}, the nonsymmetric gravitational theory (NGT) \cite{Mof96} and Modified Gravity (MOG) \cite{mog},
an extra Yukawa term besides the Newtonian potential naturally arise, see ref. \cite{NAP} for a review.
To our best knowledge, such form of non-Newtonian gravity was first proposed by Fujii \cite{Fujii71,Fuj2,Fujii91}. The non-Newtonian
gravitational potential between two objects of mass $m_1$ and $m_2$ can be written as
\begin{equation}
V(r)=-\frac{Gm_{1}m_{2}}{r}(1+\alpha e^{-r/\lambda})
\end{equation}
where $\alpha$ is a dimensionless strength parameter, $\lambda$ is
the length scale, and $G$ is the gravitational constant. In fact, it has long been proposed that
Newtonian gravity has to be modified due to either the geometrical effect of the extra
space-time dimensions predicted by string theories and/or the
exchange of weakly interacting bosons \cite{Fayet} in the super-symmetric
extension of the Standard Model. We refer the reader to refs. \cite{Adel03,Fis99,Rey05,New09,Uzan03} for reviews on this topic.
While still under debate, the Yukawa-type potential has been used successfully in explaining the flat galaxy rotation
curves \cite{Mof96,San84} and the Bullet Cluster 1E0657-558 observations~\cite{Bro07} without invoking dark matter.

In the boson exchange picture, the strength of the Yukawa potential is
\begin{equation}
\alpha=\pm g^2/(4\pi Gm_b^{2})
\end{equation}
where the $\pm$ sign is for scalar/vector bosons, $m_b$ is the baryon
mass and $g$ is the boson-baryon coupling constant. While its range is $\lambda=1/\mu$ (in natural units) where $\mu$ is the boson mass.
Much efforts have been devoted to setting upper limits on the strength $\alpha$ and range $\lambda$ from femtometer range in nuclei to the visible universe
using different experiments and observations. Shown in the left window of Fig. \ref{WFig1} are several recent constraints \cite{Nes08,Kam08,Jun13,BARB75,POKO06}
in the range of $\lambda \approx 10^{-15}-10^{-8}$ m where one has very roughly $g^{2}/\mu^{2}\approx 40-50$~GeV$^{-2}$ \cite{Wlin14}.  Extensive reviews on the constraints at larger ranges can be found in refs. \cite{Adel03,Fis99}. The extra energy density due to the Yukawa term is \cite{Long03,Kri09}
\begin{equation}\label{EDUB}
\varepsilon_ {_{\textrm{\scriptsize{UB}}}}= \frac{1}{2V}\int
\rho(\vec{x}_{1})\frac{g^{2}}{4\pi}\frac{e^{-\mu
r}}{r}\rho(\vec{x}_{2})d\vec{x}_{1}d\vec{x}_{2}=\frac{1}{2}\frac{g^{2}}{\mu^{2}}\rho^{2},
\end{equation}
where $V$ is the normalization volume. Assuming a constant
boson mass independent of the density, one obtains an additional
pressure of
\begin{equation}
P_{\textrm{\scriptsize{UB}}}=\frac{1}{2}\frac{g^{2}}{\mu^{2}}\rho^{2}.
\end{equation}

 \begin{figure*}
\begin{center}
  \resizebox{0.48\textwidth}{!}{
\includegraphics[width=8cm,height=6.5cm]{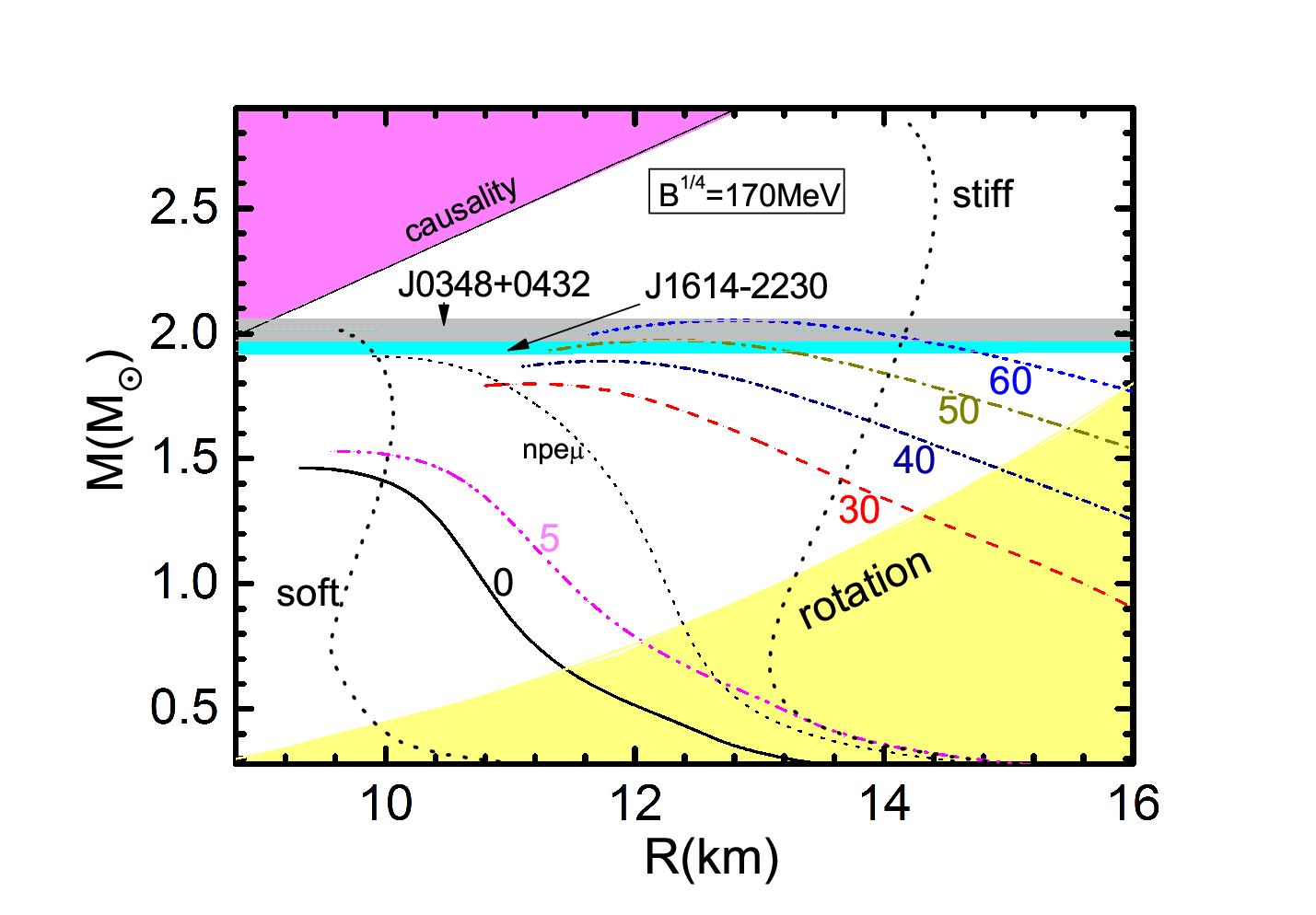}
}
  \resizebox{0.48\textwidth}{!}{
\includegraphics[width=8cm,height=6.5cm]{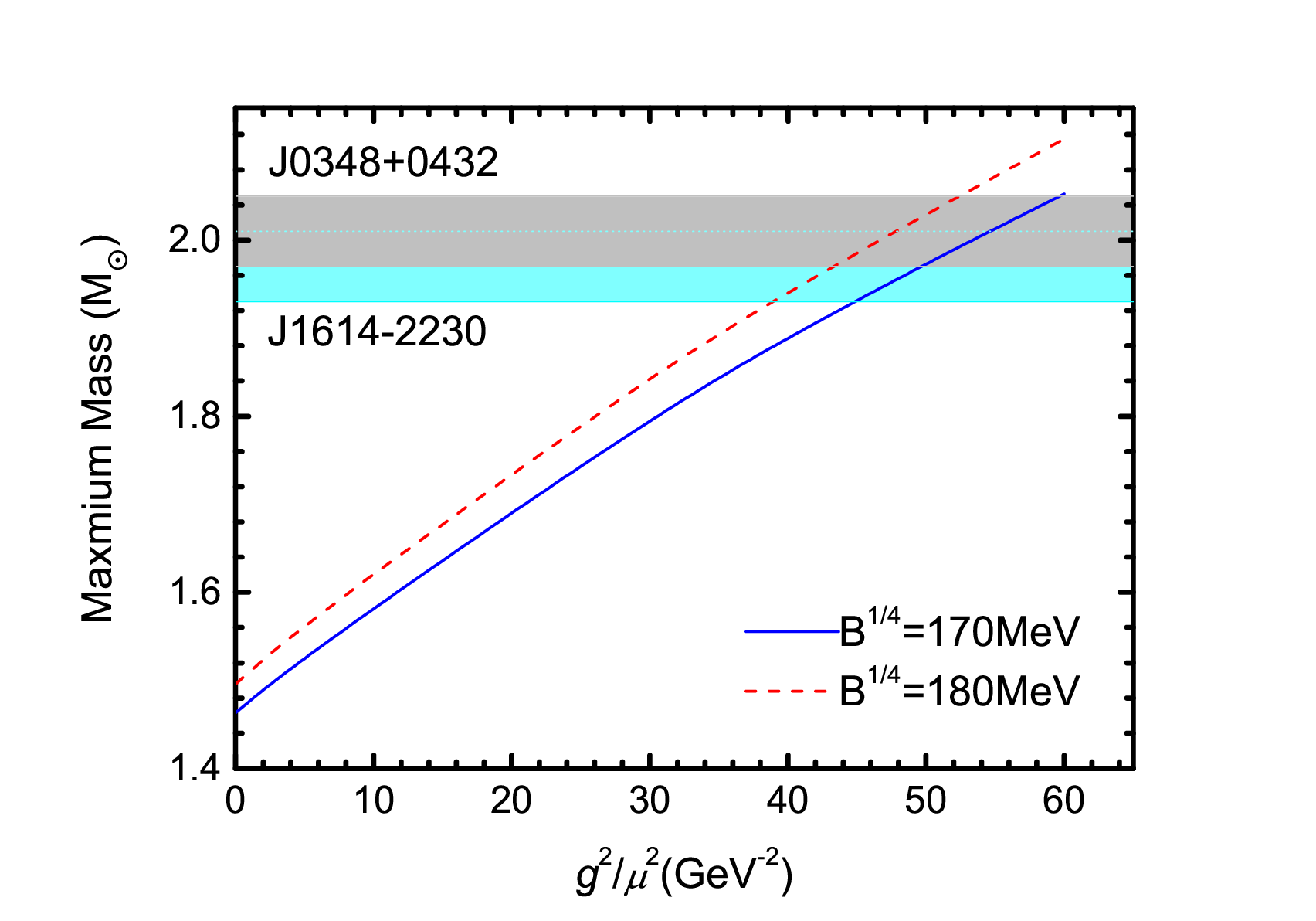}
}
\caption{\label{WFig2} Left: The mass-radius relation
of static neutron stars with $B^{1/4}$=170 MeV and several values of
$g^{2}/\mu^{2}$ in units of GeV$^{-2}$. Right: The maximum mass of neutron stars as a function of
$g^{2}/\mu^{2}$ with $B^{1/4}$=170 and 180 MeV, respectively. Taken from ref. \cite{Wlin14}.}
\end{center}
\end{figure*}
\begin{figure*}
\begin{center}
  \resizebox{0.48\textwidth}{!}{
\includegraphics{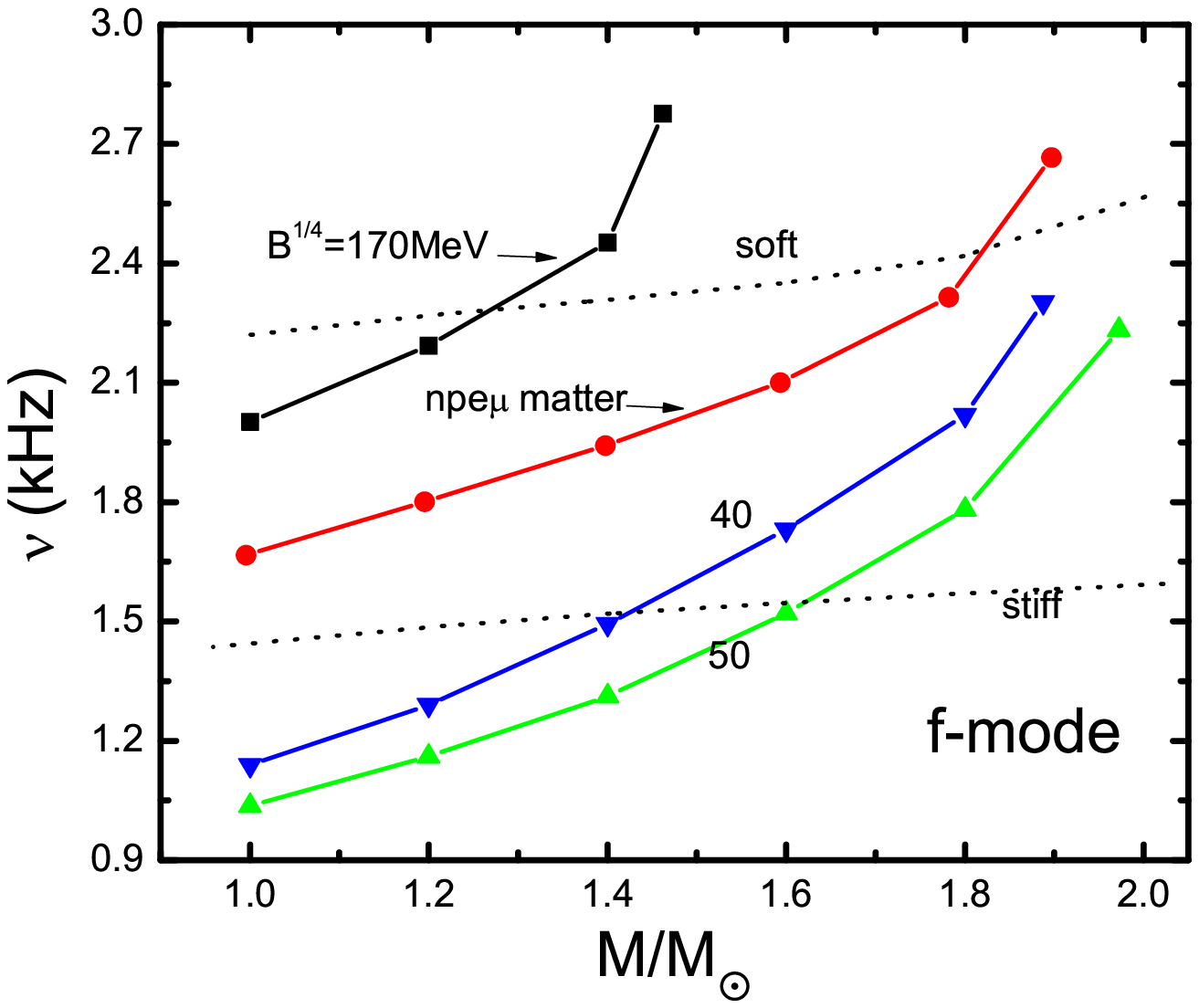}
}
  \resizebox{0.47\textwidth}{!}{
\includegraphics{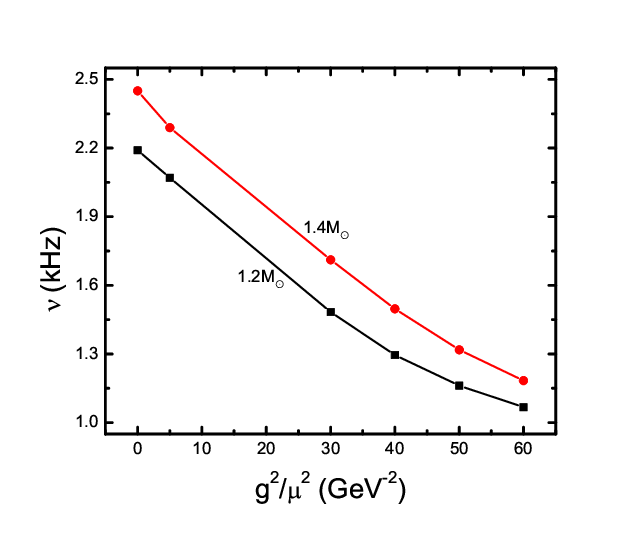}
}
\caption{\label{WFig3}(Color online) Left: The $f$-mode frequency
versus the NS mass. The upper two curves are results without
including the Yukawa potential while the lower two are obtained with
$g^2/\mu^2=40$ and $50$ GeV$^{-2}$ (for hybrid stars with $B^{1/4}$=170 MeV).  Right: The $f$-mode frequency
versus $g^2/\mu^2$ for neutron stars of mass $1.4 M_{\odot}$ and $1.2 M_{\odot}$, respectively. Taken from ref. \cite{Wlin14}.}
\end{center}
\end{figure*}
In the EiBI gravity, the resulting differential equations can be casted into the same TOV equations but with an apparent EOS including effects from both the nuclear EOS and modified gravity, reflecting the EOS-gravity degeneracy \cite{Sham2014}. The same is true for the Yukawa-type non-Newtonian gravity. Namely, a negative Yukawa term can be considered effectively either as an anti-gravity \cite{San84}
or a stiffening of the nuclear EOS \cite{Kri09}, while Fujii \cite{Fuj2} indicated that the Yukawa term is simply a part of the matter system in GR. Thus, one can study effects of the non-Newtonian gravity
by solving the TOV equation using the total pressure $P=P_{\textrm{\scriptsize{NU}}}+P {_{\textrm{\scriptsize{UB}}}}$
where $P_{\textrm{\scriptsize{NU}}}$ denotes the nuclear pressure inside neutron stars. Using this approach, Lin {\it et al.} \cite{Wlin14} constructed several typical EOSs with or without a quark core described by the MIT bag model. While the EOS for the $npe\mu$ matter is modeled by using the MDI interaction. The Gibbs construction was adopted for modeling the hadron-quark phase transition.
Shown in the right panel of Fig.~\ref{WFig1} are the typical EOSs.  Here the MDI parameter $x=0$ is used, leading to an \esym increase approximately linearly with density.
By construction, the following two sets of parameters, $B^{1/4}=170$ MeV and $g^{2}/\mu^{2}=$ 50~GeV$^{-2}$
or $B^{1/4}=180$ MeV and $g^{2}/\mu^{2}=$ 40~GeV$^{-2}$, lead to approximately the same total pressure. For comparisons, the soft and stiff EOSs for $npe$ matter constructed by Hebeler {\it et al.} \cite{Heb13} are also shown. Should one use a super-soft \esym that is decreasing with increasing density above certain high-densities, larger values of $g^{2}/\mu^{2}$ would be
required to obtain the same maximum mass \cite{wen}.

As examples, shown in the left window of Fig. \ref{WFig2} is the
mass-radius relation of hybrid stars with the bag constant
$B^{1/4}$=170 MeV and several values of $g^{2}/\mu^{2}$, while shown in the right  window are the maximum masses as a function of
$g^{2}/\mu^{2}$ with $B^{1/4}=170$ MeV and $B^{1/4}=180$ MeV,
respectively. As one expects, increasing the strength of the Yukawa term effectively stiffens the nuclear EOS, leading to
higher neutron star masses. These results reinforce the suggestion that the maximum mass of neutron stars can not rule out any EOS of dense stellar matter
before gravity is well understood \cite{DHW11}. Multiple observables may help break the EOS-gravity degeneracy. Interestingly, as shown in Fig.\ \ref{WFig3} the f-mode frequency depends sensitively on both the stiffness of the nuclear EOS and the strength $g^2/\mu^2$ of the Yukawa-type non-Newtonian gravity.
The upper two curves in the left panel of Fig. \ref{WFig3} are obtained without using the Yukawa potential while the lower two are obtained with
$g^2/\mu^2=40$ and $50$ GeV$^{-2}$, respectively. Shown in the right
panel are the $f$-mode frequency versus the strength
$g^2/\mu^2$ for a fixed mass $1.4 M_{\odot}$ and $1.2 M_{\odot}$, respectively.
Obviously, if both the mass and the $f$-mode frequency are measured, the EOS-gravity
degeneracy is readily broken. Moreover, similar EOS-gravity degeneracies were found in studying the frequencies of
the $p_1$, $p_2$, and $w_I$-modes \cite{Wlin14}. Thus, it is very hopeful that the multi-messengers approach will eventually lead to the breaking of the EOS-gravity degeneracy.

In summary of this section,  both the EOS of dense neutron-rich nuclear matter especially its symmetry energy term and the strong-field gravity are poorly known. While the GR has been successfully tested in the solar system, it has not been fully tested yet in the strong-field regime. There are reputable alternative theories for strong-field gravity. Neutron stars are among the densest objects with
strong gravity. It provides a natural testing bed for both dense matter EOS and the strong-field gravity. However, through several examples, we have seen that most of the known observables suffer from the EOS-gravity degeneracy. Realistically, to make a decisive conclusion about either one of them will require the precise knowledge of the other. Fortunately, in the era of multi-messengers astronomy,
several neutron stars observables are expected to be measured accurately. It is thus very hopeful that the EOS-gravity degeneracy of the dense neutron-rich nuclear matter in the strong-gravity field will be broken in the near future.

\section{Concluding remarks and outlook}
The density dependence of nuclear symmetry energy is the most uncertain part of the EOS of dense neutron-rich nucleonic matter. It is still poorly known especially at supra-saturation densities
 mostly because of our poor knowledge about the isovector nuclear interactions and correlations in dense medium besides the longstanding challenges in treating nuclear many-body problems.  For realizing the shared goal of determining the \esym more precisely, in parallel with the continuous efforts of predicting the \esym more accurately using various state-of-the-art theories and probing it using new experiments in terrestrial nuclear laboratories, significant efforts have also be devoted to predicting astrophysical effects of the \esym and extracting information about it from observations of neutron stars using both X-rays and gravitational waves. In this review, we focused on understanding astrophysical effects of nuclear symmetry energy.  We presented our observations and comments on some of the studies by many people in the community besides some of our own work in this area. Limited by our knowledge in this rather multi-disciplinary field, our selection of research issues and discussions made might be biased and incomplete. Without repeating too much the summaries we have given at the end of each section, we emphasize the following among the many interesting physics we have learned:
 \begin{itemize}

 \item The spin-isospin dependence of the three-body force, the tensor force induced isospin dependence of short-range nucleon-nucleon correlations as well as the finite-range part of the two-body interaction and the associated momentum-dependence of the isovector single-nucleon potential remain as the key but poorly known physics ingredients determining the high-density behaviour of nuclear symmetry energy.

\item The long-waited and well-welcomed flood of interesting papers triggered by GW170817 about the radii from analyzing the tidal deformability of neutron stars have led to a fiducial value of $R_{1.4}=12.42$ km with the lower and upper limit of $10.95$ km and $13.23$ km, respectively. This value is consistent with the pre-GW170817 prediction of 11.5 km $\leq R_{1.4}\leq13.6$ km using the \esym from analyzing terrestrial laboratory data. It is also consistent with the finding of $10.62 \leq R_{\rm{1.4}}\leq 12.83$ km from analyzing the X-ray data taken mostly by Chandra and/or XMM-Newton observatories.

\item The radii of canonical neutron stars are known to be sensitive to the \esym around $2\rho_0$. Using a new numerical technique of solving the inverse-structure problem of neutron stars in a multidimensional EOS parameter space, the extracted fiducial value of radii, observed tidal deformability from GW170817 and the earlier observed maximum mass of 2.01 M$_{\odot}$ together with the causality condition, the $E_{\rm sym}(2\rho_0)$ has been constrained to $46.9\pm10.1$ MeV. The \esym remains very uncertain at higher densities. While more precise measurements of neutron star radii and the tidal deformability in the inspiraling phase of neutron star mergers are expected to help further narrow down the nuclear symmetry energy around $2\rho_0$, new observables, such as neutrinos from the core of neutron stars, signals from the merging phase of two colliding neutron stars in space or two colliding heavy-nuclei in terrestrial laboratories are needed to probe the \esym at higher densities.

\item An absolutely maximum mass of $2.4$ M$_\odot$ independent of the EOS was predicted. Several post-GW170817 analyses using various approaches extracted a fiducial maximum mass of about $2.24$ M$_{\odot}$ for the remanent of GW170817.  It remains an interesting question and outstanding challenge to pin down the mass boundary between massive neutron stars and black holes.

\item The \esym has been predicted to have significant effects on the moment of inertia, ellipticity and the associated strain amplitude of continuous gravitational waves from isolated pulsars.

\item The frequencies and damping times of various oscillations, especially the g-modes of the core and the torsional modes of the crust, carry important information about nuclear symmetry energy.
The frequency and damping time of the f-mode were found to scale with the tidal deformability. The reported tidal deformability of neutron stars involved in GW170817 has provided useful limits for the f-mode frequency and its damping time.

\item The \esym also plays important roles in reaching the goal of breaking the EOS-gravity degeneracy. Besides Einstein's well-known GR theory of gravity, there are some reputable alternatives under debate. A better knowledge about the \esym at supra-saturation densities will help understand the nature of strong-field gravity and further test Einstein's GR using massive neutron stars or their collisions in space.

 \end{itemize}

Thanks to the great efforts of many people in the community, much progresses have been made while great challenges and new opportunities are ahead of us. While advanced X-ray observatories and gravitational wave detectors are being launched and built, new experiments in terrestrial laboratories especially those proving high-energy radioactive beams are underway.  Combining multi-messengers from multiple experiments and observations in both nuclear physics and astrophysics will certainly enable us to determine much more precisely the symmetry energy of dense neutron-rich nucleonic matter hopefully in the near future.

\section{Acknowledgement}
We would like to thank Bao-Jun Cai, Lie-Wen Chen, Farrooh Fattoyev, Frank Hall, Wei-Zhou Jiang, Xiao-Tao He, William G. Newton, Andrew Steiner, Bin Qi, S.Y. Wang, Wen-Jie Xie, Chang Xu and Jun Xu for collaborations and helpful discussions on the issues reviewed here. This work is supported in part by the U.S. Department of Energy, Office of Science, under Award Number DE-SC0013702, the CUSTIPEN (China-U.S. Theory Institute for Physics with Exotic Nuclei) under the US Department of Energy Grant No. DE-SC0009971, the China Postdoctoral Science Foundation (No. 2019M652358), the Fundamental Research Funds of Shandong University, and the National Natural Science Foundation of China under Grant Nos. 11275073, 11675226 and 11722546.



\end{document}